\documentclass[preprint]{aastex}
\usepackage{graphicx}
\usepackage{longtable}
\usepackage{amssymb}
\usepackage{color}
\usepackage{natbib}
\usepackage{slashbox}
\usepackage{hyperref}
\usepackage{lscape}

\makeatletter

\newcommand{\Rmnum}[1]{\expandafter\@slowromancap\romannumeral #1@}
\makeatother

\begin{document}

\title{INVISIBLE ACTIVE GALACTIC NUCLEI. \Rmnum{2}. RADIO MORPHOLOGIES \& FIVE NEW H\,I 21cm ABSORPTION LINE DETECTIONS}
\author{Ting Yan, John T. Stocke, Jeremy Darling}
\affil{Center for Astrophysics and Space Astronomy, UCB 389,
University of Colorado, Boulder, CO 80309-0389, USA;}
\affil{Department of Astrophysical and Planetary Sciences, UCB 391,
University of Colorado, Boulder, CO 80309-0391, USA;}
\author{Emmanuel Momjian}
\affil{National Radio Astronomy Observatory, P.O. Box O, Socorro, NM 87801, USA;} 
\author{Soniya Sharma}
\affil{Research School of Astronomy \& Astrophysics, The Australian National University, Mt Stromlo Observatory, ACT~2611, AU}
\author{Nissim Kanekar}
\affil{National Centre for Radio Astrophysics, TIFR, Post Bag 3, Ganeshkhind, Pune 411 007, India;}

\begin{abstract}

  This is the second paper directed towards finding new highly-redshifted atomic and 
molecular absorption lines at radio frequencies. To this end, we have selected a sample of 80
  candidates for obscured radio-loud active galactic nuclei and presented their basic
  optical/near-infrared (NIR) properties in Paper \Rmnum{1} \citep{yan:2012}. In this
  paper, we present both high-resolution radio continuum images for all of these sources and 
H\,I 21cm absorption spectroscopy for a few selected sources in this sample. A-configuration 
4.9 and 8.5 GHz Very Large Array (VLA) continuum observations find that 52 sources are
  compact or have substantial compact components with size $<$ 0\farcs5 and flux
  density $>$ 0.1 Jy at 4.9 GHz. The most compact 36 sources were then
  observed with the Very Long Baseline Array (VLBA) at 1.4 GHz. One definite and 10 candidate 
Compact Symmetric Objects (CSOs) are newly identified, a detection rate of CSOs $\sim$ three times
  higher than the detection rate previously found in purely flux-limited samples. Based on
  possessing compact components with high flux densities, 60 of these sources are good
  candidates for absorption-line searches. Twenty seven sources were observed for H\,I 21cm absorption
at their photometric or spectroscopic redshifts with only 6 detections made (one detection is tentative). 
However, 5 of these were from a small subset
of 6 CSOs with pure galaxy optical/NIR spectra (i.e., any AGN emission is obscured) and for which
accurate spectroscopic redshifts place the redshifted 21cm line in a radio frequency intereference (RFI)-
free spectral ``window''; i.e., the percentage of H\,I 21cm absorption-line detections could be as high as $\sim$ 80\%
in this sample. It is likely that the presence of ubiquitous RFI and the absence of accurate 
spectroscopic redshifts preclude H\,I detections in similar sources (only one detection out of the remaining 22
sources observed, 14 of which have only photometric redshifts); i.e., H\,I absorption may well be present but is masked by the RFI.
Future searches for highly-redshifted H\,I and molecular absorption can easily find more distant
CSOs among bright, ``blank field'' radio sources but will be severely hampered by an inability to 
determine accurate spectroscopic redshifts for them due to their lack of rest-frame UV continuum.

\end{abstract}

 
\section{Introduction}

Finding highly-redshifted absorption lines at radio frequencies is an
important yet challenging work. Up to now there have been $\sim$ 80 H I absorption systems
detected at cosmological distances ($z >$ 0.1), among
which about half are associated absorption systems, i.e.,
$z_{abs} = z_{em}$. The detection rate is 30\% in compact radio
sources \citep {vermeulen:2003} and can be as high
as 50\% in $z< 1$ CSO/GPS sources \citep [e.g.,][] {pihlstrom:2003, gupta:2006}. 
Except for two H I absorbers found at $z >$ 1.5,
various surveys have failed to find more associated ($z_{abs} = z_{em}$) absorption systems
at high-z \citep [e.g.,][] {curran:2008, curran:2011a}. Recently, a few 
intervening absorption systems have been found at high redshift by targeting damped Ly$\alpha$ (DLA) absorbers \citep [DLA;][]
{york:2007, kanekar:2007, kanekar:2013, kanekar:2014} but none of these include
molecular absorbers \citep {curran:2004, curran:2006}.
Meanwhile, there are only 5 known molecular
absorption systems at $z >$ 0.1 (one detected only
in OH), all at $z <$ 1 \citep{wiklind:1994, wiklind:1995, wiklind:1996, wiklind:1996a, kanekar:2005}. 
More detections are, in general, important for
understanding the cosmic evolution in the local physical properties 
of the cold dense gas out of which stars form. 

Another potentially important 
application is to use atomic and molecular radio absorption
lines to determine whether there are spacial or temporal variations
in fundamental physical parameters such as the fine-structure constant ($\alpha$)
and the proton-to-electron mass ratio. In this latter application, radio
frequency absorption-line measurements provide distinct advantages over optical techniques
due to the precision with which frequencies can be calibrated, the narrowness of the absorption
features that arise in cold gas (e.g., OH, CO and formaldehyde)
and the physics of the OH molecule's ``satellite lines''
\citep {darling:2003, darling:2004, chengalur:2003}. In contrast to claims based on
optical spectroscopy that changes in the value of $\alpha$ have been detected 
at $z>$ 2 \citep{murphy:2001, murphy:2008, webb:1999, webb:2011}, radio spectroscopy has not found any evidence for a 
change in $\alpha$ but only out to $z\approx$ 0.75 \citep{kanekar:2012}. However,
the absence of known molecular absorbers at $z>$ 1 means that the current set of radio observations cannot
rule out the claims made from optical spectroscopy at higher redshifts. Discovery of $z>$1 
molecular absorption systems must be made to confirm or challenge these optical claims.  
However, the more distant AGN are also typically much more luminous, which may
preclude the detection of neutral and molecular gas \citep{curran:2010, curran:2012}. 

To this end we have embarked on
a systematic search for obscured radio-loud active galactic nuclei (AGNs). In
Paper \Rmnum{1} of this work, we described the selection techniques for the sample
and presented our near-infrared (NIR) photometry for the sample. The selection method used the overlap
region between the Faint Images of the Radio Sky at 20cm (FIRST) survey and the Sloan Digital
Sky Survey (SDSS) to identify strong radio galaxies (f$_{20cm} >$ 0.3 Jy) whose optical 
morphologies are diffuse, likely late-type or interacting galaxies. The purpose of the
NIR imaging was to verify the optical classifications and to provide a broader baseline
in wavelength for better spectral energy distribution (SED) characterization and more
accurate photometric redshift determination. After confirming
that most of the sample objects are best-fit by pure galaxy SEDs requiring any AGN emission to be
absent or heavily obscured (i.e., a typical non-thermal AGN continuum is not present), 
in this paper we identify sources that are
compact enough at radio frequencies for absorption to be detectable. By this
we mean that a source dominated by a compact component will not have
its absorption diluted by extended flux which comes to us along unabsorbed sightlines.
We also present a preliminary search for redshifted H\,I 21cm and OH molecular absorption lines 
in which 6 new H\,I detections and no molecular detections were made.

Most radio-loud AGN are not good candidates for discovering highly-redshifted absorption since
numerous studies of radio-loud AGN host galaxies find that they are almost
exclusively ellipticals \citep [e.g.,][] {urry:1995, best:2005}
with little cold gas. But recent large surveys
make it possible to find rare cases that do not obey this
rule. In Paper 1 we selected a sample of 82 objects that make up only $\sim$ 5\%
of the $\sim$ 1500 sources with flux
densities $>$300 mJy in the April, 2003 version of the FIRST
survey \citep{becker:1995}. Among the $\sim$ 50\% of these which have coincident optical galaxies detected
by the SDSS Data Release 5 \citep[DR5;][]{adelman:2007}, the majority have typical elliptical morphology, i.e., a De Vaucouleurs
radial profile and a large concentration index \citep{strateva:2001} or are classified by SDSS as point sources. 
Only 82 optical counterparts have late-type, low central concentration, morphologies,
suggesting they could be late-type galaxies, interacting galaxies or
occasionally intervening systems for which the detected optical object is a foreground spiral galaxy 
not related physically to the radio source. In any of these cases, the nuclear source could be
obscured by dust in the gas-rich optical counterpart, becoming ``invisible'' AGN in the optical.

In Paper~\Rmnum{1}, we presented near-infrared (NIR) J, H and K$_{short}$ observations for this sample obtained using 
the Astrophysical Research Consortium 3.5-m telescope at the Apache Point Observatory (APO), which is also the
site of the SDSS 2.5-m telescope. Our NIR observing conditions typically were
better than the SDSS optical observing conditions allowing a good test of the SDSS optical
morphology classifications for these faint sources. The morphology is
largely consistent between the NIR images and optical images; with very few exceptions the diffuse
morphology of the hosts for these luminous radio sources is confirmed, 
suggesting that these can be gas-rich systems containing radio-loud AGN. Analysis
of the optical-NIR spectral energy distributions (SEDs) in
Paper~\Rmnum{1} indicates that only a small number of sources have
pure elliptical or un-reddened quasar-like SEDs. Many sources show
late-type galaxy SEDs or significantly reddened quasar-like SEDs. Most of these optical-NIR 
SEDs are consistent with pure galaxy SEDs with heavily-obscured nuclei; in several cases 
NIR turnups are present confirming this hypothesis. 
Mid-IR photometry from the {\it Spitzer Space Telescope} ``warm mission'' and from the 
{\it Wide-field Infrared Survey Explorer} (WISE) mission is available to
determine which are the most heavily-obscured sources in this sample (Truebenbach et al. in prep).

From the radio source perspective, this sample may catch a unique evolutionary phase or subclass of radio
galaxies. The study of the host galaxies and multi-wavelength
properties of the central AGN can shed light on the formation and
evolution of radio galaxies. The radio structure, particularly the size of the most compact components, can
also provide unique insights to the earliest evolutionary phases of radio galaxy evolution and can test the 
``breakout'' paradigm central to prevailing pictures of AGN feedback.

In this paper we continue the investigation of this uniquely-selected sample of obscured radio-loud AGN by
presenting Very Large Array (VLA) and Very Long Baseline Array
(VLBA) images of a large sub-sample of these sources. These maps are used both to identify the most compact sources in this sample
and to find unusual individual cases which are the best candidates for possessing cold gas absorption.   
Additionally, we present the results for redshifted H\,I 21cm absorption-line spectroscopy for 27 sources in this sample
out of which 5 definite and 1 tentative H\,I detections were made. No molecular detections were made for any of the H\,I detections.

The sample selection from Paper 1 is briefly reviewed in Section~\ref{sec:sam}. In
Section~\ref{sec:continuumobs} we describe our radio-continuum observing strategy, data
reduction process and basic observational results. The radio maps and source fluxes are presented in
an Appendix. Statistical relations between radio morphologies, radio spectral types, and
optical$+$NIR SEDs in this sample are studied and individual sources of interest (e.g., 
intervening systems, possible gravitational lens systems) are described in this same Section. In 
Section~\ref{sec:HIobs} we describe the limited radio spectroscopy which has been
attempted for this sample and present 5 H\,I 21cm detections. The overall results from this work are discussed 
and summarized in Section~\ref{sec:dis}. Future progress in the search for high-$z$ atomic and molecular absorbers
will require precise spectroscopic redshifts for the associated galaxies in this sample so that optimal
search strategies (e.g., telescope and RFI environment) can be employed.

Throughout this paper, we assume a $\mathrm{\Lambda}$CDM cosmology
with $H_0=71~\mathrm{km}~\mathrm{s}^{-1}~\mathrm{Mpc}^{-1}$,
$\Omega_\mathrm{\Lambda}=0.73$, and $\Omega_\mathrm{M}=0.27$.

\section{The Sample}\label{sec:sam}

The specific selection criteria for this sample presented in Paper~\Rmnum{1}
include the presence of an SDSS $r$-band optical counterpart within 1.5\arcsec\ of
a bright ($\geq$ 0.5 Jy) FIRST radio source position and characteristics of the optical counterpart
which favor it being a late-type galaxy; i.e., low central compactness index and a 
surface brightness profile more similar to an exponential disk \citep{strateva:2001} than an $r^{1/4}$-law. 
Of the 82 objects originally selected (Paper~\Rmnum{1}), 2 are nearby
galaxies at $z < 0.05$ (J1352$+$3126, J1413$-$0312) and are excluded
from our discussion in this paper. Among the remainder of the 80 objects,
only 3 have angular sizes larger than 5\arcsec\ in their SDSS images
which corresponds to 16 kpc at $z=0.2$; these are J1347+1217 at
$z=0.122$, J0920+2714 at $z=0.206$, and J0901+0304 at $z=0.287$. We
expect most of other objects to have redshifts between approximately 0.2 and
1.0 due to the detection limits of the SDSS photometric survey (except for
a few high-$z$ quasars, which may be mistakenly present in the sample).
The photometric and spectroscopic redshifts for individual optical counterparts 
in this sample support this conclusion. A listing of the full sample of sources and their basic
parameters can be found in Paper~\Rmnum{1} and \citet{yan:2013}

A non-elliptical optical morphology could indicate a late-type galaxy,
an interacting system, or a gravitational lens system. We find
examples already identified in our sample for each of the three
possibilities. J1415$+$1320 (a.k.a. PKS~1413+135) is an edge-on spiral galaxy at $z=0.247$
and the first object with molecular absorption lines detected at
cosmological distances \citep{wiklind:1994}. J0751+2716 is a
gravitationally-lensed system with a background quasar at $z$ = 3.2 and
a foreground galaxy at $z$ = 0.349; strong CO emission lines are found
in the high-$z$ quasar because of the high magnification factor due to a
strong lensing effect \citep{barvainis:2002}. J1347$+$1217 is an
interacting system at $z=0.122$ with \ion{H}{1} absorption lines detected
\citep{mirabel:1989}. There are also two sources, J1120$+$2327 and J1348$+$2415,
which are known alignment-effect radio galaxies in which the optical
morphology is dominated by extended emission-line atmoic gas \citep {miley:2008}.

In Paper \Rmnum{1}, we re-evaluated all 80 objects using SDSS Data
Release 8 and found 5 objects likely to be misclassified early-type
galaxies. Because the possibility of being late-type cannot be ruled
out completely and to retain a complete sample selected from SDSS DR5
and the April, 2003 version of FIRST, we still include these 5 objects
in this paper. The optical counterpart of
J0834$+$1700 is coincident with one lobe of a double-lobe radio
galaxy. We excluded it in Paper \Rmnum{1}, but it actually could be an
intervening system with potential absorption and remains in the
sample for this paper.

\section{Radio Continuum Imaging}\label{sec:continuumobs}

\subsection{Observations and Data Reduction}\label{sec:obs}

Except when archive data were available, we observed all the sources
at 4.9 GHz and 8.5 GHz using the VLA of NRAO
\footnote{The National Radio Astronomy Observatory (NRAO) is a facility of
  the National Science Foundation (NSF) operated under cooperative agreement
  by Associated Universities, Inc.} 
in its most extended
A-configuration (VLA-A; project \# AY0052). Overall, we obtained data for 35 sources at
both frequencies, 14 at 4.9 GHz only, and 8 at 8.5 GHz only. Our data
were taken during June 2007 to August 2007 in the transitional period
from VLA to Expanded VLA (EVLA; currently the Karl G. Jansky Very
Large Array). The observations were made with an effective bandwidth
of 100 MHz in full polarization centered at 4.860 GHz and 8.460
GHz. Typical on-source observing time was 4--5 minutes for each object
at one frequency. We bracketed each target or a group of close-by
targets with a single phase calibrator.

We initially reduced all VLA data using the Common Astronomy Software
Applications (CASA) Version 2.3.1; then we found the centers of many
images were shifted from the pointing center to the brightest point
probably due to problems in the self-calibration process. Reduction
was rectified using the Astronomical Imaging Processing System (AIPS)
for 4.9 GHz data but was not possible for the 8.5 GHz data. This problem 
does not affect our conclusions in this paper because the relative structure of
the 8.5 GHz images remains correct. However, caution must be exercised
in identifying individual component correlations between
frequencies. For this reason we display the 8.5 GHz images only in
relative coordinates. During the transitional period, there were
problems in VLA-EVLA baselines including phase jumps and closure
errors. Among the 27 antennas, 10 were EVLA antennas in June and July
2007, and 11 in August 2007. EVLA antennas were excluded for the 4.9 GHz
reduction in AIPS. Therefore the sensitivity is reduced by a factor of
$\sim$ 1.6.

We used the VLBA to obtain 1.4 GHz images
for sources that are unresolved or have compact components in VLA-A
images (project \# BY0020). Overall, 36 sources were observed in December 2009 and January
2010. The observations were centered at 1.446 GHz with an effective
bandwidth of 32 MHz in full polarization. For each object, $\sim$
15-minute observing scans were conducted at three well-separated hour
angles over a time span of 3--10 hours to achieve best
$uv$-coverage. The target source was phase-referenced every 4--5
minutes by a phase calibrator within 5 degrees on the sky. The data
reduction of the VLBA observations was carried out in AIPS following
standard calibration and imaging procedures. The flux density scales
of these data were set using the on-line measurements of the antenna
gains and system temperatures.

The observing log is shown in Table~\ref{tab:log} together with brief
comments about individual observations. For the archive data, if an image
is available in the literature, the reference paper is cited;
otherwise we processed the archive data ourselves and cite the
project ID as a reference in Table~\ref{tab:log}. The flux density calibrator
is completely flagged or not observed for 3 objects at 4.9 GHz and 2
objects at 8.5 GHz. In these cases, a rough flux density calibration is done by
setting the flux density of the phase calibrator to its flux density
reported in the VLA Calibrator Manual. One object at 4.9 GHz and five
objects at 1.4 GHz are not detected, which means either that the object is too
faint/extended to be detected, or that a few of the shortest baselines have
signal but they are not sufficient to make a synthesized image. The
data for one object at 4.9 GHz are completely flagged.
These sources are identified in the Tables in the Appendix. 

We have an almost complete set of VLA-A 4.9 GHz images from our own
data, the literature, and archive data processed by us. A few
exceptions without VLA-A images are objects that have been well-
studied or proven to be unresolved in VLA-A 8.5 GHz images and so are
likely unresolved at 4.9 GHz as well. The 8.5 GHz data are summarized in
Table~\ref{tab:log} including our own data, and data from the literature
or archives that would help us understand the nature of the
source. That is to say, if a source was of little interest for this project based on
its VLA-A 4.9 GHz image (for example, consisting of two well-separated
lobes with an optical/NIR source right between them), we did not
process its archive data.

\begin{deluxetable}{lcccccccl}
\tablewidth{0pt}
  \tablecaption{Observing Summary
    \label{tab:log}} \tablehead{ \colhead{Object} & \colhead{Radio
      Name} &\multicolumn{2}{|c|}{VLA 4.9 GHz} &
    \multicolumn{2}{|c|}{VLA 8.5 GHz} & \multicolumn{2}{|c|}{VLBI} &
    \colhead{Notes\tablenotemark{a}} \\ \hline \colhead{} & \colhead{}
    & \colhead{Date} & \colhead{Ref.} & \colhead{Date} &
    \colhead{Ref.}  & \colhead{Date} & \colhead{Ref.} & \colhead{} }

\startdata
	J0000$-$1054	&	PKS 2358-111	&	08/12/07	&	[1]	&	08/12/07	&	[1]	&	\nodata	&	\nodata	&	\nodata	\\
	J0003$-$1053	&	PKS 0001-111	&	08/12/07	&	[1]	&	08/12/07	&	[1]	&	\nodata	&	\nodata	&	\nodata	\\
	J0134$+$0003	&	4C -00.11	&	08/11/07	&	[1]	&	\nodata	&	\nodata	&	\nodata	&	\nodata	&	N1	\\
	J0249$-$0759	&	PKS 0247-08	&	08/11/07	&	[1]	&	08/11/07	&	[1]	&	\nodata	&	\nodata	&	N1,N2	\\
	J0736$+$2954	&	TXS 0733+300	&	07/08/07	&	[1]	&	\nodata	&	\nodata	&	\nodata	&	\nodata	&	N3	\\
	J0747$+$4618	&	4C +46.16	&	\nodata	&	[13]	&	\nodata	&	[13]	&	\nodata	&	\nodata	&	\nodata	\\
	J0749$+$2129	&	TXS 0746+216	&	06/16/07	&	[1]	&	06/16/07	&	[1]	&	\nodata	&	\nodata	&	\nodata	\\
	J0751$+$2716	&	B2 0748+27	&	\nodata	&	[15]	&	\nodata	&	[15]	&	\nodata	&	\nodata	&	\nodata	\\
	J0759$+$5312	&	TXS 0755+533	&	06/16/07	&	[1]	&	06/16/07	&	[1]	&	12/10/09	&	[1]	&	\nodata	\\
	J0805$+$1614	&	PKS 0802+16	&	09/13/87	&	[3]	&	07/08/07	&	[1]	&	12/08/09	&	[1]	&	\nodata	\\
	J0807$+$5327	&	TXS 0803+536	&	06/16/07	&	[1]	&	06/16/07	&	[1]	&	12/10/09	&	[1]	&	\nodata	\\
	J0824$+$5413	&	TXS 0820+543	&	06/16/07	&	[1]	&	06/16/07	&	[1]	&	12/10/09	&	[1]	&	N4	\\
	J0834$+$1700	&	4C +17.45	&	\nodata	&	\nodata	&	07/08/07	&	[1]	&	12/08/09	&	[1]	&	\nodata	\\
	J0839$+$2403	&	4C +24.18	&	07/08/07	&	[1]	&	\nodata	&	\nodata	&	\nodata	&	\nodata	&	\nodata	\\
	J0843$+$4215	&	B3 0840+424A	&	\nodata	&	\nodata	&	\nodata	&	[16]	&	\nodata	&	[14]	&	\nodata	\\
	J0901$+$0304	&	PKS 0859+032	&	07/06/07	&	[1]	&	\nodata	&	\nodata	&	01/03/10	&	[1]	&	\nodata	\\
	J0903$+$5012	&	4C +50.28	&	06/16/07	&	[1]	&	06/16/07	&	[1]	&	\nodata	&	\nodata	&	\nodata	\\
	J0905$+$4128	&	B3 0902+416	&	\nodata	&	[13]	&	\nodata	&	[13]	&	\nodata	&	[12]	&	\nodata	\\
	J0907$+$0413	&	4C +04.30	&	12/17/84	&	[8]	&	\nodata	&	\nodata	&	\nodata	&	\nodata	&	\nodata	\\
	J0910$+$2419	&	4C +24.19	&	07/08/07	&	[1]	&	\nodata	&	\nodata	&	\nodata	&	\nodata	&	\nodata	\\
	J0915$+$1018	&	TXS 0912+105	&	02/07/85	&	[8]	&	\nodata	&	\nodata	&	\nodata	&	\nodata	&	\nodata	\\
	J0917$+$4725	&	B3 0914+476	&	06/16/07	&	[1]	&	06/16/07	&	[1]	&	12/10/09	&	[1]	&	\nodata	\\
	J0920$+$1753	&	4C +18.29	&	07/08/07	&	[1]	&	\nodata	&	\nodata	&	12/08/09	&	[1]	&	N4	\\
	J0920$+$2714	&	B2 0917+27B	&	06/17/07	&	[1]	&	06/17/07	&	[1]	&	12/08/09	&	[1]	&	\nodata	\\
	J0939$+$0304	&	PKS 0937+033	&	12/17/84	&	[8]	&	\nodata	&	\nodata	&	01/03/10	&	[1]	&	\nodata	\\
	J0945$+$2640	&	B2 0942+26	&	01/26/01	&	[9]	&	\nodata	&	\nodata	&	12/08/09	&	[1]	&	\nodata	\\
	J0951$+$1154	&	TXS 0948+121	&	02/08/85	&	[8]	&	07/06/07	&	[1]	&	\nodata	&	\nodata	&	\nodata	\\
	J1008$+$2401	&	B2 1005+24	&	07/08/07	&	[1]	&	\nodata	&	\nodata	&	12/08/09	&	[1]	&	N4	\\
	J1010$+$4159	&	B3 1007+422	&	\nodata	&	[13]	&	\nodata	&	[13]	&	\nodata	&	[12]	&	\nodata	\\
	J1019$+$4408	&	B3 1016+443	&	\nodata	&	[13]	&	\nodata	&	[13]	&	\nodata	&	[12]	&	\nodata	\\
	J1023$+$0424	&	TXS 1021+046	&	04/22/86	&	[4]	&	07/06/07	&	[1]	&	01/03/10	&	[1]	&	\nodata	\\
	J1033$+$3935	&	B2 1030+39	&	\nodata	&	\nodata	&	\nodata	&	[16]	&	\nodata	&	[14]	&	\nodata	\\
	J1034$+$1112	&	TXS 1031+114	&	06/17/07	&	[1]	&	06/17/07	&	[1]	&	\nodata	&	\nodata	&	\nodata	\\
	J1043$+$0537	&	4C +05.45	&	07/06/07	&	[1]	&	\nodata	&	\nodata	&	01/03/10	&	[1]	&	N4	\\
	J1045$+$0455	&	TXS 1043+051	&	06/17/07	&	[1]	&	06/17/07	&	[1]	&	\nodata	&	\nodata	&	\nodata	\\
	J1048$+$3457	&	4C +35.23	&	03/08/86	&	[2]	&	\nodata	&	\nodata	&	12/08/09	&	[1]	&	\nodata	\\
	J1120$+$2327	&	4C +23.27	&	\nodata	&	\nodata	&	\nodata	&	\nodata	&	\nodata	&	\nodata	&	\nodata	\\
	J1125$+$1953	&	PKS 1123+201	&	07/06/07	&	[1]	&	\nodata	&	\nodata	&	12/08/09	&	[1]	&	\nodata	\\
	J1127$+$5743	&	TXS 1124+579	&	06/16/07	&	[1]	&	06/16/07	&	[1]	&	12/10/09	&	[1]	&	\nodata	\\
	J1129$+$5638	&	TXS 1126+569	&	06/16/07	&	[1]	&	06/16/07	&	[1]	&	12/10/09	&	[1]	&	\nodata	\\
	J1142$+$0235	&	TXS 1139+028	&	02/08/85	&	[8]	&	07/08/07	&	[1]	&	01/03/10	&	[1]	&	N4	\\
	J1147$+$4818	&	4C +48.33	&	06/17/07	&	[1]	&	06/17/07	&	[1]	&	12/10/09	&	[1]	&	\nodata	\\
	J1148$+$1404	&	TXS 1145+143	&	04/08/86	&	[4]	&	07/06/07	&	[1]	&	01/03/10	&	[1]	&	\nodata	\\
	J1202$+$1207	&	TXS 1200+124	&	06/17/07	&	[1]	&	06/17/07	&	[1]	&	01/03/10	&	[1]	&	\nodata	\\
	J1203$+$4632	&	B3 1200+468	&	07/08/07	&	[1]	&	\nodata	&	\nodata	&	\nodata	&	\nodata	&	\nodata	\\
	J1207$+$5407	&	4C +54.26	&	06/16/07	&	[1]	&	06/16/07	&	[1]	&	\nodata	&	\nodata	&	\nodata	\\
	J1215$+$1730	&	4C +17.54	&	\nodata	&	\nodata	&	\nodata	&	[16]	&	\nodata	&	[14]	&	\nodata	\\
	J1228$+$5348	&	4C +54.28	&	06/16/07	&	[1]	&	06/16/07	&	[1]	&	\nodata	&	\nodata	&	\nodata	\\
	J1238$+$0845	&	TXS 1235+090	&	06/17/07	&	[1]	&	06/17/07	&	[1]	&	01/03/10	&	[1]	&	\nodata	\\
	J1300$+$5029	&	TXS 1258+507	&	\nodata	&	\nodata	&	\nodata	&	[17]	&	\nodata	&	[14]	&	\nodata	\\
	J1312$+$1710	&	TXS 1310+174	&	04/08/86	&	[4]	&	07/08/07	&	[1]	&	12/08/09	&	[1]	&	\nodata	\\
	J1315$+$0222	&	TXS 1312+026	&	12/17/84	&	[8]	&	07/08/07	&	[1]	&	\nodata	&	\nodata	&	\nodata	\\
	J1341$+$1032	&	4C +10.36	&	06/17/07	&	[1]	&	06/17/07	&	[1]	&	\nodata	&	\nodata	&	\nodata	\\
	J1345$+$5846	&	4C +59.20	&	06/17/07	&	[1]	&	06/17/07	&	[1]	&	12/10/09	&	[1]	&	\nodata	\\
	J1347$+$1217	&	4C +12.50	&	\nodata	&	\nodata	&	\nodata	&	\nodata	&	\nodata	&	\nodata	&	\nodata	\\
	J1348$+$2415	&	4C +24.28	&	\nodata	&	[11]	&	\nodata	&	\nodata	&	\nodata	&	[10]	&	\nodata	\\
	J1354$+$5650	&	4C +57.23	&	06/17/07	&	[1]	&	06/17/07	&	[1]	&	\nodata	&	\nodata	&	\nodata	\\
	J1357$+$0046	&	PKS 1355+01	&	06/17/07	&	[1]	&	06/17/07	&	[1]	&	01/03/10	&	[1]	&	\nodata	\\


	J1410$+$4850	&	TXS 1408+490	&	07/15/07	&	[1]	&	07/15/07	&	[1]	&	12/10/09	&	[1]	&	\nodata	\\
	J1413$+$1509	&	TXS 1411+154	&	07/15/07	&	[1]	&	\nodata	&	\nodata	&	\nodata	&	\nodata	&	\nodata	\\
	J1414$+$4554	&	B3 1412+461	&	\nodata	&	\nodata	&	\nodata	&	[17]	&	\nodata	&	[18]	&	\nodata	\\
	J1415$+$1320	&	PKS 1413+135	&	\nodata	&	\nodata	&	\nodata	&	\nodata	&	\nodata	&	[19]	&	\nodata	\\
	J1421$-$0246	&	4C -02.60	&	07/10/07	&	[1]	&	07/10/07	&	[1]	&	\nodata	&	\nodata	&	\nodata	\\
	J1424$+$1852	&	4C +19.47	&	07/10/07	&	[1]	&	07/10/07	&	[1]	&	\nodata	&	\nodata	&	\nodata	\\
	J1502$+$3753	&	B2 1500+38	&	07/15/07	&	[1]	&	07/15/07	&	[1]	&	\nodata	&	\nodata	&	\nodata	\\
	J1504$+$5438	&	TXS 1503+548	&	\nodata	&	\nodata	&	\nodata	&	[16]	&	\nodata	&	[14]	&	\nodata	\\
	J1504$+$6000	&	TXS 1502+602	&	06/17/07	&	[1]	&	06/17/07	&	[1]	&	01/11/10	&	[1]	&	\nodata	\\
	J1523$+$1332	&	4C +13.54	&	07/10/07	&	[1]	&	07/10/07	&	[1]	&	01/11/10	&	[1]	&	\nodata	\\
	J1527$+$3312	&	B2 1525+33	&	07/15/07	&	[1]	&	07/26/91	&	[6]	&	01/11/10	&	[1]	&	N5	\\
	J1528$-$0213	&	PKS 1525-020	&	07/10/07	&	[1]	&	07/10/07	&	[1]	&	\nodata	&	\nodata	&	\nodata	\\
	J1548$+$0808	&	TXS 1545+082	&	07/15/07	&	[1]	&	07/15/07	&	[1]	&	01/11/10	&	[1]	&	\nodata	\\
	J1551$+$6405	&	TXS 1550+642	&	\nodata	&	\nodata	&	\nodata	&	[16]	&	\nodata	&	[14]	&	\nodata	\\
	J1559$+$4349	&	4C +43.36	&	06/23/99	&	[7]	&	\nodata	&	\nodata	&	01/11/10	&	[1]	&	\nodata	\\
	J1604$+$6050	&	TXS 1603+609	&	06/17/07	&	[1]	&	06/17/07	&	[1]	&	01/11/10	&	[1]	&	\nodata	\\
	J1616$+$2647	&	PKS 1614+26	&	07/15/07	&	[1]	&	05/04/90	&	[5]	&	01/11/10	&	[1]	&	N2	\\
	J1625$+$4134	&	4C +41.32	&	\nodata	&	\nodata	&	\nodata	&	[16]	&	\nodata	&	[14]	&	\nodata	\\
	J1629$+$1342	&	4C +13.60	&	07/15/07	&	[1]	&	07/15/07	&	[1]	&	\nodata	&	\nodata	&	\nodata	\\
	J1633$+$4700	&	4C +47.43	&	07/15/07	&	[1]	&	07/15/07	&	[1]	&	01/11/10	&	[1]	&	\nodata	\\
	J1724$+$3852	&	B2 1722+38	&	07/15/07	&	[1]	&	07/15/07	&	[1]	&	01/11/10	&	[1]	&	\nodata	\\
	J2203$-$0021	&	4C -00.79	&	08/11/07	&	[1]	&	\nodata	&	\nodata	&	\nodata	&	\nodata	&	N1	\\
\enddata

\tablerefs{[1] This work (VLA project AY0052 and VLBA project BY 0020); [2] VLA project AA052; [3] VLA project
  AA073; [4] VLA project AB375; [5] VLA project AB568; [6] VLA project
  AB611; [7] VLA project AB922; [8] VLA project AH167; [9] VLA project
  AS704; [10] \citet{cai:2002}; [11] \citet{chambers:1996}; [12]
  \citet{dallacasa:2002}; [13] \citet{fanti:2001}; [14]
  \citet{helmboldt:2007}; [15] \citet{lehar:1997}; [16]
  \citet{myers:2003}; [17] \citet{patnaik:1992}; [18]
  \citet{peck:2000}; [19] \citet{perlman:2002}.}

\tablenotetext{a}{N1--N5 in this column means: \\ N1, no flux calibration on the VLA-A 4.9 GHz image; \\ N2, no flux calibration on the VLA-A 8.5 GHz image;\\ N3, not detected on the VLA-A 4.9 GHz image; \\ N4, not detected on the VLBA 1.4 GHz image; \\ N5, all data are flagged on the VLA-A 4.9 GHz image.}

\end{deluxetable}

\subsection{Observational Results}\label{sec:res}

The basic observational results of this study are detailed in the 
Appendix, including high resolution and dynamic range VLA and VLBA maps of these sources
as well as measurements of the component fluxes, sizes and locations. In this Section we use the
basic data from the Appendix to investigate systematic relationships between the radio
source sizes, morphologies and spectral indices, and the optical-NIR morphologies and SEDs.

\subsubsection{Morphological Classification of Radio Images}\label{sec:mor}

We find that 52 out of the 80 sources are compact or have compact
components with size $<$ 0\farcs5 and flux density $>$ 0.1 Jy at 4.9
GHz. Twenty-nine objects are unresolved in VLA-A images and their
morphological types are determined only by the VLBA observations.

In Table~\ref{tab:type}, we compile the optical-NIR properties of each source together 
with the basic radio data from the Appendix. For each source we list the spectroscopic
or revised photometric redshift if available, the optical$+$NIR SED types and Hubble types in columns (2), (3)
and (4) respectively. The optical$+$NIR
SEDs were divided into four classes in Paper~\Rmnum{1}: quasar ($Q$),
quasar with extinction signatures in the bluer bands ($Q+abs$), galaxy
with a quasar signature in the bluer bands ($G+Q$), and pure galaxy
($G$). We fit $G$-type objects with template spectra of five Hubble
types (E, S0, Sa, Sb, and Sc), and give the best-fit Hubble-types in column (4) (see Paper~\Rmnum{1} for the
details of the procedure used). Also in Paper~\Rmnum{1} we used the SDSS photometry and our newly-acquired APO NIR
photometry to determine more accurate photometric redshifts for $G$ sources only; these revised $z_{phot}$ values 
are listed in column (2) with asterisks. Radio continuum flux densities are listed at three frequencies in columns (5), (6) \& (7)
followed by the low frequency (between 365 MHz and 1.4 GHz) spectral index in column (8) and the high frequency
spectral index (between 1.4 and 4.9 GHz) in column (9).
Column (10) lists the abbreviated spectral type of the source (SS = steep spectrum; USS = ultra-steep spectrum; 
GPS = Giga-Hertz Peaked Source; FS = flat spectrum). Each object's largest angular size (LAS) in arcsecs is listed in
Column (11) and its largest linear size (LLS) in kpc in Column (12). Because radio source size
is frequency and resolution dependent, we also list the observing
frequency and telescope/configuration in Column (13) for these measurements. 
When an object's spectroscopic redshift is not available, its
photometric redshift is used to estimate its linear size. In Paper
\Rmnum{1} we fit photometric redshifts and Hubble types from
optical$+$NIR SEDs for objects with galaxy-like SEDs; in these cases 
the redshift uncertainties are $\pm$ 0.1-0.2. For objects with a
quasar-like signature in their optical$+$NIR SEDs, due to the
uncertain amount of quasar contribution and extent of obscuration, it
is not possible to derive a reliable photometric redshift. In this
case, an upper limit on LLS is set by using the maximum angular scale
possible over cosmic time, 8.6 kpc/\arcsec at $z \sim 1.6$. 

{
\begin{landscape}
\begin{deluxetable}{llllrrrrrlcclll}
\tabletypesize{\tiny}
\tablecaption{Summary of Radio Galaxy Properties \label{tab:type}}
\tablewidth{0pt}
\tablehead{ \colhead{Object} &
        \colhead{$z$} & \colhead{SED} & \colhead{$G$} &
        \colhead{$S_{365}$} & \colhead{$S_{1.4}$} &
        \colhead{$S_{4.9}$} & \colhead{$\alpha_\mathrm{l}$} &
        \colhead{$\alpha_\mathrm{h}$} & \colhead{Spec.} &
        \colhead{LAS} & \colhead{LLS} & \colhead{Freq.(GHz)/} &
        \colhead{Morph.} & \colhead{EVAL}\\
        \colhead{} & \colhead{} & \colhead{Type} & \colhead{Type} &
        \colhead{(Jy)} & \colhead{(Jy)} & \colhead{(Jy)} & \colhead{}
        & \colhead{} & \colhead{Type} & \colhead{($\arcsec$)} &
        \colhead{(kpc)} &
        \colhead{Telescope} & \colhead{} & \colhead{} \\
        \colhead{(1)} & \colhead{(2)} & \colhead{(3)} & \colhead{(4)}
        & \colhead{(5)} & \colhead{(6)} & \colhead{(7)} &
        \colhead{(8)} & \colhead{(9)} & \colhead{(10)} &
        \colhead{(11)} & \colhead{(12)} & \colhead{(13)} &
        \colhead{(14)} & \colhead{(15)} }

\startdata
	J0000$-$1054		&	\nodata		&	Q$+$abs	&	\nodata	&	$		2.20	$	&	0.40	&	0.07	&	$		-1.28	$	&	-1.41	&	USS	&	1.5	&	$	<	12.9	$	&	4.9	/	VLA-A	&	MSO	&	Y			\\
	J0003$-$1053		&	1.474	\tablenotemark{a}	&	Q$+$abs	&	\nodata	&	$		1.47	$	&	0.40	&	0.14	&	$		-0.97	$	&	-0.84	&	SS	&	2.3	&	$		20.0	$	&	4.9	/	VLA-A	&	CPLX	&	Y	,	I:	\\
	J0134$+$0003		&	0.879		&	G	&	Sa	&	$		1.11	$	&	0.89	&	0.46	&	$		-0.17	$	&	-0.53	&	GPS	&	0.018	&	$		0.140	$	&	2.3	/	VLBI	&	CSO:	&	Y			\\
	J0249$-$0759		&	\nodata		&	G$+$Q	&	\nodata	&	$		2.34	$	&	0.62	&	0.19	&	$		-0.99	$	&	-0.95	&	SS	&	1.9	&	$	<	16.3	$	&	4.9	/	VLA-A	&	LSO	&	N			\\
	J0736$+$2954		&	\nodata		&	G$+$Q	&	\nodata	&	$		0.35	$	&	0.49	&	0.52	&	$		0.26	$	&	0.05	&	FS	&	0.014	&	$	<	0.118	$	&	5	/	VLBI	&	CJ	&	Y			\\
	J0747$+$4618		&	2.926		&	Q$+$abs	&	\nodata	&	$		2.00	$	&	0.52	&	0.13	&	$		-1.00	$	&	-1.09	&	USS	&	1.3	&	$		10.6	$	&	4.9	/	VLA-A	&	MSO	&	N			\\
	J0749$+$2129		&	0.52	*	&	G	&	S0	&	$		1.13	$	&	0.42	&	0.12	&	$		-0.73	$	&	-1.05	&	SS	&	2.6	&	$	\sim	16.1	$	&	4.9	/	VLA-A	&	CPLX	&	Y			\\
	J0751$+$2716		&	3.200	\tablenotemark{b}	&	G	&	Sc	&	$		1.47	$	&	0.60	&	0.21	&	$		-0.66	$	&	-0.83	&	SS	&	0.80	&	$		6.14	$	&	8.5	/	VLA-A	&	GA	&	Y	,	I	\\
	J0759$+$5312		&	\nodata		&	G$+$Q	&	\nodata	&	$		1.12	$	&	0.33	&	0.15	&	$		-0.91	$	&	-0.63	&	SS	&	1.4	&	$	<	11.9	$	&	4.9	/	VLA-A	&	MSO	&	Y			\\
	J0805$+$1614		&	0.632	:	&	G	&	Sc	&	$		1.78	$	&	0.63	&	0.20	&	$		-0.77	$	&	-0.94	&	SS	&	0.45	&	$	:	3.10	$	&	8.5	/	VLA-A	&	MSO	&	Y			\\
	J0807$+$5327		&	\nodata		&	Q$+$abs	&	\nodata	&	$		0.83	$	&	0.33	&	0.11	&	$		-0.69	$	&	-0.92	&	SS	&	1.1	&	$	<	9.15	$	&	4.9	/	VLA-A	&	MSO	&	Y	,	I:	\\
	J0824$+$5413		&	0.639		&	G$+$Q	&	\nodata	&	$		1.12	$	&	0.39	&	0.15	&	$		-0.79	$	&	-0.77	&	SS	&	1.8	&	$		12.4	$	&	4.9	/	VLA-A	&	MSO	&	Y			\\
	J0834$+$1700		&	\nodata		&	Q$+$abs	&	\nodata	&	$		5.64	$	&	1.64	&	0.58	&	$		-0.92	$	&	-0.84	&	SS	&	0.11	&	$	<	0.966	$	&	1.4	/	VLBI	&	SL	&	Y	,	I	\\
	J0839$+$2403		&	\nodata		&	G$+$Q	&	\nodata	&	$		1.77	$	&	0.66	&	0.23	&	$		-0.73	$	&	-0.84	&	SS	&	4.1	&	$	<	35.3	$	&	4.9	/	VLA-A	&	LSO	&	N			\\
	J0843$+$4215		&	0.60	*	&	G	&	Sb	&	$		2.50	$	&	1.46	&	0.58	&	$		-0.40	$	&	-0.74	&	SS$\checkmark$	&	0.021	&	$	\sim	0.143	$	&	5	/	VLBI	&	CJ	&	Y			\\
	J0901$+$0304		&	0.287		&	G	&	Sc	&	$		0.56	$	&	0.38	&	0.26	&	$		-0.28	$	&	-0.31	&	GPS$\checkmark$	&	0.011	&	$		0.047	$	&	1.4	/	VLBI	&	PS	&	Y			\\
	J0903$+$5012		&	\nodata		&	G$+$Q	&	\nodata	&	$		2.78	$	&	0.95	&	0.31	&	$		-0.80	$	&	-0.90	&	SS	&	3.4	&	$	<	29.0	$	&	4.9	/	VLA-A	&	LSO	&	N			\\
	J0905$+$4128		&	\nodata		&	G$+$Q	&	\nodata	&	$		1.10	$	&	0.48	&	0.16	&	$		-0.61	$	&	-0.90	&	SS	&	0.33	&	$	<	2.82	$	&	1.66	/	VLBI	&	MSO	&	Y			\\
	J0907$+$0413		&	1.10	*	&	G	&	Sc	&	$		1.85	$	&	0.70	&	0.19	&	$		-0.72	$	&	-1.04	&	SS	&	2.6	&	$	\sim	21.4	$	&	4.9	/	VLA-A	&	LSO	&	N			\\
	J0910$+$2419		&	\nodata		&	G$+$Q	&	\nodata	&	$		2.72	$	&	0.83	&	0.21	&	$		-0.89	$	&	-1.12	&	SS	&	4.2	&	$	<	36.0	$	&	4.9	/	VLA-A	&	LSO	&	N			\\
	J0915$+$1018		&	\nodata		&	G$+$Q	&	\nodata	&	$		0.97	$	&	0.35	&	0.18	&	$		-0.76	$	&	-0.56	&	SS	&	1.9	&	$	<	16.5	$	&	4.9	/	VLA-A	&	LSO	&	Y			\\
	J0917$+$4725		&	\nodata		&	G$+$Q	&	\nodata	&	$		1.21	$	&	0.35	&	0.13	&	$		-0.91	$	&	-0.81	&	SS	&	0.58	&	$	<	4.96	$	&	4.9	/	VLA-A	&	SL	&	N	,	I	\\
	J0920$+$1753		&	0.68	*	&	G	&	Sc	&	$		3.45	$	&	1.08	&	0.27	&	$		-0.86	$	&	-1.11	&	SS	&	0.42	&	$	\sim	2.97	$	&	4.9	/	VLA-A	&	MSO	&	Y			\\
	J0920$+$2714		&	\nodata	\tablenotemark{c}	&	G	&	S0	&	$		1.60	$	&	0.46	&	0.13	&	$		-0.93	$	&	-1.04	&	SS	&	2.9	&	$	<	24.7	$	&	4.9	/	VLA-A	&	LSO	&	Y	,	I	\\
	J0939$+$0304		&	0.34	*	&	G	&	S0	&	$		0.75	$	&	0.47	&	0.23	&	$		-0.35	$	&	-0.56	&	FS:	&	0.23	&	$	\sim	1.11	$	&	1.4	/	VLBI	&	CJ	&	Y			\\
	J0945$+$2640		&	\nodata		&	Q	&	\nodata	&	$		1.56	$	&	0.57	&	0.20	&	$		-0.74	$	&	-0.86	&	SS	&	5.3	&	$	<	45.7	$	&	4.9	/	VLA-A	&	LSO	&	N			\\
	J0951$+$1154		&	0.65	*	&	G	&	Sc	&	$		1.25	$	&	0.37	&	0.11	&	$		-0.90	$	&	-0.99	&	SS	&	2.9	&	$	\sim	20.1	$	&	4.9	/	VLA-A	&	LSO	&	N			\\
	J1008$+$2401		&	0.62	*	&	G	&	Sc	&	$		1.36	$	&	0.43	&	0.12	&	$		-0.85	$	&	-1.05	&	SS	&	2.0	&	$	\sim	13.6	$	&	4.9	/	VLA-A	&	MSO	&	Y			\\
	J1010$+$4159		&	0.71	*	&	G	&	Sc	&	$		1.40	$	&	0.42	&	0.14	&	$		-0.90	$	&	-0.88	&	SS	&	0.27	&	$	\sim	1.90	$	&	1.66	/	VLBI	&	MSO	&	Y			\\
	J1019$+$4408		&	\nodata		&	Q$+$abs	&	\nodata	&	$		1.04	$	&	0.34	&	0.07	&	$		-0.82	$	&	-1.27	&	SS	&	0.16	&	$	<	1.33	$	&	1.66	/	VLBI	&	CSO	&	Y			\\
	J1023$+$0424		&	0.71	*	&	G	&	Sc	&	$		0.90	$	&	0.33	&	0.11	&	$		-0.75	$	&	-0.89	&	SS	&	0.56	&	$	\sim	4.03	$	&	8.5	/	VLA-A	&	MSO	&	Y			\\
	J1033$+$3935		&	1.095	:	&	G	&	Sc	&	$		0.82	$	&	0.41	&	0.65	&	$		-0.52	$	&	0.37	&	FS	&	0.005	&	$		0.039	$	&	5	/	VLBI	&	CJ	&	Y			\\
	J1034$+$1112		&	\nodata		&	G$+$Q	&	\nodata	&	$		4.59	$	&	1.21	&	0.36	&	$		-0.99	$	&	-0.98	&	SS	&	4.9	&	$	<	41.7	$	&	4.9	/	VLA-A	&	LSO	&	N			\\
	J1043$+$0537		&	0.42	*	&	G	&	E	&	$		2.67	$	&	0.67	&	0.20	&	$		-1.03	$	&	-0.97	&	USS	&	6.6	&	$	\sim	36.5	$	&	4.9	/	VLA-A	&	LSO	&	N			\\
	J1045$+$0455		&	\nodata		&	Q$+$abs	&	\nodata	&	$		1.18	$	&	0.38	&	0.14	&	$		-0.84	$	&	-0.82	&	SS	&	2.9	&	$	<	24.8	$	&	4.9	/	VLA-A	&	LSO	&	N			\\
	J1048$+$3457		&	1.594		&	Q$+$abs	&	\nodata	&	$		2.44	$	&	1.05	&	0.38	&	$		-0.63	$	&	-0.81	&	SS	&	0.045	&	$		0.385	$	&	1.4	/	VLBI	&	CPLX	&	Y			\\
	J1120$+$2327		&	1.819		&	Q	&	\nodata	&	$		5.48	$	&	1.38	&	0.36	&	$		-1.02	$	&	-1.09	&	USS	&	4.5	&	$		38.4	$	&	4.9	/	VLA-A	&	LSO	&	N			\\
	J1125$+$1953		&	\nodata		&	Q	&	\nodata	&	$		1.60	$	&	0.43	&	0.15	&	$		-0.98	$	&	-0.83	&	SS	&	0.47	&	$	<	4.02	$	&	4.9	/	VLA-A	&	CJ	&	Y			\\
	J1127$+$5743		&	0.49	*	&	G	&	Sb	&	$		1.70	$	&	0.65	&	0.19	&	$		-0.71	$	&	-1.00	&	SS	&	0.49	&	$	\sim	2.95	$	&	8.5	/	VLA-A	&	MSO	&	Y			\\
	J1129$+$5638		&	0.892		&	G	&	Sa	&	$		1.25	$	&	0.50	&	0.12	&	$		-0.68	$	&	-1.16	&	SS	&	0.080	&	$		0.622	$	&	1.4	/	VLBI	&	CSO:	&	Y			\\
	J1142$+$0235		&	0.37	*	&	G	&	Sa	&	$		0.89	$	&	0.38	&	0.20	&	$		-0.63	$	&	-0.50	&	SS$\checkmark$	&	1.5	&	$	\sim	7.67	$	&	4.9	/	VLA-A	&	MSO	&	Y			\\
	J1147$+$4818		&	\nodata		&	Q	&	\nodata	&	$		1.89	$	&	0.50	&	0.12	&	$		-0.99	$	&	-1.18	&	SS	&	1.0	&	$	<	8.32	$	&	4.9	/	VLA-A	&	CJ/GA	&	Y	,	I:	\\
	J1148$+$1404		&	0.54	*	&	G	&	Sc	&	$		1.00	$	&	0.33	&	0.10	&	$		-0.84	$	&	-0.97	&	SS	&	1.5	&	$	\sim	9.29	$	&	4.9	/	VLA-A	&	MSO	&	Y			\\
	J1202$+$1207		&	\nodata		&	G$+$Q	&	\nodata	&	$		1.01	$	&	0.37	&	0.16	&	$		-0.75	$	&	-0.69	&	SS	&	0.098	&	$	<	0.840	$	&	1.4	/	VLBI	&	CSO:	&	Y			\\
	J1203$+$4632		&	0.36	*	&	G	&	E	&	$		0.52	$	&	0.42	&	0.20	&	$		-0.17	$	&	-0.58	&	GPS	&	0.040	&	$	\sim	0.200	$	&	5	/	VLBI	&	CSO:	&	Y			\\
	J1207$+$5407		&	0.61	*	&	G	&	Sc	&	$		2.58	$	&	0.60	&	0.16	&	$		-1.09	$	&	-1.05	&	USS	&	1.4	&	$	\sim	9.74	$	&	4.9	/	VLA-A	&	MSO	&	Y			\\
	J1215$+$1730		&	0.268		&	Q$+$abs	&	\nodata	&	$		2.40	$	&	1.03	&	0.62	&	$		-0.63	$	&	-0.41	&	SS$\checkmark$	&	0.14	&	$		0.551	$	&	5	/	VLBI	&	CSO:	&	Y			\\
	J1228$+$5348		&	\nodata		&	G$+$Q	&	\nodata	&	$		1.94	$	&	0.53	&	0.15	&	$		-0.96	$	&	-1.02	&	SS	&	4.9	&	$	<	41.7	$	&	4.9	/	VLA-A	&	LSO	&	N			\\
	J1238$+$0845		&	0.65	*	&	G	&	Sc	&	$		1.25	$	&	0.41	&	0.08	&	$		-0.82	$	&	-1.32	&	SS	&	3.2	&	$	\sim	22.0	$	&	4.9	/	VLA-A	&	LSO	&	N			\\
	J1300$+$5029		&	1.561		&	Q	&	\nodata	&	$		0.67	$	&	0.38	&	0.39	&	$		-0.43	$	&	0.03	&	FS	&	0.10	&	$		0.855	$	&	4.9	/	VLA-B	&	CJ	&	Y			\\
	J1312$+$1710		&	0.63	*	&	G	&	Sc	&	$		0.95	$	&	0.34	&	0.11	&	$		-0.78	$	&	-0.90	&	SS	&	0.056	&	$	\sim	0.385	$	&	1.4	/	VLBI	&	CSO:	&	Y			\\
	J1315$+$0222		&	0.85	*	&	G	&	Sc	&	$		1.33	$	&	0.52	&	0.16	&	$		-0.70	$	&	-0.93	&	SS	&	2.4	&	$	\sim	18.2	$	&	4.9	/	VLA-A	&	LSO	&	Y			\\
	J1341$+$1032		&	\nodata		&	Q$+$abs	&	\nodata	&	$		2.62	$	&	0.69	&	0.17	&	$		-1.00	$	&	-1.12	&	USS	&	1.0	&	$	<	8.72	$	&	4.9	/	VLA-A	&	MSO	&	Y	,	I:	\\
	J1345$+$5846		&	\nodata		&	G$+$Q	&	\nodata	&	$		1.81	$	&	0.42	&	0.12	&	$		-1.08	$	&	-1.00	&	USS	&	2.3	&	$	<	19.3	$	&	4.9	/	VLA-A	&	LSO	&	N			\\
	J1347$+$1217		&	0.122		&	G	&	Sa	&	$		8.31	$	&	4.86	&	3.09	&	$		-0.40	$	&	-0.36	&	FS	&	0.10	&	$		0.217	$	&	1.66	/	VLBI	&	CSO	&	Y			\\
	J1348$+$2415		&	2.879		&	Q	&	\nodata	&	$		2.74	$	&	0.56	&	0.14	&	$		-1.19	$	&	-1.14	&	USS	&	3.0	&	$		23.8	$	&	4.9	/	VLA-A	&	LSO	&	N			\\
	J1354$+$5650		&	0.65	*	&	G	&	Sc	&	$		1.86	$	&	0.72	&	0.28	&	$		-0.71	$	&	-0.76	&	SS	&	2.5	&	$	\sim	17.4	$	&	4.9	/	VLA-A	&	LSO	&	N			\\
	J1357$+$0046		&	\nodata		&	G$+$Q	&	\nodata	&	$		4.74	$	&	2.00	&	0.49	&	$		-0.64	$	&	-1.11	&	SS	&	0.086	&	$		0.647	$	&	1.4	/	VLBI	&	CSO:	&	Y			\\



	J1410$+$4850		&	0.592		&	G$+$Q	&	\nodata	&	$		0.76	$	&	0.33	&	0.11	&	$		-0.63	$	&	-0.86	&	SS	&	0.043	&	$		0.284	$	&	1.4	/	VLBI	&	CSO:	&	Y			\\
	J1413$+$1509		&	0.22	*	&	G	&	S0	&	$	<	0.25	$	&	0.50	&	0.41	&	$	>	0.52	$	&	-0.16	&	GPS$\checkmark$	&	0.015	&	$	\sim	0.051	$	&	5	/	VLBI	&	CSO	&	Y			\\
	J1414$+$4554		&	0.38	*	&	G	&	S0	&	$		0.40	$	&	0.41	&	0.21	&	$		0.02	$	&	-0.53	&	GPS	&	0.031	&	$	\sim	0.162	$	&	5	/	VLBI	&	CSO	&	Y			\\
	J1415$+$1320		&	0.247		&	G	&	Sb	&	$		2.74	$	&	1.18	&	0.84	&	$		-0.63	$	&	-0.27	&	FS	&	0.12	&	$		0.450	$	&	1.66	/	VLBI	&	CSO	&	Y			\\
	J1421$-$0246		&	0.49	*	&	G	&	E	&	$		1.73	$	&	0.55	&	0.16	&	$		-0.85	$	&	-0.97	&	SS	&	3.5	&	$	\sim	21.2	$	&	4.9	/	VLA-A	&	LSO	&	N			\\
	J1424$+$1852		&	\nodata		&	Q$+$abs	&	\nodata	&	$		2.69	$	&	0.70	&	0.19	&	$		-1.00	$	&	-1.06	&	USS	&	1.6	&	$	<	13.8	$	&	4.9	/	VLA-A	&	MSO	&	Y			\\
	J1502$+$3753		&	\nodata		&	Q$+$abs	&	\nodata	&	$		1.11	$	&	0.34	&	0.12	&	$		-0.89	$	&	-0.81	&	SS	&	2.0	&	$	<	17.5	$	&	4.9	/	VLA-A	&	LSO	&	N			\\
	J1504$+$5438		&	0.621		&	G	&	Sc	&	$		0.78	$	&	0.38	&	0.17	&	$		-0.53	$	&	-0.64	&	SS	&	0.002	&	$		0.014	$	&	5	/	VLBI	&	PS	&	Y			\\
	J1504$+$6000		&	1.024		&	Q	&	\nodata	&	$		5.22	$	&	1.55	&	0.44	&	$		-0.90	$	&	-1.01	&	SS	&	0.62	&	$		4.98	$	&	8.5	/	VLA-A	&	MSO	&	Y			\\
	J1523$+$1332		&	\nodata		&	Q$+$abs	&	\nodata	&	$		1.59	$	&	0.35	&	0.11	&	$		-1.13	$	&	-0.92	&	USS	&	3.7	&	$	<	32.0	$	&	4.9	/	VLA-A	&	LSO	&	N			\\
	J1527$+$3312		&	0.32	*	&	G	&	Sb	&	$		0.75	$	&	0.32	&	0.11	&	$		-0.64	$	&	-0.89	&	SS	&	0.13	&	$	\sim	0.601	$	&	8.5	/	VLA-A	&	CPLX	&	Y			\\
	J1528$-$0213		&	0.54	*	&	G	&	E	&	$		1.06	$	&	0.47	&	0.16	&	$		-0.61	$	&	-0.85	&	SS	&	1.4	&	$	\sim	8.79	$	&	4.9	/	VLA-A	&	MSO	&	Y			\\
	J1548$+$0808		&	0.51	*	&	G	&	E	&	$		1.79	$	&	0.63	&	0.21	&	$		-0.78	$	&	-0.89	&	SS	&	0.25	&	$	\sim	1.51	$	&	1.4	/	VLBI	&	MSO	&	Y			\\
	J1551$+$6405		&	\nodata		&	G$+$Q	&	\nodata	&	$		0.59	$	&	0.68	&	0.21	&	$		0.10	$	&	-0.96	&	GPS	&	0.074	&	$	<	0.630	$	&	5	/	VLBI	&	CJ	&	Y			\\
	J1559$+$4349		&	1.232		&	Q$+$abs	&	\nodata	&	$		2.08	$	&	0.75	&	0.19	&	$		-0.76	$	&	-1.12	&	SS	&	0.79	&	$		6.62	$	&	4.9	/	VLA-A	&	MSO/GA	&	Y	,	I:	\\
	J1604$+$6050		&	0.559		&	G	&	S0	&	$		0.64	$	&	0.59	&	0.20	&	$		-0.06	$	&	-0.88	&	GPS	&	0.014	&	$		0.089	$	&	1.4	/	VLBI	&	CSO:	&	Y			\\
	J1616$+$2647		&	0.755		&	G	&	Sc	&	$		1.71	$	&	1.41	&	0.70	&	$		-0.14	$	&	-0.56	&	GPS	&	0.12	&	$		0.915	$	&	1.4	/	VLBI	&	CSO	&	Y			\\
	J1625$+$4134		&	2.550		&	Q	&	\nodata	&	$		2.47	$	&	1.72	&	1.36	&	$		-0.27	$	&	-0.19	&	FS	&	0.006	&	$		0.049	$	&	5	/	VLBI	&	CJ	&	Y			\\
	J1629$+$1342		&	0.74	*	&	G	&	Sc	&	$		1.95	$	&	0.71	&	0.23	&	$		-0.75	$	&	-0.92	&	SS	&	5.1	&	$	\sim	37.4	$	&	4.9	/	VLA-A	&	LSO	&	N			\\
	J1633$+$4700		&	\nodata		&	Q$+$abs	&	\nodata	&	$		1.46	$	&	0.46	&	0.15	&	$		-0.87	$	&	-0.91	&	SS	&	5.2	&	$	<	44.5	$	&	4.9	/	VLA-A	&	LSO	&	N			\\
	J1724$+$3852		&	1.542		&	Q$+$abs	&	\nodata	&	$		0.88	$	&	0.37	&	0.14	&	$		-0.64	$	&	-0.77	&	SS	&	0.040	&	$		0.340	$	&	1.4	/	VLBI	&	CSO:	&	Y			\\
	J2203$-$0021		&	0.729		&	G	&	Sc	&	$		1.71	$	&	0.61	&	0.18	&	$		-0.77	$	&	-0.98	&	SS	&	0.88	&	$		6.38	$	&	4.9	/	VLA-A	&	MSO	&	Y			\\
      \enddata

      \tablecomments{Columns list: (1) object name; (2) redshift, where suffix (*) indicates a photometric redshift fit by its optical$+$NIR SED in Paper~\Rmnum{1}, 
and the suffix ``:'' means the redshift is uncertain; (3) the optical$+$NIR SED type from Paper~\Rmnum{1}; (4) the Hubble type for pure ``G''-type objects in column (3); 
(5) the flux density at 365 MHz; (6) the flux density at 1.4 GHz; (7) the flux density at 4.9 GHz; (8) the low-frequency spectral index ($\alpha_l$) between 365 MHz and 1.4 GHz; 
(9) the high-frequency spectral index ($\alpha_h$) between 1.4 GHz and 4.9 GHz; 
(10) the radio spectral classification, where the check mark $\checkmark$ means the classification has been modified 
using additional information to the spectral indices in column (8) and (9). 
Classifications include SS = Steep Spectrum, FS = Flat Spectrum, USS = Ultra-Steep Spectrum, and GPS =
Giga-Hertz Peak Spectrum; (11) the largest angular size (LAS) in arcsecs between major components; (12) the largest linear size (LLS) in kpc, 
where ``$\sim$'' means a photometric redshift is used and ``$<$'' means $z=1.6$ is used to obtain the largest size possible, (13) frequency and telescope with 
which the LAS and LLS are measured; (14) radio-source morphological classification, where CPLX means complex, CJ means core-jet, GA means gravitational arc, PS means point source, 
SL means single lobe, and CSO: means CSO candidates; (15) an evaluation (EVAL) whether the object remains a good candidate for an absorption line search at radio frequencies (Y or N) 
and whether it is an intervening system (I) or candidate (I:).}

      \tablenotetext{a}{The redshift listed is for the background radio source which has an SDSS detected galaxy in the foreground.}

      \tablenotetext{b}{The redshift listed is for the background radio source, which is gravitationally-lensed by a foreground galaxy at $z=0.349$.}

      \tablenotetext{c}{This is an intervening system with a foreground galaxy at $z$=0.206. The radio source redshift is unknown.}

\end{deluxetable}

\end{landscape}


Most of the objects show classic double-lobe structure with or without
a core. The separation of the two lobes ranges from tens of parsecs to
tens of kpcs. We adopt the terminology introduced by
\citet{fanti:1995} to indicate an LLS of $>$ 15 kpc, 15 kpc to 1
kpc, and $<$ 1 kpc with the terms Large Symmetric Object (LSO),
Medium-size Symmetric Object (MSO), and Compact Symmetric Object (CSO)
respectively. In general, 1 kpc is considered a typical size for the
narrow-line emission regions around AGNs, and 15 kpc is a
typical size for the visible portions of the faint optical host galaxies 
in this sample. We find 24 LSOs, 22 MSOs,
and 16 CSOs in the sample. LSOs are classic FR\Rmnum{1} or FR\Rmnum{2} radio galaxies
\citep{fanaroff:1974} with extended radio emission that expand to scales much
larger than their host galaxies. These objects are not likely to be
intrinsic absorbers due to their extended structure and large scale,
unless a strong core is present. In contrast, MSOs and CSOs are better
candidates because of their compact structure and small scale close
to the gas-rich galaxy with which they are identified. However, from
its morphology at one-band alone, a CSO can hardly be discriminated
from a core-jet object with multiple jet components if small-scale 
spectral index data is lacking. Therefore, 10 of the CSOs
remain candidates as multi-wavelength and multi-epoch observations are
necessary to characterize ``bona-fide'' CSOs. CSOs are discussed in detail in Section~\ref{sec:cso}.

If the jet of a radio source axis is oriented toward us, the radiation
close to the core could be strongly boosted due to relativistic
beaming of the jet. Radio structure of these sources generally
consists of a compact core and some extended jet components aligned on
one side of the core (core-jet [CJ] morphology). Sometimes a weak
counterjet is also detected. These are also good candidates for
absorption-line searches given their very compact structure.

Morphological classifications are listed in Column~(14) of
Table~\ref{tab:type}. Except for the double-lobe and core-jet types
mentioned above, there are four minor types: single-lobe [SL],
gravitational-arc [GA], complex [CPLX], and point-source [PS] objects. J0834$+$1700 and
J0917$+$4725 have a single, extended component in their VLA-A
images. We classify them as SL objects since they are very likely
one of the two lobes in a classic double-lobe structure. J0751$+$2716
is a GA system with an arcuate shape in its radio image
\citep{lehar:1997}. We find two more objects showing potential gravitational
arcs, J1147$+$4818 and J1559$+$4349. Four objects are classified as
CPLX objects for their atypical morphologies. All the above sources
will be discussed in detail in subsequent sections. The point-source (PS)
objects, J0901$+$0304 and J1504$+$5438, are unresolved in their VLBA
images and most likely core-jet objects without the jet components
detected separately.

Column~(15) is our overall evaluation (``EVAL'') concerning the likelihood for a radio absorption line
to be present based primarily upon the compactness of the radio emission. Approximately 70\%
of these sources appear to be good candidates (``Y'' in column ~(15)) 
for absorption searches once spectroscopic redshifts are in-hand. Intervening (``I'') and
possibly intervening (``I:'') galaxies are also good candidates for absorption-line searches.

\subsubsection{Compact Symmetric Objects}\label{sec:cso}

CSOs are luminous radio galaxies which have symmetrical,
 double-lobe structure fully contained within 1 kpc. Rather
than ``frustrated'' jets obstructed by a dense inter-stellar medium
(ISM), CSOs are now believed to be young objects with age $\lesssim
10^4$ yr that will evolve into classical FR\Rmnum{1} or FR\Rmnum{2} radio
galaxies \citep{fanti:1995,begelman:1996,readhead:1996}. The hotspot
advance speed of CSOs is found to be sub-luminal
\citep{owsianik:1998} with nearly equal-brightness extended lobes, 
in sharp contrast to the core-jet objects. CSOs also have
low flux variability and low polarization. All of these are
consistent with the speculation that the jets are propagating nearly
perpendicular to the line of sight \citep{readhead:1996}. While CSO candidates are often selected
based on their morphologies in one band, it is preferred that a
flat-spectrum or inverted core between two symmetric lobes is identified using
multi-wavelength observations. Since the cores of CSOs are often
weak or undetected, the advance speed of hotspots and
jet components can be measured to make sure there is no strong projection
effect (i.e., no apparent superluminal motion as in the core-jet sources). The
detection rate of \ion{H}{1} 21 absorption is exceptionally high in
CSOs, 30--50\% (e.g., \citealt{pihlstrom:2003,gupta:2006}), which may be due, at
least in part, to their very compact radio structure \citep {curran:2010}. One of
the two known intrinsic atomic and molecular absorbers, J1415$+$1320,
is a CSO \citep{perlman:1996,perlman:2002}. The high frequency of occurrence of 
H\,I absorption in CSOs may be due to them having physical sizes (10 pc to 1 kpc) 
which are comparable in size to the absorbing screen \citep{curran:2013}. Thus, 
identification of CSO candidates in this sample is an important step towards
discovering new H\,I and molecular absorption systems. 

There are 5 previously known CSOs in our sample (see Table~\ref{tab:cso};
previously known CSOs are listed below the solid line in this Table). 
J1019$+$4408 was observed by
\citet{dallacasa:2002a} at 1.67 GHz and exhibits all the typical CSO
features in morphology, i.e., symmetric, edge-brightened, and S-shaped
lobes. J1347$+$1217 is an ultra-luminous infrared galaxy at $z=0.122$
in an ongoing merger. It was observed using VLBI by
\citet{stanghellini:1997} at 5 GHz, and the presence of a core was confirmed by
\citet{stanghellini:2001} at 15 GHz and \citet{xiang:2002} at 1.66 GHz
although luminosities of the two lobes are quite asymmetrical especially
at higher frequencies. \citet{lister:2003} found that J1347$+$1217 has
atypical CSO features including superluminal motions and large linear
polarization, suggesting a small angle relative to the line-of-sight.
\ion{H}{1} 21-cm absorption is detected towards the weak lobe
\citep{morganti:2004}. 

J1413$+$1509 was classified by \citet{helmboldt:2007} as a CSO candidate at 
5 GHz and later confirmed by \citet{tremblay:2011} using VLBI observations 
at 4.8 GHz, 8.3 GHz, and 15 GHz. J1414$+$4554 also has typical CSO morphology;
\citet{gugliucci:2005} did not detect motions between the two hotspots
from 1997 to 2002 and set an upper limit of 0.014 mas yr$^{-1}$. This
indicates a relative projected velocity of $\leq$ 0.24 $c$ assuming a
photometric redshift $z=0.38$ (Paper~\Rmnum{1}). J1415$+$1320 is an
atomic and molecular absorber at radio frequencies. \citet{perlman:1996} 
identified the core using multi-frequency VLBI observations. \citet{gugliucci:2005} measured the
motion of the hotspots and estimate an age as young as 130$\pm$47 yr for this CSO.

We have identified one new CSO (J1616$+$2647 with typical CSO
morphology, an extended double-lobed morphology with at best a very weak core; see
its VLBA map in the Appendix) 
and 10 new CSO candidates (CSO: in Column~15 in Table~\ref{tab:type} and
all listings above the solid line in Table~\ref{tab:cso}). All of
them are compact in 8.5 GHz VLA-A images, i.e., there is no extended
flux detected at or larger than the $\sim$ 0\farcs1 level. For 9 of the 10
objects, 70\%--85\% of the flux density detected by the FIRST survey
is accounted for in our 1.4 GHz VLBA images. J1312$+$1710 has only
52\% of its FIRST flux detected, indicating it may have a second set
of weak lobes further out from the core.

Since we lack spatially-resolved multi-frequency VLBA images and/or very accurate optical-NIR/radio
astrometry we cannot identify the core component in these sources unambiguously either by its flat 
or inverted spectral index or its position with respect to the host galaxy. This means that 
in most cases (all but J1616$+$2647) a CSO morphology cannot be distinguished from a
CJ morphology. Until better VLBA maps and/or multi-epoch VLBA maps become available, the
10 sources listed above the line in Table~\ref{tab:cso} remain CSO candidates.

Three of our CSO candidates were previously observed with VLBA. J0134$+$0003
was observed in the VLBA Calibrator Survey by \citet{beasley:2002}. 
J1215$+$1730 and J1203+4632 were selected by
\citet{helmboldt:2007} as CSO candidates in 5 GHz VLBA maps. 
\citet{tremblay:2011} claim there is no structure seen in the
counterjet of J1215$+$1730 and refute it as a CSO candidate. But we
still consider it a candidate based on our CSO-selection criteria.

Once all 10 CSO candidates are confirmed, the frequency of CSO occurrence in this sample 
will be $\sim$ 20\%, nearly three higher than the 7.5\% identification rate found by
\citet{polatidis:1999} in a flux-limited sample with flux density
$\geqslant$ 0.7 Jy at 5 GHz. While selecting at high frequencies tends
to include more compact object than at low frequencies, our
selection frequency is 1.4 GHz compared to 5 GHz in
\citet{polatidis:1999}. We will show in Section~\ref{sec:sta} that most
of the CSO candidates have multi-band properties that are consistent
with typical CSOs, implying a very high success rate of CSO selection. The high
CSO detection rate might be caused by our selection of non-elliptical
galaxies, consistent with the fact that disturbed optical structure is
often seen in nearby CSOs \citep{perlman:2001}. 
Once confirmation of these CSO
candidates is obtained, our sample will have returned a much higher frequency
of CSOs than other selection techniques.

\begin{deluxetable}{llrrllc}
  \tablecaption{Summary of CSOs and CSO
    Candidates\label{tab:cso}}\tablewidth{0pt}
 \tablehead{ \colhead{Object} &
    \colhead{$z$} & \colhead{LAS} & \colhead{LLS} & \colhead{$z$} &
    \colhead{CSO} &
    \colhead{Status} \\
    \colhead{} & \colhead{} & \colhead{(mas)} &
    \colhead{(pc)} & \colhead{Ref.} & \colhead{Ref.} & \colhead{}\\
    \colhead{(1)} & \colhead{(2)} & \colhead{(3)} & \colhead{(4)} &
    \colhead{(5)} & \colhead{(6)} & \colhead{(7)} }

\startdata
J0134$+$0003	&	0.879	&	18	\tablenotemark{a}	&		140	&	[5]	&	[3]	&	?	\\
J1129$+$5638	&	0.892	&	80		&		622	&	[1]	&This work		&	?	\\
J1202$+$1207	&	\nodata	&	98		&	$<$	840	&	\nodata	&This work		&	?	\\
J1203$+$4632    &       0.36*   &       40              &       $\sim$  200	&	[2]	&[12]		&	?	\\
J1215$+$1730	&	0.268	&	135	\tablenotemark{b}	&		551	&	[12]	&	[12]	&	?	\\
J1312$+$1710	&	0.63*	&	56		&	$\sim$	385	&	[2]	&This work		&	?	\\
J1357$+$0046	&	\nodata	&	86		&	$<$	738	&	\nodata	&This work		&	?	\\
J1410$+$4850	&	0.592	&	43		&		284	&	[1]	&This work		&	?	\\
J1604$+$6050	&	0.559	&	14		&		89	&	[1]	&This work		&	?	\\
J1724$+$3852	&	1.542	&	40		&		340	&	[1]	&This work		&	?	\\ \hline
J1019$+$4408	&	\nodata	&	155	\tablenotemark{c}	&	$<$	1326	&	\nodata	&	[4]	&	Y	\\
J1347$+$1217$^{\dagger}$	&	0.122	&	100	\tablenotemark{d}	&		217	&	[6]	&	[10], [13], [8]	&	Y	\\
J1413$+$1509	&	0.22*	&	14	\tablenotemark{e}	&	$\sim$	49	&	[2]	&	[12]	&	Y	\\
J1414$+$4554	&	0.38*	&	31	\tablenotemark{f}	&	$\sim$	160	&	[2]	&	[7]	&	Y	\\
J1415$+$1320$^{\dagger}$	&	0.247	&	117	\tablenotemark{g}	&		450	&	[11]	&	[9], [7]	&	Y	\\ 
J1616$+$2647	&	0.755	&	124		&		915	&	[1]	&This work		&	Y	\\
\enddata

\tablecomments{The two sections separated by a
  horizontal line summarize the 10 CSO candidates (at top) and
  the 6 previously-known/confirmed CSOs (at bottom). Columns list: (1)
  object name; (2) redshift, where suffix ``*'' indicates photometric
  redshift fit by optical-NIR SED in Paper \Rmnum{1}; (3) largest
  angular size (LAS) between major components in milli-arcsecs; (4) largest 
  physical size (LLS) in parsecs,
  where ``$\sim$'' means a photometric redshift is used and ``$<$''
  means $z=1.6$ is assumed because no good redshift estimate in available; (5) redshift
  reference; (6) CSO reference; (7) status, where ``?'' indicates a CSO
  candidate and ``Y'' means confirmed CSO.}

\tablerefs{[1] Spectroscopic redshift (Paper~\Rmnum{1}); [2] photometric redshift
  derived by fitting the optical--NIR SED (Paper~\Rmnum{1});
  [3] \citet{beasley:2002}; 
  [4] \citet{dallacasa:2002a};
  [5] \citet{drinkwater:1997};
  [6] \citet{grandi:1977}; 
  [7] \citet{gugliucci:2005}; 
  [8] \citet{lister:2003}; 
  [9] \citet{perlman:1996};
  [10] \citet{stanghellini:2001}; 
  [11] \citet{stocke:1992};   
  [12] \citet{tremblay:2011}; 
  [13] \citet{xiang:2002}.
}

\tablenotetext{a}{Measured at 2.3 GHz \citep{beasley:2002}.}
\tablenotetext{b,e,f}{Measured at 5 GHz \citep{helmboldt:2007}.}
\tablenotetext{c}{Measured at 1.66 GHz \citep{dallacasa:2002a}.}
\tablenotetext{d}{Measured at 1.66 GHz \citep{xiang:2002}.}
\tablenotetext{g}{Measured at 1.66 GHz \citep{perlman:1996}.}

\end{deluxetable}

\subsubsection{Radio SED types}\label{sec:sed}

To classify radio SEDs, flux densities at three frequency bands are
used: 365 MHz, 1.4 GHz, and 5 GHz. The Texas Survey of Radio Sources
at 365 MHz \citep{douglas:1996} detected all objects in the sample,
except J1413$+$1509 for which the flux density limit of the survey, 25
mJy, is used as an upper limit. Flux densities at 1.4 GHz from the
FIRST survey are compared with those from other lower-resolution
(45\arcsec) surveys such as the NRAO VLA Sky Survey
\citep{condon:1998} to make sure all the flux is accounted
for by FIRST. In fact, the differences are all within $\sim$ 10\%,
comparable to the uncertainties. At 5 GHz, all the sources with
declination higher than 2\fdg5 are in the 87GB catalog of radio
sources \citep{gregory:1991}. For those lower than 2\fdg5, flux
densities at 5 GHz are cited from the Parkes-MIT-NRAO surveys
\citep{griffith:1994,griffith:1995} or the Parkes Catalog
\citep{wright:1990}. Flux densities of the three bands are listed in
Columns~(5)--(7) of Table~\ref{tab:type}.

We define $\alpha_l$ (``l'' for low) as the spectral index between 365 MHz and 1.4
GHz, and $\alpha_h$ as between 1.4 GHz and 5 GHz ($S_{\nu} \propto
\nu^{\alpha}$). While slightly different criteria are adopted by
different authors, the following empirical criteria are commonly used
at the above listed frequencies: flat-spectrum (FS) objects ($\alpha_h
\geqslant -0.5$; e.g., \citealt{healey:2007}), steep-spectrum (SS)
objects ($\alpha_h < -0.5$ and $\alpha_l < -0.5$),
gigahertz-peaked-spectrum (GPS) objects ($\alpha_h < -0.5$, $\alpha_l
\geqslant -0.5$, and $\alpha_l-\alpha_h \geqslant 0.3$;
e.g. \citealt{odea:1998}), and ultra-steep-spectrum (USS) objects
($\alpha_l \leqslant -1.0$; e.g. \citealt{roettgering:1994}). In
Table~\ref{tab:type}, $\alpha_l$ and $\alpha_h$ are shown in
Column~(8) and Column~(9) respectively, with the inferred radio SED type in
Column~(10).

Our spectral classification scheme described above can be interpreted in
terms of turnover frequency: 
$\nu_m \lesssim 400$ MHz for SS objects and $\nu_m \gtrsim $ 400 MHz
for GPS objects. A small fraction of SS objects are USS with $\alpha <
-1.0$ due to extreme energy loss probably caused by electron depletion
or inverse-Compton radiation. It has been found that USS objects are
excellent candidates for finding high-$z$ radio galaxies
\citep{miley:2008}. When the radiation becomes opaque at a range of physical sizes, the spectrum
has a shallower and often bumpy character, which makes it an FS object.

We have double-checked the radio SED types using flux densities
measured from other radio frequencies, with help from the NASA/IPAC
Extragalactic Database (NED). Two FS objects, J0901$+$0304 and J1413$+$1509,
are re-classified as GPS objects because their turnover frequencies
($\nu_\mathrm{m}$) are between 5 GHz and 8.5 GHz. Two FS objects,
J1215$+$1730 and J1142$+$0235, are re-classified as SS because their
SEDs are straight from 365 MHz to 8.5 GHz with $\alpha
\simeq-0.5$. One GPS object, J0843$+$4215, is re-classified as SS
because $\nu_m \leqslant$ 500 MHz with $\alpha_l=-0.4$ and
$\alpha_h=-0.7$. One object, J0939$+$0304, does not fall into any of
the four categories listed above. It has $\alpha_h < -0.5$, $\alpha_l
> -0.5$, and $(\alpha_l-\alpha_h) < 0.3$. Flux densities at higher or
lower frequencies are needed to pin down the SED types of objects like
J0939$+$0304. It could be either FS or GPS as there is no flux density
available $>$ 5 GHz, so it is marked as ``FS:''.

Overall, there are 7 FS objects, 8 GPS objects, 47 SS objects, 9 USS
objects, and one uncertain object (FS or GPS ?). The reliability of the
original classification method is 60\% for FS objects and 83\% for GPS
objects. A spectral index between $\sim$ 1 GHz and $\sim$ 10 GHz might
be a better choice for $\alpha_h$ to classify FS and GPS objects than
the spectral index between $\sim$ 1 GHz and $\sim$ 5 GHz used here. For SS and
USS objects, 100\% of them are indeed steep spectrum, although the
critical spectral index $-$1.0 is rather arbitrarily chosen. The radio
spectral types are shown in Column (10) of Table~\ref{tab:type}.


\subsection{Statistical Properties of the Radio Continuum Emission}\label{sec:sta}

In this section, we study the statistical properties of our sample
including radio morphologies, radio SEDs, and optical$+$NIR SEDs in
order to identify correlations between different types of sources which
would make eventual redshifted H\,I 21cm absorption searches more effective. 
To study a homogeneous sample with only intrinsic obscuration (and 
possible related absorption),
systems with intervening galaxies need to be excluded. Column~(15) of
Table~\ref{tab:type} lists 4 intervening systems and 5 suspect
objects (see Sections~\ref{sec:pos}--\ref{sec:oth} for more
details). Excluding these 9 objects gives us a statistical 
sample of 71 objects with which to study source systematics.

%
%
%
%
%
%

\subsubsection{Radio Morphologies and SEDs}\label{sec:morspe}

We summarize the number of objects with each radio morphological type
and SED type in Table~\ref{tab:morspe}. The following statistical
trends are clearly shown in the data.

\begin{deluxetable}{r|r|cccccc}
  \tablecaption{Statistics of Morphological Types and Radio Spectral
    Types\label{tab:morspe}} \tablewidth{0pt}\tablehead{ \colhead{\backslashbox{Radio
        SED}{Morphology}} & \colhead{} & \colhead{CJ} & \colhead{CSO}
    & \colhead{MSO} & \colhead{LSO} & \colhead{PS} & \colhead{CPLX} }

\startdata
	&		&	8	&	16	&	19	&	23	&	2	&	3	\\	\hline
FS	&	7	&	5	&	2	&	0	&	0	&	0	&	0	\\	
GPS	&	8	&	1	&	6	&	0	&	0	&	1	&	0	\\	
SS	&	47	&	2	&	8	&	15	&	18	&	1	&	3	\\	
USS	&	9	&	0	&	0	&	4	&	5	&	0	&	0	\\	
\enddata

\tablecomments{This table shows the number of objects with a
  particular morphological type and radio spectral type. The row
  header lists morphological types and the column header lists
  spectral types. The total number of objects for each type are shown in
  the first row and the first column.}

\end{deluxetable}

\begin{itemize}
\item There is not a single MSO or LSO among the FS and GPS
  objects. This implies that $\nu_m \gtrsim$ 1 GHz almost guarantees a
  small-scale core-jet object, CSO, or unresolved source. On the
  other hand, 75\% of the core-jet objects and 50\% of the CSOs are FS
  or GPS objects.
\item FS objects are mostly or all core-jet objects (at least 5 out of
  7).
\item Most of the GPS objects (6 out of 8) are CSOs. As previously noted,
GPS objects are an excellent source for finding CSOs
  \citep{stanghellini:1997,liu:2007}.
\item Steep spectrum (SS) CSOs have a larger median size (622 pc) than
  GPS CSOs (162 pc). Meanwhile, most of the SS CSOs have $\nu_m \sim$
  365~MHz to 1.4~GHz, which means that $\nu_m$ varies inversely with
  the size. This correlation has been found previously, both in a sample
of CSOs and MSOs \citep{fanti:1990} and in a sample of Compact Steep Spectrum 
(CSS) and GPS sources \citep{odea:1997}. While this trend might also apply to 
MSOs and LSOs, their inferred $\nu_m$ shifts to frequencies lower 
than the 365 MHz band and so cannot be verified using current data.

\item Double-lobe sources in the sample have $\nu_m <$ 10 GHz
and their sizes are inversely proportional to $\nu_m$, while FS
objects have no turnover frequency up to at least 10 GHz
\citep{odea:1998}. It is also worth noting that the intervening
systems/candidates are exclusively SS objects, i.e., the radio sources
in these systems are of the most common variety, unlike those with
intrinsic obscurations.

\end{itemize}

The presence in the current sample of the same trends found in more traditionally-defined 
samples means that these radio sources are all examples of previously known types, not some new radio
morphology associated only with gas-rich systems. However, some of these types (e.g., CSOs) 
are greatly favored by the novel selection method used here (i.e., selecting radio sources 
by their optical morphology).

\subsubsection{Radio Morphologies and Optical$+$NIR
  SEDs}\label{sec:morsed}

Similar to Table~\ref{tab:morspe}, Table~\ref{tab:morsed}
summarizes the number of objects with a particular radio morphology
and optical$+$NIR SED type (see Paper~\Rmnum{1} for a listing of these
types and details of their determination).
We find the following statistical tendencies in the sample:

\begin{deluxetable}{r|r|cccccc}
  \tablecaption{Statistics of Radio Morphologies and Optical$+$NIR
    SEDs\label{tab:morsed}} \tablewidth{0pt}\tablehead{ \colhead{\backslashbox{OIR
        SED}{Morphology}} & \colhead{} & \colhead{CJ} & \colhead{CSO}
    & \colhead{MSO} & \colhead{LSO} & \colhead{PS} & \colhead{CPLX} }

\startdata
	&		&	8	&	16	&	19	&	23	&	2	&	3	\\	\hline
G	&	37	&	3	&	10	&	12	&	8	&	2	&	2	\\	
G+Q	&	16	&	2	&	3	&	3	&	8	&	0	&	0	\\	
Q+abs	&	11	&	0	&	3	&	3	&	4	&	0	&	1	\\	
Q	&	7	&	3	&	0	&	1	&	3	&	0	&	0	\\	
\enddata

\tablecomments{This table shows the number of objects with a
  particular radio morphology and optical$+$NIR SED. The row header
  lists radio morphological types and the column header lists
  optical$+$NIR SED types. The total number of objects for each type are
  shown in the first row and the first column.}

\end{deluxetable}

\begin{itemize}
\item Core-jet (CJ) objects have a high percentage ($\sim$ 60\%) of quasar-type SEDs (
  $Q$, $Q+abs$, and $G+Q$). Since our sample-selection
  criteria were meant to exclude sources with stellar counterparts, the actual
  fraction of CJ sources which are quasars is much higher in traditionally-selected
  samples. As expected this indicates that CJ objects are predominantly
  unobscured quasars; the few found in this sample are due either to misclassification
of the SDSS images or to a partial obscuration of the nuclear regions.

\item CSOs and MSOs have a high percentage ($>$ 80\%) of galaxy-type SEDs ($G$
  and $G+Q$) implying that their AGN are mostly obscured in the optical and even in 
the NIR in some cases. Use of mid-IR survey work by the {\it Spitzer and WISE} satellites should
find large numbers of CSOs and MSOs by cross-correlation of bright radio and mid-IR sources.  

\item The fraction of $G$-type objects decreases and that of $Q$-type objects increases with 
increasing size (CSOs, MSO, LSOs). This suggests that
  optical obscuration diminishes with increasing radio size and thus with AGN age
\citep{fanti:1995, begelman:1996}. This result also suggests that, while
highly-obscured radio-loud AGN may exist in abundance at high-$z$, they are exceptionally
difficult to identify and confirm (i.e., obtain redshifts) because the counterparts 
rapidly become invisible optically with increasing redshift.

\end{itemize}



\subsubsection{Radio SEDs and Optical$+$NIR SEDs}\label{sec:spesed}

Table~\ref{tab:sedspe} summarizes the number of objects with a
particular radio SED type and optical$+$NIR SED type. Following the
discussions in Section~\ref{sec:morspe} and \ref{sec:morsed}, we find
that GPS objects are associated with $G$ or $G+Q$-type objects, while
2 out of the 7 FS objects show $Q$-type optical-NIR SEDs. Here again
we see that the fraction of quasars in FS objects is much lower than in
traditionally-selected samples. We conclude that there is a good
correspondence between radio morphological types, radio SED types and
optical-NIR SED types: CSO-GPS-G, MSO-CSS-G, LSO-SS-G, CJ-FS-Q, where
``CSS'' stands for Compact Steep Spectrum sources, a name that
combines morphological and SED features \citep{fanti:1990}. Although
the radio SED types require fewer observations to classify, they do not
specify a radio morphology; e.g., CSOs can be FS, GPS,
or SS objects depending on the source's turnover frequency. Therefore, high-resolution
radio images are crucial to reveal the nature of these objects.

\begin{deluxetable}{r|r|cccc}
\tablewidth{0pt}
 \tablecaption{Statistics of Optical$+$NIR SEDs and Radio Spectral
    Types\label{tab:sedspe}} \tablehead{
    \colhead{\backslashbox{OIR SED}{Radio SED}} & \colhead{} &
    \colhead{FS} & \colhead{GPS} & \colhead{SS} & \colhead{USS}}

\startdata
	&		&	7	&	8	&	47	&	9	\\	\hline
G	&	37	&	4	&	7	&	24	&	2	\\	
G+Q	&	16	&	1	&	1	&	13	&	1	\\	
Q+abs	&	11	&	0	&	0	&	7	&	4	\\	
Q	&	7	&	2	&	0	&	3	&	2	\\	
\enddata

\tablecomments{This table shows the number of objects with a
  particular optical-NIR SED type and radio SED type. The row header
  lists radio SED types and the column header lists optical-NIR SED
  types. The total number of objects for each type are shown in the first
  row and the first column.}

\end{deluxetable}

Among the 9 USS objects, 7 have quasar-type SEDs; additionally, all of them are
MSOs or LSOs. Redshifts for 3 USS objects are available, all at high-$z=$
1.8--2.9. For comparison, the other two objects with $z \geqslant
1.8$ are the gravitational-lens system J0751$+$2716 and an FS core-jet
object J1625$+$4134. In objects with $\alpha_l < -0.9$, only
J0920$+$2714 has a $G$-type SED, and we will show in
Section~\ref{sec:pos} that this is an intervening absorption system
with a foreground galaxy at $z=0.206$ much brighter than the AGN host
galaxy in the optical$+$NIR. The radio source host remains undetected
on our deepest optical ($r>$23) and NIR (K$_s >$ 19) images (see 
Paper \Rmnum{1}). Our findings in this sample are consistent with many 
of the USS objects being at high-$z$ \citep [e.g.][]{roettgering:1994}.

\subsection{Individual Sources of Interest}\label{sec:individuals}

\subsubsection{Positional Offsets}\label{sec:pos}

In Paper~\Rmnum{1}, we detected a NIR counterpart for every optical
counterpart, i.e., the offset between the NIR and optical centroid positions is
within 5$\sigma$ of their combined uncertainties. The measurement errors 
of the optical/NIR objects are of order $\sim$ 0\farcs1--0\farcs2 and their 
observed angular sizes are $\lesssim$ 2\arcsec\ except for a
few larger objects ($\sim$ 5\arcsec). However, the diffuse, asymmetrical morphology
of some of these galaxies make their centroid determination more uncertain than
the measurement errors. Also, the resolution of the
FIRST survey is $\sim$ 5\arcsec, insufficient to determine a precise
offset between the optical/NIR and radio positions. With our new
high-resolution radio images we scrutinized each object's radio core positions 
to see whether there is a significant offset which might indicate unusual systems
such as intervening or interacting galaxies. In Figure 5 in the Appendix the optical
and NIR centroids are located on the VLA and VLBA maps. These positions are
seen most easily on the 4.9 GHz maps as blue (optical) and red (NIR) crosses. When
the centroid positions are inside the green boxes marking the VLBA map field-of-view,
the centroids are also located on the 1.4 GHz VLBA maps although typically the positional
errors are much larger than the component sizes, making firm identifications of the VLBI core
difficult with current data in most cases.

We discuss here all sources with significant radio/optical-NIR positional offsets. 
The radio and NIR images and component solutions for all of these sources can be found in 
the Appendix.

\textbf{J0003$-$1053} has four components in its 4.9 GHz and 8.5 GHz
images. Its NIR centroid position is coincident with the ``C''
component, which has an inverted spectral index. All other components
have steep spectral indices $\alpha < -0.5$. However, the optical centroid
position is close to the ``B'' component. A redshift of
1.474$\pm$0.001 is obtained from emission lines in the NIR (see
Paper~\Rmnum{1}), but the photometric redshift given by SDSS DR8
is $0.57 \pm 0.12$. The optical-NIR SED is very red with
$(r-K_\mathrm{s})=5.5\pm0.3$, redder that typical elliptical galaxies,
and classified as $Q+abs$. We suggest that the object at $z=1.474$ is
heavily obscured either intrinsically or, more likely, by the foreground
galaxy that may or may not have associated radio emission. The co-linearity 
of components B, C and D strongly favors component B being a part of the
background radio source at $z$=1.474, not emission from the foreground galaxy.
Either situation makes this source favorable for the detection of absorption lines
at $z$=0.5-0.7.

\textbf{J0807$+$5327} has a double-lobe radio structure. However, the
entire radio structure is to the south of the optical and NIR centroid
positions by $\sim$ 1.5\arcsec, which makes it possible that the radio
source is not associated with the optical$+$NIR object we detect. If this is a
core-jet object with a core in the ``A'' component, the core is still
$\sim$ 0.5\arcsec\ away from the NIR centroid position. The VLBA detected
source appears similar morphologically to a hot spot in a single radio lobe, identified
as component ``A'' in Figure 5 in the Appendix. 
However, there is no bright source to the NE of this position
which would be the second associated lobe of this source. Regardless of the overall
nature of the source, the coincidence on the sky of a diffuse galaxy and a $\sim$ 100 mJy
compact radio source is favorable for the detection of absorption lines.
 
\textbf{J0917$+$4725} shows extended structure in its 4.9 GHz and 8.5
GHz images, and the radio center is 1\farcs3 away from the NIR
center. We overlay a larger field-of-view FIRST field with an $r$-band SDSS image in
Figure~\ref{fig:J0917}. The maps in Figure 5 in Appendix I are at higher resolution than the FIRST
image and show a morphology consistent with being a single radio lobe with its hot spot detected
by the VLBA. A second radio lobe is seen in the northeast
$\sim$ 40\arcsec\ away, making it unlikely that the NIR object shown 
in Figure~\ref{fig:J0917} is the host galaxy. It is unknown which is the optical counterpart
to the radio source. Absorption lines are unlikely to be detected due
to the positional offset between the compact source seen with VLBA and the NIR galaxy
even if the optical$+$NIR object is in the foreground of the radio source.

\begin{figure}[htdp]
\includegraphics[scale=0.5]{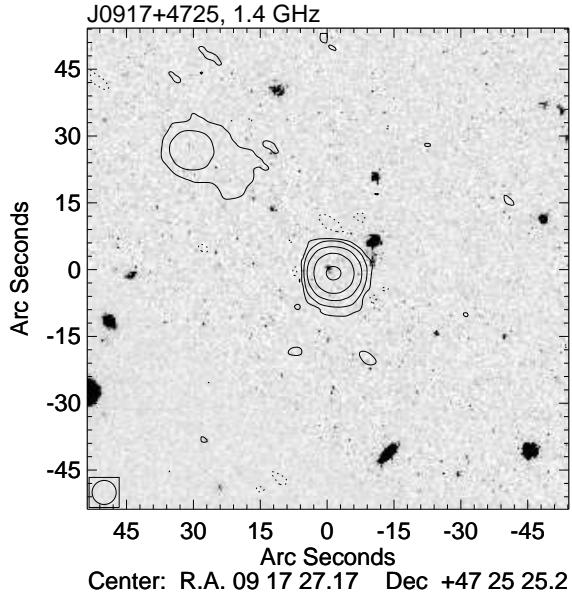}
\caption[The 1.4 GHz contour of the field of J0917$+$4725]{The 1.4 GHz FIRST flux
  contours of the field of J0917$+$4725 overlaid with an optical $r$-band SDSS image. 
The radio contour levels are $-$0.45, 0.45, 2.3, 11,
  54, and 267 mJy. \label{fig:J0917}}
\end{figure}

\textbf{J0920$+$2714} is an LSO with the brighter lobe overlapping with
but not centered on an elliptical galaxy at $z=0.209$ (1\farcs5 away
from optical center). An \ion{H}{1} absorption line has been detected at the redshift of the
elliptical galaxy (see Section~\ref{sec:HIobs}. We conclude that the 
optical--NIR galaxy is not the counterpart to the radio source; the radio source 
counterpart is too faint and/or too obscured to be seen on our deepest 
optical and NIR images. The USS for this radio source and the absence of a strong
core suggest that this radio source is likely at high-$z$.

\textbf{J1341$+$1032} is also an MSO which is offset from its
optical$+$NIR center similar to J0807$+$5327 and J0920$+$2714. This
makes it a good candidate to detect absorption lines at the redshift
of the offset foreground galaxy, which is very diffuse and has an SDSS $z_{phot} \approx$ 0.4.
A possible detection of the optical/NIR counterpart to the radio source can be seen between
components A \& B in the NIR image in the Appendix.

\subsubsection{Gravitational Lens Candidates}\label{sec:lens}

If well-aligned, the foreground galaxy in an intervening system can
substantially bend the light from the background source and form a
gravitationally-lensed arc. In this case the optical/NIR object we identify may be
a foreground galaxy with a positional offset from the radio source core.
Therefore, an arc-like structure to
the radio source can help us to discover candidates for lensing
systems. There are three objects showing arc-like features in the
sample including J0751$+$2716, a previously known gravitational lens
system \citep{lehar:1997}, which will not be discussed further here.

\textbf{J1559$+$4349} has a NIR redshift of 1.232$\pm$0.001
(Paper~\Rmnum{1}) and an optical photometric redshift of
0.42$\pm$0.09 according to SDSS DR8 \citep{eisenstein:2011}. We classify its optical$+$NIR SED
as $Q+abs$ because it is bluer than an Sc-type
galaxy. Figure~\ref{fig:J1559} shows that the $K$-band center (red cross) is close
to the radio core position. If the radio structure is part of an Einstein ring
with radius 0\farcs64, the center of the ring is close to the $r$-band
center (blue cross). Therefore, we suggest that the foreground lens dominates 
the optical light, while the background AGN host at $z=1.232$ is dominant in
the NIR. J1559$+$4349 has two components in its VLA-A 4.9 GHz and 8.5
GHz images with spectral indices $-1.0$ and $-1.2$
respectively. However, extrapolated to 1.4 GHz, only $\sim$ 50\% of
the total flux density of ``A'' and $\sim$ 20\% of ``B'' is detected
in the VLBA image.

\begin{figure}[htdp]
\includegraphics[scale=0.5]{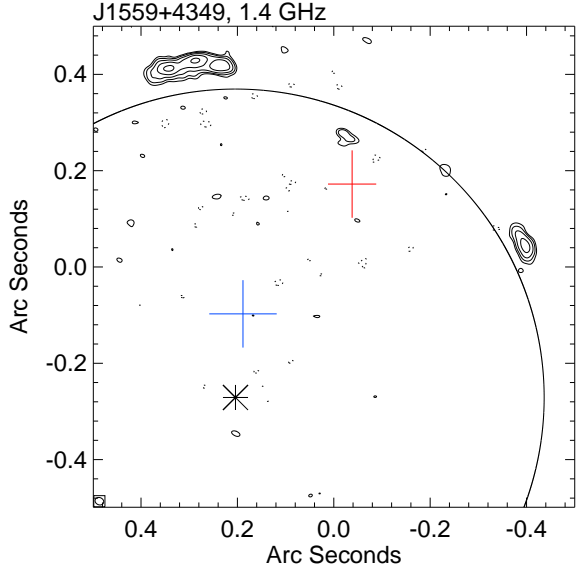}
\caption[The 1.4 GHz contour of J1559$+$4349.]{The 1.4 GHz VLBA contours of
  J1559$+$4349. The black arc is part of a circle with radius 0\farcs64
  centered at the position labeled with an asterisk. The NIR centroid is shown as a red cross
while the optical $r$-band centroid is shown as a blue cross. (This plot is
  otherwise the same as the 1.4 GHz plot in the Appendix.)\label{fig:J1559}}
\end{figure}

\textbf{J1147$+$4818} is classified as a CJ object based on its
VLA-A images. No emission from component ``B'' (as is seen on the VLA-A 4.9 and 8.5 GHz maps
in the Appendix) is detected in the 1.4
GHz VLBA image. We show in Figure~\ref{fig:J1147} that the arc-like string of
VLBA components lies on a circle with radius 0\farcs20. The optical and
NIR centroid positions are coincident and both are in the vicinity of
the arc although their uncertainties are large. J1147$+$4818 is a $Q$-type object but DR8 gives a photometric
redshift of $z=0.17\pm0.05$ by assuming it is a galaxy. Compared to
J1559$+$4349, J1147$+$4818 is a less-likely gravitational arc system due
to the absence of evidence for two distinct optical and NIR objects, as in the previous case. Plus, radio
sources often show modest intrinsic curvature of core-jet components as seen in
Figure~\ref{fig:J1147}. We conclude that while J1559$+$4349 almost
certainly possesses
arcuate radio structure due to a foreground galaxy, J1147$+$4818 is a much
less-likely gravitationally-lensed background source.

\begin{figure}[htdp]
\includegraphics[scale=0.5]{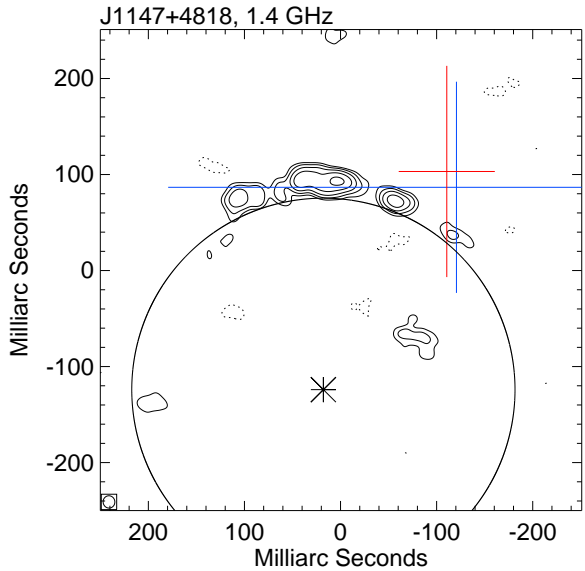}
\caption[The 1.4 GHz contour of J1147$+$4818.]{The 1.4 GHz VLBA contours of
  J1147$+$4818. The black arc is part of a circle with radius 0\farcs20
  centered at the position labeled with an asterisk. The blue and red crosses are the SDSS
$r$-band and NIR centroids respectively. (This plot is
  otherwise the same as the 1.4 GHz plot in the Appendix.)\label{fig:J1147}}
\end{figure}

\subsubsection{Other Objects of Interest}\label{sec:oth}

\textbf{J0834$+$1700} is one lobe in a double-lobe structure
25\arcsec\ apart (see Figure~\ref{fig:J0834}), similar to another
single-lobe object J0917$+$4725. The radio source is so well-aligned
with the optical$+$NIR object that no significant positional offset is
present, unlike in the case of J0917$+$4725. Absorption lines may be
detected if the radio source is in the background, although the large
amount of extended radio flux makes this a poor candidate for finding
intervening absorption.

\begin{figure}[htdp]
\includegraphics[scale=0.5]{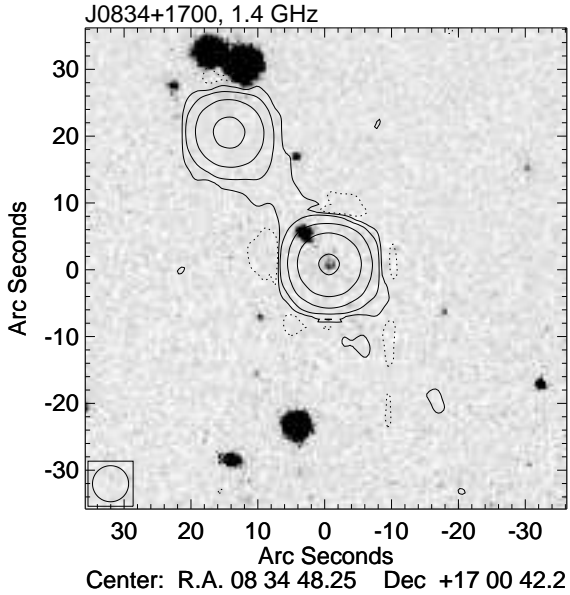}
\caption[The 1.4 GHz FIRST contours of the field of J0834$+$1700.]{The 1.4 GHz
  contour of the field of J0834$+$1700 overlaid with optical image in
  the SDSS $r$-band. The radio contour levels are $-$0.9, 0.9, 5.5, 33, 203,
  and 1239 mJy.\label{fig:J0834}}
\end{figure}

\textbf{J1048$+$3457} is a likely C~IV broad-absorption-lines (BAL) QSO at
$z=1.6$ \citep{willott:2003}. Without spectral information for the VLBA components,
we cannot distinguish a core-jet structure from a CSO type structure in this source. 
The multi-component sub-arcsec scale structure might indicate interactions with a 
dense ISM at an early stage of its radio activity
\citep{kunert-bajraszewska:2007, kunert-bajraszewska:2010}.

\textbf{J1527$+$3312} is unresolved in the 8.5 GHz VLA-A image, but
is very extended in the higher-resolution 1.4 GHz VLBA image for which
only 25\% of the total flux density is accounted. While further
observations are needed to detect the diffuse radio emission, this
``missing flux'' makes J1527$+$3312 a poor candidate for having H\,I absorption.

\section{REDSHIFTED H\,I 21cm SPECTROSCOPY}\label{sec:HIobs}

In this Section we report the results of an incomplete search for redshifted
H\,I 21cm and OH absorption at sub-GHz frequencies. Due to the presence of
substantial, site-specific RFI the most successful portion of our attempted radio spectroscopy is
for those 6 sources where accurate spectroscopic redshifts were available and where those 
redshifts move the H\,I 21cm absorption to a spectral region relatively clear of
RFI. Seven sources with good spectroscopic redshifts have significant RFI at the frequency of
the redshifted H\,I 21cm, precluding a sensitive search for absorption. 
Four of the 6 sources which have usable data at their spectroscopic redshifts
have clear H\,I detections with a 5th showing a tentative detection. Optical spectroscopy
supporting these observations was presented in Paper 1 although, since that publication,
four new spectroscopic redshifts were obtained at the {\it Gemini North Observatory}
using the multi-object spectrograph (GMOS) in a single-object mode. Based on the detection and
measurement of  strong, narrow emission lines in these spectra, these redshifts appear in 
Table~\ref{tab:line_log} with a reference of ``Gemini''. 

\begin{landscape}
\begin{deluxetable}{lllclcl}
\tabletypesize{\tiny}
\tablewidth{0pt}
  \tablecaption{Observing Log.\label{tab:line_log}} 
\tablehead{
    \colhead{Object} & \colhead{$z$}  & \colhead{$z$ ref}	&
    \colhead{Line(s)}	&	\colhead{Telescope/Receiver}
    &	\colhead{Date}
    &	\colhead{Line(s)} \\
   \colhead{} & \colhead{} & \colhead{} & \colhead{searched} & \colhead{} &
   \colhead{} & \colhead{detected} \\
   \colhead{(1)} & \colhead{(2)} & \colhead{(3)} & \colhead{(4)} & \colhead{(5)} &
   \colhead{(6)} & \colhead{(7)}} 

\startdata
J0003$-$1053  \tablenotemark{a}	&	*0.42$\pm$0.02 	&	\citep{yan:2013}	&	HI	&	GBT/PF2	&	28 \& 30-Sep-09	&		\\
J0003$-$1053  \tablenotemark{a}	&	1.474$\pm$0.001	&	\citep{yan:2013}	&	HI	&	GMRT	& 21-Jan-10 \& 19-Oct-13	&		\\
J0134$+$0003	&	0.879	&	NED	&	HI	&	Arecibo/800-MHz	&	09-Oct-09	&		\\
J0134$+$0003	&	0.879	&	NED	&	HI, OH	&	GBT/PF1 680--920MHz	&	07-Feb-13	&		\\
J0751$+$2716  \tablenotemark{b}	&	0.34937 \& 3.2	&	NED	&	HI	&	GBT/PF2	&	08-Oct-09	&		\\
J0805$+$1614	&	:0.632$\pm$0.002	&	\citep{yan:2013}	&	HI	&	GBT/PF1 680--920MHz	&	10-Feb-13	&		\\
J0805$+$1614	&	:0.632$\pm$0.002	&	\citep{yan:2013}	&	OH	&	GBT/PF2	&	08-Jun-09	&		\\
J0805$+$1614	&	:0.632$\pm$0.002	&	\citep{yan:2013}	&	OH	&	GBT/PF2	&	13-Oct-09	&		\\
J0824$+$5413	&	0.6385$\pm$0.0003	&	\citep{yan:2013}	&	HI	&	GBT/PF1 680--920MHz	&	21-May-09	&		\\
J0839$+$2403	&	*0.88$\pm$0.19 	&		&	HI	&	GBT/PF1 680--920MHz	&	22-May-09	&		\\
J0839$+$2403	&	*0.88$\pm$0.19 	&		&	HI	&	Arecibo/800-MHz	&	14-Mar-11	&		\\
J0839$+$2403	&	*0.88$\pm$0.19 	&		&	HI	&	Arecibo/800-MHz	&	15-Mar-11	&		\\
J0901$+$0304	&	0.2872$\pm$0.0001	&	\citep{yan:2013}	&	HI	&	GBT/PF2	&	09-Oct-09	&	HI	\\
J0901$+$0304	&	0.2872$\pm$0.0001	&	\citep{yan:2013}	&	HI	&	VLA	&	28-Dec-09	&	HI	\\
J0901$+$0304	&	0.2872$\pm$0.0001	&	\citep{yan:2013}	&	HI	&	GMRT	&	28-Jan-10	&	HI	\\
J0910$+$2419	&	*0.70$\pm$0.06 	&		&	HI	&	GBT/PF1 680--920MHz	&	21-May-09	&		\\
J0920$+$2714	&	0.2064$\pm$0.0002	&	\citep{yan:2013}	&	HI	&	GBT/L-band	&	10-Sep-06	&	HI	\\
J0939$+$0304	&	*0.47$\pm$0.03 	&		&	HI	&	GBT/PF2	&	08-Jun-09	&		\\
J0951$+$1154	&	*0.62$\pm$0.26 	&		&	HI	&	GBT/PF1 680--920MHz	&	01-Jul-09	&		\\
J1008$+$2401	&	*0.88$\pm$0.22 	&		&	HI	&	GBT/PF1 680--920MHz	&	01-Jul-09	&		\\
J1023$+$0424	&	*0.63$\pm$0.15 	&		&	HI	&	GBT/PF1 680--920MHz	&	22-May-09	&		\\
J1023$+$0424	&	*0.63$\pm$0.15 	&		&	HI	&	Arecibo/800-MHz	&	15-Mar-11	&		\\
J1033$+$3935	&	1.095$\pm$0.002	&	NED	&	HI	&	GBT/PF1 510--690 MHz	&	03-Nov-09	&		\\
J1043$+$0537	&	*0.87$\pm$0.29 	&		&	HI	&	GBT/PF1 680--920MHz	&	22-May-09	&		\\
J1043$+$0537	&	*0.87$\pm$0.29 	&		&	HI	&	Arecibo/800-MHz	&	14-Mar-11	&		\\
J1129$+$5638	&	0.892$\pm$0.002	& Gemini	&	HI	&	GBT/PF1 680--920MHz	&	15-Apr-12	&	HI	\\
J1238$+$0845	&	*0.83$\pm$0.14 	&		&	HI	&	GBT/PF1 680--920MHz	&	22-May-09	&		\\
J1238$+$0845	&	*0.83$\pm$0.14 	&		&	HI	&	Arecibo/800-MHz	&	14-Mar-11	&		\\
J1238$+$0845	&	*0.83$\pm$0.14 	&		&	HI	&	Arecibo/800-MHz	&	15-Mar-11	&		\\
J1315$+$0222	&	*0.55$\pm$0.48 	&		&	HI	&	GBT/PF1 680--920MHz	&	01-Jul-09	&		\\
J1357$+$0046	&	*1.00$\pm$0.17 	&		&	HI	&	GBT/PF1 680--920MHz	&	22-May-09	&	HI	\\
J1357$+$0046	&	*1.00$\pm$0.17 	&		&	HI	&	GBT/PF1 680--920MHz	&	01-Jul-09	&	HI	\\
J1357$+$0046	&	*1.00$\pm$0.17 	&		&	OH	&	GBT/PF2	&	08-Jun-09	&		\\
J1410$+$4850	&	0.592$\pm$0.001	& Gemini	&	HI	&	GBT/PF1 680--920MHz	&	08-Feb-13	&		\\
J1414$+$4554    &       :0.186$\pm$0.002        &       NED     &       HI      &       GBT/L-band      &       12-Mar-09       &               \\
J1414$+$4554	&	:0.186$\pm$0.002	&	NED	&	HI	&	GMRT/L-band    	&	29 Jan-10	&		\\
J1421$-$0246	&	*0.53$\pm$0.02 	&		&	HI	&	GBT/PF2	&	05-Jun-09	&		\\
J1424$+$1852	&	*0.53$\pm$0.02 	&		&	HI	&	GBT/PF1 680--920MHz	&	22-May-09	&		\\
J1504$+$5438	&	0.62244$\pm$0.00013	& SDSS	&	HI	&	GBT/PF1 680--920MHz	&	21-May-09	&		\\
J1604$+$6050	&	0.559$\pm$0.001	& Gemini	&	HI	&	GBT/PF1 680--920MHz	&	08-Feb-13	&	HI:	\\
J1616$+$2647	&	0.755$\pm$0.001	& Gemini	&	HI	&	GBT/PF1 680--920MHz	&	15-Apr-12	&	HI	\\
J2203$-$0021	&	0.729$\pm$0.001	&	\citep{yan:2013}	&	HI	&	Arecibo/800-MHz	&	09-Oct-09	&		\\
J2203$-$0021	&	0.729$\pm$0.001	&	\citep{yan:2013}	&	OH	&	GBT/PF2	&	01-Oct-09	&		\\
\enddata

\pagebreak

\tablecomments{Columns list: (1) Object name in IAU convention; (2) the redshift used for the line search, either (a) a spectroscopic 
redshift obtained by us or found in the literature, or (b) a photometric redshift from SDSS DR6, (``*''). Uncertain redshifts 
are indicated by a colon; (3) redshift reference; ``Gemini'' indicates a new redshift obtained by us at Gemini North and 
reported here for first time; (4) line(s) searched; (5) telescope/receiver used; (6) observing date; (7) line(s) detected.} 

\noindent \tablenotetext{a} {J0003$-$1053 was observed at the foreground photometric redshift at the GBT (first line entry above) and at the background
spectroscopic redshift at GMRT (second line entry).} 

\noindent \tablenotetext{b} {J0715$+$2716 was observed only at its foreground photometric redshft at GBT, not
at its $z=3.2$ intrinsic AGN redshift.} 

\end{deluxetable}

\end{landscape}

But because we lacked optical and/or NIR spectroscopy for most of the sources 
in this sample, we attempted radio spectroscopy for 14 additional sources at their 
photometric redshifts making only one detection. While these sources were chosen using the
same optical/NIR photometric classifications (i.e., pure ``G''- and ``G$+$Q'' types preferred) and radio 
continuum observations (i.e., CSOs and GPS preferred) as guides for which are the most
likely to have observable H\,I and molecular absorption, the rather coarse accuracy of the
photo-$z$s ($\sim \pm$ 0.1-0.2, which translates into a frequency uncertainty of $\pm$35-70 MHz at $z$=1)
made these searches mostly unsuccessful.
Given the strength and ubiquity of the RFI over these very broad observing bands, it is not
possible even to establish viable detection limits for the 14 sources observed with only photometric redshifts.
 
\subsection{OBSERVATIONS AND DATA REDUCTION}\label{sec:HIobsred}

Observations were carried out at several radio telescopes, each using specific receivers
and observing modes to access the location in frequency of the redshifted H\,I 21cm line. 
The observing log shown in Table~\ref{tab:line_log} indicates the telescope/receiver combinations
used for these various observations. 

At the Robert C. Byrd Green Bank Telescope (GBT) of NRAO observations were conducted with the
two prime focus receivers PF1 ($\sim$ 510--690 or $\sim$ 680--920 MHz)
and PF2 ($\sim$ 901--1230 MHz) at the appropriate frequency bands
for the redshifted H\,I hyper-fine transition. The spectrometer was set to the
narrow-bandwidth, high resolution mode with dual linear polarizations
and multiple spectral windows taking spectra simultaneously. Usually 4
spectral windows were used with each having a 50 MHz bandwidth and
4096 channels. Position-switched scans were performed on the target
and sky reference positions with 5 min durations and 2 sec integration time. 
For a typical object of
0.8 Jy at 900 MHz, 5 position-switched pairs were carried out to
detect a $\tau = 0.01$ absorption line at 5$\sigma$ significance. The
total observing time for each target ranged from $\sim$ 20 minutes to
$\sim$ 2 hours due to the large flux density range of our targets. Flux densities are
calibrated using the noise diode calibrator, which is turned on half of
the time for each integration. The uncertainty of this ``standard''
flux calibration is $\pm$ 10\%. Nonetheless, only relative fluxes
(optical depths) are of interest to this work. The data were reduced
using GBTIDL, an interactive package for the reduction and analysis of
spectral line data taken with GBT. Each integration and polarization
was calibrated separately and then averaged together. However, the
RFI, while either intermittent or persistent, is always strong and
variable. Therefore, a median spectrum of all integrations for each
polarization is calculated to eliminate sporadic RFIs not present in
most integrations. The noise of a median spectrum is $\pi/2$ larger
than that of a mean spectrum if the frequency range is RFI-free or only
slightly contaminated. If a possible absorption line is identified,
each integration is carefully inspected to flag those with the strongest RFI. A
final spectrum is obtained by averaging all the remaining integrations.

At the Arecibo Observatory (AO), 6 objects were observed with the 800
MHz receiver ($\sim$ 700--800 MHz) in dual polarization mode. 
J0134$+$0003 was searched for H I at its spectroscopic redshift, but an upper limit was obtained
with less sensitivity compared to our GBT observations.
Additionally, four objects were searched at their photometric redshifts. These four objects were
also searched at GBT using the 680-920 MHz receiver.
No detection or RFI-clear upper limits were obtained.

At the AO the Mock Spectrometer contains 14 independent boxes, which were configured to
have 8192 channels and a bandwidth of 12 MHz with some overlapping to
cover the entire 100 MHz frequency range of the
receiver. Double position-switched scans were performed to cancel
residual standing waves, which resulted in an observing sequence containing
four scans: on-target, off-target, on-reference, and
off-reference. A reference source was observed close to each target
source in space and time. Each scan was 3-5 minutes long, and the total
observing time was 1-2 hours for each source. The data were reduced by
IDL routines provided by AO. The reference sources are also used as
flux calibrators. The AO observations took advantage of a period
which was relatively RFI-free just prior to the inception of digital
television in Puerto Rico. This observing window has now been closed.

Observations to confirm the GBT detection of the redshifted HI 21 cm 
\ion{H}{1} absorption lines toward J0901$+$0304 were
carried out with the Very Large Array (VLA) of NRAO on 28 December 2009
The array was in the D-configuration, and the total observing time was 1 hr. These
observations used 21 antennas that were converted to the
Expanded VLA (EVLA) standards. The data were correlated using the old
VLA correlator. The aliasing that was known to affect the lower 0.5
MHz of the bandwidth for EVLA data which were correlated with the old
VLA correlator was avoided by properly placing the frequencies of the
two redshifted HI 21 cm lines within a more restrictive 3.125 MHz
band. The calibrator source 3C147 (J0542$+$4951) was used to set the
absolute flux density scale, while the compact source J0925+0019 was
used as the complex gain calibrator. The editing, calibration, and
imaging of the VLA data were carried out using {\sc AIPS}. The HI line-free
channels were split, and the continuum emission from J0901$+$0304 was
self-calibrated in both phase and amplitude in a succession of
iterative cycles. The final phase and amplitude solutions were then
applied on the spectral-line data set. The continuum emission was
subtracted in the UV-domain, and a Stokes I image cube was made with a
synthesized beam width of $97\arcsec \times 67\arcsec$ (PA=-26.60).

The Giant Metrewave Radio Telescope (GMRT) was also used to confirm the detection of
\ion{H}{1} 21cm absorption at $z = 0.2872$ in J0901+2734, as well as to search
for absorption at $z = 0.186$ towards J1414+4554 and $z = 1.474$ towards J0003-1053.
The initial GMRT observations of the three sources were carried out in January 2010,
using $23-25$ working antennas, the GMRT hardware correlator as the backend, two
circular polarizations, and the GMRT L-band (J0901+2734 and J1414+4554) and 610-MHz
band (J0003-1053) receivers. A bandwidth of 4 MHz was used for all observations, sub-divided
into 128 channels (L-band) or 256 channels (610-MHz band); this yielded a velocity resolution
of $\approx 8-8.5$~km~s$^{-1}$ and a total velocity coverage of $\approx 1100$~km~s$^{-1}$ (L-band)
and $\approx 2100$~km~s$^{-1}$ (610-MHz band). Observations of one of the standard calibrators
3C48, 3C147 or 3C286 were used to calibrate the flux density scale and the system passband,
while nearby compact sources were used for secondary calibration.

The initial GMRT observations of J0003-1053 resulted in a tentative ($\approx 5\sigma$)
detection of a weak absorption feature. We hence re-observed this source with the GMRT
in October 2013, using the GMRT Software Backend, two circular polarizations, and a bandwidth
of 4.167~MHz sub-divided into 512 channels, yielding a velocity resolution of
$\approx 4.3$~km~s$^{-1}$ and a velocity coverage of $\approx 2175$~km~s$^{-1}$. No detection was made.

All GMRT data were analysed in ``classic'' AIPS, using standard data editing and calibration
procedures to obtain the antenna-based complex gains by interpolating between the values
derived from observations of the secondary calibrator. An iterative self-calibration procedure
was then used to determine more accurate gains on the target field, and finally, to obtain a
continuum image of the field. 3-D imaging techniques were used for all sources, sub-dividing
the field into 13 (L-band) or 19 (610-MHz band) facets to correct for the non-coplanarity
of the GMRT. The target flux densities were measured via a single Gaussian fit to the central region of
the final image with the AIPS task {\sc jmfit}. We obtained flux densities of $(402.39 \pm 0.92)$~mJy
(J0901+2734, at $\approx 1104.2$~MHz), $(385.09 \pm 0.78) $~mJy (J1414+4554, at $\approx 1198.5$~MHz),
and  $\approx (753.0 \pm 1.6)$~mJy (J0003-1053, at $\approx 574.2$~MHz). The final continuum image
was then subtracted out from the calibrated visibility data, using the AIPS task {\sc uvsub}, after
which a first-order polynomial was fit to the visibility spectra on each interferometer baseline and
subtracted. For each source, the residual UV data were then shifted to the barycentric frame and
imaged in all channels to obtain the final spectral cube. A cut through this cube at the target
location yielded the \ion{H}{1} 21cm spectrum. The root mean square (RMS) noise values were measured
(in off-line regions of the spectra) to be $\approx 0.0012$ per 17.0~km~s$^{-1}$ channel (J0901+2734),
$\approx 0.0028$ per 15.6~km~s$^{-1}$ channel (J1414+4554) and $\approx 0.0034$ per 16.3~km~s$^{-1}$
channel (J0003-1053, in 2010) and $\approx 0.0030$ per 8.5~km~s$^{-1}$ (J0003-1053, in 2013), all
in units of optical depth. We note that the weak feature seen in the original spectrum of J0003-1053
was not detected in the deeper spectrum of 2013, and is hence likely to have arisen from radio frequency
interference (RFI).

At the GBT, twelve objects were searched at their
spectroscopic redshift, which led to 5 detections (the absorption
in J0901$+$0304 was contaminated by RFI, but
confirmed later by VLA and GMRT observations) and one
tentative detection (J1357$+$0046). Seven are in RFI-rich regions, precluding
a sensitive search at $z_{spec}$. A 3$\sigma$ upper limit of $\tau <$ 0.01 (N$_{HI} <$ 2.0 $\times$ 10$^{20}$ cm$^{-2}$)
was obtained for J0134$+$0003 from an observation free of RFI at the spectroscopic redshift 
of the source. These GBT observations targeted all sources in this sample 
with spectroscopic redshifts (see Table 1 in Paper \Rmnum{1})
which placed the redshifted H\,I 21cm line in an observable band relatively free of
intense RFI. Since the publication of Paper \Rmnum{1} three new spectroscopic redshifts were
obtained for sources in this program (see below). The targets observed at GBT were mostly those
with either point-like or CSO radio continuum emission, pure galaxy SEDs (``G-type''; see  Paper \Rmnum{1})
and GPS SEDs in the radio. The detection rate for ``G-type''/GPS/CSO type sources is at least 67\% (6 of 9)
with only one firm non-detection (J0134$+$0003) completely clear of RFI at its spectroscopic redshift.
Therefore, the actual detection rate in this class of source could be much higher.

Fourteen objects were searched at GBT with wide bandwidth for 21 cm absorption lines at their photometric redshifts.
These observations led to only one detection, in the CSO, ``G$+$Q''-type source J1357+0046. Interestingly, the 
detected H\,I absorption lines were found at $z=$ 0.7971, 0.7962, quite far away from the SDSS 
$z_{phot}$=0.57$\pm$0.17 for the optical host galaxy.
While the selection criteria for observing these sources were similar to the group with spectroscopic redshifts, a much
smaller detection rate was obtained: 1/14 = 7\% for sources with $z_{phot}$ only versus 5/13 = 38\% for sources
with accurate $z_{spec}$ values. And since only 6 of the $z_{spec}$ sources were in regions clear of RFI, the
detection rate is actually 5/6 = 83\% for our sample. We infer that redshifted H\,I likely is present in most 
of the 14 sources we observed, but its presence was hidden from detection by RFI. 

\begin{landscape}
\tabletypesize{\tiny}

\begin{deluxetable}{lcccrrrrrr}
\tablewidth{22cm}
  \tablecaption{\ion{H}{1}\ Detections. \label{tab:line_h1}} 
\setlength{\tabcolsep}{0.05in}
\tablehead{
    \colhead{Object} & \colhead{Comp.} & \colhead{Resolution} &
    \colhead{$rms$} & \colhead{Center} & \colhead{FWHM} &
    \colhead{$\tau_p$} & \colhead{$z$} & \colhead{$\int\tau\mathrm{d}v$} & \colhead{$N_{\mathrm{H\Rmnum{1}}}$}\\
    \colhead{} & \colhead{ID} & \colhead{(kHz)} &
    \colhead{(mJy)} & \colhead{(MHz)} & \colhead{(km
      s$^{-1}$)} & \colhead{} & \colhead{} & \colhead{(km s$^{-1}$)} &
    \colhead{($10^{20} \mathrm{cm}^{-2}$)} \\
    \colhead{(1)} & \colhead{(2)} & \colhead{(3)} & \colhead{(4)} &
    \colhead{(5)} & \colhead{(6)} & \colhead{(7)} & \colhead{(8)} &
    \colhead{(9)} & \colhead{(10)} }

\startdata
J0901$+$0304	\tablenotemark{a}	&	A	&	24	&	1.5	&	1102.277	$(	2	$)	&	46	$(	2	$)	&	-0.0775	$(	23	$)	&	0.288611	$(	3	$)	&	3.79	$\pm$	0.17	&	6.82	$\pm$	0.31	\\
		&	B	&		&		&					&					&					&					&				&				\\
		&	Int\tablenotemark{b}	&		&		&					&					&					&					&	3.82	$\pm$	0.10	&	6.88	$\pm$	0.18	\\
J0901$+$0304	\tablenotemark{c}	&	A	&	65	&	2.0	&	1102.267	$(	8	$)	&	54	$(	5	$)	&	-0.0560	$(	47	$)	&	0.288623	$(	10	$)	&	4.70	$\pm$	0.22	&	8.46	$\pm$	0.39	\\
		&	B	&		&		&	1103.486	$(	15	$)	&	92	$(	10	$)	&	-0.0387	$(	36	$)	&	0.287199	$(	18	$)	&	5.52	$\pm$	0.31	&	9.94	$\pm$	0.56	\\
		&	Int	&		&		&					&					&					&					&	10.22	$\pm$	0.38	&	18.40	$\pm$	0.68	\\
J0901$+$0304	\tablenotemark{d}	&	A	&	48	&	2.0	&	1102.341	$(	6	$)	&	34	$(	4	$)	&	-0.0608	$(	62	$)	&	0.288536	$(	7	$)	&	3.04	$\pm$	0.17	&	5.47	$\pm$	0.30	\\
		&	B	&		&		&	1103.625	$(	13	$)	&	68	$(	9	$)	&	-0.0401	$(	44	$)	&	0.287036	$(	15	$)	&	4.03	$\pm$	0.22	&	7.26	$\pm$	0.39	\\
		&	Int	&		&		&					&					&					&					&	7.07	$\pm$	0.27	&	12.73	$\pm$	0.49	\\
J0920$+$2714		&	A	&	24	&	1.8	&	1177.131	$(	2	$)	&	13	$(	1	$)	&	-0.0819	$(	107	$)	&	0.206668	$(	2	$)	&	1.10	$\pm$	0.19	&	1.97	$\pm$	0.34	\\
		&	B	&	24	&	1.8	&	1177.072	$(	13	$)	&	33	$(	5	$)	&	-0.0405	$(	47	$)	&	0.206728	$(	13	$)	&	1.42	$\pm$	0.26	&	2.56	$\pm$	0.47	\\
		&	Int	&		&		&					&					&					&					&	2.61	$\pm$	0.12	&	4.70	$\pm$	0.21	\\
J1129$+$5638		&	A	&	12	&	5.0	&	750.764	$(	15	$)	&	104	$(	12	$)	&	-0.0276	$(	68	$)	&	0.891947	$(	38	$)	&	3.07	$\pm$	0.83	&	5.52	$\pm$	1.49	\\
		&	B	&	12	&	5.0	&	751.033	$(	193	$)	&	170	$(	109	$)	&	-0.0063	$(	24	$)	&	0.891270	$(	487	$)	&	1.13	$\pm$	0.85	&	2.04	$\pm$	1.53	\\
		&	Gfit\tablenotemark{e}	&	12	&	5.0	&	750.788	$(	6	$)	&	127	$(	6	$)	&	-0.0285	$(	11	$)	&	0.891886	$(	15	$)	&	3.85	$\pm$	0.22	&	6.93	$\pm$	0.40	\\
		&	Int	&		&		&					&					&					&					&	4.16	$\pm$	0.14	&	7.49	$\pm$	0.24	\\
J1357$+$0046	\tablenotemark{f}	&	A	&	24	&	2.8	&	790.764	$(	5	$)	&	80	$(	5	$)	&	-0.0130	$(	6	$)	&	0.796246	$(	11	$)	&	1.12	$\pm$	0.08	&	2.01	$\pm$	0.15	\\
		&	B	&	24	&	2.8	&	790.396	$(	7	$)	&	71	$(	6	$)	&	-0.0088	$(	7	$)	&	0.797081	$(	16	$)	&	0.66	$\pm$	0.08	&	1.19	$\pm$	0.14	\\
		&	Int	&		&		&					&					&					&					&	1.82	$\pm$	0.04	&	3.28	$\pm$	0.07	\\
J1357$+$0046	\tablenotemark{g}	&	A	&	24	&	3.7	&	790.767	$(	5	$)	&	90	$(	4	$)	&	-0.0140	$(	6	$)	&	0.796239	$(	10	$)	&	1.34	$\pm$	0.08	&	2.41	$\pm$	0.15	\\
		&	B	&	24	&	3.7	&	790.410	$(	8	$)	&	59	$(	7	$)	&	-0.0068	$(	7	$)	&	0.797049	$(	18	$)	&	0.42	$\pm$	0.07	&	0.76	$\pm$	0.12	\\
		&	Int	&		&		&					&					&					&					&	1.74	$\pm$	0.05	&	3.14	$\pm$	0.09	\\
J1604$+$6050		&	Gfit	&	6	&	3.0	&	911.025	$(	2	$)	&	8	$(	2	$)	&	-0.0142	$(	25	$)	&	0.559130	$(	4	$)	&	0.12	$\pm$	0.03	&	0.22	$\pm$	0.06	\\
		&	Int	&		&		&					&					&					&					&	0.14	$\pm$	0.02	&	0.25	$\pm$	0.04	\\
J1616$+$2647		&	Gfit	&	49	&	3.4	&	809.133	$(	34	$)	&	447	$(	30	$)	&	-0.0084	$(	5	$)	&	0.755467	$(	73	$)	&	3.98	$\pm$	0.35	&	7.16	$\pm$	0.63	\\
		&	Int	&		&		&					&					&					&					&	4.00	$\pm$	0.16	&	7.21	$\pm$	0.28	\\
\enddata

\tablecomments{Columns list: (1) object name in IAU convention; (2) ID
  of the absorbing component, where ``Gfit'' means there is only one
  component and ``Int'' means an integrated value for all components; (3)
  spectral resolution of the spectrum; (4) $rms$ measured from
  continuum-subtracted background; (5) (6) (7) (8) Gaussian fit
  results of the line center in frequency, FWHM in rest-frame
  velocity, peak of optical depth, and redshift of line center; (9)
  integrated optical depth $\int\tau\mathrm{d}v$ either from a Gaussian
  fit or in the case of ``Int'', directly integrated from the spectrum non-parametrically; 
  (10) the column density of the absorbing component assuming an 100\% covering
  factor and a spin temperature of 100 K.}  
\tablenotetext{a}{GBT
  observation.}  \tablenotetext{b}{``A'' component only.}
\tablenotetext{c}{GMRT observation.}  \tablenotetext{d}{VLA
  observation.}  \tablenotetext{e}{Gaussian fit of single compoent.}
\tablenotetext{f}{GBT observation on May 22, 2009.}
\tablenotetext{g}{GBT observation on July 01, 2009.}

\end{deluxetable}

\end{landscape}


Details of the H\,I detections made in this program are shown in Table~\ref{tab:line_h1} including the
optical depth $\tau$ and inferred column density \ion{H}{1} assuming a spin temperature $T_s =$ 100 K
and a 100\% fraction of the source covered by the absorbing gas (i.e., $f$ = 1).
By integrating the optical depth $\tau$ over the velocity width (v) of the H\,I absorption line, 
we find that the \ion{H}{1} column density of the absorbing gas is:

\begin{equation}
  N_\mathrm{HI} = 1.8\times10^{18} \mathrm{cm^{-2}} \frac{T_s}{f} \int \tau \mathrm{d} v,
\end{equation}


Assuming this same formalism stated above our observations made clear non-detections
only when: (1) an accurate spectroscopic redshift was available and (2)
observations of the redshifted \ion{H}{1} 21cm line frequency were made in 
an RFI clear region at the telescope of choice. Sources with only photometric redshift estimates have redshift errors 
so large that RFI covers a significant percentage of the range of possible redshifts. This renders our
observations incapable of determining non-detections without doubt in most cases. So only three non-detections are firm. 
Details of the H\,I non-detections made in this program are shown in Table~\ref{tab:line_h1non}.

In summary, of the 15 sources with spectroscopic redshifts observed at redshifted \ion{H}{1}, only 5 detections (1 tentative) 
were made, 3 observations netted clear non-detections and 7 objects were indeterminant due to RFI. Five of our six total 
\ion{H}{1} detections (one detection has only a photometric redshift) are CSOs while only one (J0134$+$003) of our 
three non-detections is a CSO. Thus, the detection rate of CSOs in our sample is very high ($83\pm17$\%) although obviously our sample 
size is quite small.

\begin{deluxetable}{lrrrrc}
\tablewidth{0pt}
  \tablecaption{\ion{H}{1}\ Upper Limits.\label{tab:line_h1non}} \tablehead{
    \colhead{Object} & \colhead{Search Freq.} &
    \colhead{$rms$}  & \colhead{$\tau_{3\sigma}$}  & \colhead{$N_{\mathrm{H\Rmnum{1}}}$} &  \\
    \colhead{} & \colhead{(MHz)} & \colhead{(mJy)} & \colhead{} &
    \colhead{($10^{20} \mathrm{cm}^{-2}$)} & \colhead{} }

\startdata 
J0003$-$1053  \tablenotemark{a}         &       574.13  &       2.6     &       $<      0.010   $  	&       $<      0.80    $       &              \\
J0134$+$0003  \tablenotemark{b}  	&	755.94	&	11.6	&	$<	0.030	$	&	$<	1.8	$	&		\\
J0134$+$0003  \tablenotemark{c} 	&	755.94	&	5.4	&	$<	0.010	$	&	$<	0.61	$	&		\\
J1414$+$4554    \tablenotemark{a}       &      1197.64  &        1.5    &       $<      0.0084    $     &       $<     0.65     $       &               \\
\enddata

\tablecomments{The RMS noise and the $3\sigma$ optical depth limits are computed at velocity resolutions of 16.3~km~s$^{-1}$ (J0003$-$1053),
10~km~s$^{-1}$ (J0134$+$0003), and 15.6~km~s$^{-1}$ (J1414$+$4554). The $N_{\mathrm{H\Rmnum{1}}}$ limits assume a spin temperature 
of 100~K, a covering factor of unity, and a FWHM of 100~km~s$^{-1}$ for the absorption line. These calculations use the RMS noise after 
smoothing the spectra to a velocity resolution of 100~km~s$^{-1}$.}
\tablenotetext{a}{GMRT observation.}  
\tablenotetext{b}{AO observation.}
\tablenotetext{c}{GBT observation.}
\end{deluxetable}

No OH main or satellite absorption line has been detected in any 
source in our sample although only 6 sources have usable data for this
purpose. We obtained good 3$\sigma$ upper limits for column density of the 1667 MHz main and satellite
lines ($N_{OH} <$ 1.6 and 16 $\times$ 10$^{14}$ cm$^{-2}$ respectively) in J2203$-$0021 and limits
on one of the two satellite lines in J0751$+$2716 and J1357$+$0046 ($<$ 20 and 6 $\times$ 10$^{14}$ cm$^{-2}$ respectively).
These limits assume an excitation temperature of 100 K, a covering factor of 100\% and a rest-frame velocity 
width of 100 km s$^{-1}$. Three other sources (J0805+1614, J1129+5638 \& J1357+0046) have the main or satellite lines 
observable in the band but have RFI present which precludes an accurate upper limit. Among these six, only 
J1129+5638 and J1357+0046 have detected redshifted H\,I 21cm absorption.

\subsection{Individual Sources}

\subsubsection{J0134$+$0003}

J0134$+$0003 is a GPS source and a VLBA calibrator which has an optical counterpart with a ``G''-type SED best-fit by an
Sa galaxy template \citep{mannucci:2001} and spectroscopic $z$=0.8790. This redshift places the potential H\,I line in a clear spectral
region but no detection was made either at GBT or AO. At a search frequency of 755.94 MHz our better GBT observations set
a 3$\sigma$ upper limit of $\tau <$ 0.010 and $N_{H\,I} <$ 2.0 $\times$ 10$^{20}$ cm$^{-2}$ (assuming a T$_s$=100K and a 100\% 
covering factor; this limit is less than all of our definite H\,I detections in the same sample. Evidently, not all 
compact/GPS/``G''-type sources have detectable H\,I absorption. \citet{curran:2006} observed this source for \ion{H}{1} absorption
at Westerbork and also reported a non-detection although the predicted redshifted 21cm location was at the edge of intense RFI.  

\subsubsection{J0901$+$0304}

See Figure~\ref{fig:0901} for the 21 cm spectra of J0901$+$0304. We
first observed this object using the GBT and found two tentative
absorption features. One feature is in a small RFI-free window while
the other is badly contaminated by RFI at Green Bank. We later confirmed both
features using the VLA and GMRT where the RFI was not as difficult an issue.

J0901$+$0304 has a pure Sc galaxy (``G''-type) optical$+$NIR SED 
\citep [based on template SEDs from][]{mannucci:2001}
and its optical and NIR images suggest that it may be an interacting system with some
diffuse structure extending to the NE of the main galaxy. The optical spectrum 
obtained by us at the APO 3.5m telescope includes moderate strength narrow emission 
lines whose line ratios suggest a LINER or weak Seyfert ionization source \citep{kewley:2006}.
J0901$+$0304 is a GPS source which is not resolved in our VLBA map with a resolution of $\sim$ 10 mas at
1.4 GHz, which means it could either be a very small-scale CSO or a core-jet
object. The GPS SED suggests that this source is likely a very small CSO.

Two \ion{H}{1} absorption lines are detected, one coincident with the optical
redshift of $z$=0.2872$\pm$0.0001, the other redshifted by a velocity of 328 km s$^{-1}$.
Since the broad line has a velocity consistent with the nucleus, we identify this absorption as 
disk gas, which leaves the narrow component as a likely infalling high-velocity cloud (HVC).
This broad-narrow absorption-line pairing is quite
similar to the H\,I 21cm absorption lines found by \citet{keeney:2011} at low-redshift
($z$=0.018) in PKS 1327$-$206. In that case the broad, shallow HI absorber is coincident in velocity with
the emission-line H\,II region gas in a spiral arm in the interacting galaxy ESO 1327$-$2041 and the narrower H\,I absorption is 
HVC gas at a velocity of $\Delta v \approx$ 250 km s$^{-1}$ with respect to the disk. The low-$z$ of ESO 1327$-$2041 allowed
a definitive determination of which H\,I absorber is disk gas and which is an HVC. With more examples
of broad and narrow H\,I absorption we may be able to use the line width as an indicator for where in the galaxy
the absorbing cloud(s) is(are) located. However, similar, very broad H\,I absorption also has been seen to 
originate from a circumnuclear disk as in 4C 31.04 \citep{mirabel:1990} and 3C 190
\citep{ishwara-chandra:2003}. 

There is also a very broad absorption blueward of the narrow absorption line in these spectra
which is almost certainly real. It is a cumulative 10$\sigma$ depression below the continuum in
the GMRT spectrum (see Figure~\ref{fig:0901} top spectrum) with a velocity width of $\sim$ 
140 km s$^{-1}$. A similar feature is seen in the GBT spectrum but it is in a frequency region possibly
affected by RFI. While this broad absorption is not obviously seen in the VLA spectrum, the continuum level to lower
frequencies is not well-defined and so it may be present in those data as well. Even though the VLA receivers have lower 
system temperatures than at the GMRT, the GMRT has $\sim$ 25\% more aperture (29 antennae to only 21 VLA antennae working for
our observation). Since these two differences largely cancel, the $>$ 4 times longer on-source integration time at GMRT makes 
this observation 2--3 times more sensitive. We conclude that the broad absorption is real and is most easily visible as the excess absorption
below the dotted line-fit in the GMRT spectrum. Given its redshift relative to the
disk absorption and to the emission line redshift for the nucleus, this feature also appears to be infalling gas.  
\begin{figure}[hdtp]
\includegraphics[width=12cm]{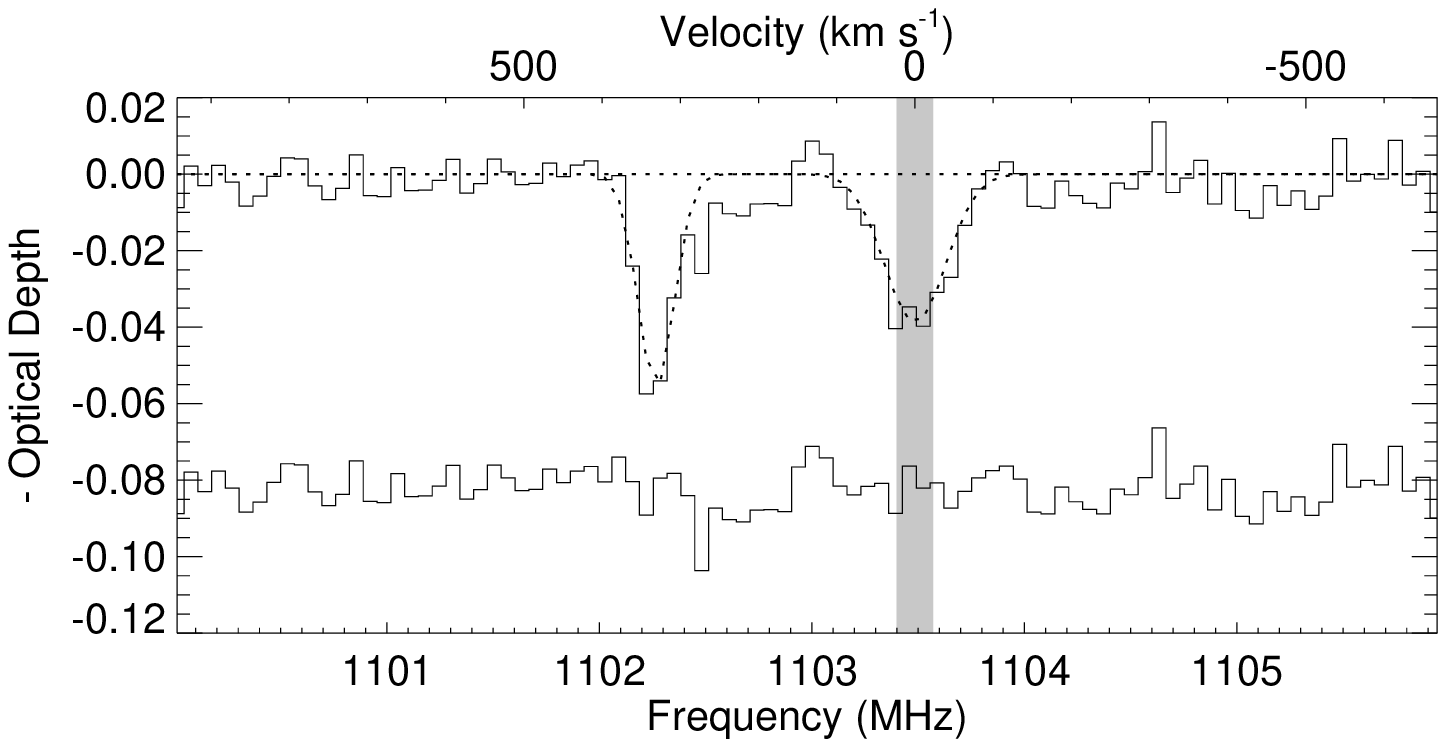}
\includegraphics[width=12cm]{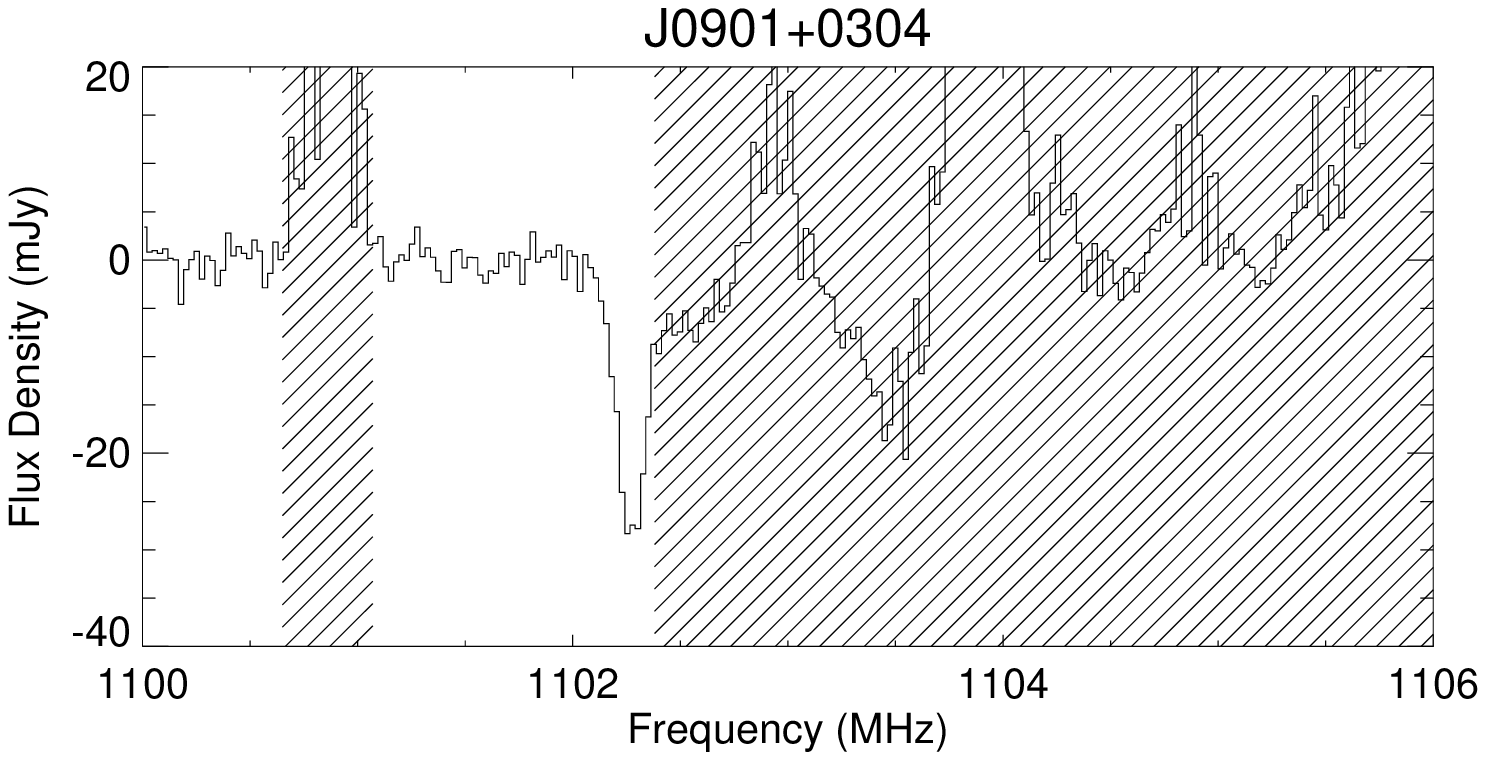}
\includegraphics[width=12cm]{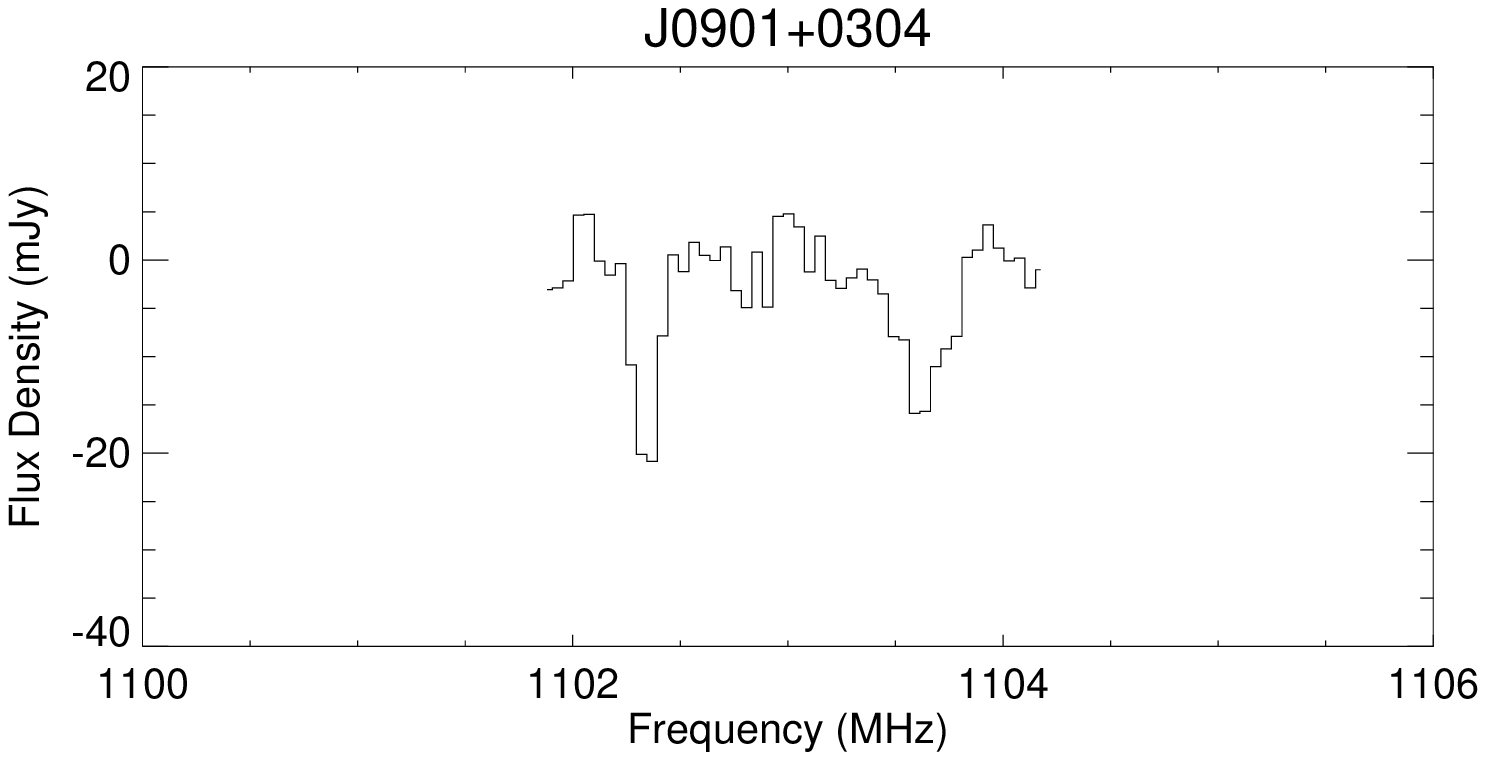}
\caption[The \ion{H}{1}\ 21 cm spectrum in J0901$+$0304.]{The
  \ion{H}{1}\ 21 cm spectrum in J0901$+$0304 as observed by GMRT (top),
  GBT (middle), and VLA (bottom). On the GMRT plot, Gaussian fits to
  the two absorption components are shown as dotted lines. Residuals
  are displayed at the bottom of this plot offset in the ``y''-coordinate for clarity. 
Zero velocity is set to the redshift of the optical galaxy at $z=0.2872\pm0.0001$, with
  redshift uncertainty shown in gray shade. The hatched regions in the GBT spectrum are
  regions contaminated by RFI.  \label{fig:0901}}
\end{figure}

\subsubsection{J0920$+$2714}

See Figure~\ref{fig:0920} for the 21 cm spectrum from the GBT where we
observed J0920$+$2714 in 2006 through an exploratory program. We later
obtained an optical redshift in 2009 and found that the broad 21 cm
absorption feature is slightly redshifted ($\sim \Delta v \approx$ 65 km s$^{-1}$)
with respect to the galaxy nucleus.

This is an ``intervening'' H\,I system in which a double-lobe, steep spectrum radio source is in the
background (unknown redshift) and is well-offset on the sky from the center of an S0 galaxy at $z=0.2064\pm0.0002$ in the foreground. 
The S0 classification is based both upon a visual inspection of the SDSS and APO 3.5m NIR images 
and the optical$+$NIR SED which is matched quite well by an S0 template from \citet{mannucci:2001}. 
An APO/DIS spectrum shows absorption lines from star light and a weak H$\alpha$ emission
line. The clear positional offset between the radio source and the optical/NIR galaxy 
is described in Section~\ref{sec:pos} and shown explicitly in the radio/NIR image overlay in the Appendix. 

We fit two Gaussians (dotted lines) to the asymmetric H\,I absorption feature shown in Figure~\ref{fig:0920}. 
The deep narrow component is not resolved in velocity and the shallower component has a FWHM of 33 km s$^{-1}$. Both are
redshifted with respect to the systemic galaxy redshift by 66 and 81 km s$^{-1}$
respectively. However, the rotation curve of the S0 galaxy is redshifted toward the southwest where the 
strongest radio source component lies. Therefore, at least one of these absorptions is consistent with being
normal disk gas in the S0.

\begin{figure}[hdtp]
\includegraphics[width=12cm]{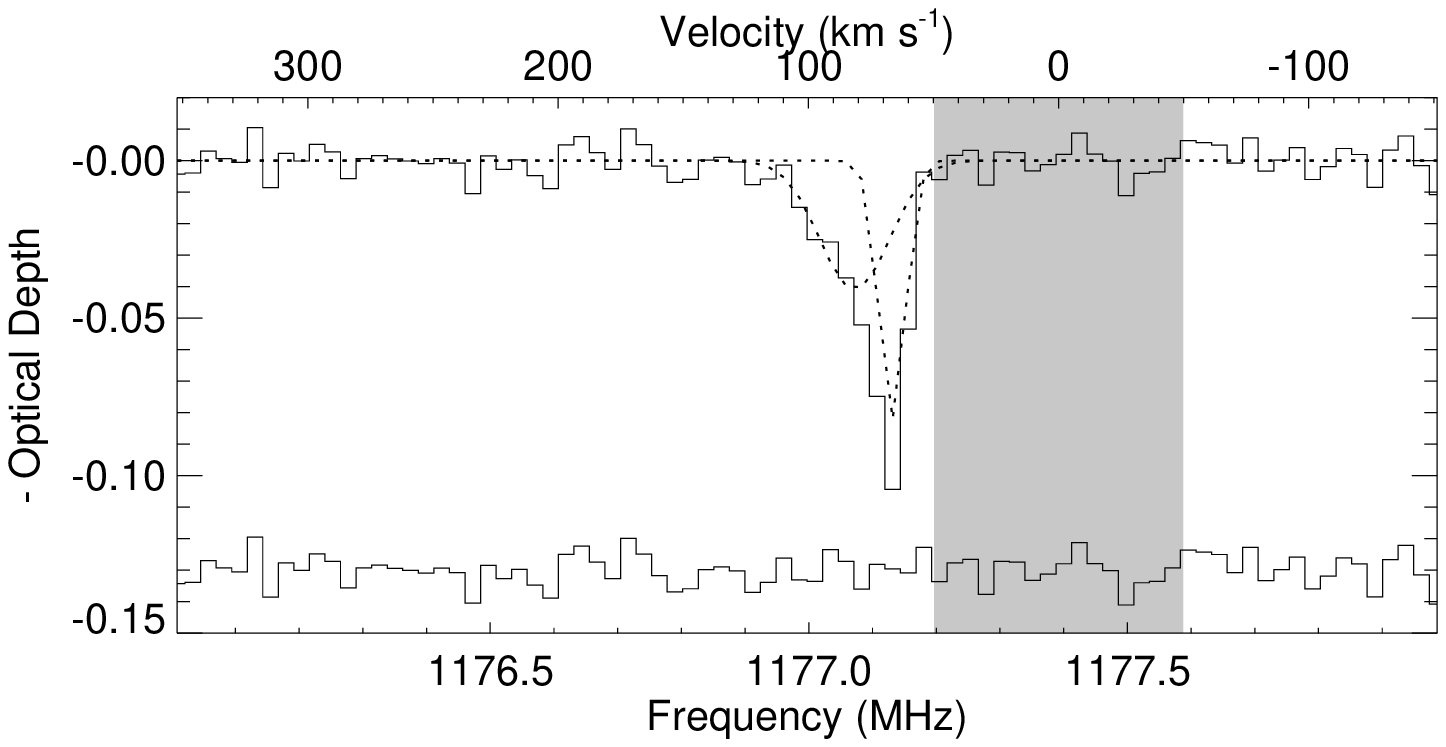}
\caption[\ion{H}{1}\ 21 cm spectrum in J0920$+$2714.]{The \ion{H}{1}\ 21 cm
  spectrum in J0920$+$2714 as observed by the GBT. We fit the absorption
  feature with two Gaussian components shown as dotted lines. The narrow component is not
  resolved (FWHM $\approx$ 2 spectral channels). Residuals between the data and the models 
are displayed with an offset in ``y'' for clarity. Zero
  velocity is set to the redshift of the optical galaxy, with
  the uncertainty in the galaxy velocity shown in gray shade. \label{fig:0920}}
\end{figure}

\subsubsection{J1129$+$5638}

See Figure~\ref{fig:1129} for the redshifted H\,I 21cm spectrum of J1129$+$5638 taken at the GBT. It
is a steep-spectrum source which is a CSO candidate according to our VLBA observations (see the 
radio images in the Appendix). Its optical/NIR SED is best fit by a pure ``G''-type SED with an
Sc-type spiral galaxy template and at ground-based resolution the optical/NIR morphology 
is amorphous (see Appendix) with a blue extension to the south that may be a companion galaxy. 
We recently used Gemini/GMOS to obtain a redshift of $0.8925\pm0.0008$ from
strong narrow emission lines (see Figure~\ref{fig:1129gmos}). The presence of the high
ionization [\ion{Ne}{5}] line classifies this source as a narrow-line radio galaxy.

The 21 cm absorption is quite broad (FWHM = 318 km~s$^{-1}$ if fit by
a single Gaussian profile) at a redshift consistent with the redshift of the 
optical emission lines to within rather broad errors ($\pm$ 120
km~s$^{-1}$). Thus, this detection is consistent either with normal disk gas or
outflowing material. However, the CSO-type radio morphology is not
consistent with an outflow in our direction. Therefore, we interpret
this detection as being disk gas in a rather massive galaxy ($\sim$ 4L$^*$ including $K$- and 
evolutionary corrections).

\begin{figure}[hdtp]
\includegraphics[width=12cm]{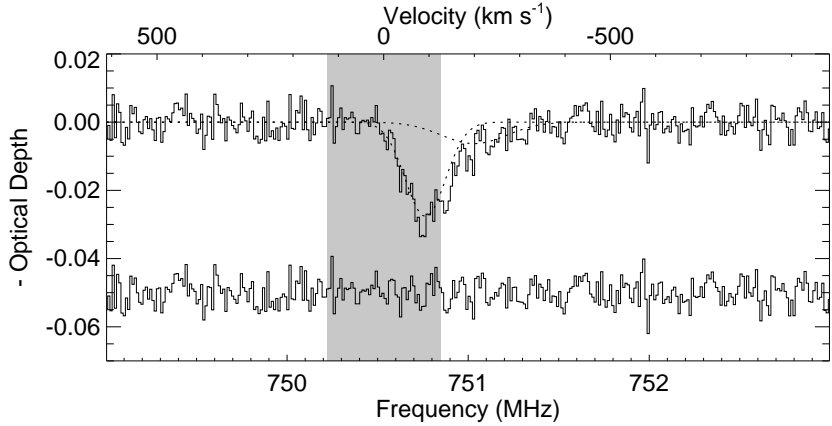}
\caption[\ion{H}{1}\ 21 cm spectrum in J1129$+$5638.]{The \ion{H}{1}\ 21
  cm spectrum in J1129$+$5638 as observed by the GBT. Although an
  asymmetry is present in the absorption line, the shallower component is not
  well detected ($< 2\sigma$ for a two-component Gaussian fit). We
  have also fit the feature with a single component which gives a
  slightly larger $\chi^2$. Gaussian fits to the two absorption
  components are shown as dotted lines. Residuals are displayed with
  an offset in ``y'' for clarity. Zero velocity is set to the redshift of
  the optical galaxy, with 1$\sigma$ velocity uncertainty shown in gray
  shade.\label{fig:1129}}
\end{figure}


%
\begin{figure}[hdtp]
  \includegraphics[]{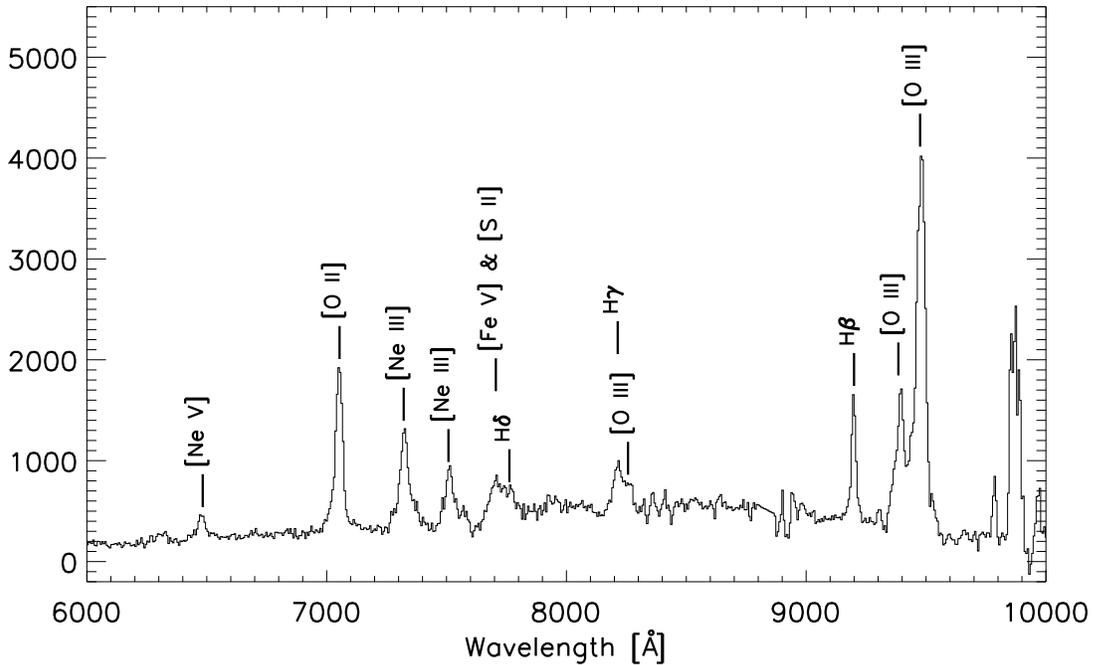}
  \caption{The Gemini/GMOS spectrum of J1129$+$5638 in counts versus wavelength (i.e., no flux
    calibration applied). Unidentified spikes are residual cosmic rays and sky
    lines. This spectrum is typical for a high-ionization, narrow-line radio galaxy. The apparent
emission line at $\sim$ 9900\AA\ is spurious. \label{fig:1129gmos}}
\end{figure}

\subsubsection{J1357$+$0046}

See Figure~\ref{fig:1357} for the redshifted 21cm GBT spectrum for this source. This is the only
successful detection made in our program without the benefit of an optical/NIR spectroscopic
redshift and was made only because the GBT PF-1 receiver has a very broad bandwidth
since the H\,I detection redshift ($z=$0.7971,0.7962) is quite different from the 
$z_{phot}=$0.57 of the optical galaxy. This large discrepancy is due almost certainly to the presence of
a significant amount of non-thermal continuum in the spectrum producing a substantial error in the photo-$z$
estimate; i.e., the optical/NIR SED classification is
``G$+$Q''. The SDSS and APO NIR images (see Appendix) are quite compact but with a significant
amount of diffuse emission around what may be a compact core. J1357$+$0046 shows an obvious 
CSO VLBA structure (see Appendix). 

The two H\,I lines have a velocity
separation of 140 km s$^{-1}$, and FWHMs of 80 km s$^{-1}$ and 70 km s$^{-1}$ so neither absorption
is the more obvious candidate for disk gas. Without an accurate spectroscopic redshift an unambiguous
assignment of these absorptions as disk gas, infall or outflow is not possible.

\begin{figure}[hdtp]
\includegraphics[width=12cm]{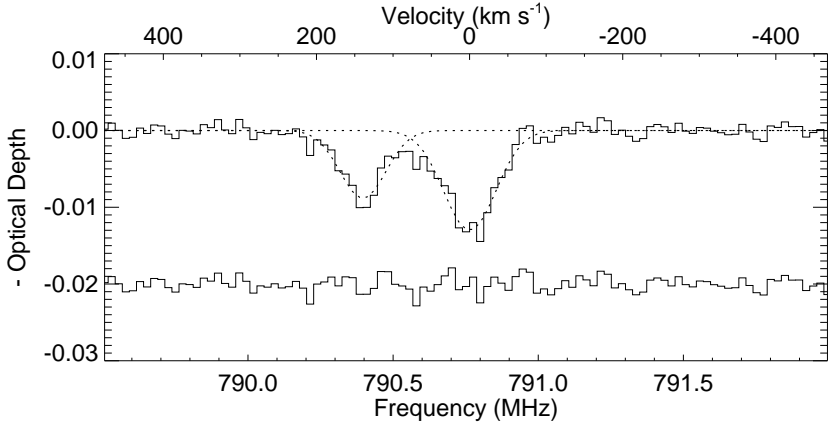}
\caption[\ion{H}{1}\ 21 cm spectra in J1357$+$0046.]{The \ion{H}{1}\ 21 cm
  spectrum for J1357$+$0046 as observed by the GBT. J1357$+$0046 does not have
  a spectroscopic redshift in the optical. Gaussian fits to the two
  absorption components are shown in dotted lines. Residuals are
  displayed with an offset in ``y'' for clarity. The zero velocity is set
  to the center of the low redshift H\,I component ($z=0.7962$) since only
  a photometric redshift is available for this source.\label{fig:1357}}
\end{figure}



\subsubsection{J1604$+$6050}

See Figure~\ref{fig:1604} for a tentative detection of a single, narrow 21 cm absorption
component at the optical redshift obtained by Gemini-GMOS (Figure~\ref{fig:1604gmos}). The optical redshift of
the weak, narrow emission lines is $0.5590\pm0.0004$ and that of the tentative H\,I absorption
line is $0.55913\pm0.00015$ ($\Delta v \approx$ 25 km s$^{-1}$). The tentative 21cm detection is an
extremely narrow line with an FWHM of only 8 km/s. J1604$+$6050 is a CSO candidate in our VLBA map
(see Appendix) with an optical$+$NIR SED best-fit by a pure ``G''-type SED with an
S0-type galaxy template. The rather amorphous optical/NIR galaxy images have centroids consistent with the
CSO centroid seen with the VLBA. Given the optical/NIR (pure ``G''-type) and radio continuum properties
(GPS and CSO) of this source, it would not be unusual to find an H\,I absorption line detection
associated with it. Longer integration times are required to confirm this tentative detection.

\begin{figure}[hdtp]
\includegraphics[width=12cm]{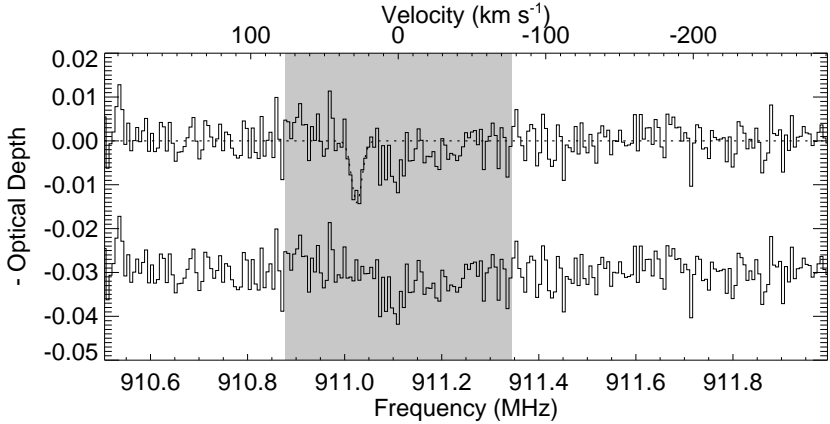}
\caption[\ion{H}{1}\ 21 cm spectrum in J1604$+$6050.]{The \ion{H}{1}\ 21
  cm spectrum in J1604$+$6050 as observed by the GBT. There is a tentative
  absorption line at 911.025 MHz. A Gaussian fit to the absorption
  component is shown in dotted lines and is almost unresolved by these observations. Residuals are displayed with an
  offset in ``y'' for clarity. Zero velocity is set to the redshift of the
  optical galaxy determined from emission lines with the 1$\sigma$ redshift uncertainty
  shown in gray shade. The apparent absorption at $\sim$ 911.03 MHz is an instrumental artifact. \label{fig:1604}}
\end{figure}

\begin{figure}[hdtp]
  \includegraphics[]{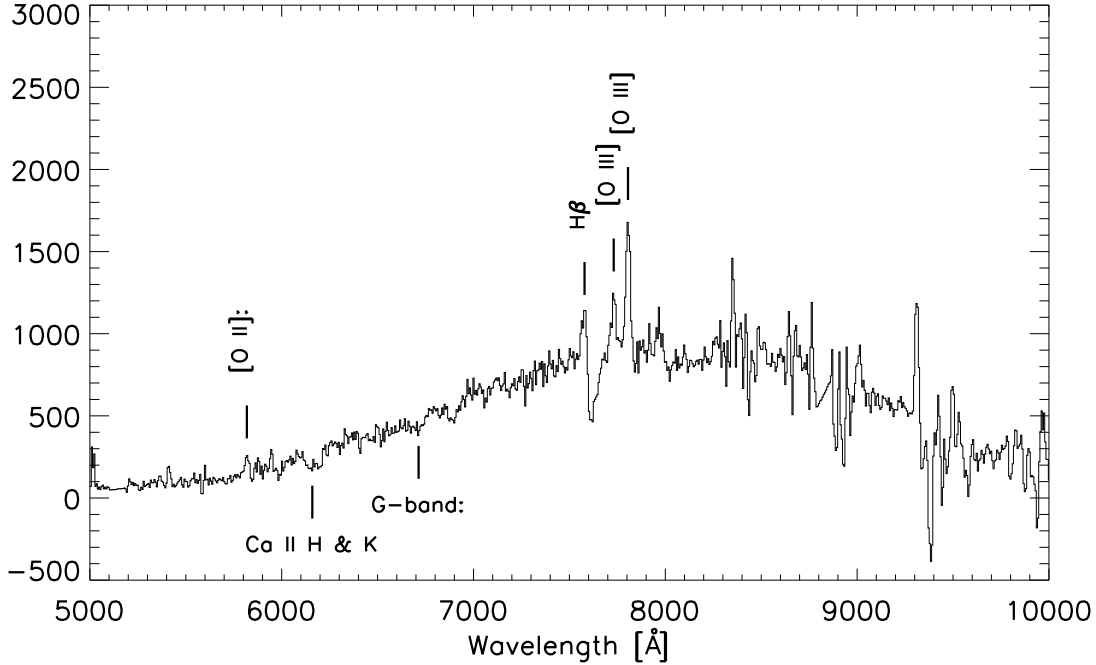}
  \caption{The GMOS spectrum of J1604$+$6050 displayed as counts versus wavelength (i.e., no flux
    calibration). Unidentified spikes are residual cosmic rays and sky
    lines. The atmospheric oxygen B- and A-bands are visible in absorption. The redshift
of $z=$0.5990$\pm$0.0004 was obtained using just the emission lines. \label{fig:1604gmos}}
\end{figure}


%

\subsubsection{J1616$+$2647}

See Figure~\ref{fig:1616} for the 21 cm GBT spectrum of J1616$+$2647 which reveals a very broad, symmetric
profile at the optical redshift. J1616$+$2647 is a
GPS source which is a CSO candidate according to our VLBA observations (see Appendix). Its
optical$+$NIR SED is best fit by an Sc-type galaxy template
(i.e., pure ``G''-type). The optical/NIR images reveal a compact structure, well-centered
at the CSO position. This source was observed by Gemini/GMOS to obtain a
redshift of $0.7553\pm0.0003$ from very strong, narrow emission lines
(Figure~\ref{fig:1616gmos}).

J1616$+$2647 has one of the broadest H\,I absorption systems detected so far
with an FWHM = 447 km s$^{-1}$. There are four objects that have 21 cm
absorption lines broader than 400 km/s, including 3C 84 and PKS
2322$-$12, both cD galaxies close to the center of galaxy clusters
\citep{inoue:1996,sarazin:1995}. While it would be natural to
associate these broad absorptions with out-flowing gas, the absorption in
J1616$+$2647 is centered at the systemic redshift and could also be
partly or largely disk gas in an extremely massive galaxy. Deep imaging to determine if a rich
cluster is present around this galaxy is needed to test this hypothesis.

\begin{figure}[hdtp]
\includegraphics[width=12cm]{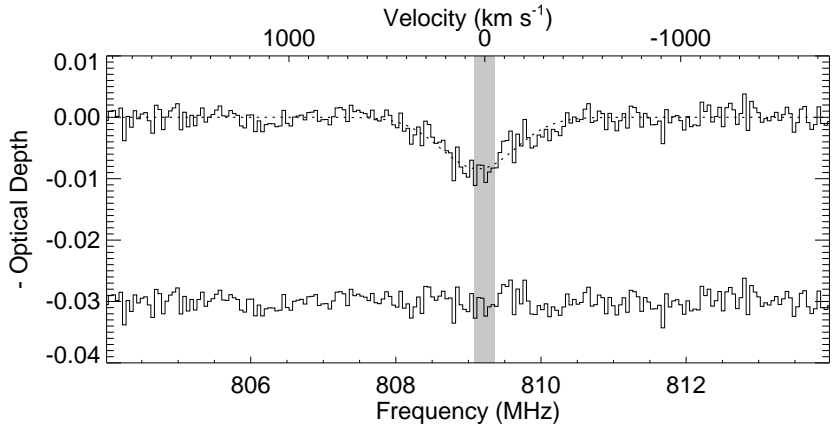}
\caption[\ion{H}{1}\ 21 cm spectrum in J1616$+$2647.]{The \ion{H}{1}\ 21
  cm spectrum in J1616$+$2647 as observed by the GBT. A Gaussian fit to the 
absorption component is shown in dotted lines and has FWHM = 447 km s$^{-1}$.
Residuals are displayed with an offset in ``y''
  for clarity. Zero velocity is set to the redshift of the optical
  galaxy with 1$\sigma$ redshift uncertainty shown in gray shade.\label{fig:1616}}
\end{figure}


%
\begin{figure}[hdtp]
  \includegraphics[]{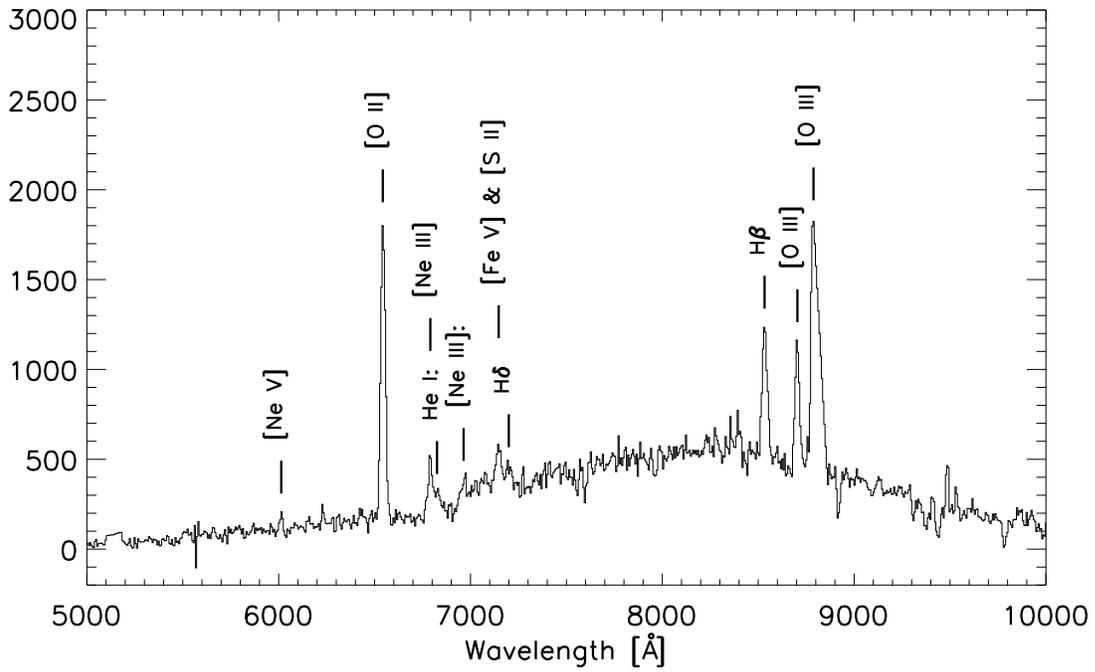}
  \caption{The Gemini/GMOS spectrum of J1616$+$2647 displayed as counts versus wavelength 
(i.e., no flux calibration). Unidentified spikes are residual cosmic rays and sky
    lines. The redshift of $z$=0.7553$\pm$0.0003 was determined from the emission lines only. 
\label{fig:1616gmos}}
\end{figure}

\section{Discussion and Conclusions}\label{sec:dis}

We have selected and studied a sample of highly-obscured radio-loud AGN with the ultimate goal
of discovering new highly-redshifted absorption systems at radio
frequencies. After demonstrating that most of the objects remain
obscured in the NIR in Paper~\Rmnum{1}, we find in this paper that
their radio properties are also promising for absorption-line searches. 
There are 52 objects that have a compact component with a size
0\farcs5, providing a small background source that can be efficiently covered by foreground, absorbing gas
that might be present in the host galaxy. We also have identified 9 potentially intervening systems
and 16 CSOs or CSO candidates (including two previously identified) in our sample. There are 20 MSOs and
8 core-jet objects, which also are good potential targets for absorption-line
searches. In fact, only the few (14) LSOs not in intervening systems or without a
bright, dominant radio core should be excluded from further searches (see Column~15 of Table~\ref{tab:type})
as unlikely to possess redshifted H\,I absorption lines in their low frequency radio spectrum. 

Using VLBA maps, the overall radio spectrum and the optical/NIR SEDs of these sources, we find a strong
correlation between compact radio source structure, a GPS radio spectrum and a pure galaxy (``G''-type)
SED. Reasoning that the absence of a non-thermal AGN spectrum in these CSO radio sources indicates
that we are viewing the central AGN through an absorbing screen, we targeted these combined classes of sources 
(CSO/GPS/pure galaxy optical/NIR SED) in our survey for redshifted H\,I absorption-line spectroscopy. 

Of the 6 CSO/GPS/``G''-type sources with accurate spectroscopic redshifts we detected H\,I absorption lines in 5 of them at
their spectroscopic redshift. This includes one tentative detection and there was one clear non-detection
in this small sample. Additionally, one absorber which is at a lower redshift than the AGN was also discovered. Using a 
different sample, \citet {vermeulen:2003} comes to similar conclusions concerning the high H\,I absorption detection rate
from the CSO/GPS class of sources ($\sim$ 33\% detection rate in their sample). Our small sample has a significantly larger
detection rate for CSOs than was found by that study. Severe RFI at the frequencies of the 
redshifted H\,I 21cm lines prevented a sensitive search in 7 other CSO/GPS/``G''-type sources with spectroscopic redshifts. 
Redshifted H\,I absorption searches in 14 other sources which had only photometric redshifts were largely frustrated by the 
presence of strong, ubiquitous RFI throughout the sub-GHz radio spectrum. Only one detection was made from this sample of 
14 but the RFI is so strong that we cannot rule out the possiblity that many of these sources actually possess 
H\,I absorption that we cannot detect from the ground (or anywhere close to human-made RFI for that matter). Clearly having 
a spectroscopic redshift in-hand is critical in the search for highly redshifted H\,I and, eventually, 
molecular absorption lines.

A higher detection rate of H\,I absorption in radio-loud galaxies compared to radio-loud quasars, 
in GPS objects compared to flat-spectrum (FS) objects and in CSOs compared to core-jet sources is broadly
consistent with unified schemes of radio-loud AGN \citep{urry:1995}. In their simplest form, unified schemes suggest that
radio-loud galaxies are viewed close to the plane of a circum-nuclear disk and torus (if present) while radio-loud quasars
are viewed close to the jet propagation direction. This suggests that the obscuration of the non-thermal light
from the AGN is strongly related to the presence of 21 cm absorption gas. If so, this predicts the low (or zero) detection rate
in high UV luminosity objects (L$_{UV} >$ 10$^{23}$ W Hz$^{-1}$) as has been seen by \citet{curran:2010}.

Taken in the light of standard unified schemes, the results of this study do not bode well for finding
higher redshift atomic and molecular absorption lines in the radio. However, we should not take the present results and this statement
to mean that such sources do not exist but rather that they will be quite difficult to find and study. 
In Paper~\Rmnum{1} we embarked on a program to identify excellent candidates for discovering highly redshifted radio absorption lines
by cross-correlating bright ($\geq$ 0.5 Jy) FIRST radio sources with SDSS counterparts which are less centrally-concentrated
than giant elliptical galaxies and point sources typical of radio galaxies and quasars. The use of the SDSS for this work 
limited this search to $z\leq$ 1.0 due to the faintness of the optical counterparts; i.e., more distant AGN similar to the
present sample are too faint to be detected in the SDSS photometric sample.
This approximate redshift limit is confirmed by the photometric redshifts found by the SDSS and by us using NIR photometry in 
conjunction with the SDSS photometry in Paper~\Rmnum{1}. This is specifically the case for those sources whose optical/NIR SEDs are 
dominated by galaxy starlight, not by AGN emission. Those few sources with quasar-like SEDs are biased towards $z>$ 1 
in this sample and thus at too high a UV rest-frame luminosity to be viable H\,I absorption candidates \citep{curran:2010}.

To locate and study higher redshift ($>$ 2) sources similar to the present sample does {\bf not} require using fainter 
radio sources than the 0.3 Jy limit used here. As shown in this paper the radio properties of such sources are easily studied with current instruments.
The limiting factors are the optical/NIR identifications of faint host or intervening galaxies associated with these bright sources and obtaining 
spectroscopic redshifts for them. Our failure to discover significant numbers of redshifted H\,I absorbers using photometric redshifts
alone, makes the measurement of a spectroscopic redshift an essential step in this process. 

The existence of sensitive IR sky surveys from the {\it Spitzer Space Telescope} and the {\it Wide-field Infrared Survey Explorer} (WISE) satellite
can be used to construct substantial samples of optically-faint (i.e., SDSS non-detects), radio and NIR bright sources. VLBA mapping of such
sources can isolate the CSO and point sources which are the best candidates for H\,I and molecular absorption-line searches. In a preview of this technique we
used {\it Spitzer} IRAC and WISE 12 and 22 $\mu$ photometry to find that of the 13 CSOs and candidate CSOs from the present survey, 10 (7 definite
and 3 probable) show mid-IR flux turnups. Of the 4 \ion {H}{1} absorption-line detected CSOs observed by WISE, 
3 show definite or probable mid-IR turnups. In this manner the best candidates for the most heavily obscured radio-loud AGN can be identified. 
Then the problem
becomes how to determine the source redshift to allow an H\,I detection in the forest of RFI confronting us at sub-GHz frequencies. 
Developing capabilities in the sub-mm wavelength regime may make such redshift determinations possible.

An alternative approach is to target directly molecular absorption at higher rest frequencies where observations would be 
uninhibited by significant RFI. But this would require a flat spectrum (FS) radio source. CSOs are generally not FS sources. Instead, FS core-jet
sources are generally dominated by luminous rest-frame UV continuum emission in the optical. The presence of the UV continuum appears to preclude
the presence of strong absorption \citep {vermeulen:2003, curran:2010}. While very rare, FS core-jet sources whose optical/NIR 
continuum is dominated by galaxy starlight 
would be promising targets for redshifted absorption-line studies. However, blind surveys for molecular absorption in such sources have
been unsuccessful up to now \citep[e.g.,][]{murphy:2003a, kanekar:2014}. 

A detailed description of all of the sources in the sample on which this work is based can be found in the dissertation of the first 
author \citep {yan:2013}.

\appendix 

\section {Appendix: Basic Radio Source Data and Maps from VLA-A and VLBA}

Table~\ref{tab:obs} lists basic observational parameters for all radio maps
made for each source in the sample including the following information: the major
and minor axis sizes in arcsecs, the position angle (P.A.; angle east of north) of the restored
beam, the map $rms$ ($\sigma$) in the vicinity of the source in mJy per beam and the maximum flux 
density in mJy per beam in the map. The median $rms$ for the maps in Figure~\ref{fig:image} 
is 0.19, 0.12, and 0.36 mJy at 4.9 GHz, 8.5
GHz, and 1.4 GHz (VLBA) respectively. Dynamic ranges of up to a few hundreds are
generally achieved through self-calibration. The median major axis
value of the restored beam is 0\farcs5, 0\farcs3, and 15 milliarc
seconds at 4.9 GHz, 8.5 GHz, and 1.4 GHz respectively. 
For a few VLBA 1.4 GHz observations, the data were tapered to
exclude long baselines with no signal.

Our VLA and VLBA images are shown in Figure~\ref{fig:image}. Each
image has five contour levels ranging from 3$\sigma$ to
0.8$S_\mathrm{m}$ and logarithmically spaced, where $S_\mathrm{m}$ is
the maximum flux density, i.e., by assuming $R=0.8S_\mathrm{m}/3\sigma$,
the contour levels are (1, $R^{1/4}$, $R^{2/4}$, $R^{3/4}$, R)$\times
3\sigma$. A negative 3$\sigma$ value is shown as a dashed line. In the
4.9 GHz and 1.4 GHz images, the optical centroid position in the
$r$-band is labeled in blue with uncertainties along the RA and Dec
axes. The red cross is the NIR centroid position (usually
$K_\mathrm{s}$-band) with uncertainties. A green rectangle in the 4.9
GHz images represents the boundary of the object's 1.4 GHz VLBA image
where available. Therefore even if an object's NIR and/or optical
position crosses are completely or partly outside of its 1.4 GHz VLBA
image, their relative positions can be inferred from the crosses and
rectangle in the 4.9 GHz images. Center coordinates of the 4.9 GHz and
1.4 GHz images are shown at the bottom of each plot. Relative
positions along RA and Dec are labeled on the ``X'' and ``Y'' axes
respectively. Please note that absolute 8.5 GHz astrometry is
problematical and, for this reason, the central coordinates of the 
8.5 GHz images are not given.

The 4.9 GHz images are overlaid with NIR images (see Paper
\Rmnum{1}). $K_\mathrm{s}$-band images are used with two
exceptions. J1125$+$1953 was not detected in the $K_\mathrm{s}$-band so its
$J_\mathrm{s}$-band image is used. J1502$+$3753 was not observed in the $K_\mathrm{s}$-band
so its $H$-band image is used. NIR images have been re-gridded and
smoothed by cubic interpolation to match the resolution of the 4.9 GHz
images. This procedure sometimes resulted in artificially fracturing diffuse
emission in the NIR image due to trying to match the radio resolution. 
Pixel counts are displayed by grayscale where 0 is white and 1 is dark, and where 1 represents the
maximum pixel count in the object and where 0 and 0.2 represent the
$-$2.5$\sigma$ and +2.5$\sigma$ noise of the sky background
respectively. Between 0 and 1 the gray shade scales
linearly with pixel counts. As can be seen in the overlaid images, the
host galaxies of these sources are often quite diffuse and irregular in
morphology as discussed in detail in Paper~\Rmnum{1}. These asymmetrical morphologies
makes the determination of a galaxy centroid quite uncertain, often with 
errors much larger than the fitting errors shown on the radio maps.

For each radio image, we fit major components with elliptical Gaussian
profiles and report the results in Tables~\ref{tab:vlac},
\ref{tab:vlax}, and~\ref{tab:vlba} for the 4.9 GHz, 8.5 GHz, and 1.4
GHz VLBA images respectively. Tables~\ref{tab:vlac} and \ref{tab:vlba} list
(1) object name, (2) component ID, (3,4) RA and Dec of the component,
(5,6) peak and integrated flux density, and (7,8,9) deconvolved major
axis, minor axis, and P.A. of the Gaussian component (angle east of north). The 8.5 GHz
results are shown in Table~\ref{tab:vlax} without coordinate
information for the reason mentioned in Section~\ref{sec:obs}. Without
reliable astrometry, we match components in the 4.9 GHz and 8.5 GHz images
by eye and mark the same component with the same letter ID. The
associations are unambiguous in all cases despite the lack of accurate 
astrometrical registration. Spectral index data for
each component in Column (8) of Table~\ref{tab:vlax} is calculated
between 4.9 GHz and 8.5 GHz. In cases in which one component at 4.9
GHz is resolved into several components at 8.5 GHz, each sub-component
is named by the letter ID at 4.9 GHz followed by a number and the component
flux densities are summed in calculating the spectral index. Note
well, that because the beam sizes differ between 4.9 GHz and 8.5 GHz
in our observations, these spectral indices should be treated as
estimates. The Gaussian profile may not be a good estimate for
resolved components with complicated structures, but we make sure the
total flux density is conserved during the fits.

\clearpage
\maxdeadcycles=1000

\begin{figure}[htdp]
\begin{tabular}{lll}

\includegraphics[scale=0.25]{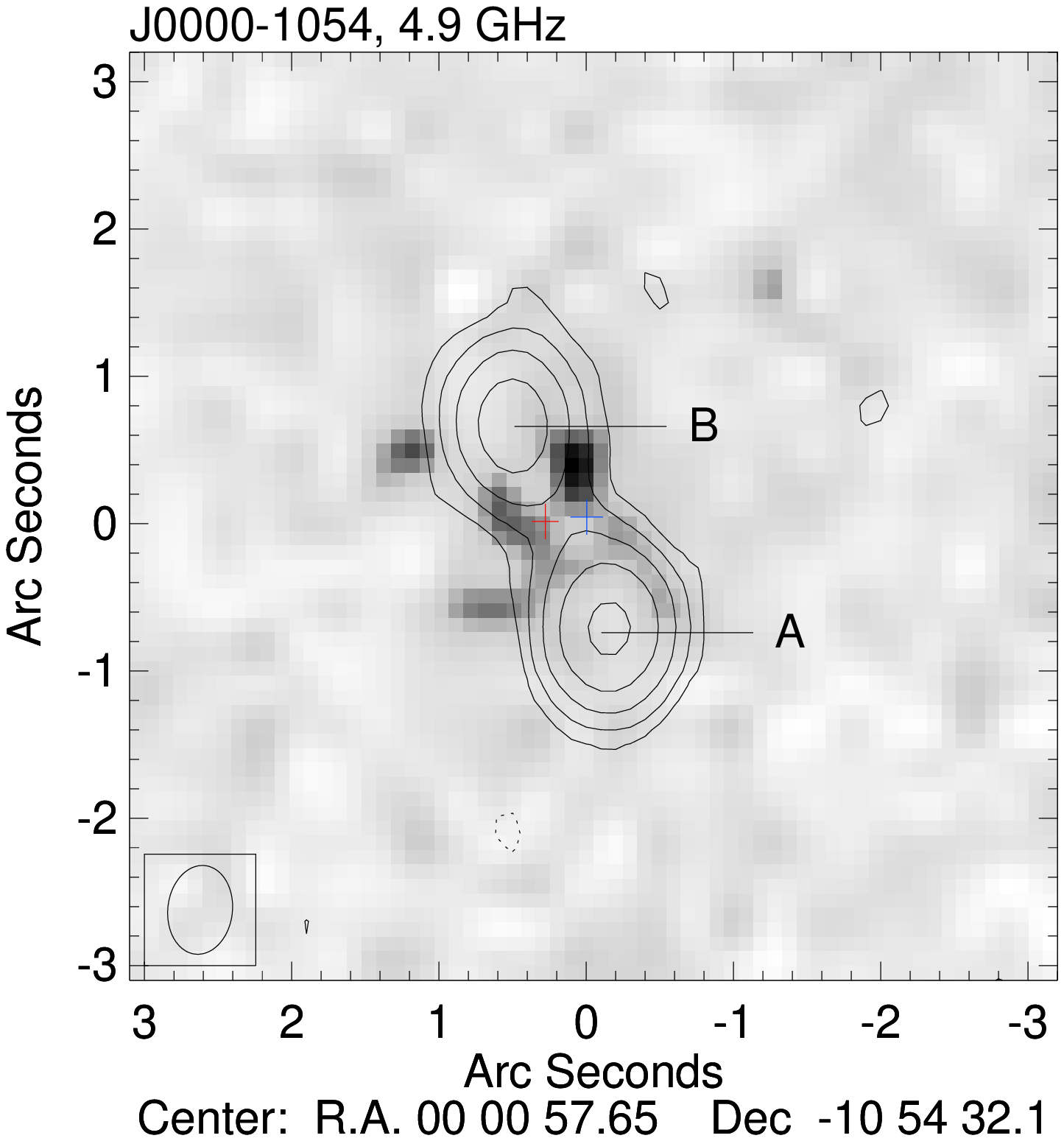}	&	\includegraphics[scale=0.25]{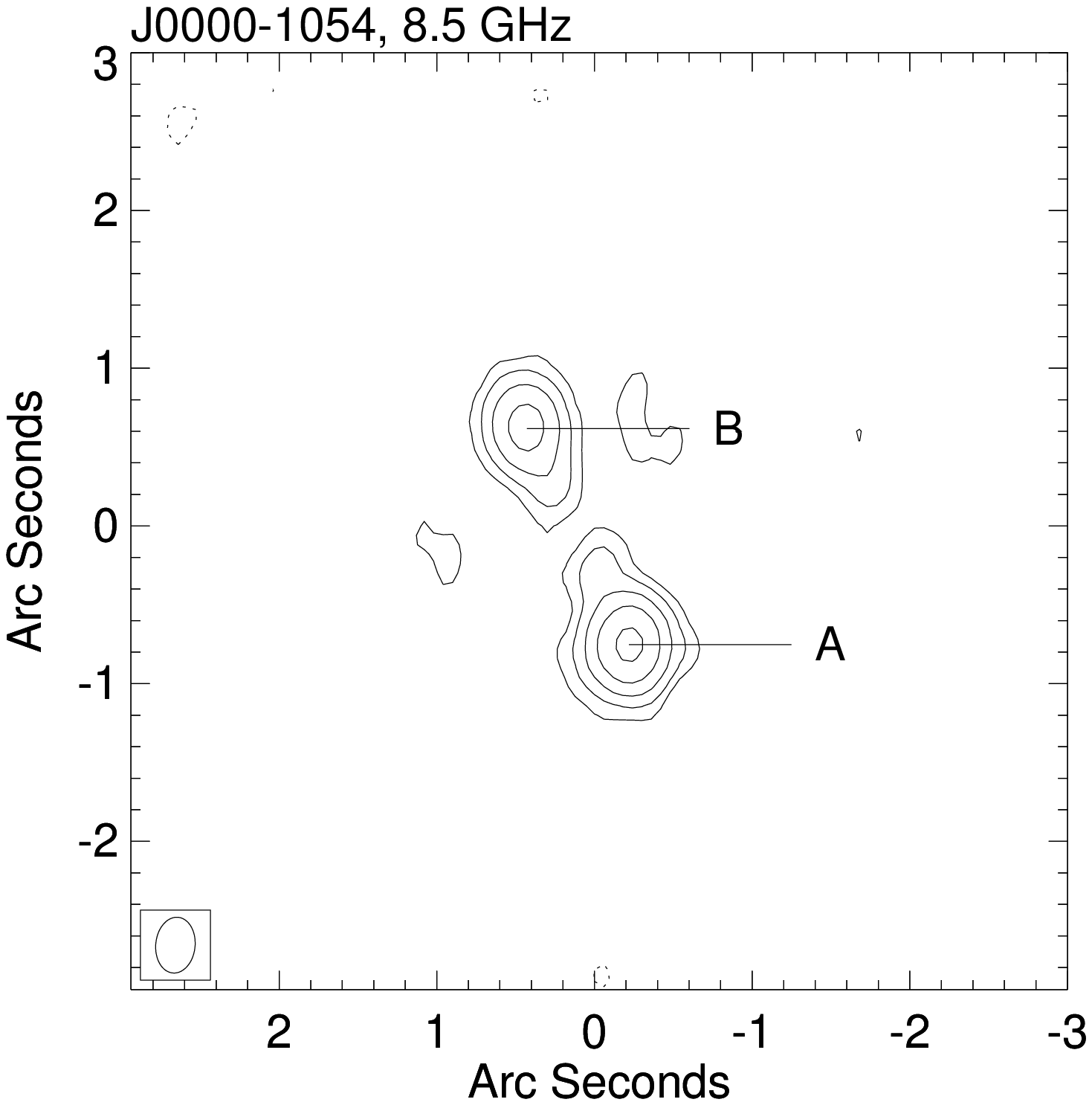}	&		\\
\includegraphics[scale=0.25]{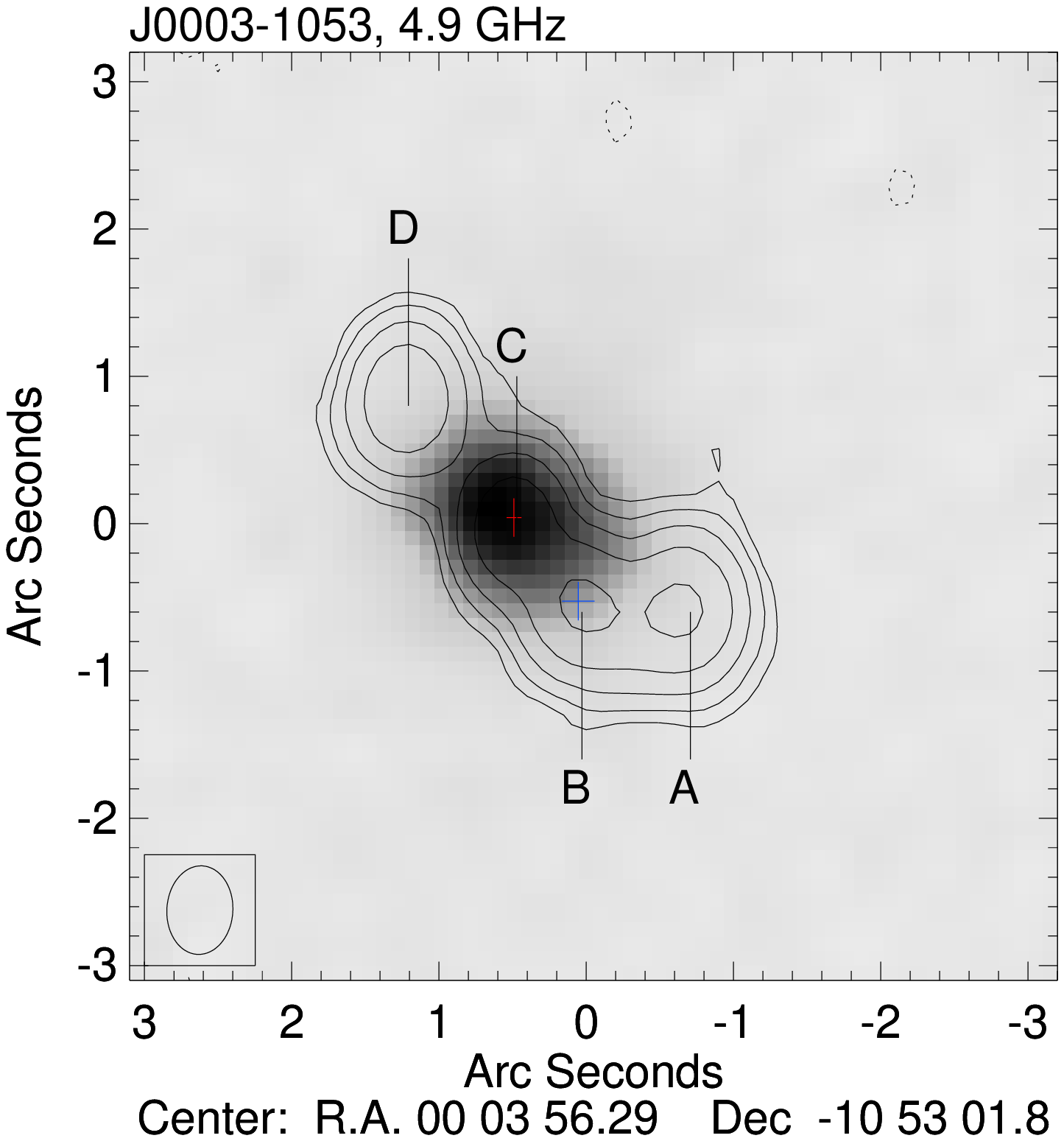}	&	\includegraphics[scale=0.25]{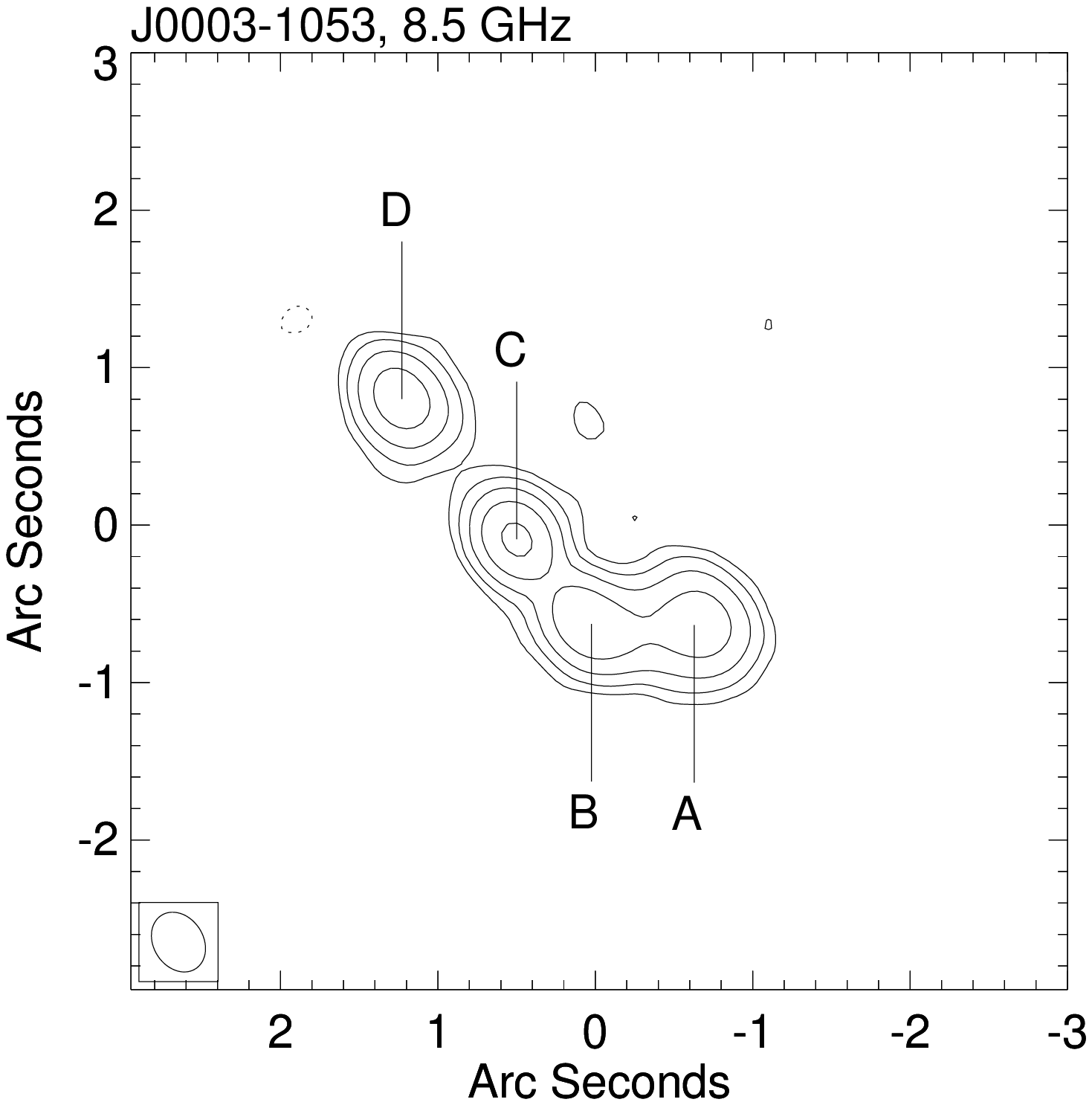}	&		\\
\includegraphics[scale=0.25]{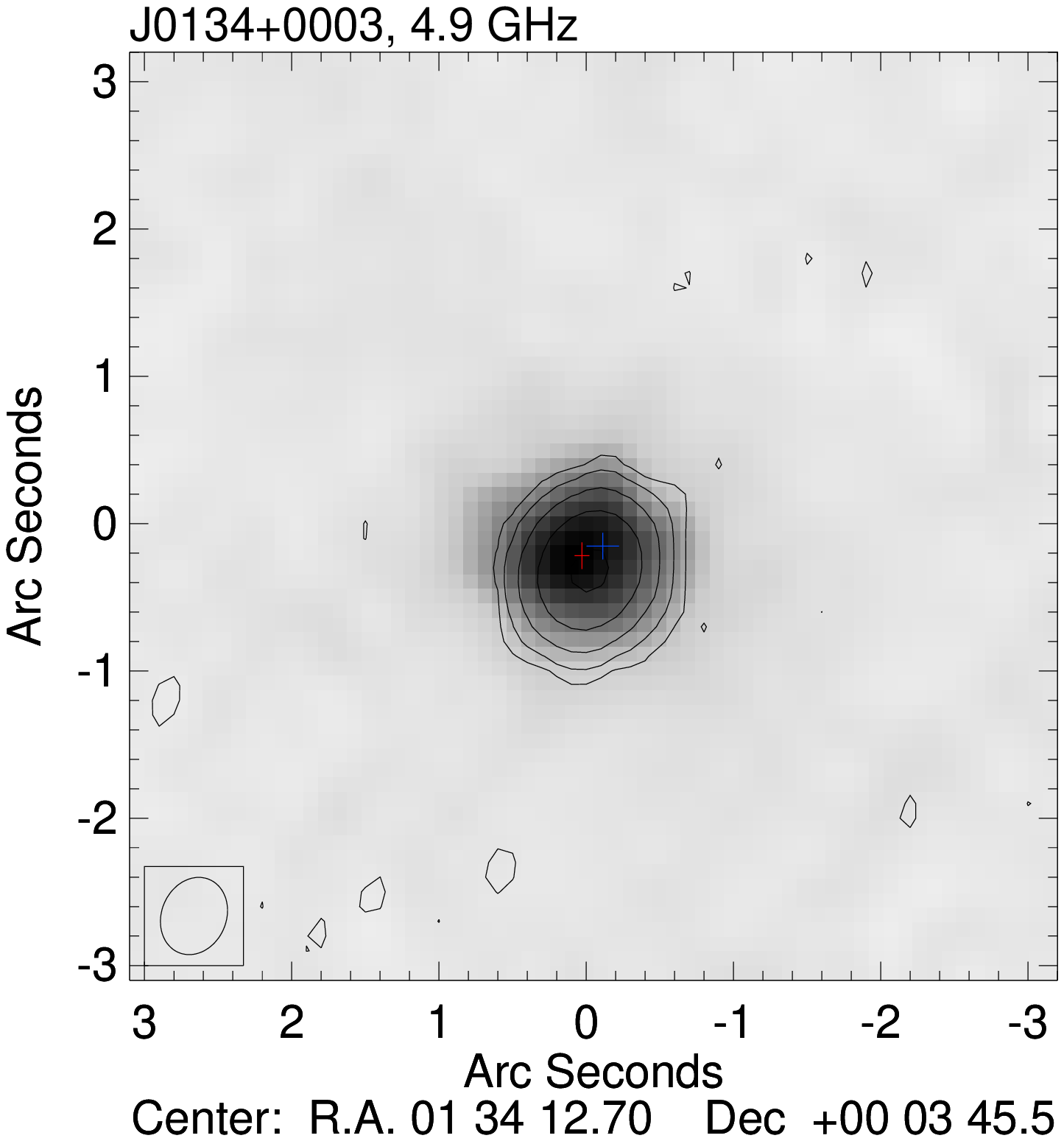}	&		&		\\
\includegraphics[scale=0.25]{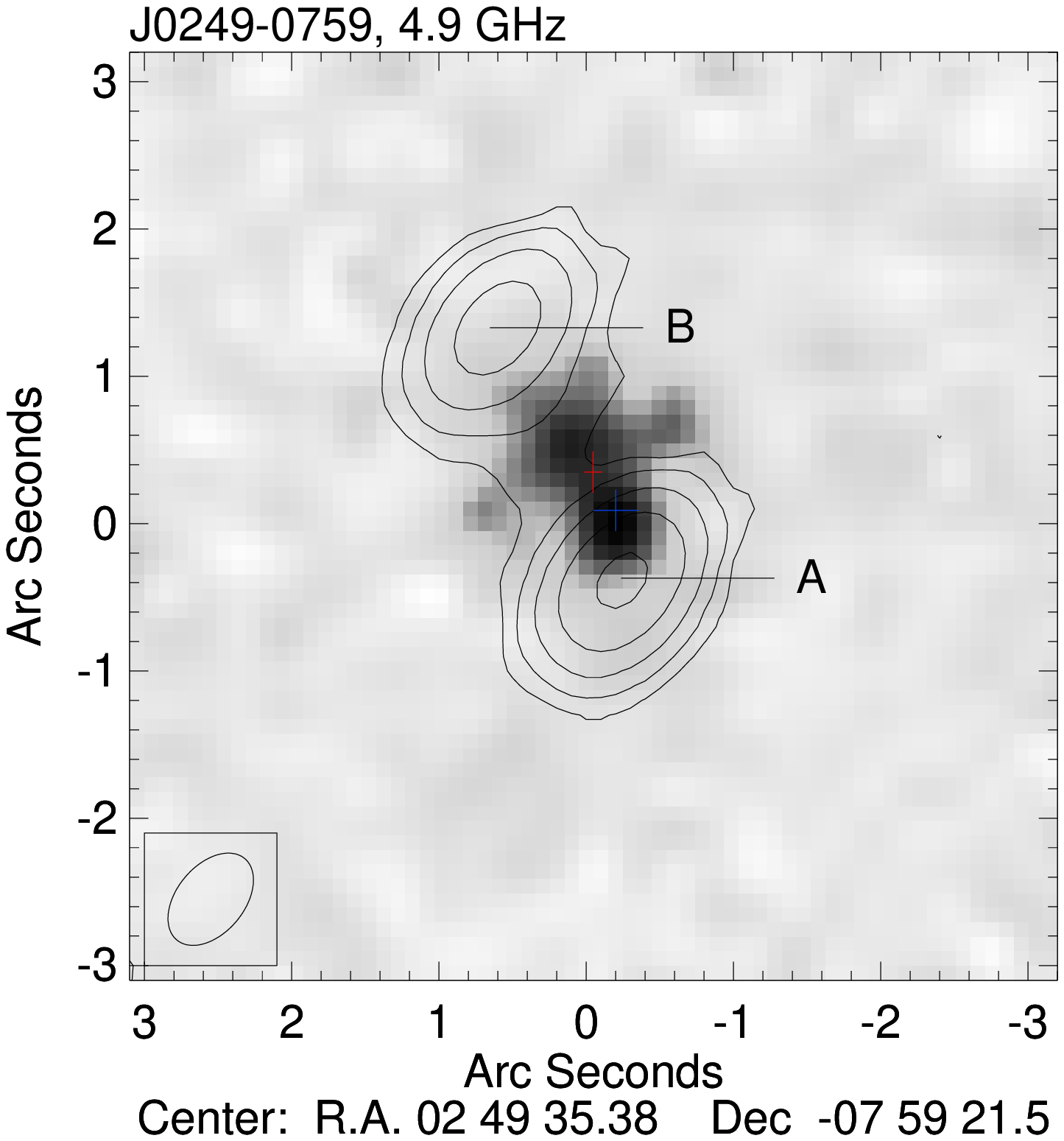}	&	\includegraphics[scale=0.25]{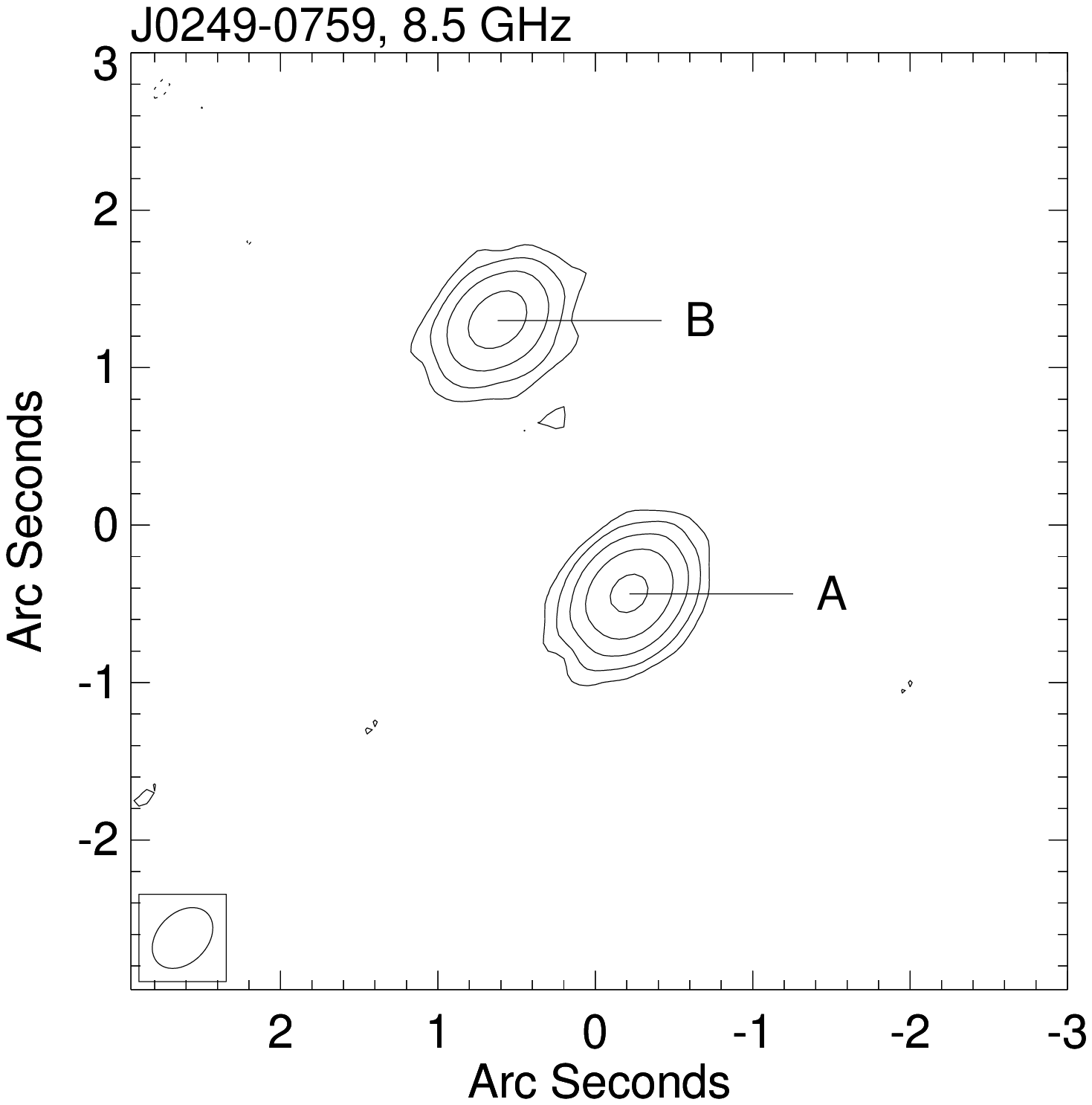}	&		\\
\includegraphics[scale=0.25]{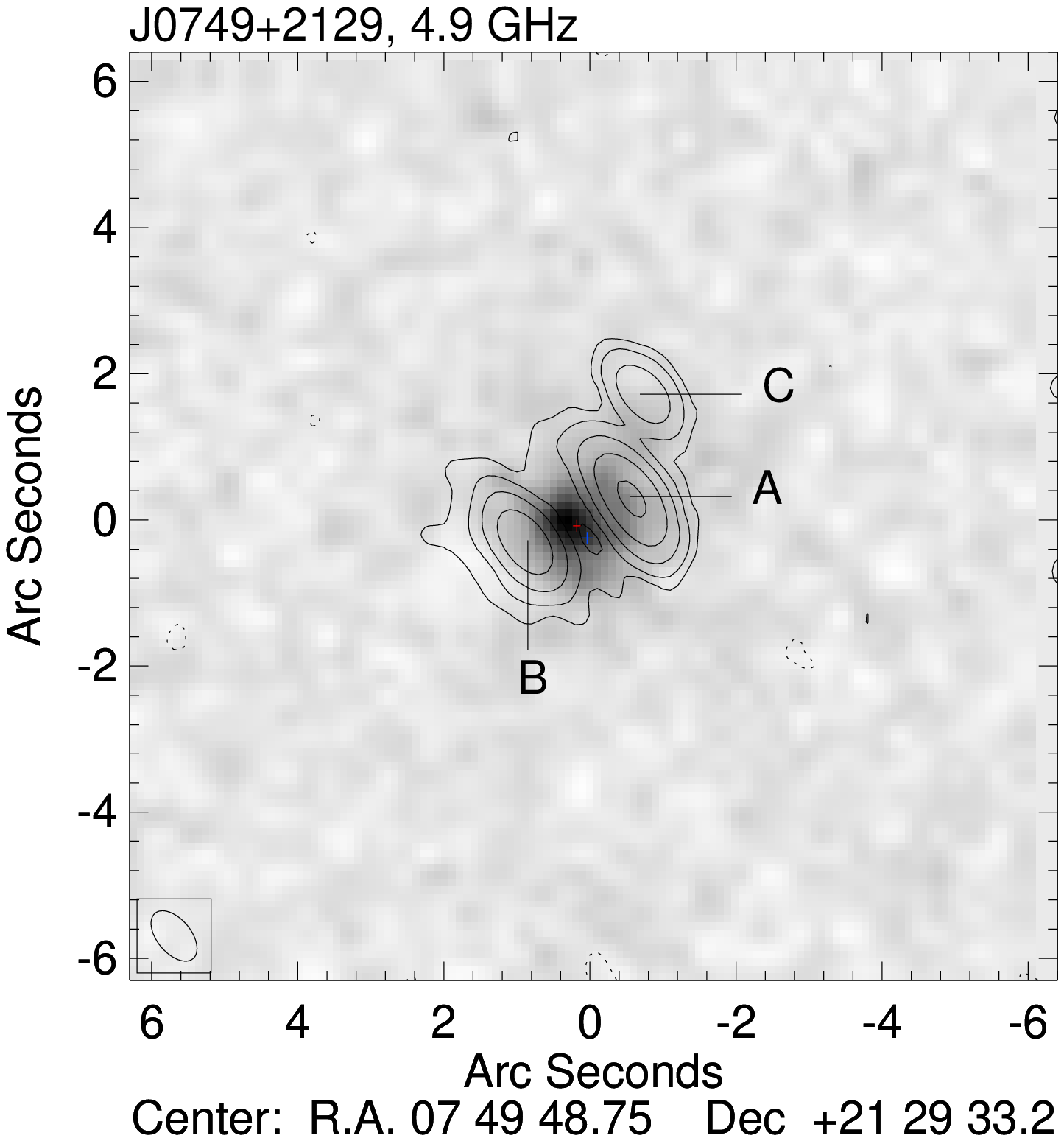}	&	\includegraphics[scale=0.25]{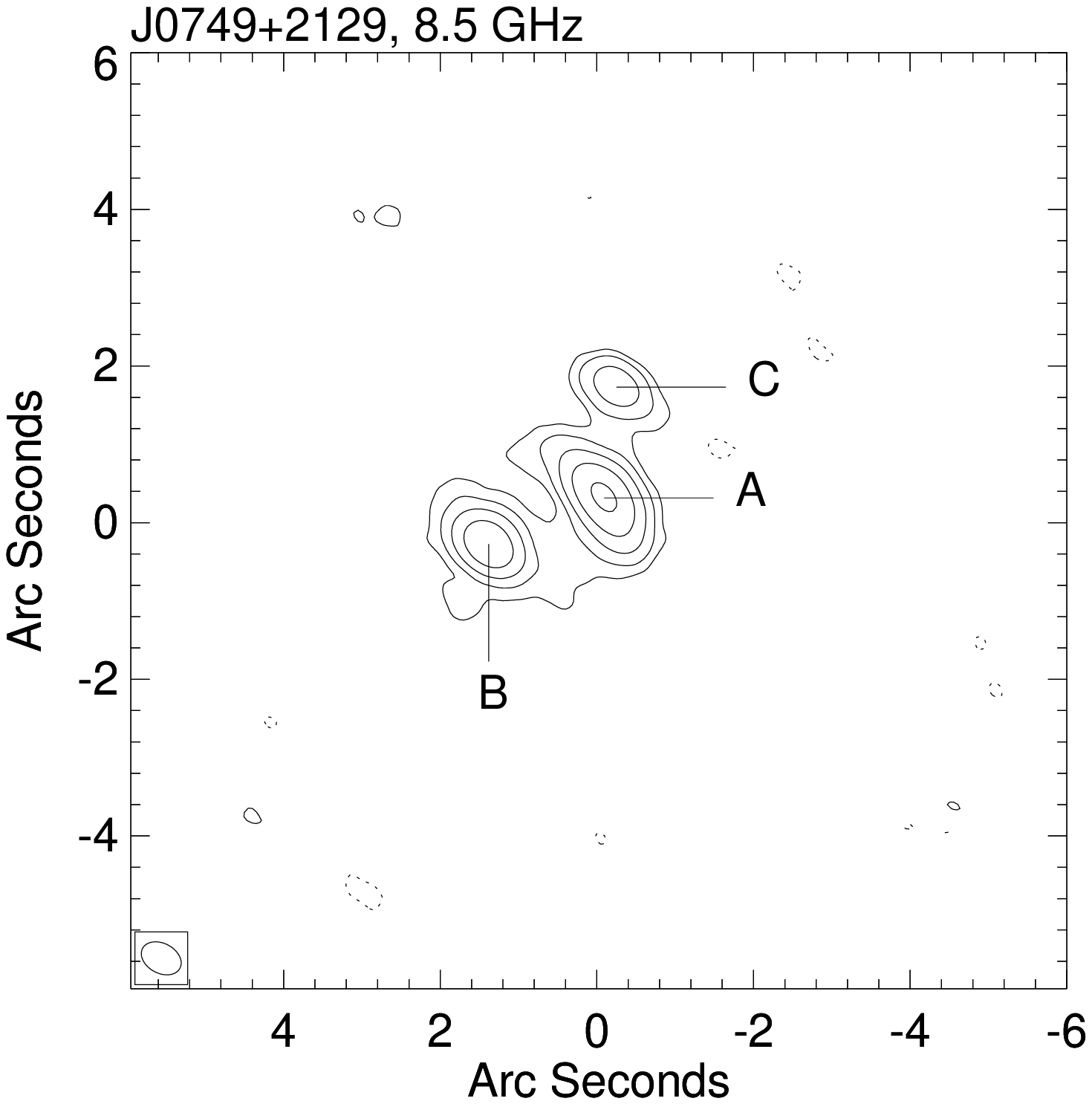}	&		\\

\end{tabular}
\end{figure}

\begin{figure}[htdp]
\begin{tabular}{lll}

\includegraphics[scale=0.25]{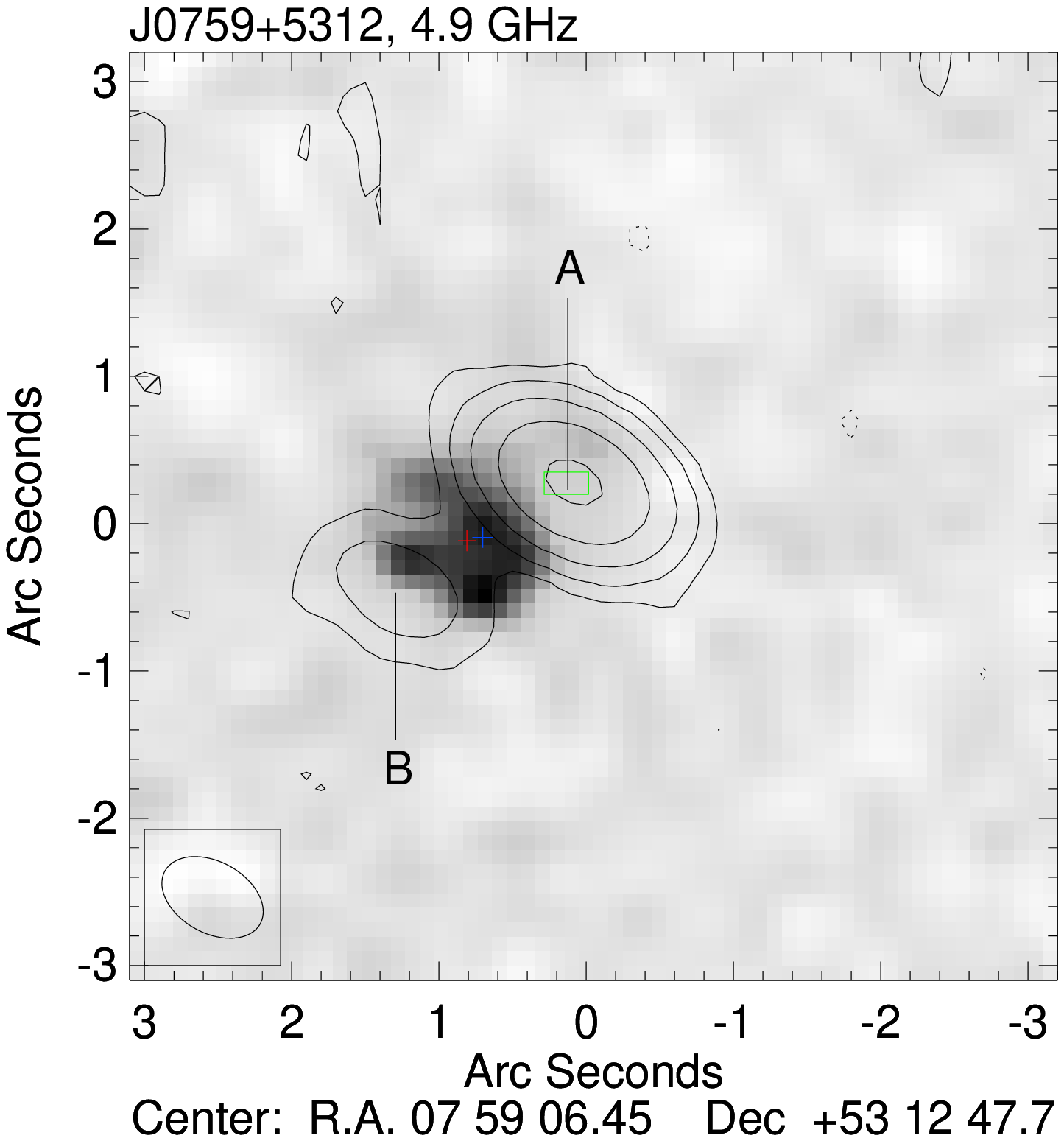}	&	\includegraphics[scale=0.25]{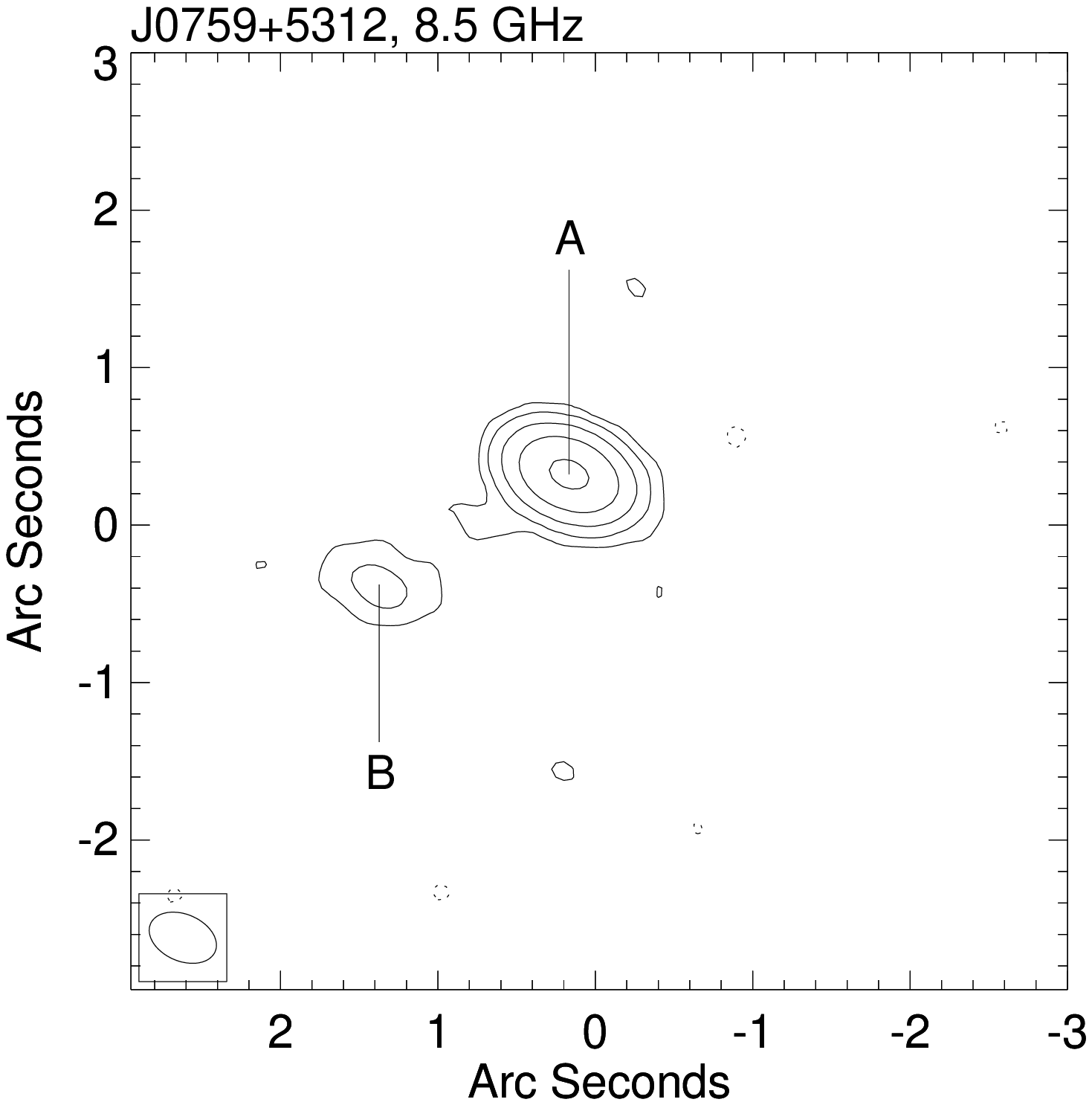}	&	\includegraphics[scale=0.3,bb=50 0 901 453]{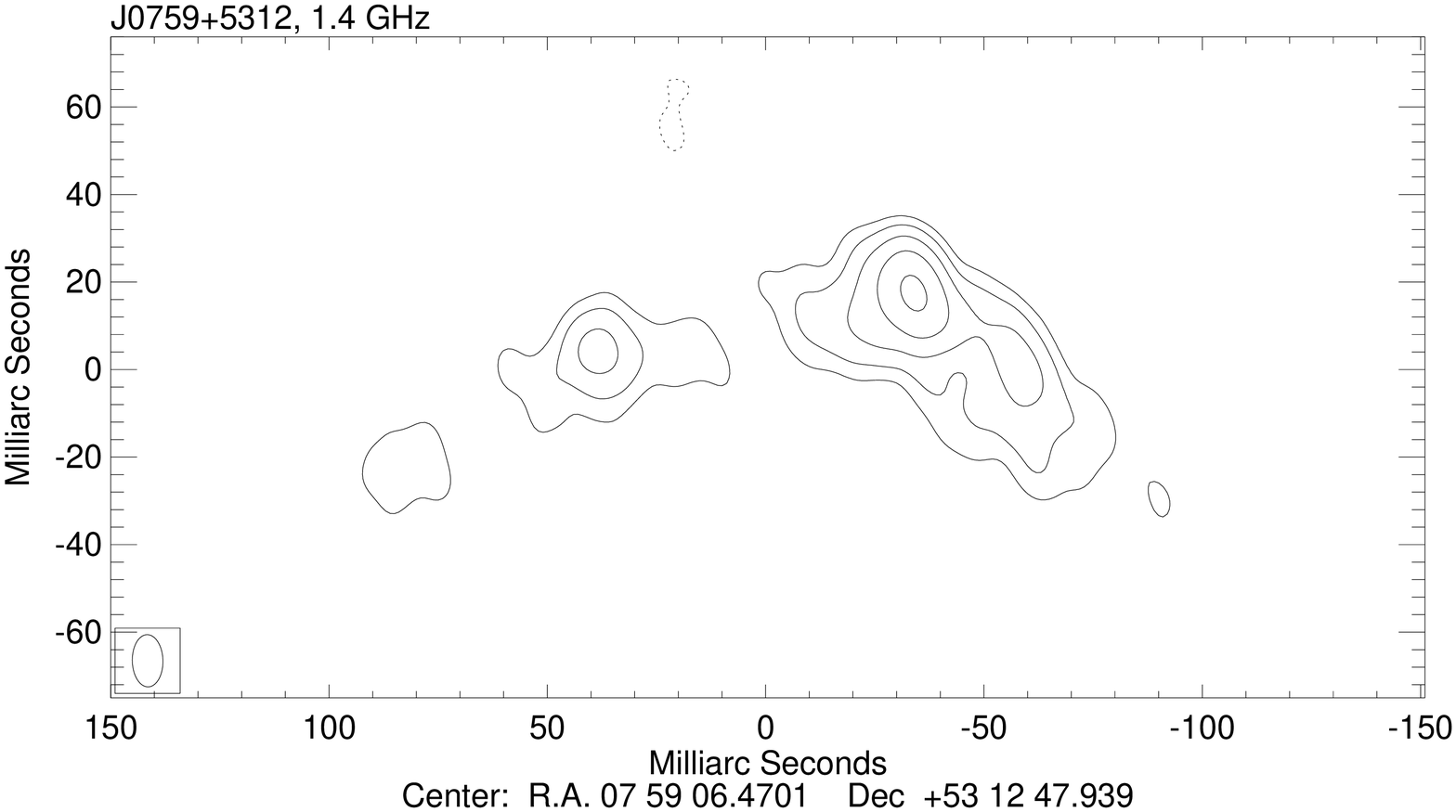}	\\
\includegraphics[scale=0.25]{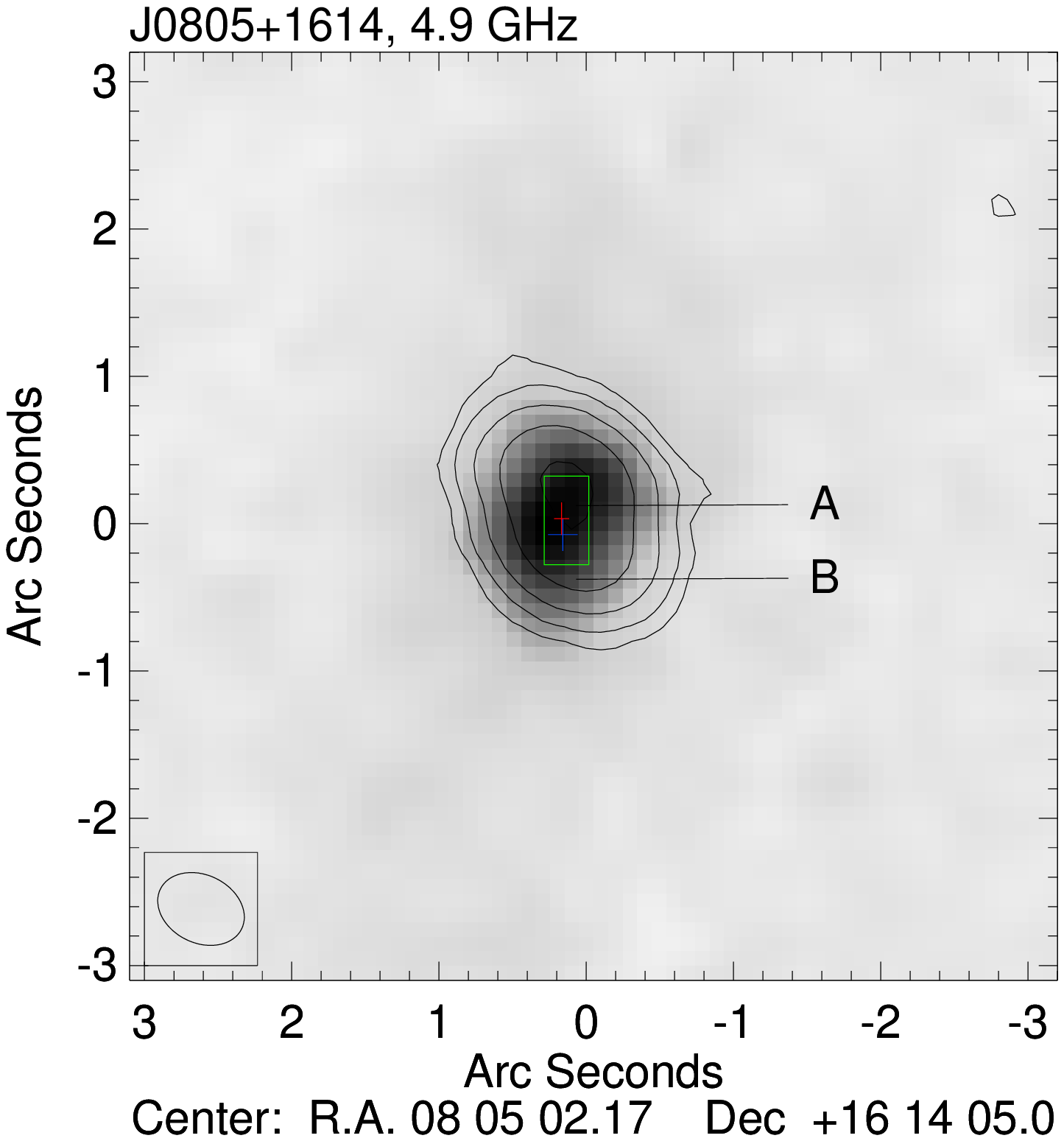}	&	\includegraphics[scale=0.25]{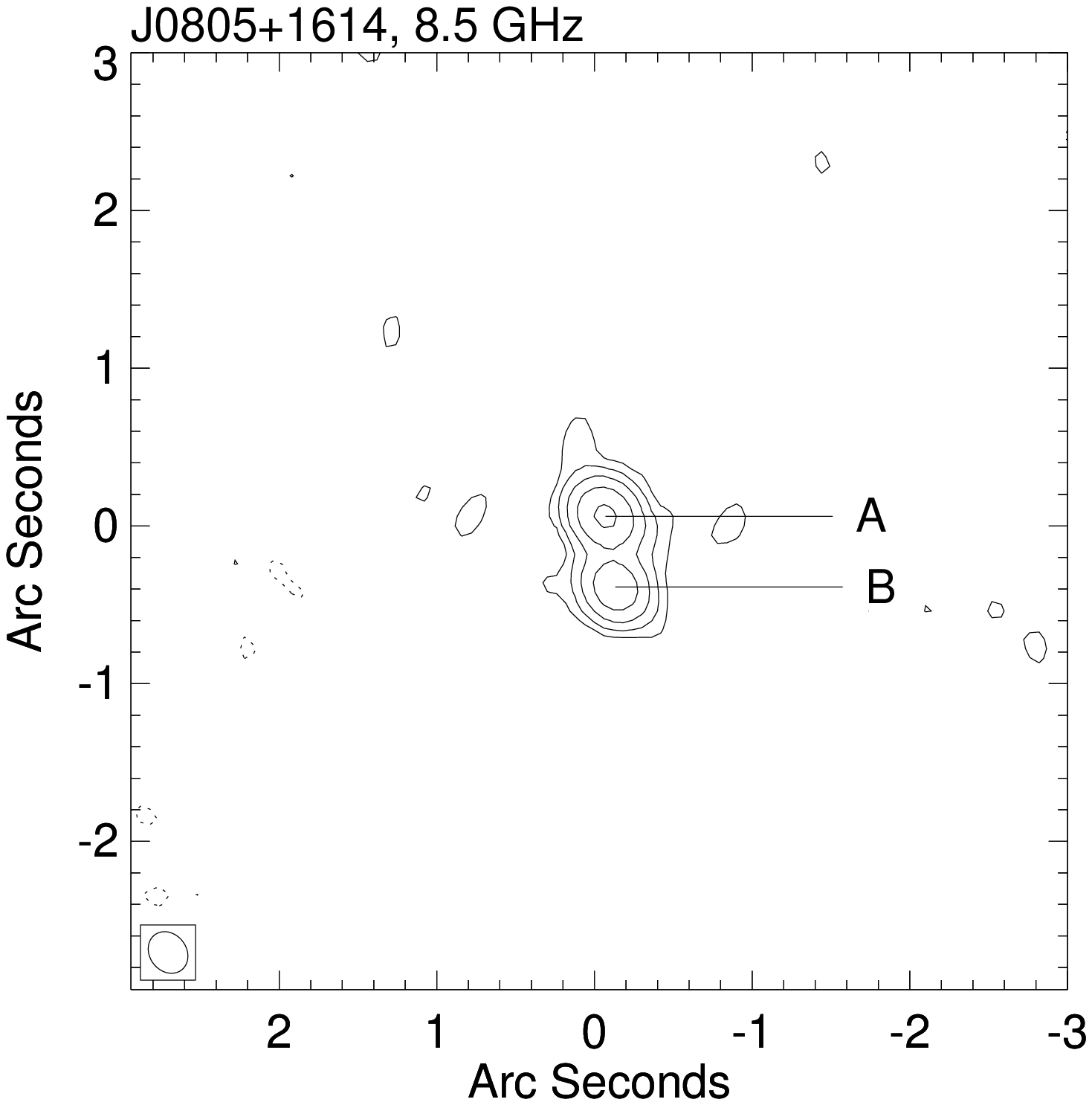}	&	\includegraphics[scale=0.3]{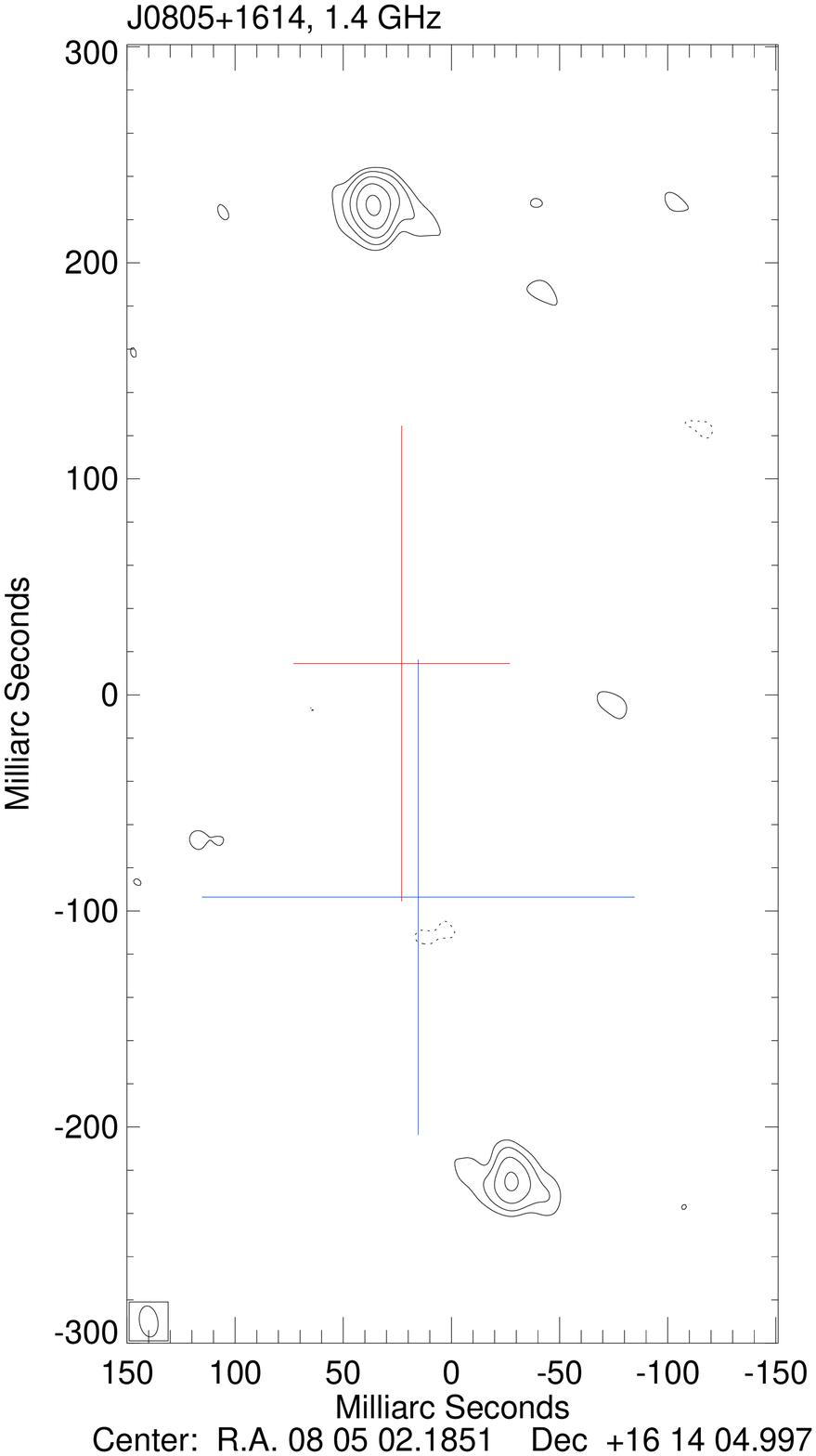}	\\
\includegraphics[scale=0.25]{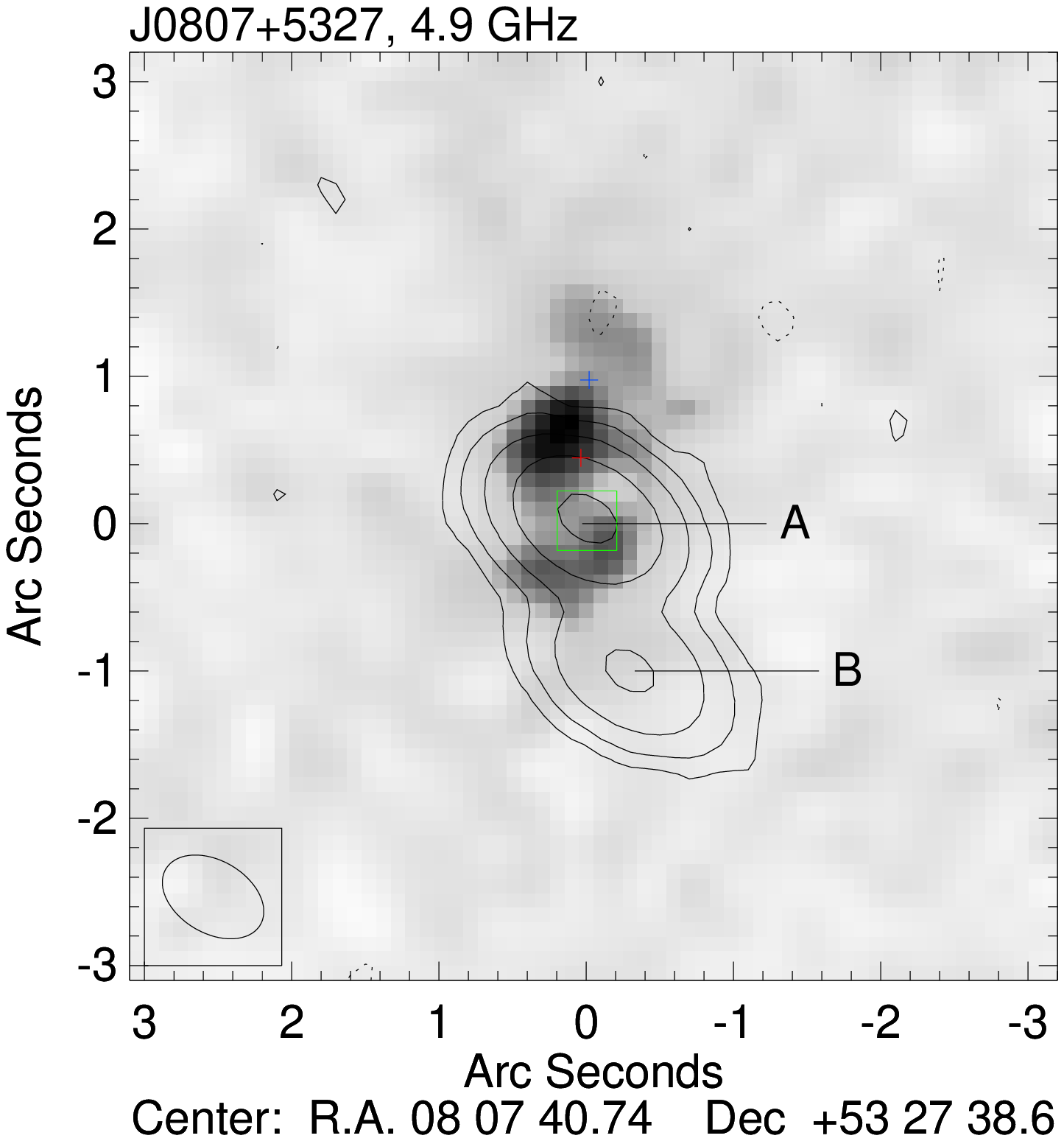}	&	\includegraphics[scale=0.25]{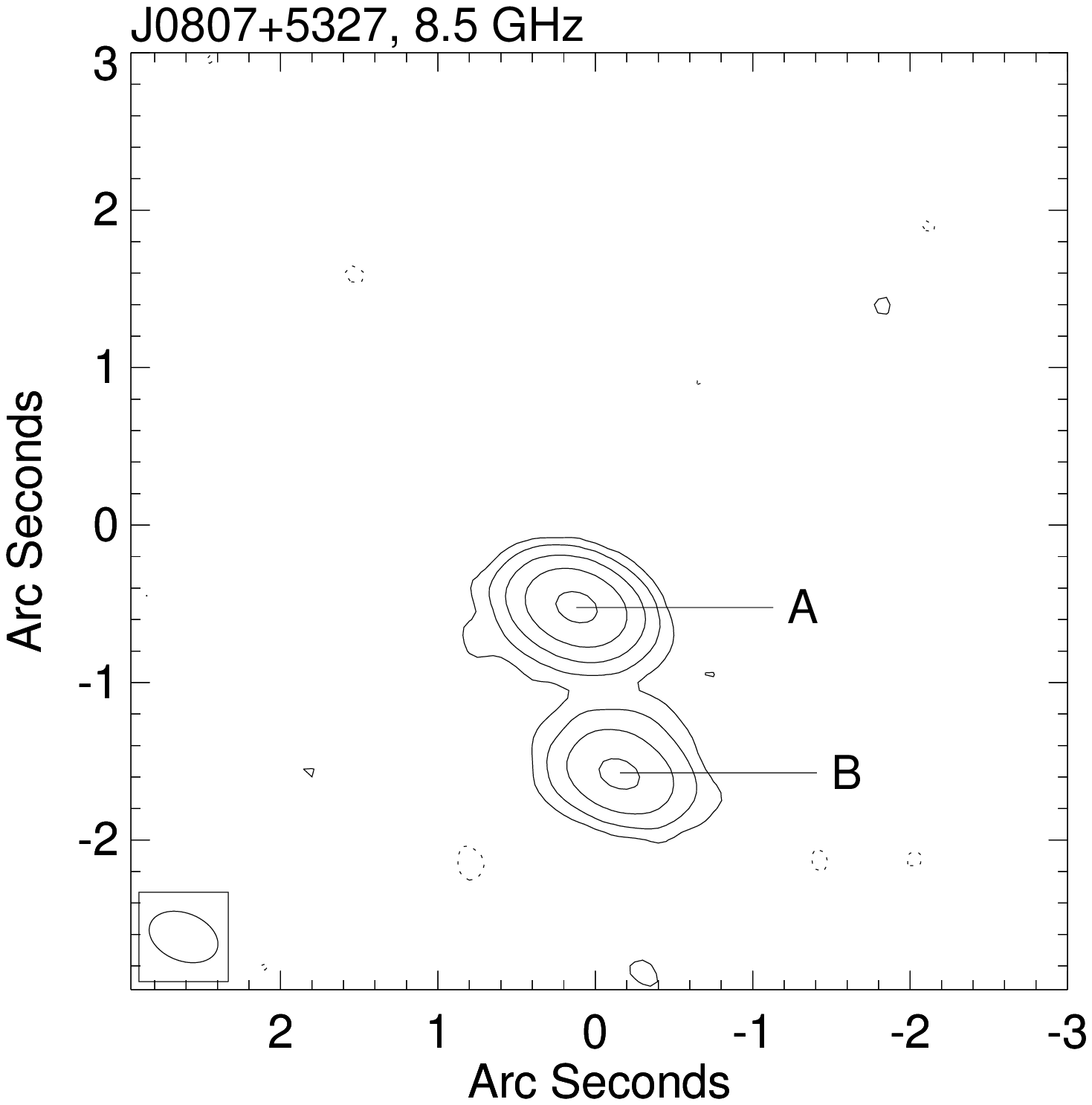}	&	\includegraphics[scale=0.3]{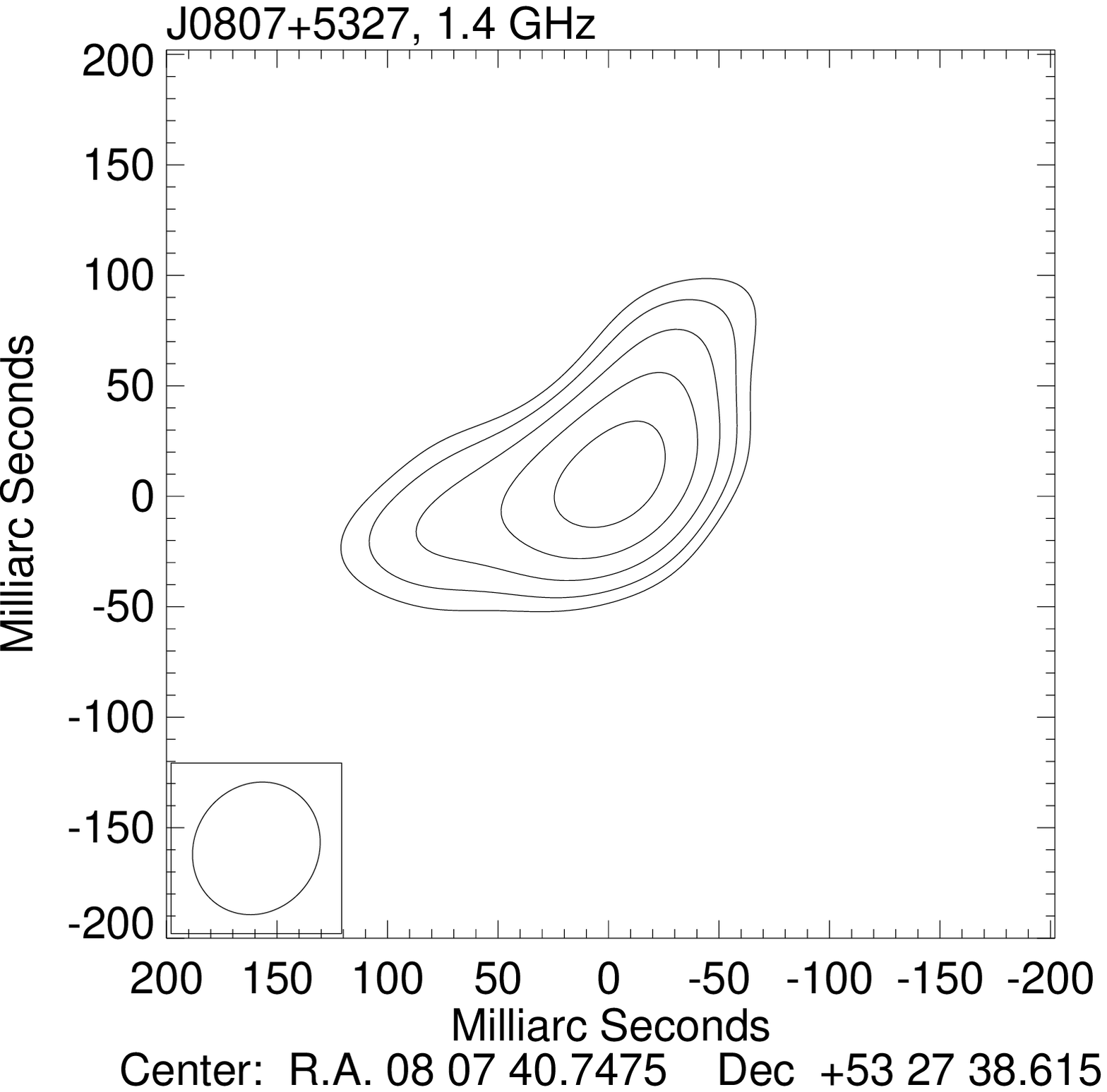}	\\
\includegraphics[scale=0.25]{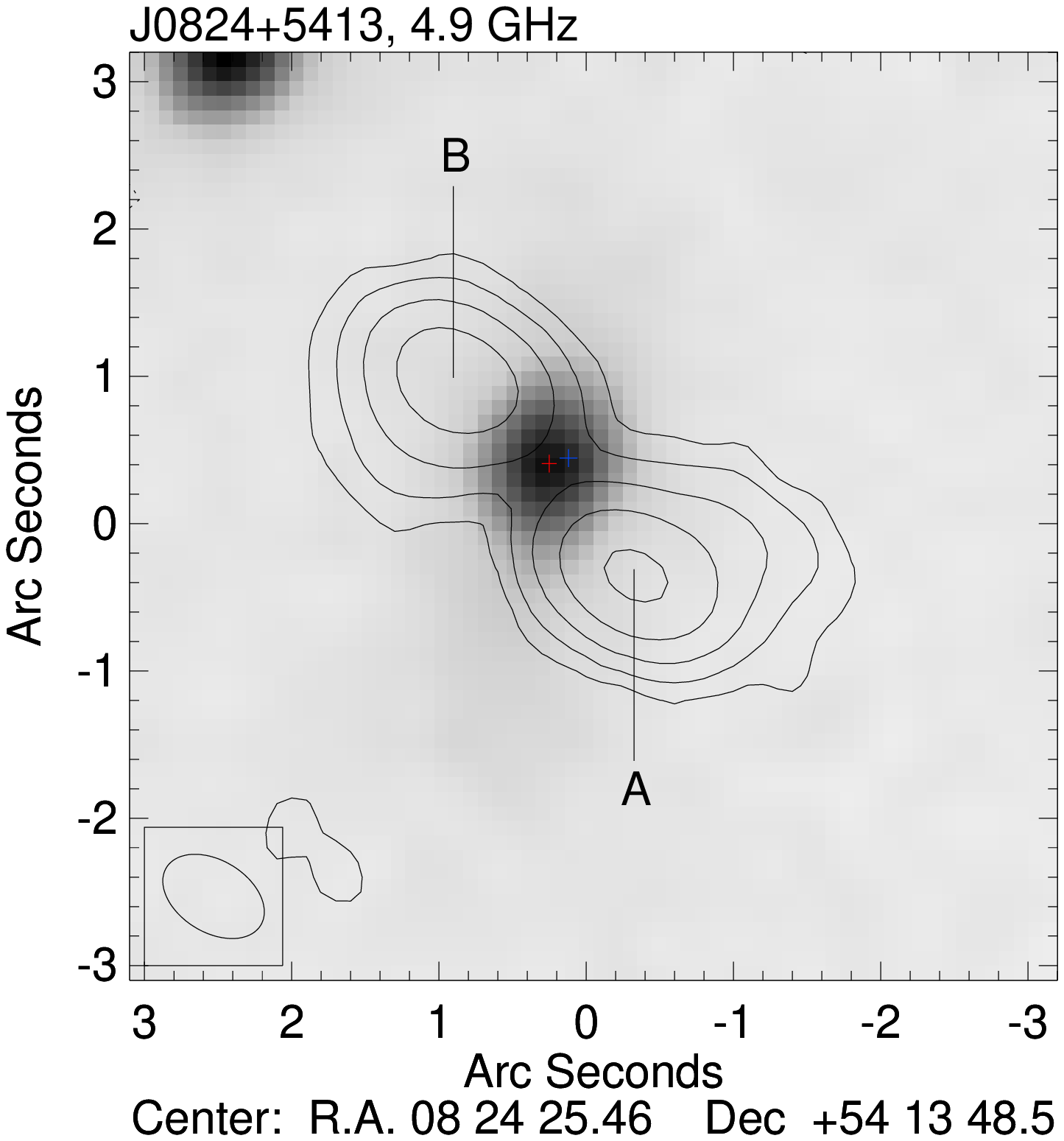}	&	\includegraphics[scale=0.25]{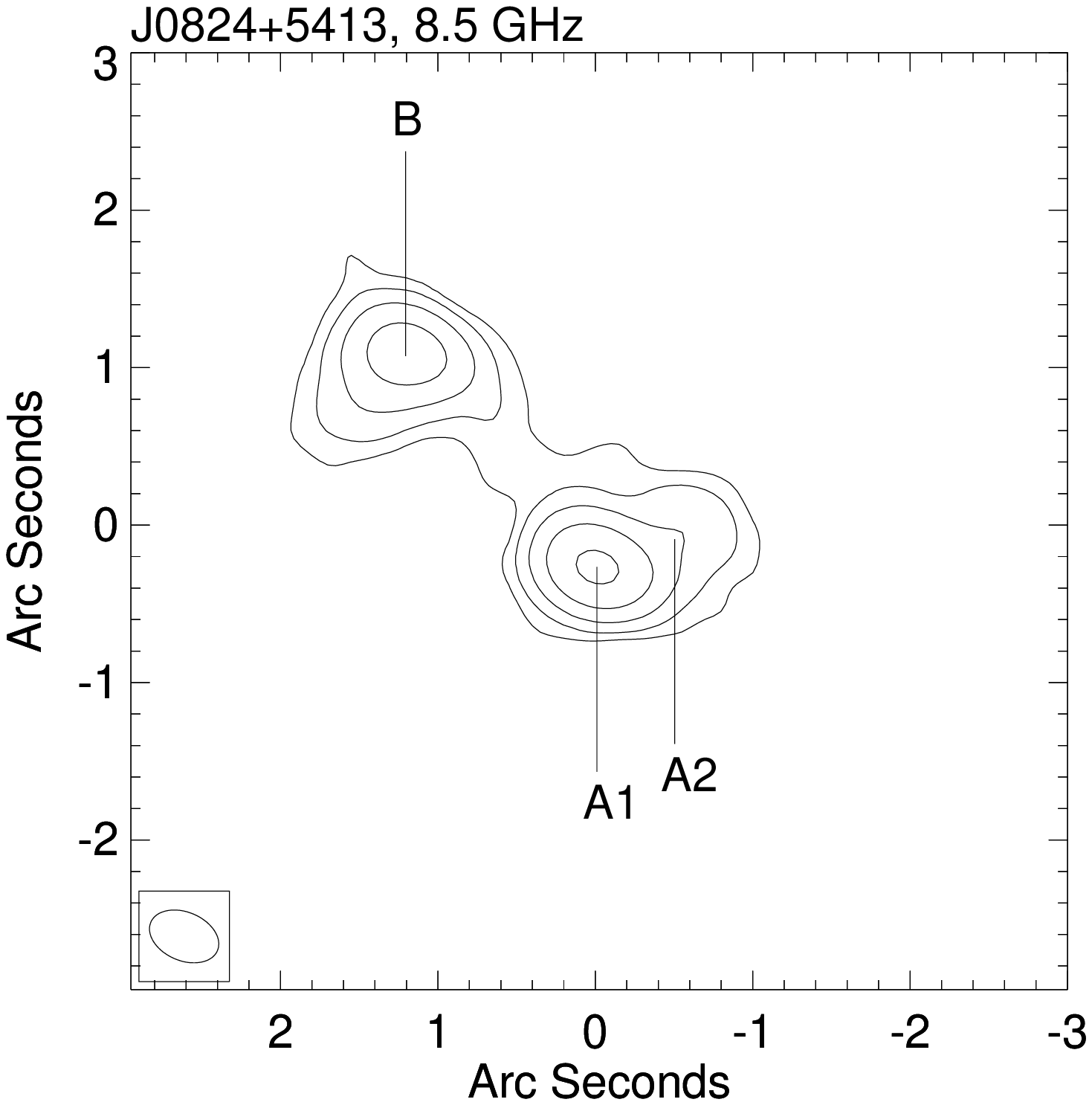}	&		\\

\end{tabular}
\end{figure}

\begin{figure}[htdp]
\begin{tabular}{lll}

	&	\includegraphics[scale=0.25]{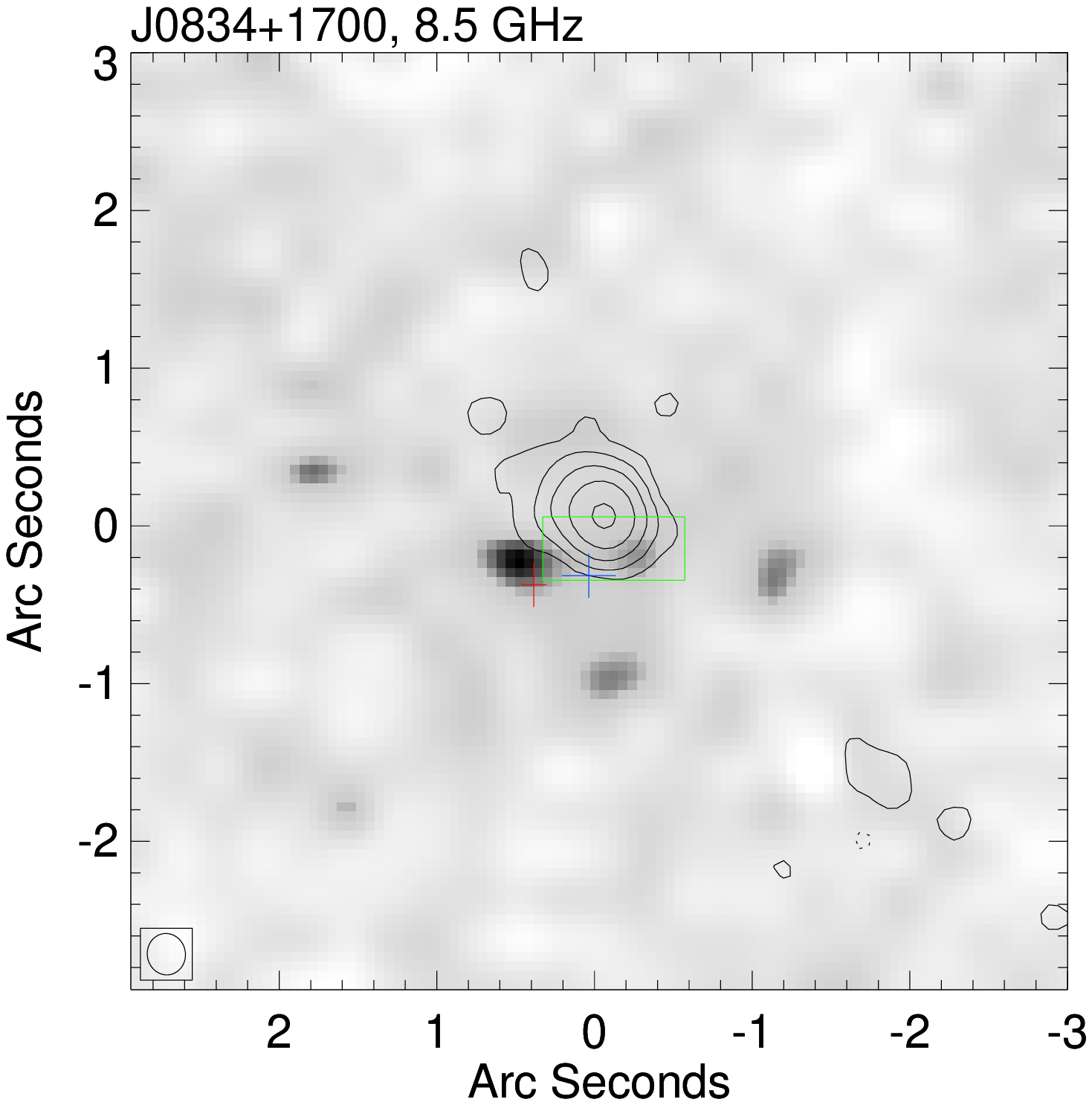}	&	\includegraphics[scale=0.25,bb=50 0 1017 453]{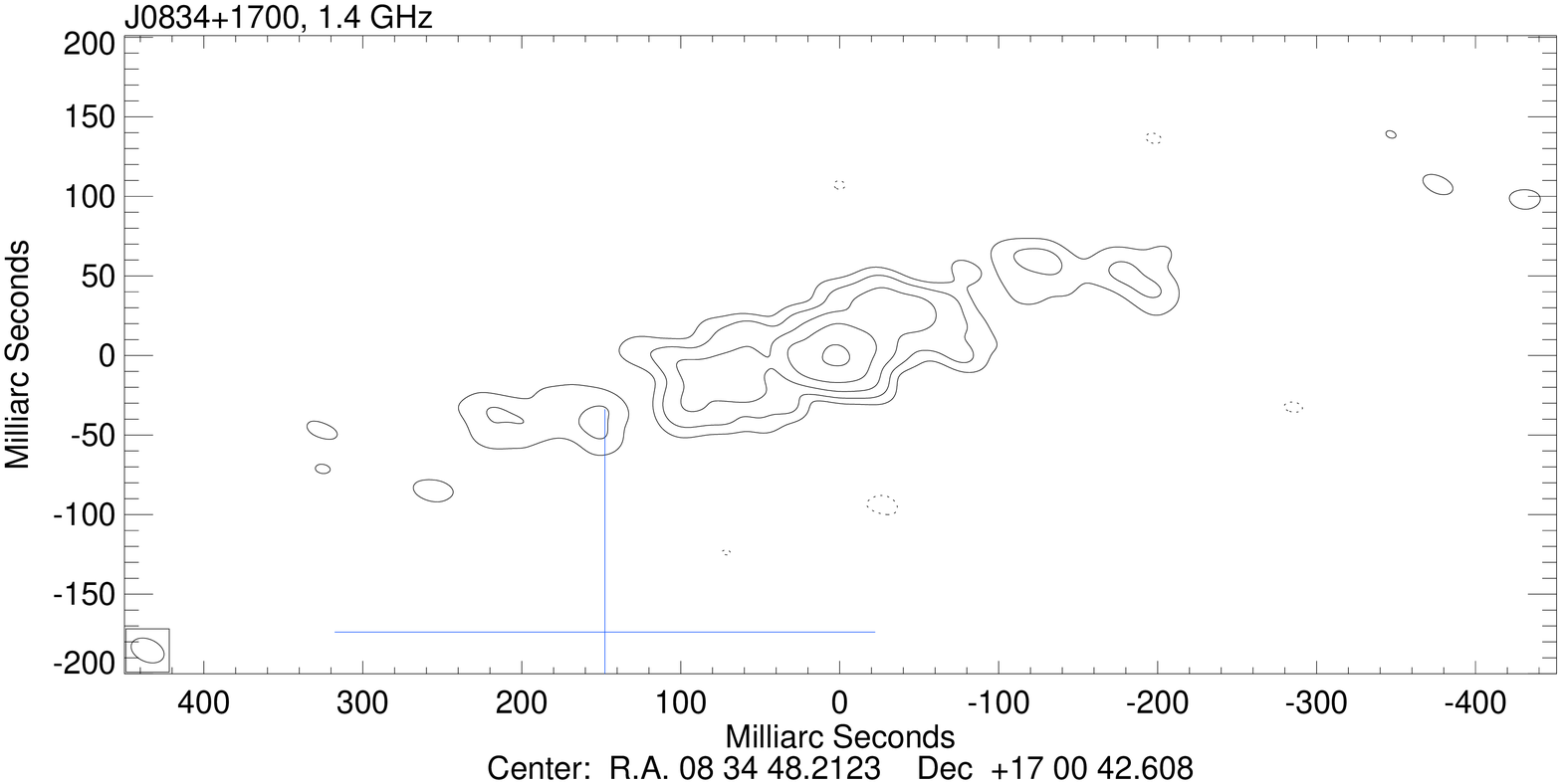}	\\
\includegraphics[scale=0.25]{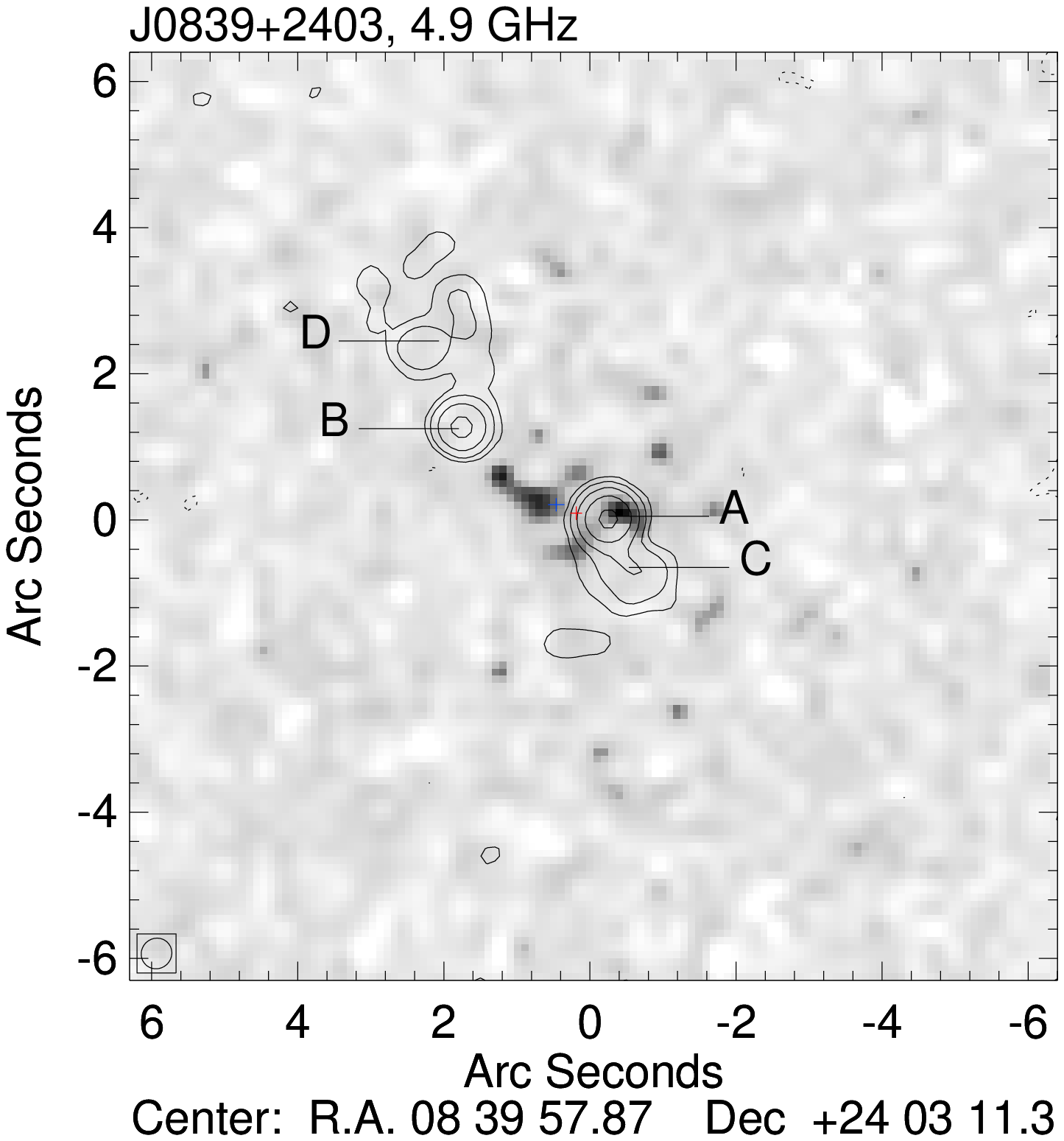}	&		&		\\
\includegraphics[scale=0.25]{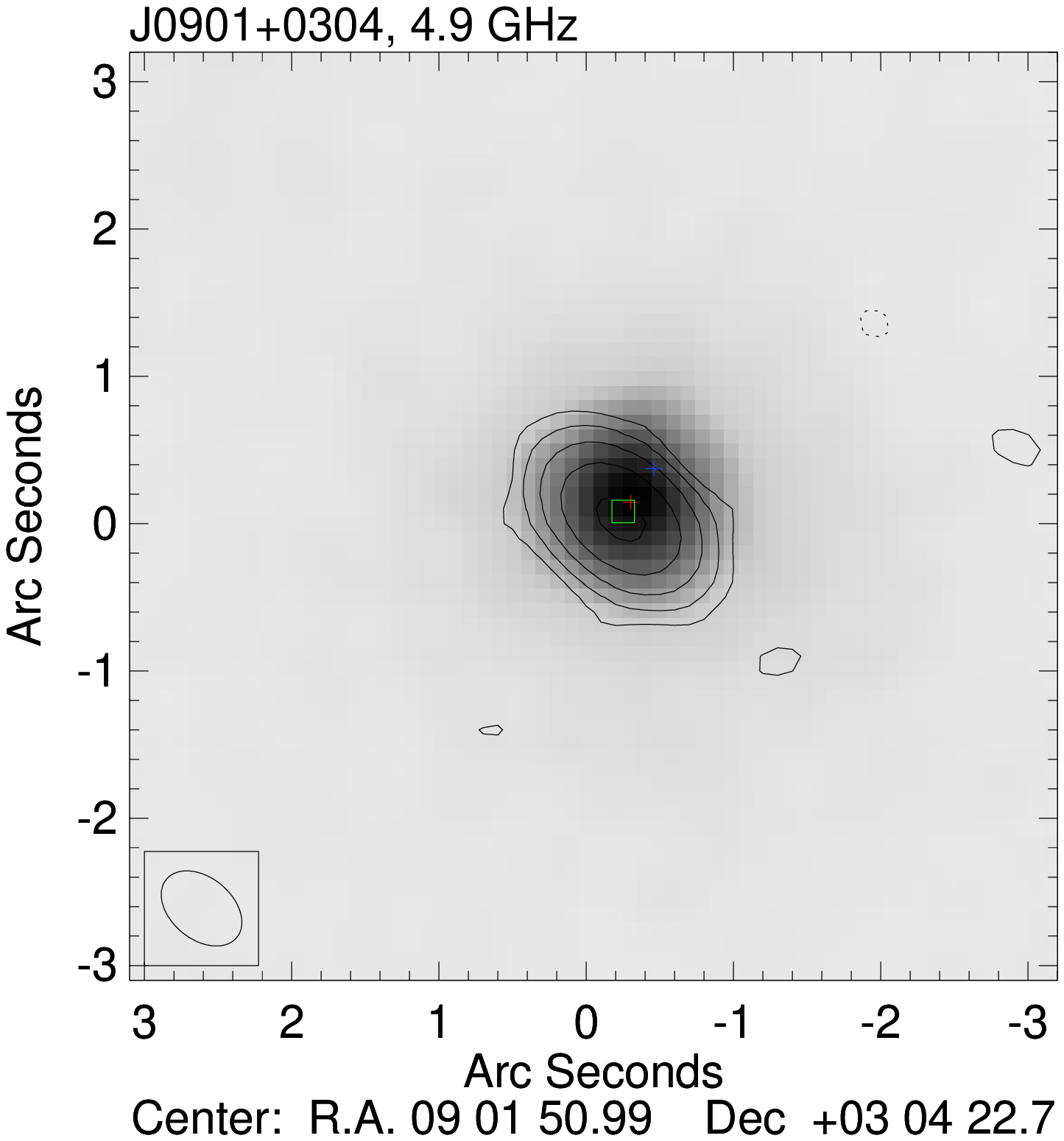}	&		&	\includegraphics[scale=0.25]{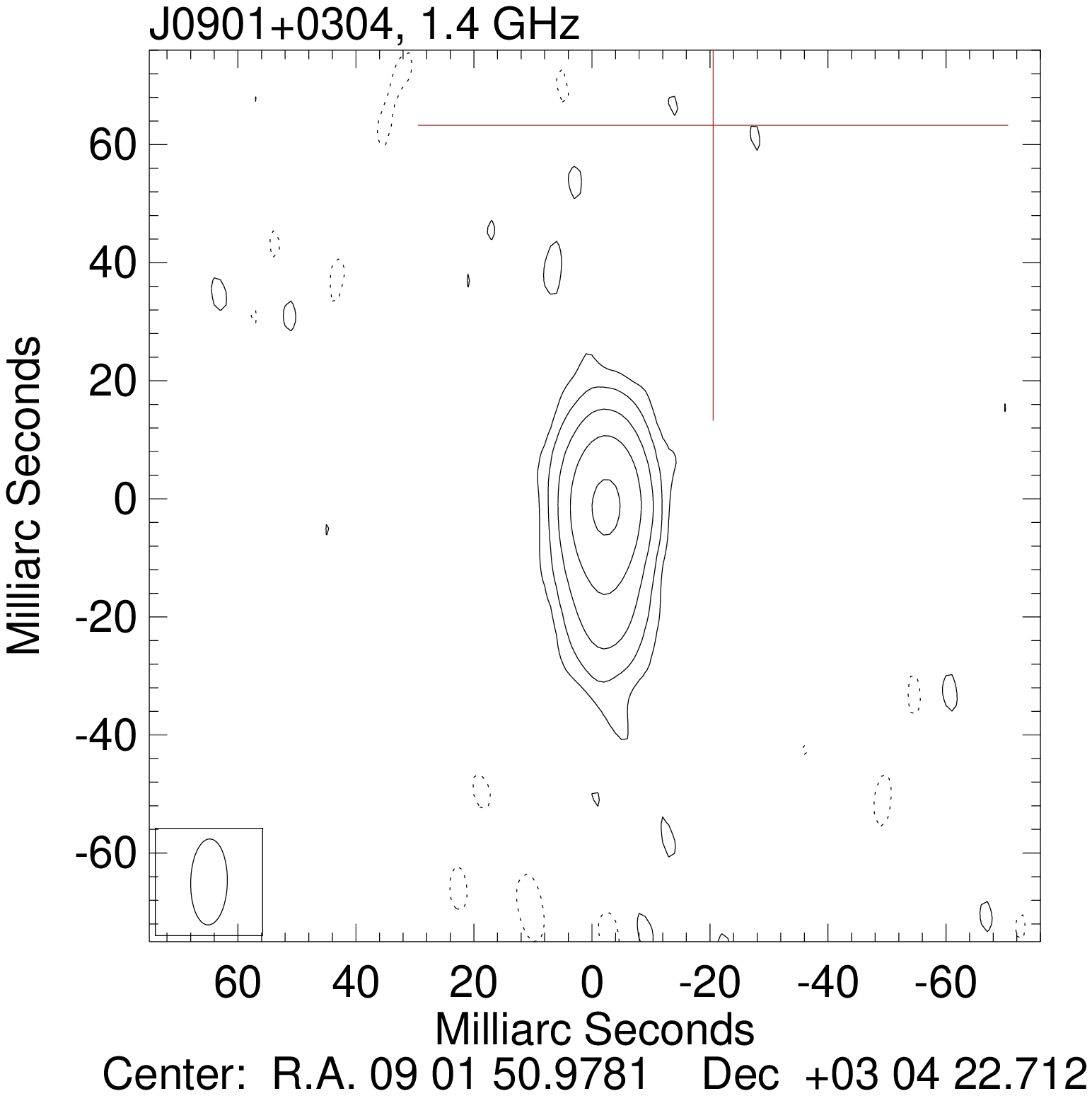}	\\
\includegraphics[scale=0.25]{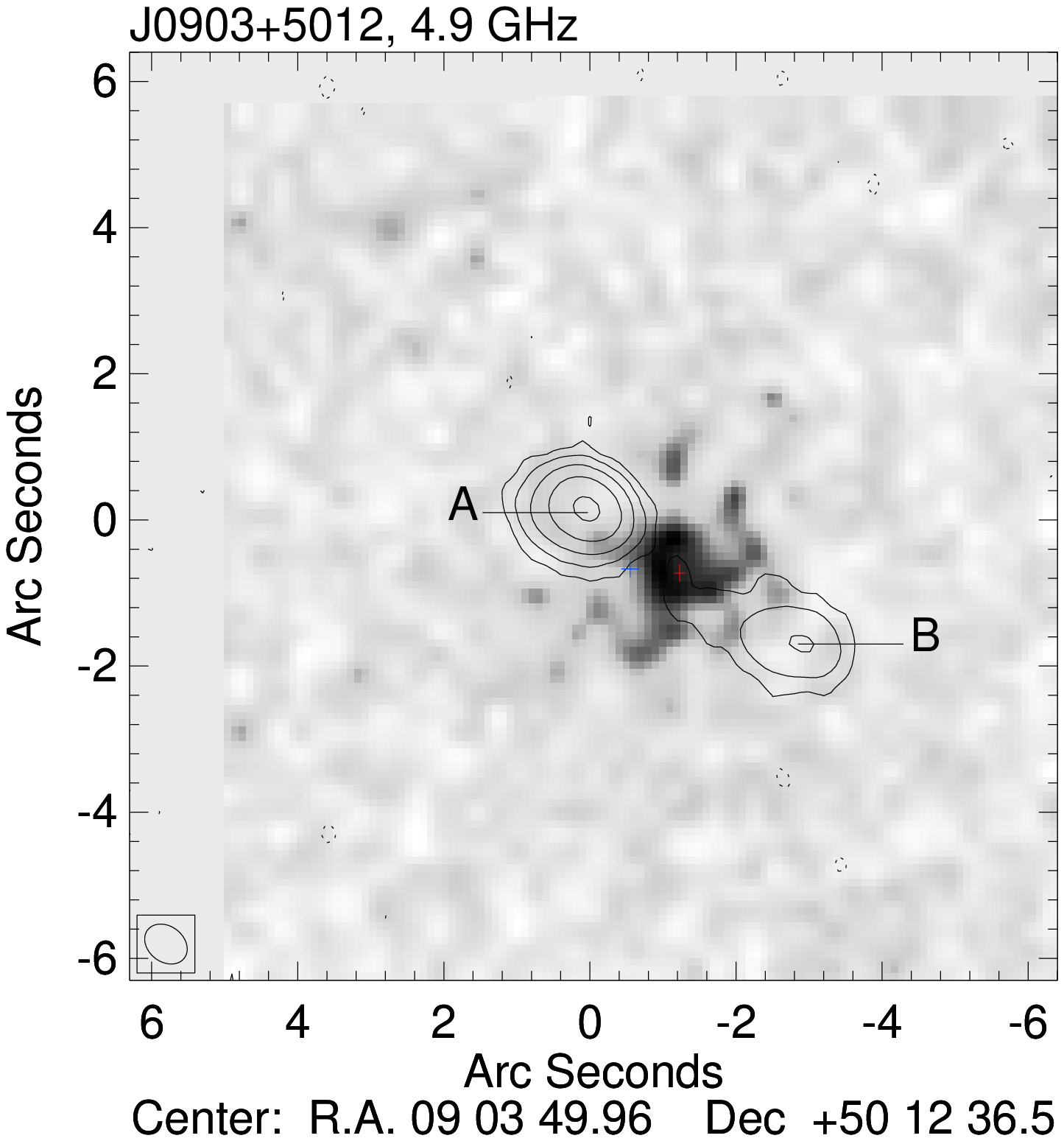}	&	\includegraphics[scale=0.25]{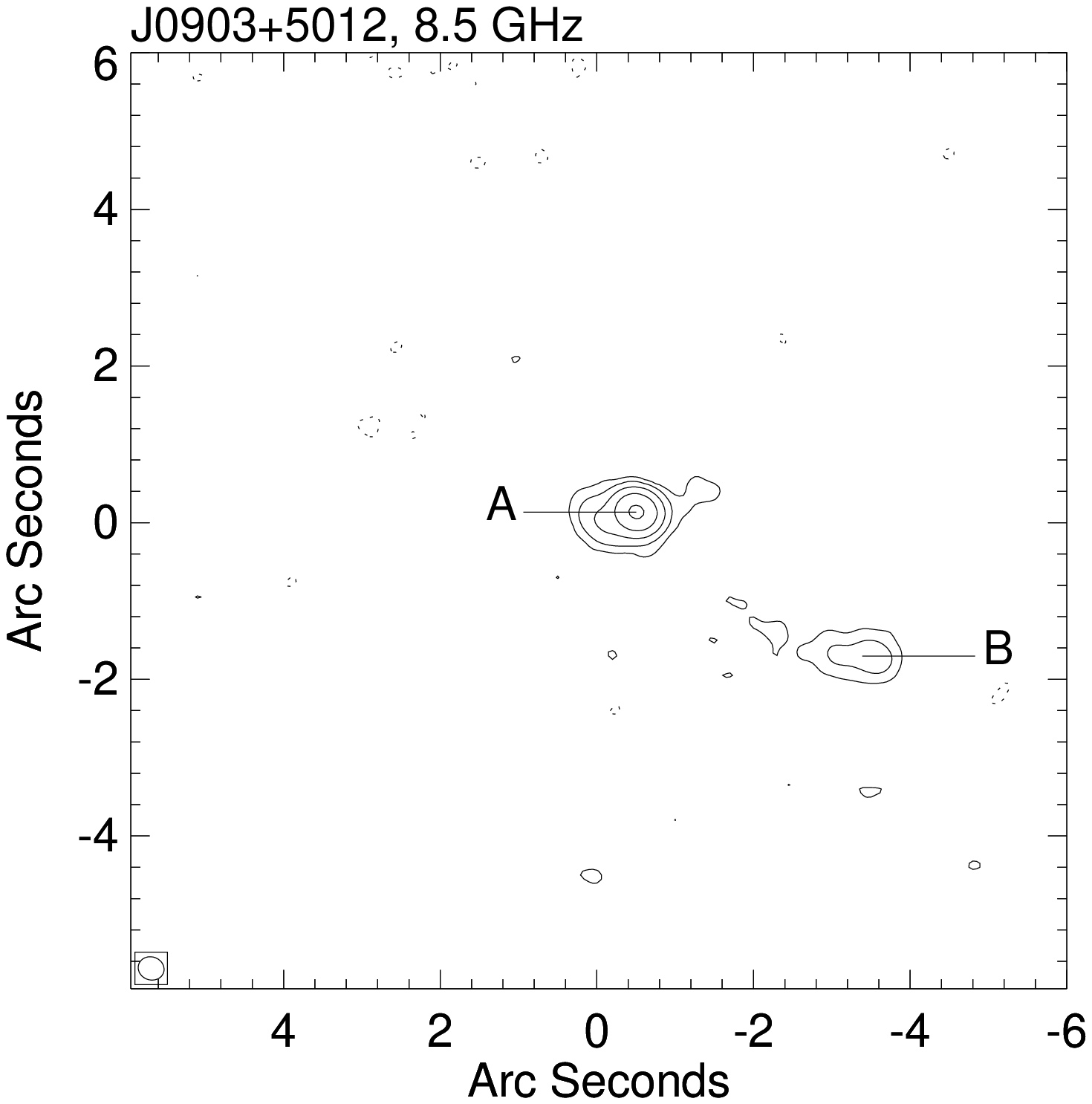}	&		\\
\includegraphics[scale=0.25]{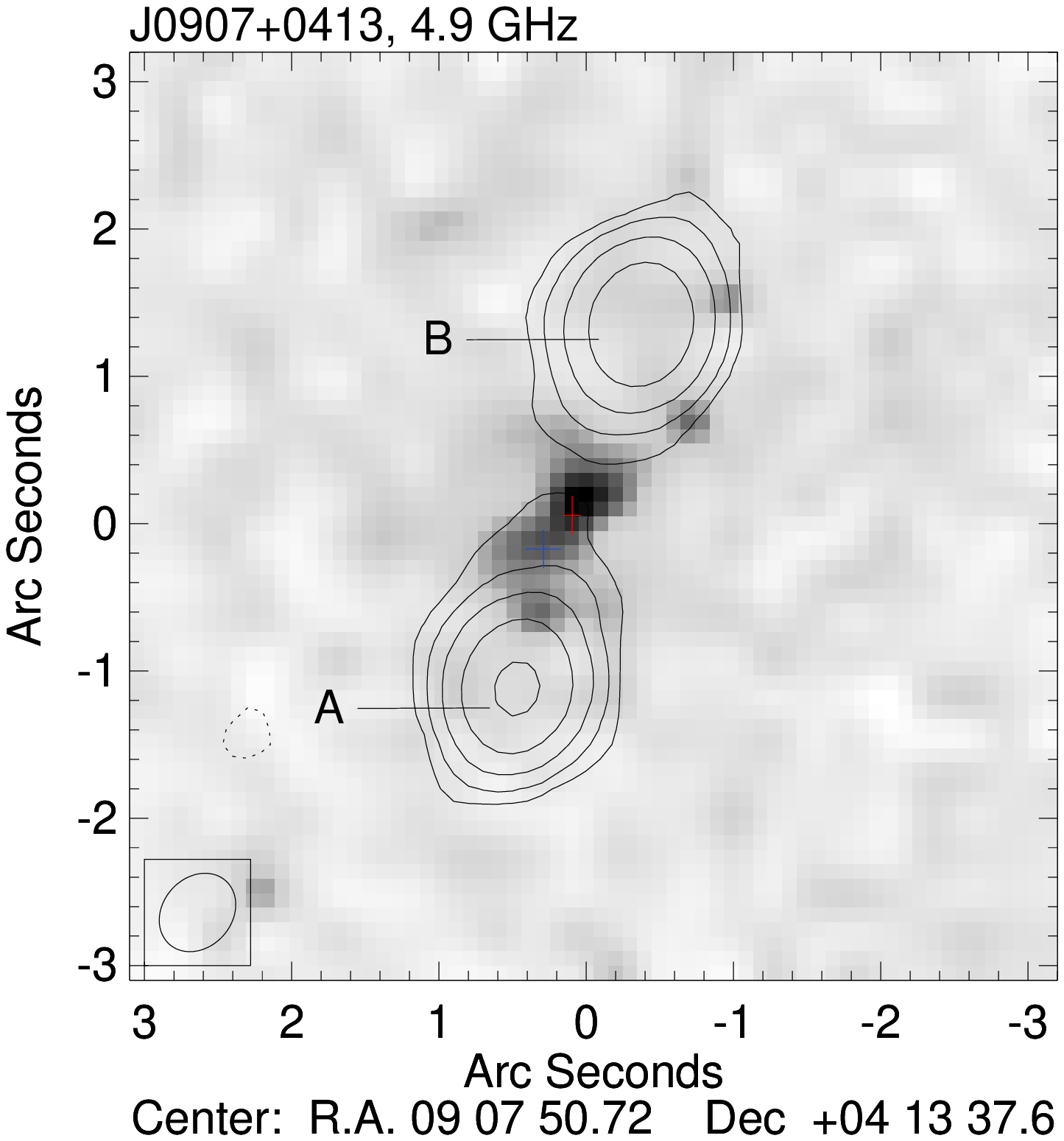}	&		&		\\

\end{tabular}
\end{figure}

\begin{figure}[htdp]
\begin{tabular}{lll}

\includegraphics[scale=0.25]{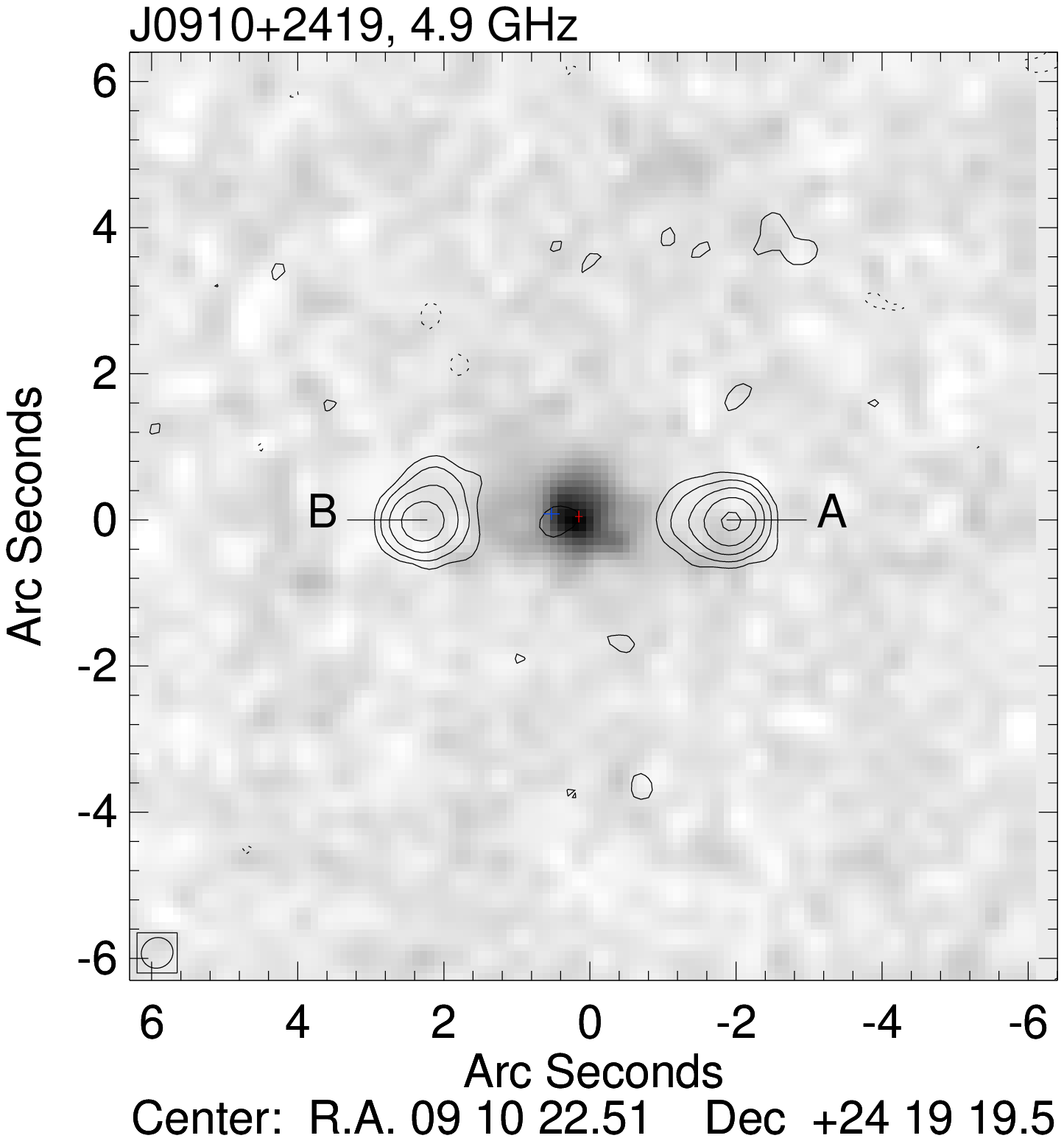}	&		&		\\
\includegraphics[scale=0.25]{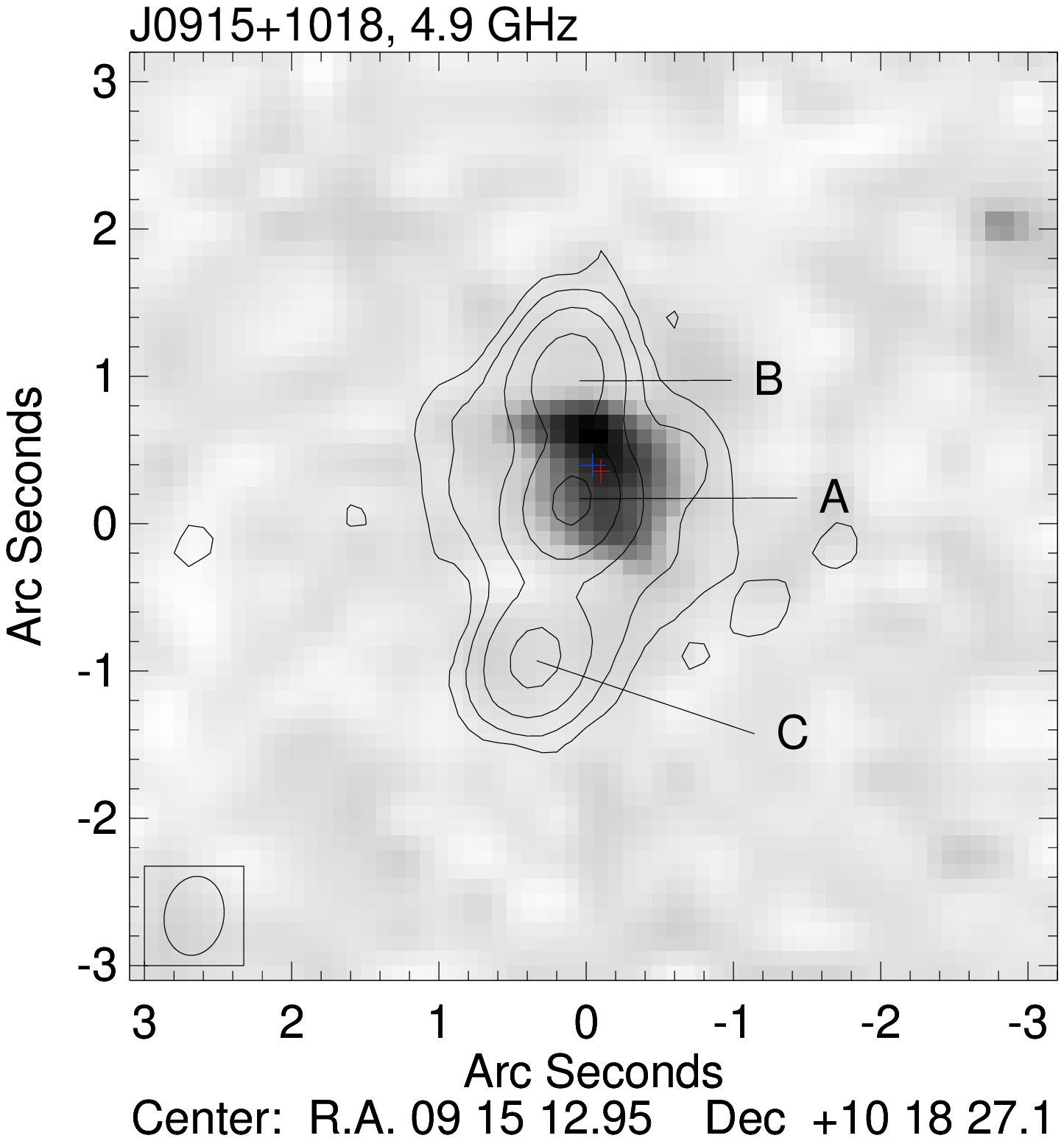}	&		&		\\
\includegraphics[scale=0.25]{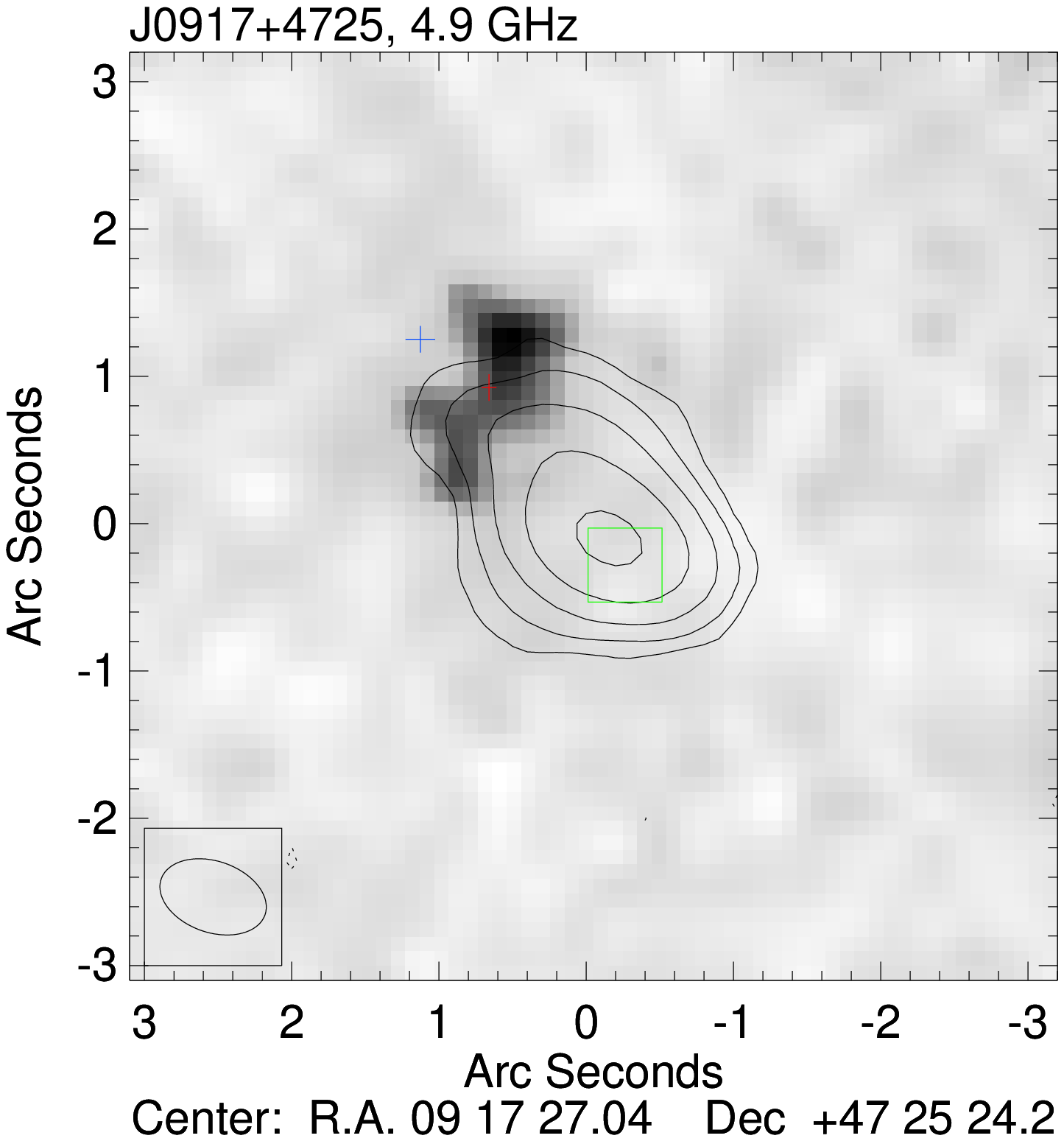}	&	\includegraphics[scale=0.25]{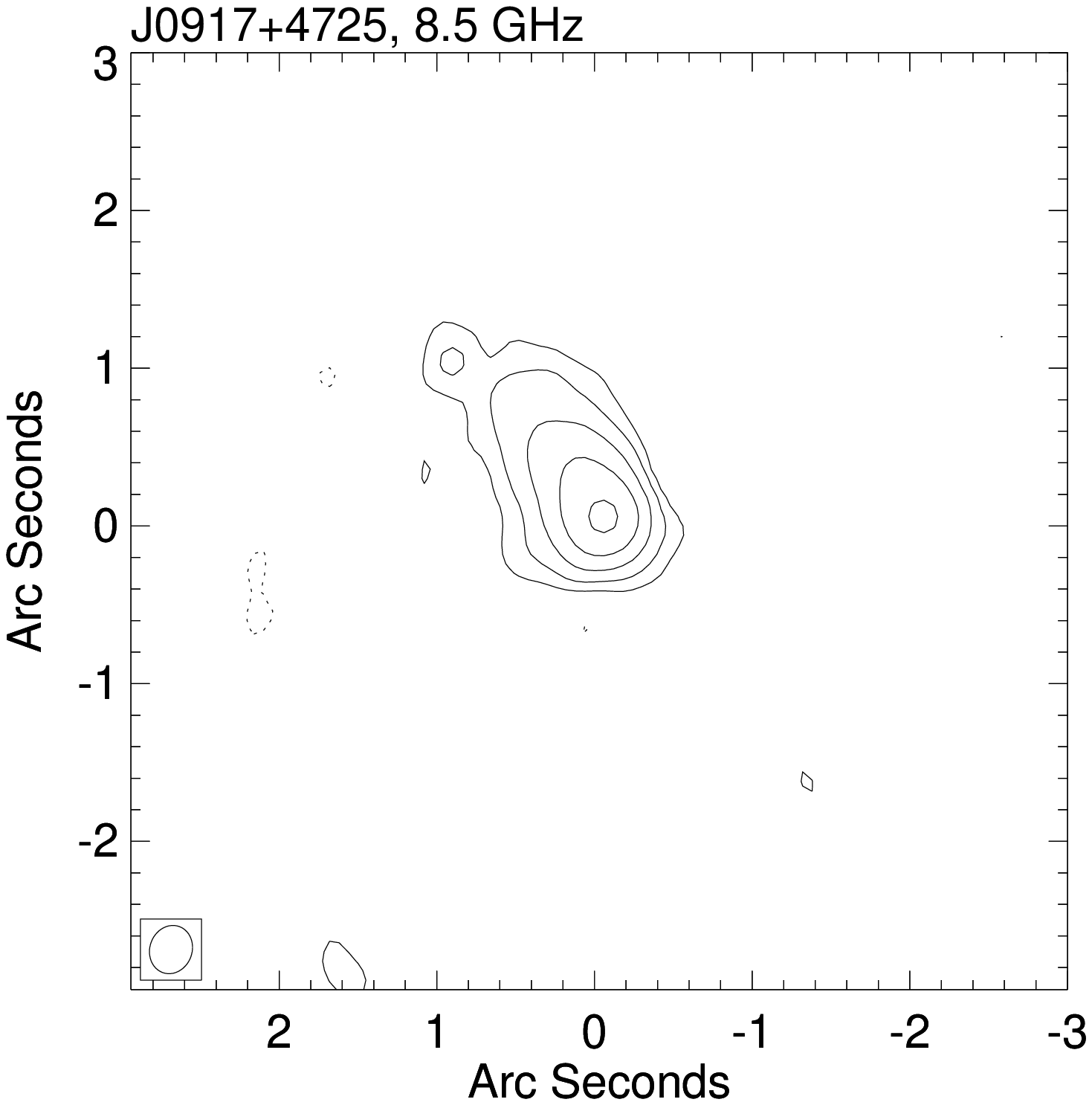}	&	\includegraphics[scale=0.25]{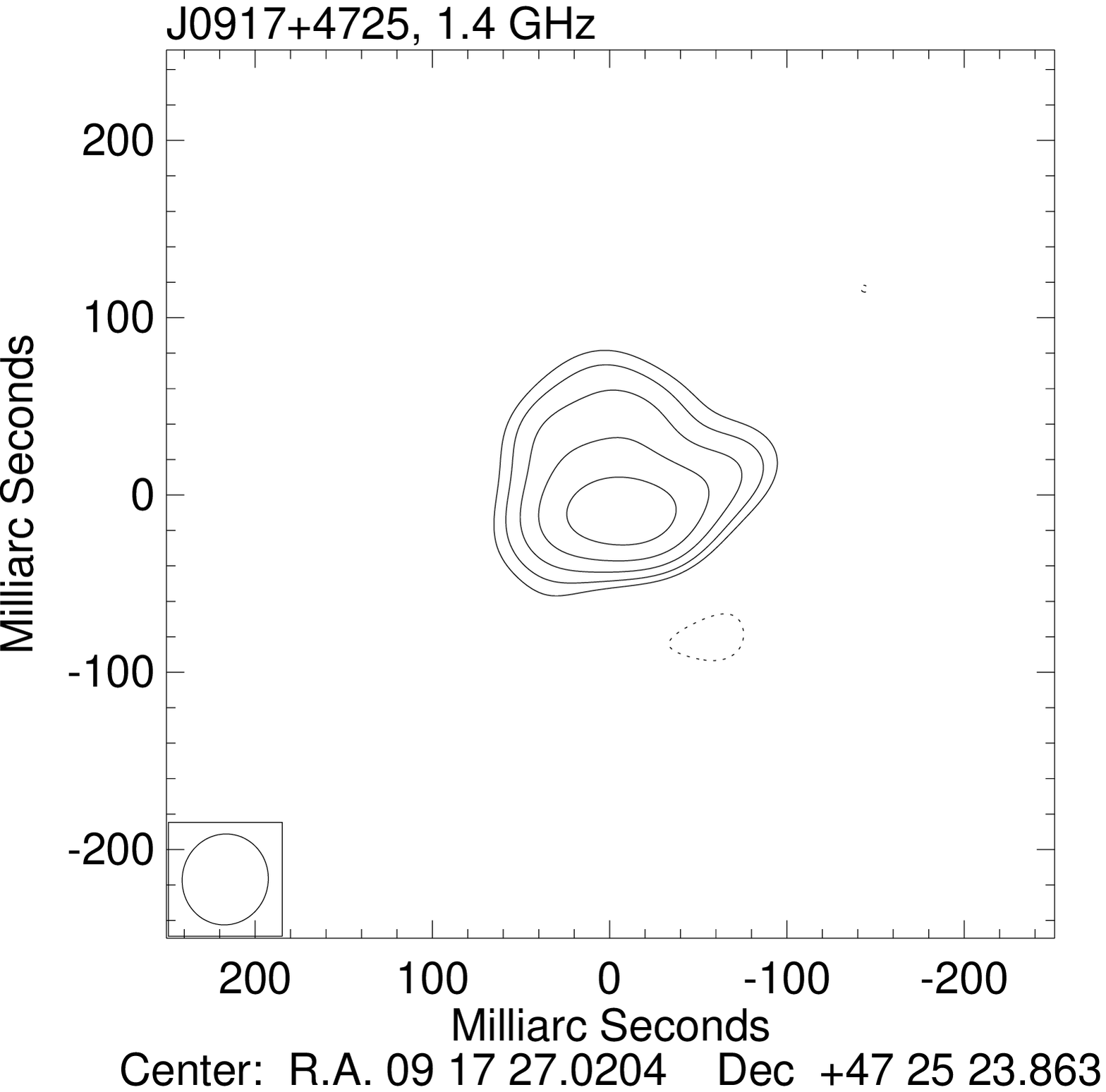}	\\
\includegraphics[scale=0.25]{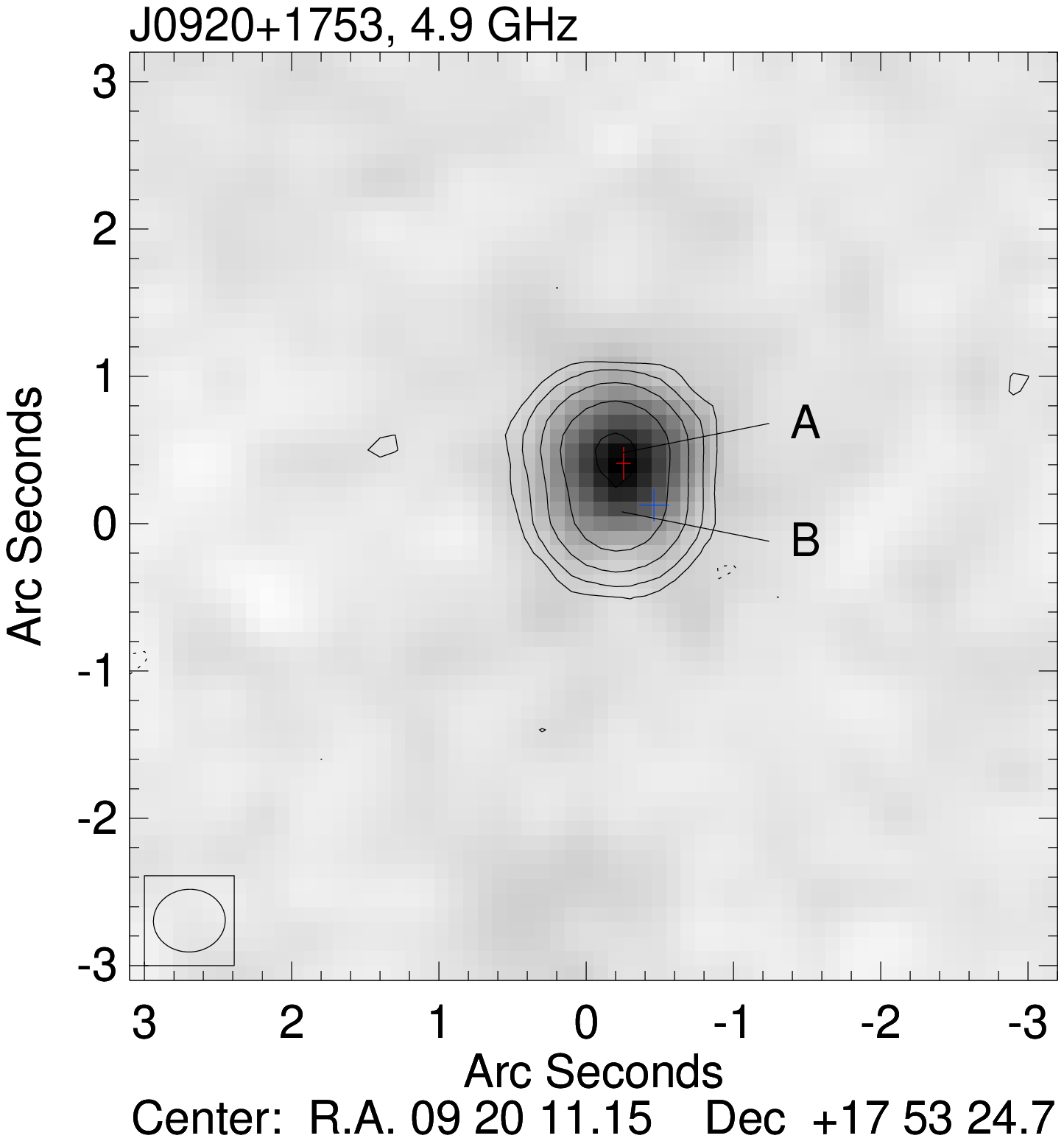}	&		&		\\
\includegraphics[scale=0.25]{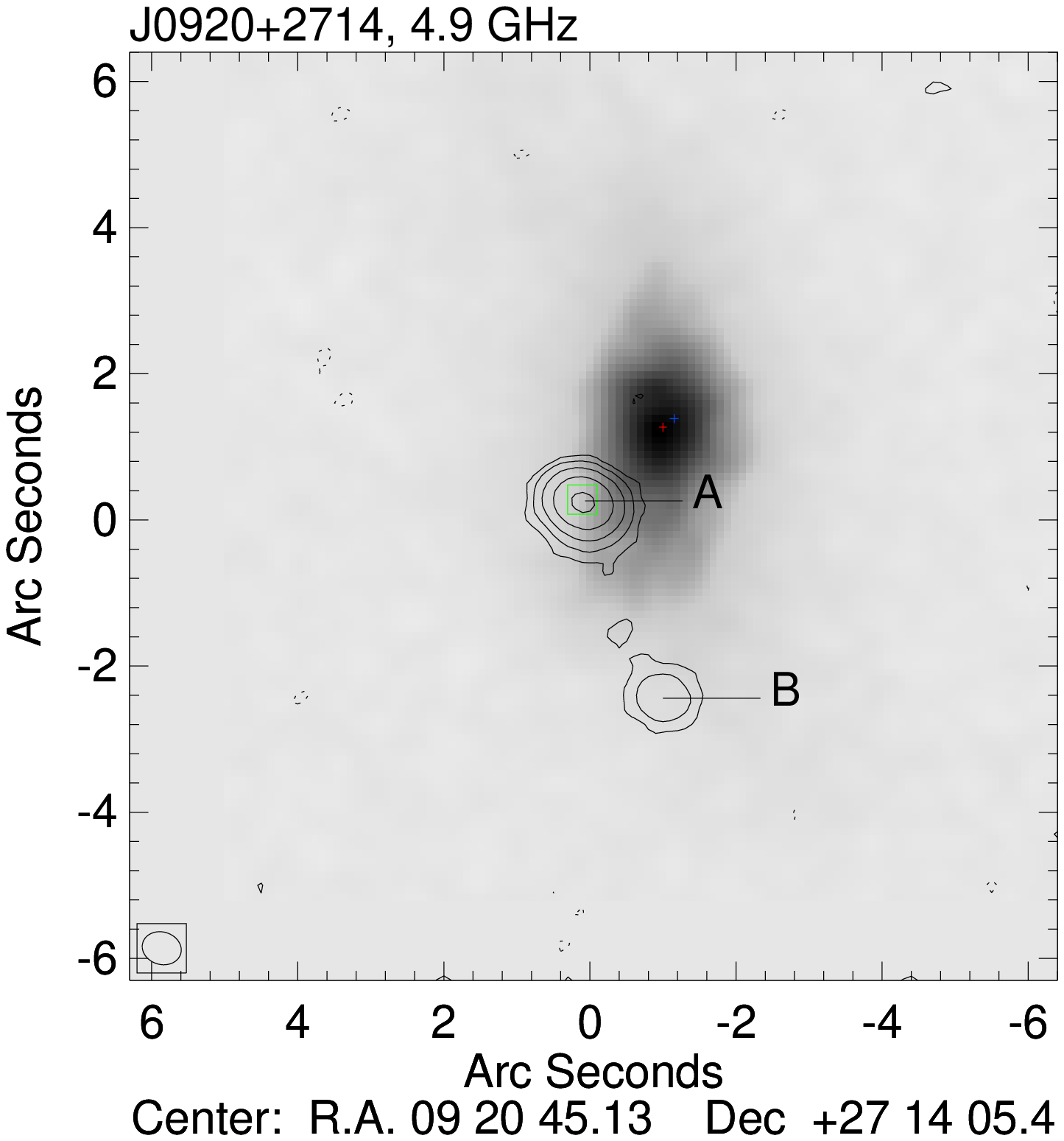}	&	\includegraphics[scale=0.25]{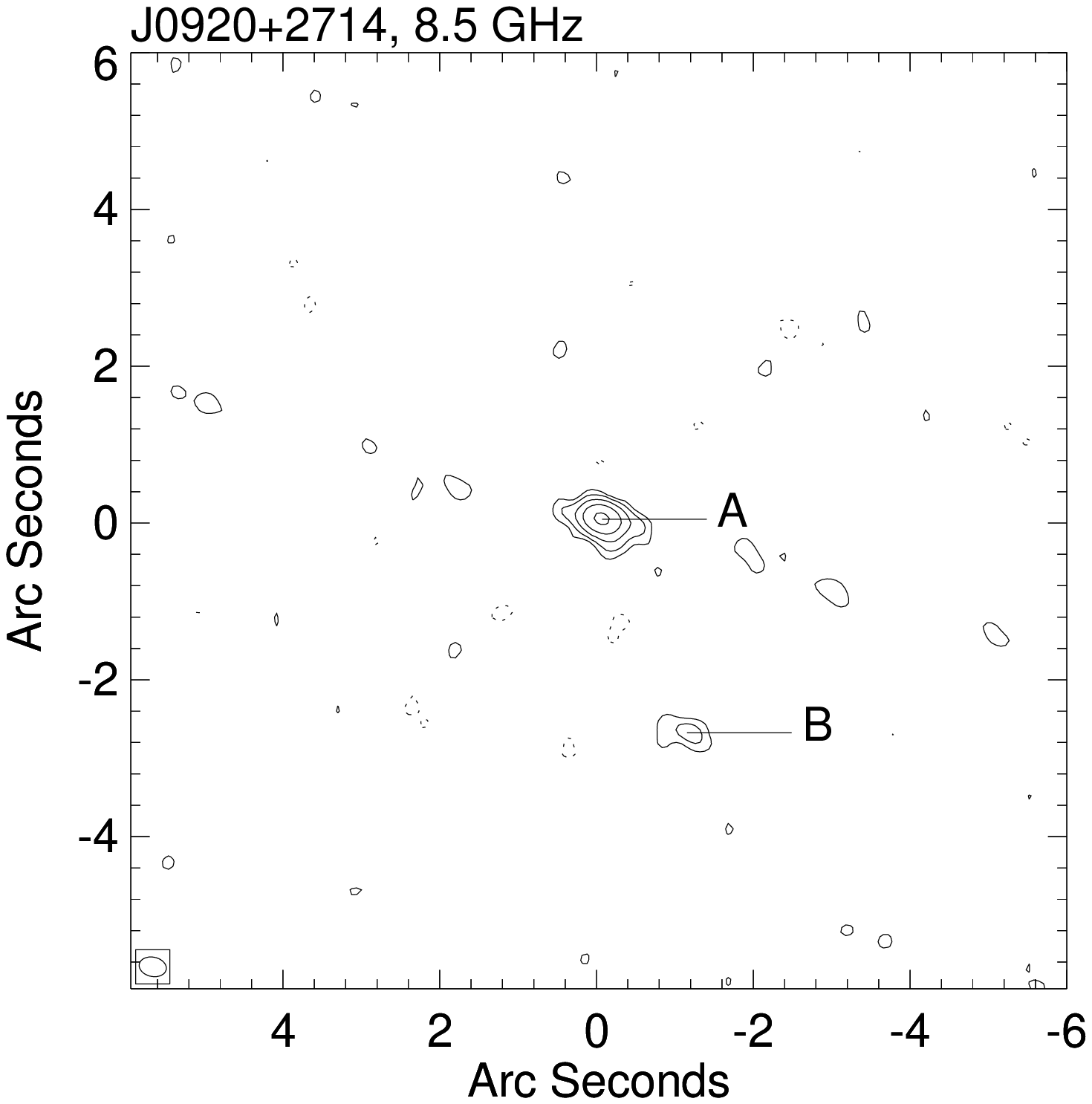}	&	\includegraphics[scale=0.25]{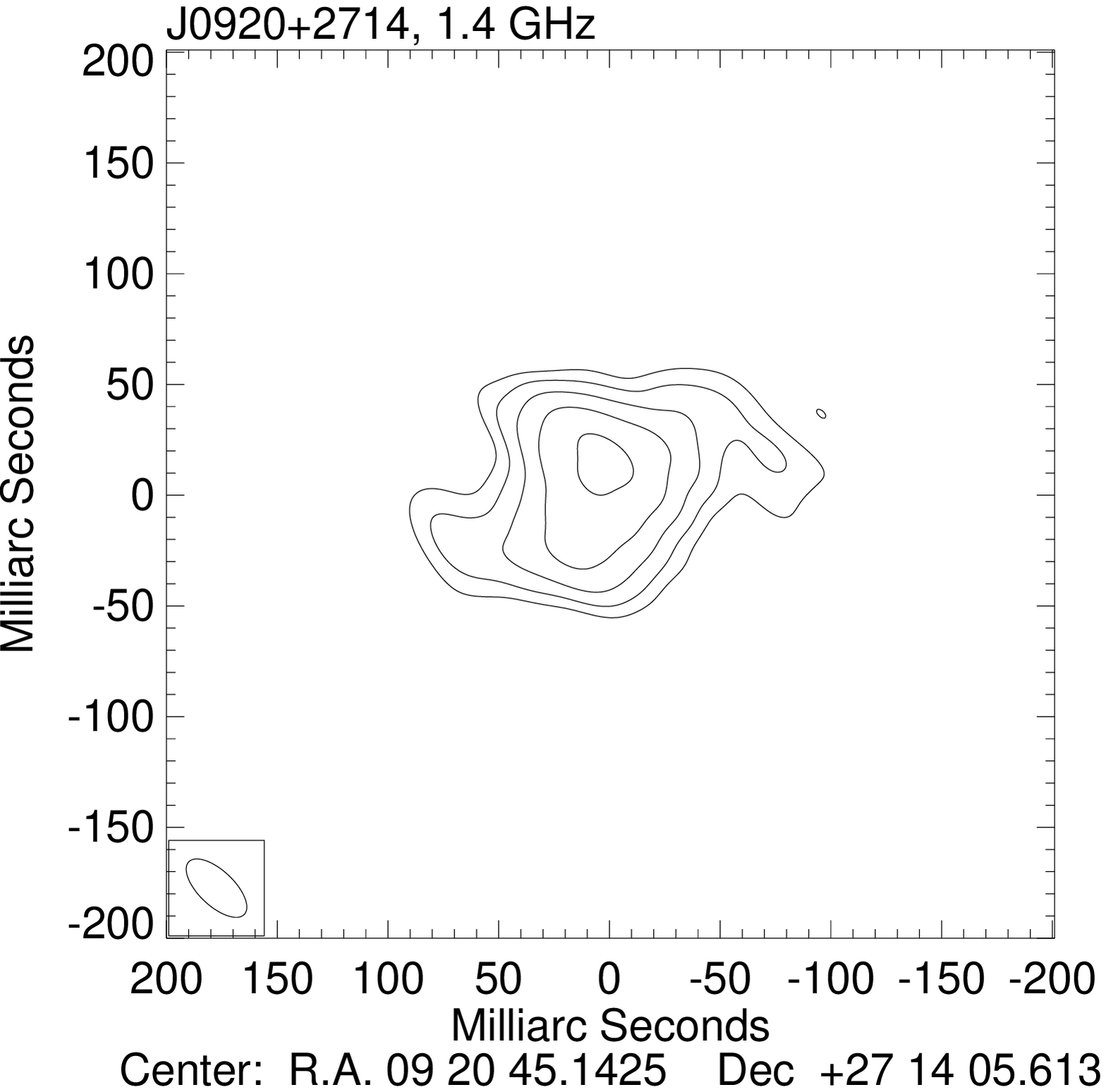}	\\

\end{tabular}
\end{figure}

\begin{figure}[htdp]
\begin{tabular}{lll}

\includegraphics[scale=0.25]{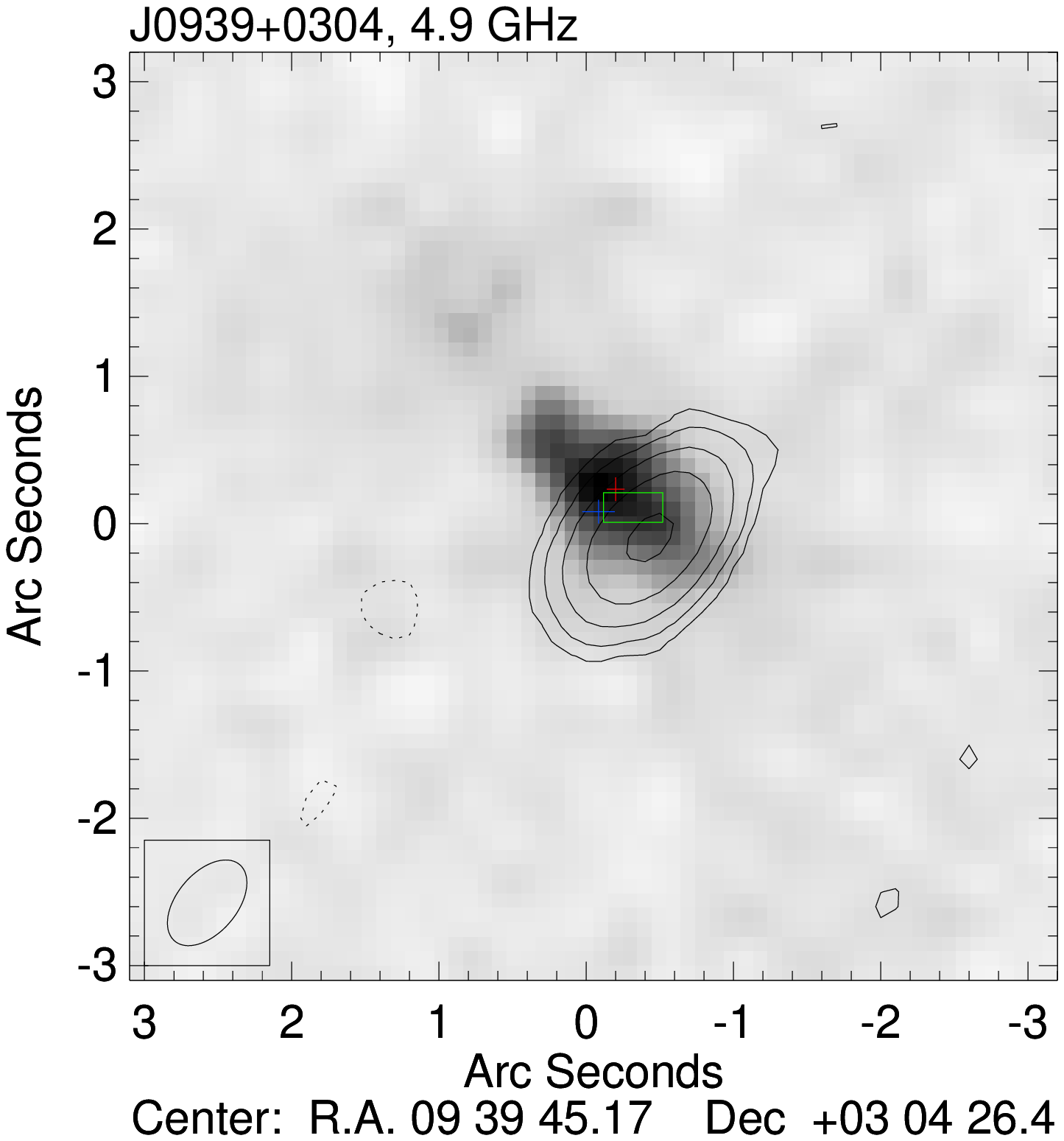}	&		&	\includegraphics[scale=0.25,bb=70 0 902 453]{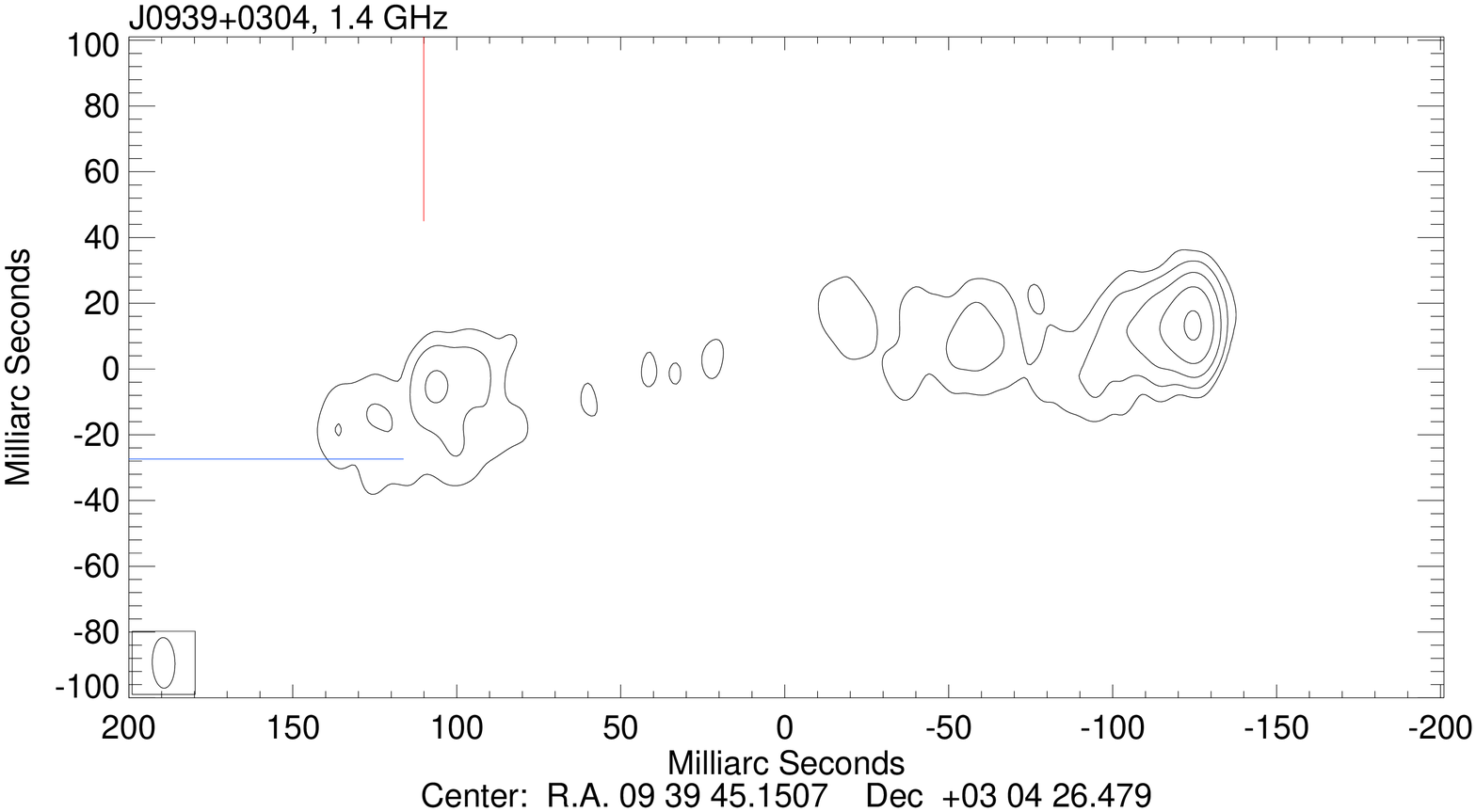}	\\
\includegraphics[scale=0.25]{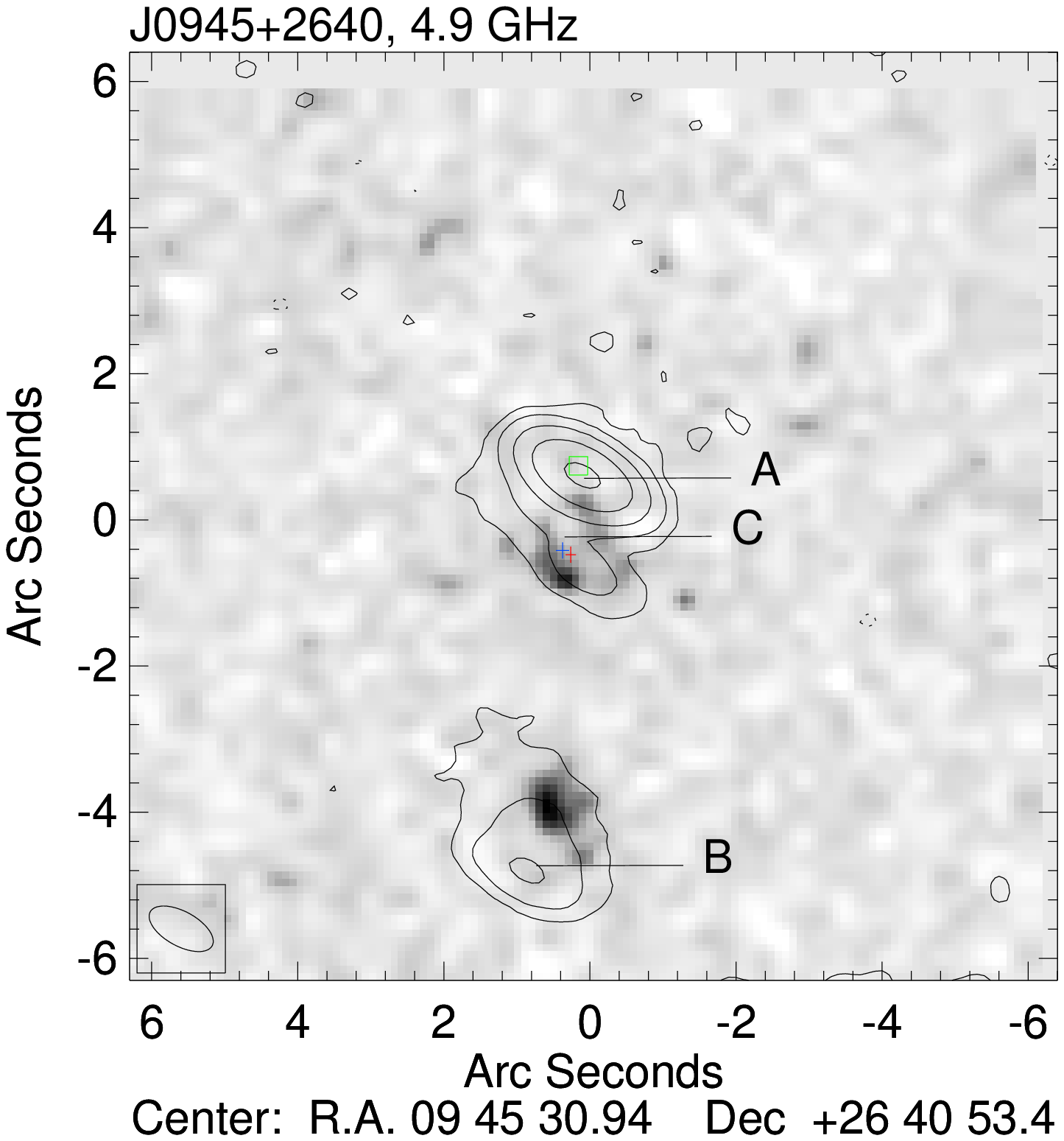}	&		&	\includegraphics[scale=0.25]{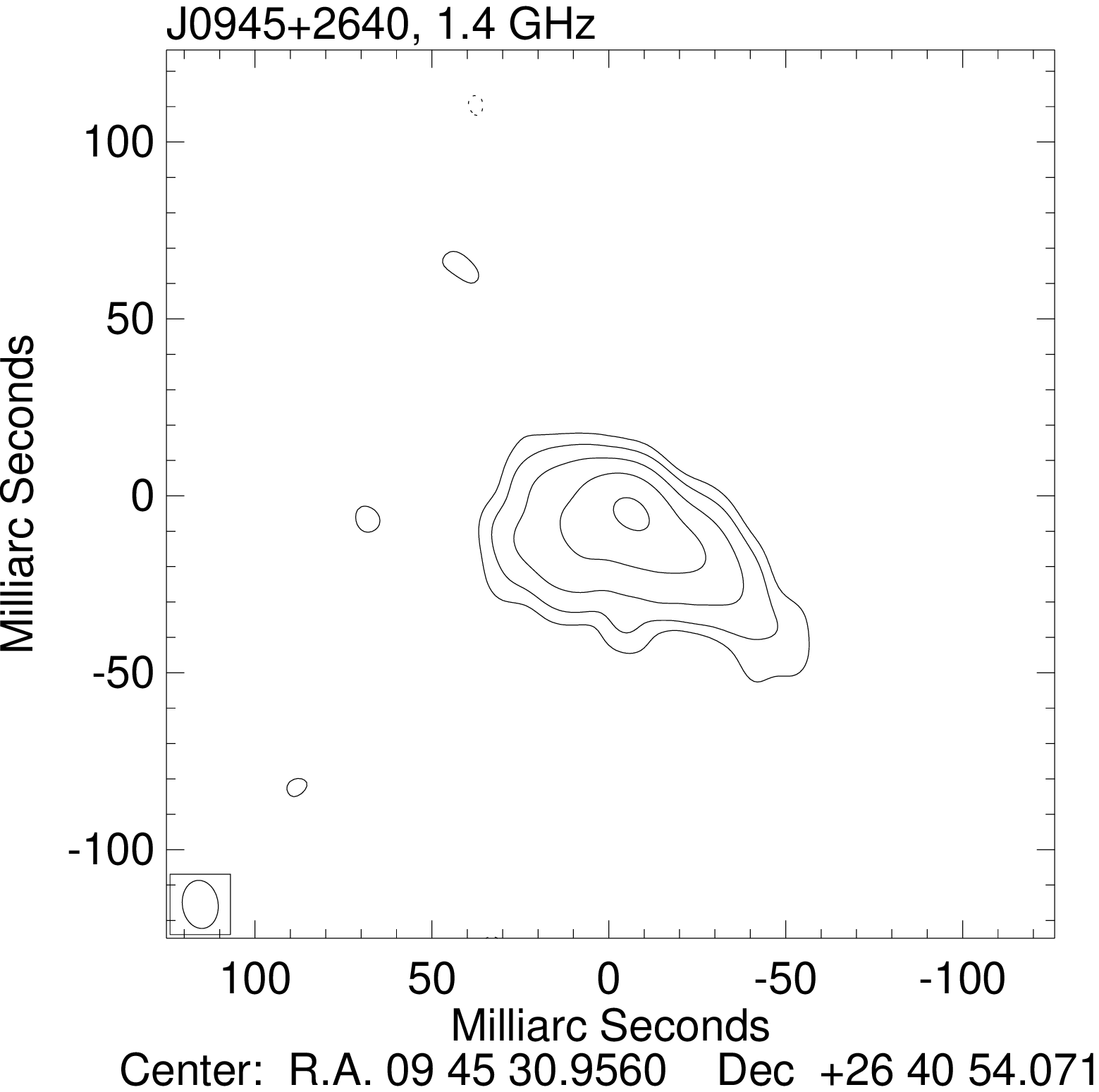}	\\
\includegraphics[scale=0.25]{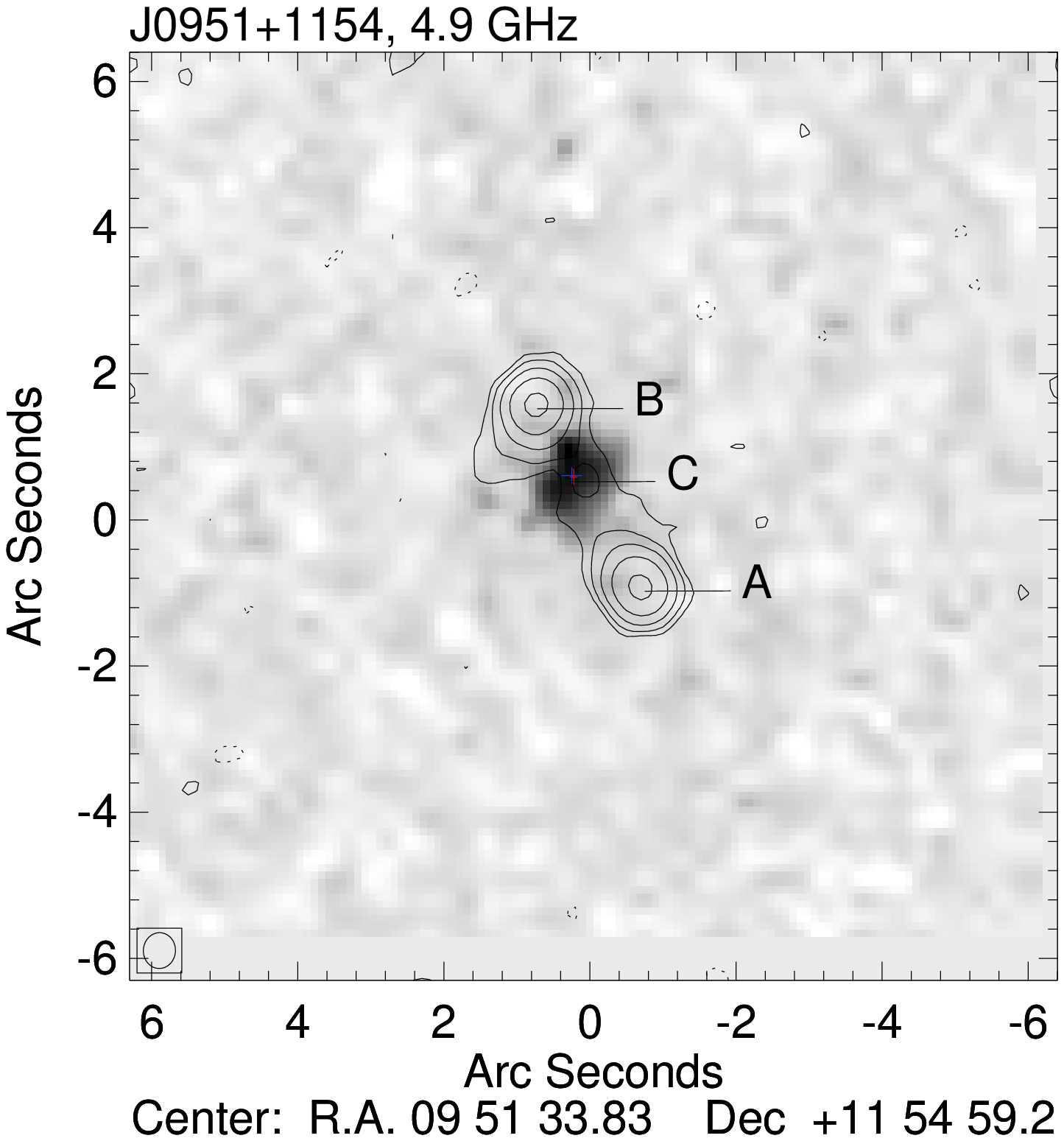}	&	\includegraphics[scale=0.25]{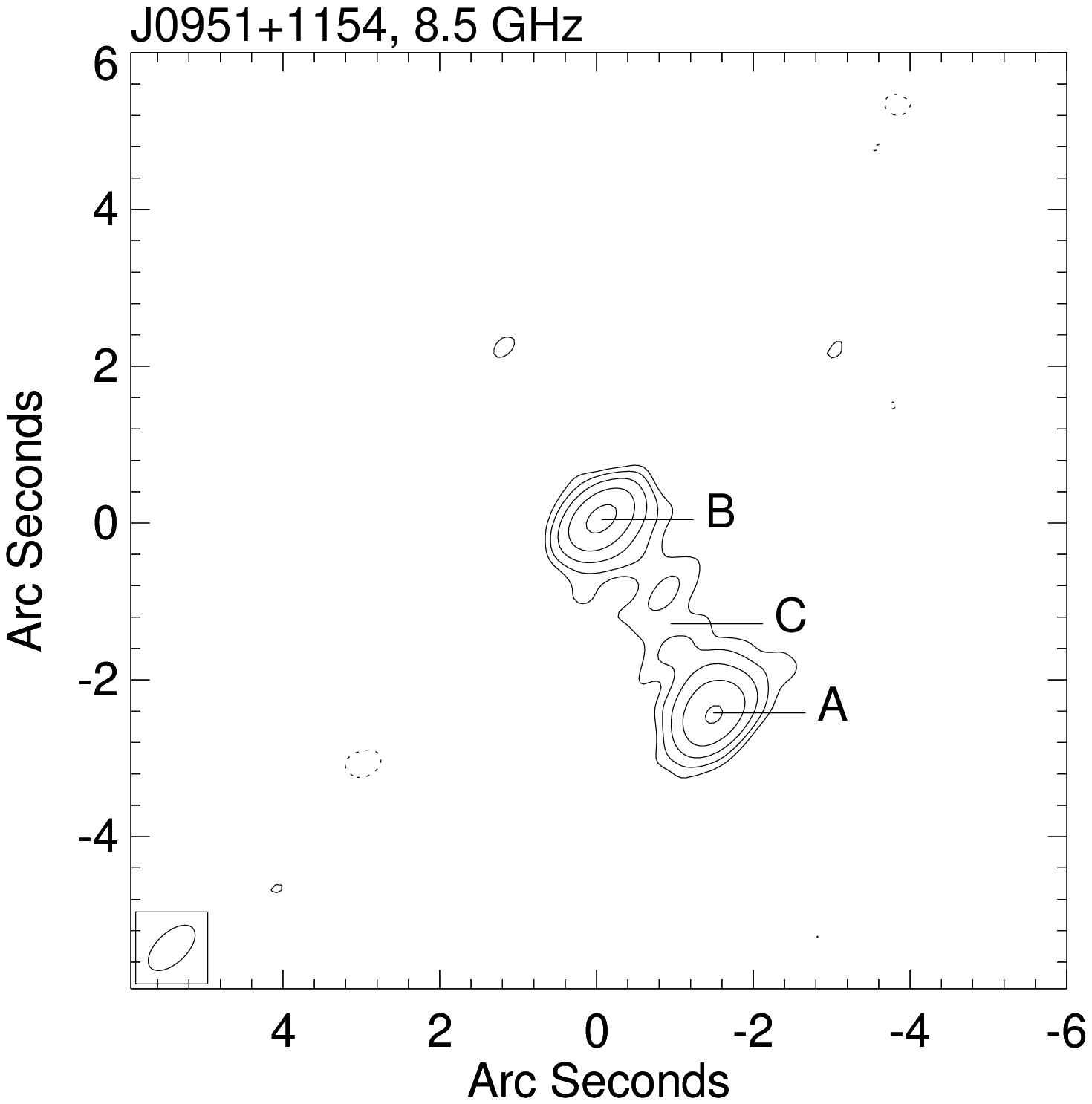}	&		\\
\includegraphics[scale=0.25]{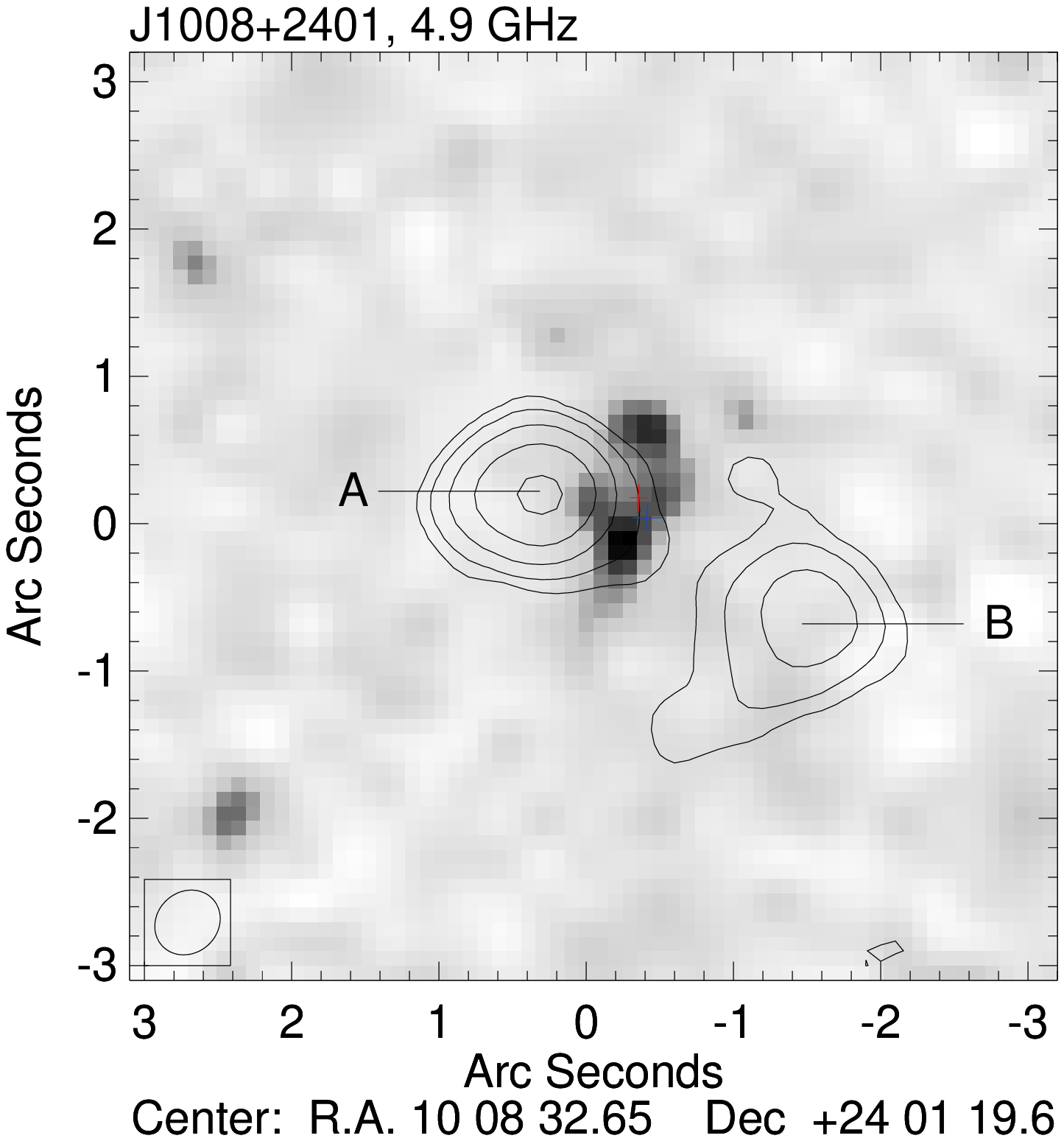}	&		&		\\
\includegraphics[scale=0.25]{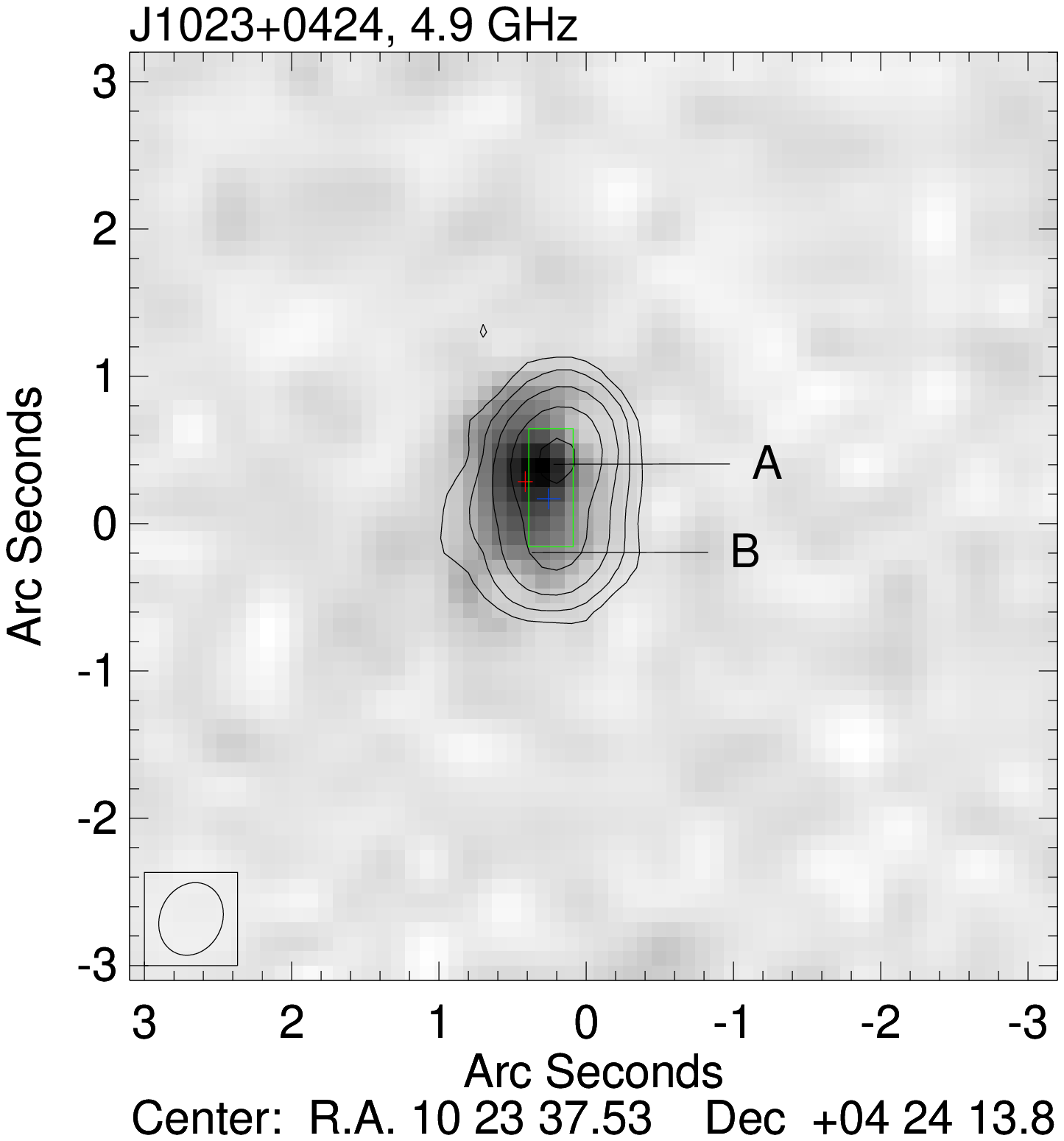}	&	\includegraphics[scale=0.25]{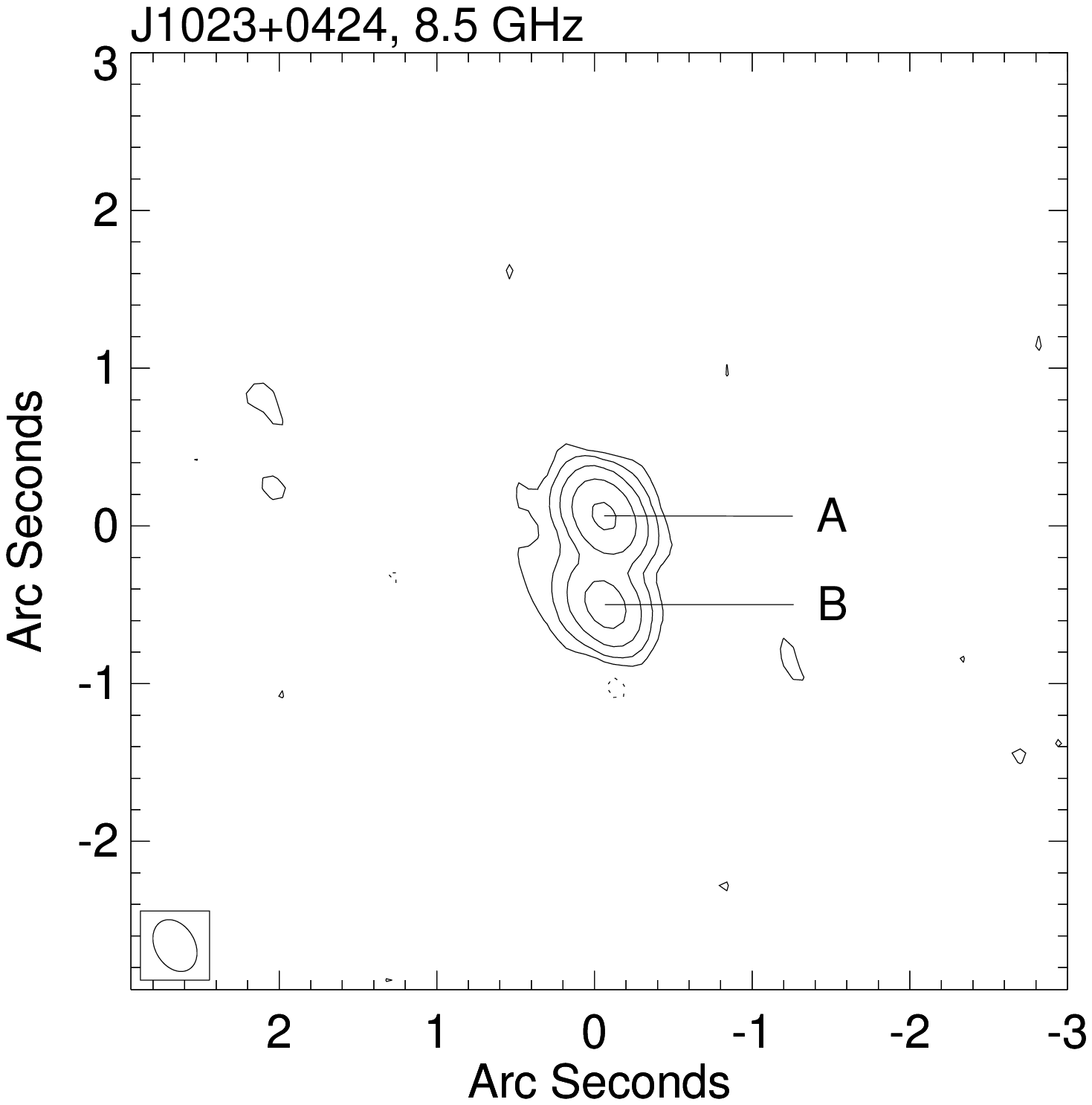}	&	\includegraphics[scale=0.25,bb=0 120 453 453]{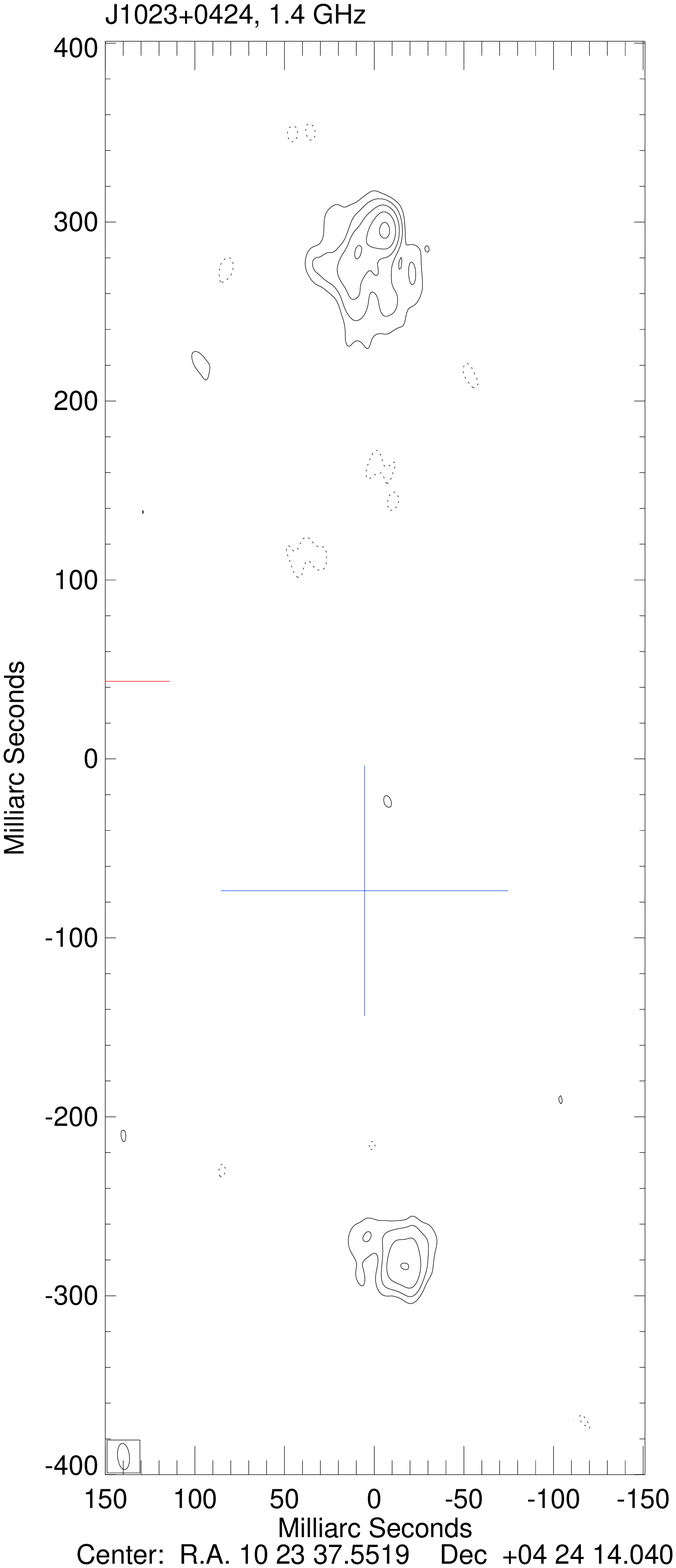}	\\

\end{tabular}
\end{figure}

\begin{figure}[htdp]
\begin{tabular}{lll}

\includegraphics[scale=0.25]{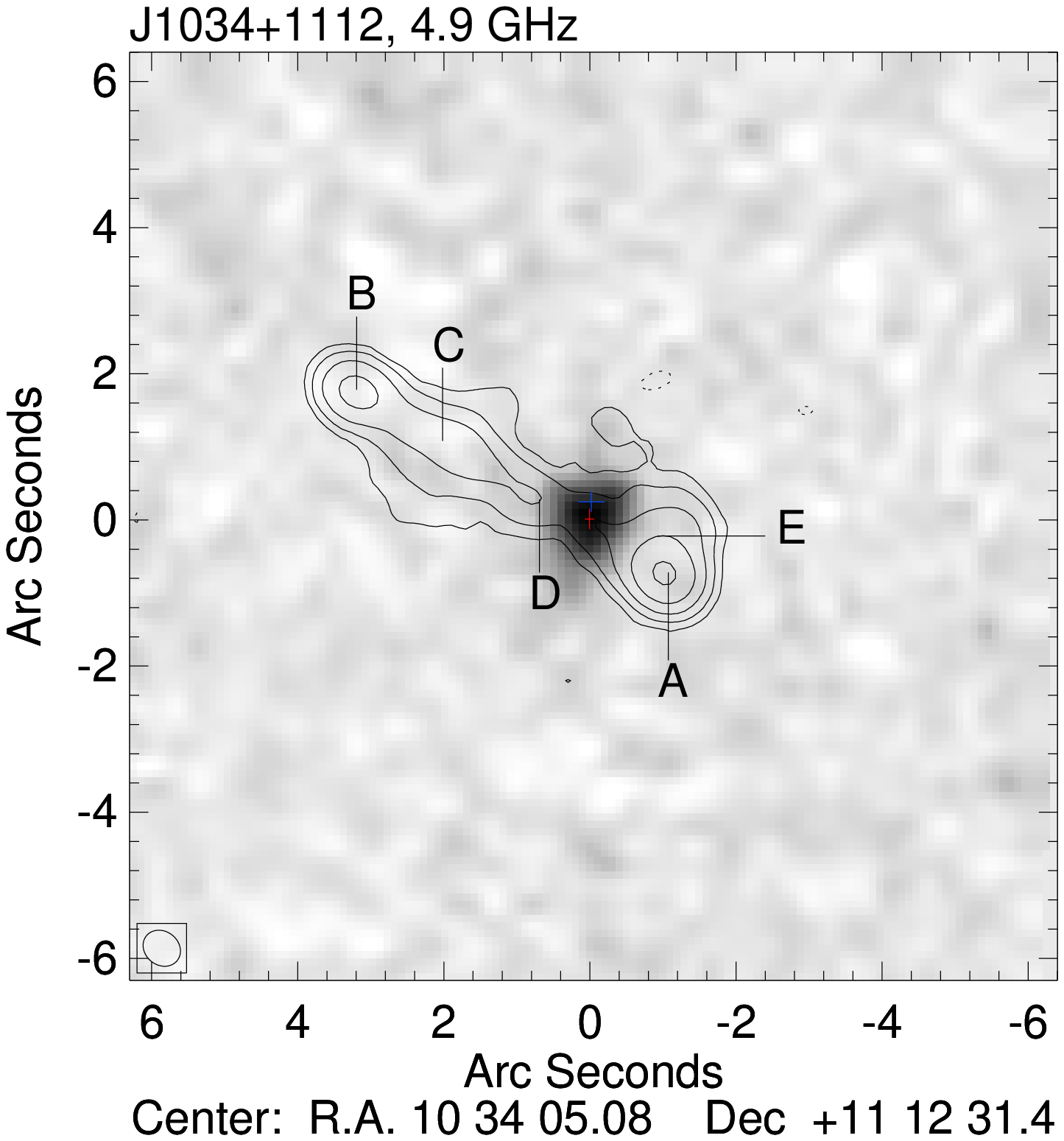}	&	\includegraphics[scale=0.25]{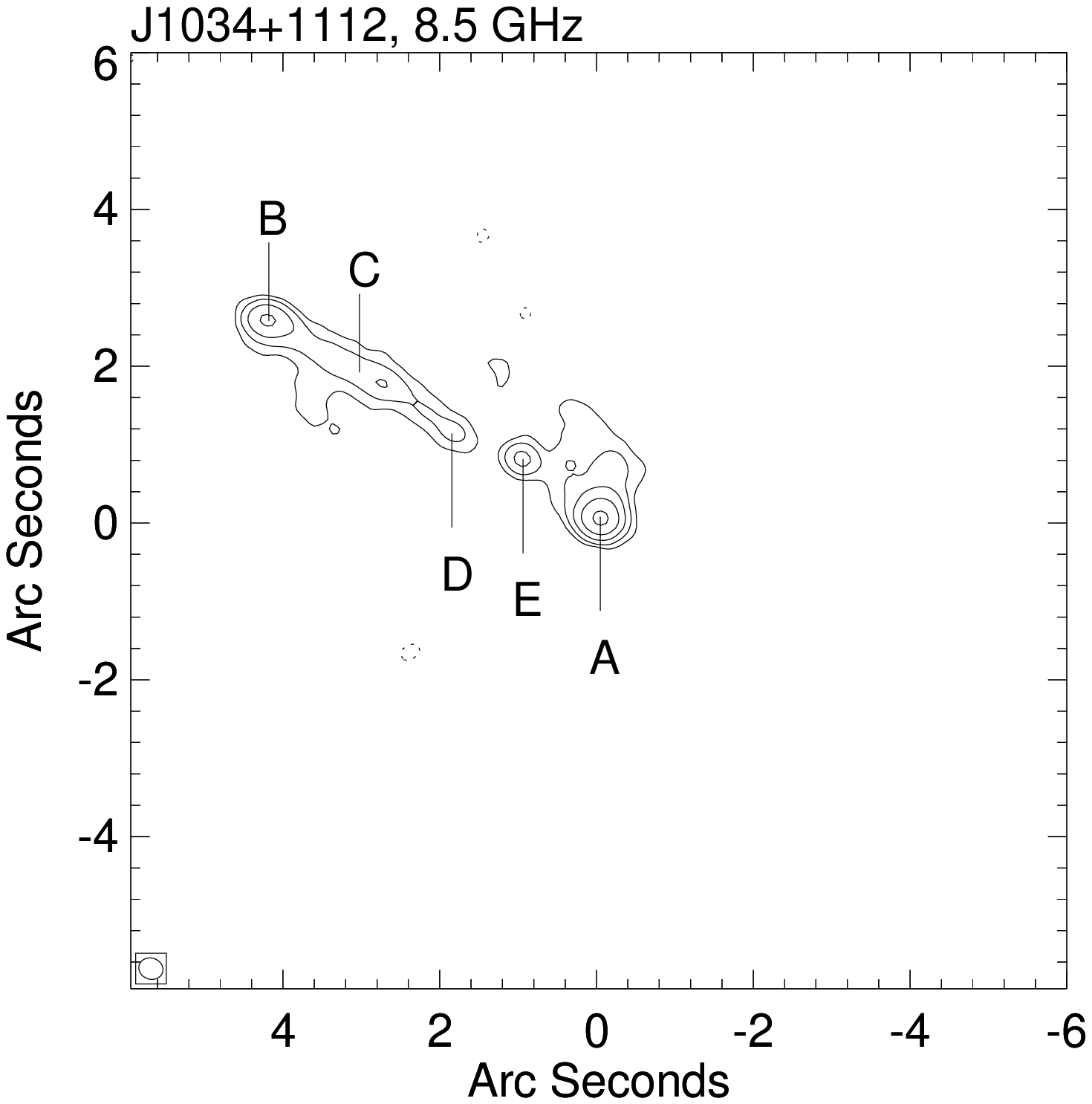}	&		\\
\includegraphics[scale=0.25]{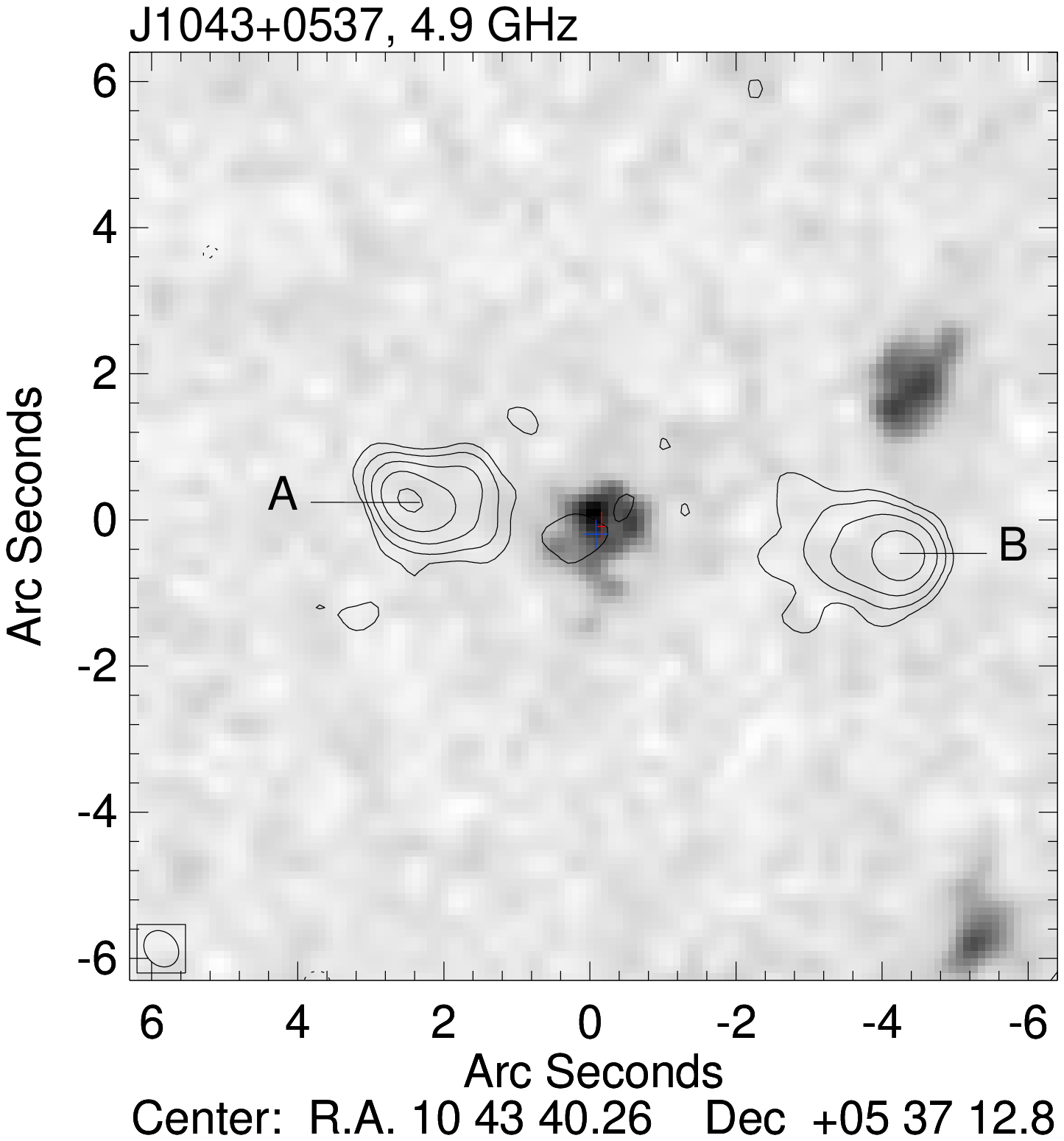}	&		&		\\
\includegraphics[scale=0.25]{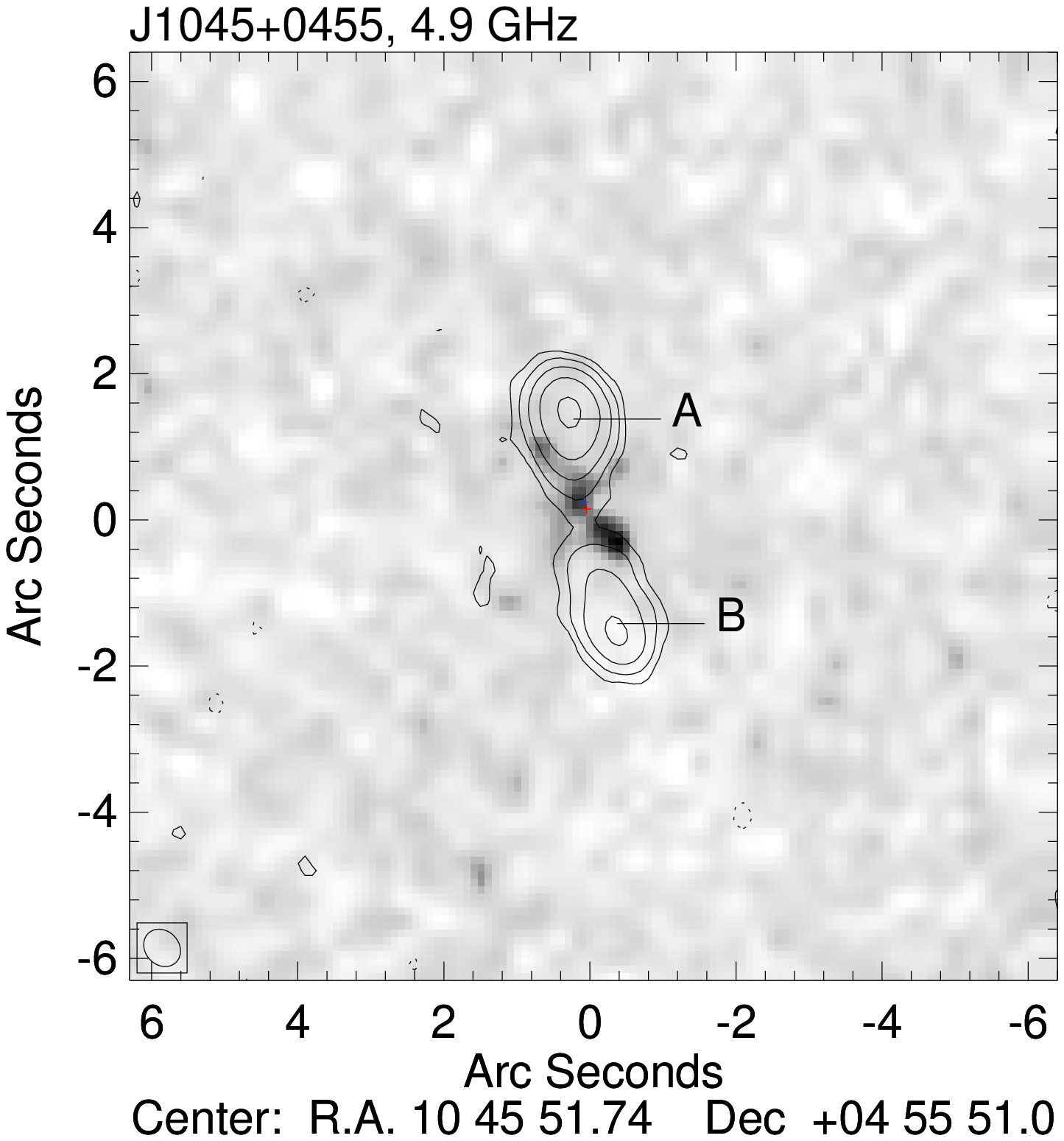}	&	\includegraphics[scale=0.25]{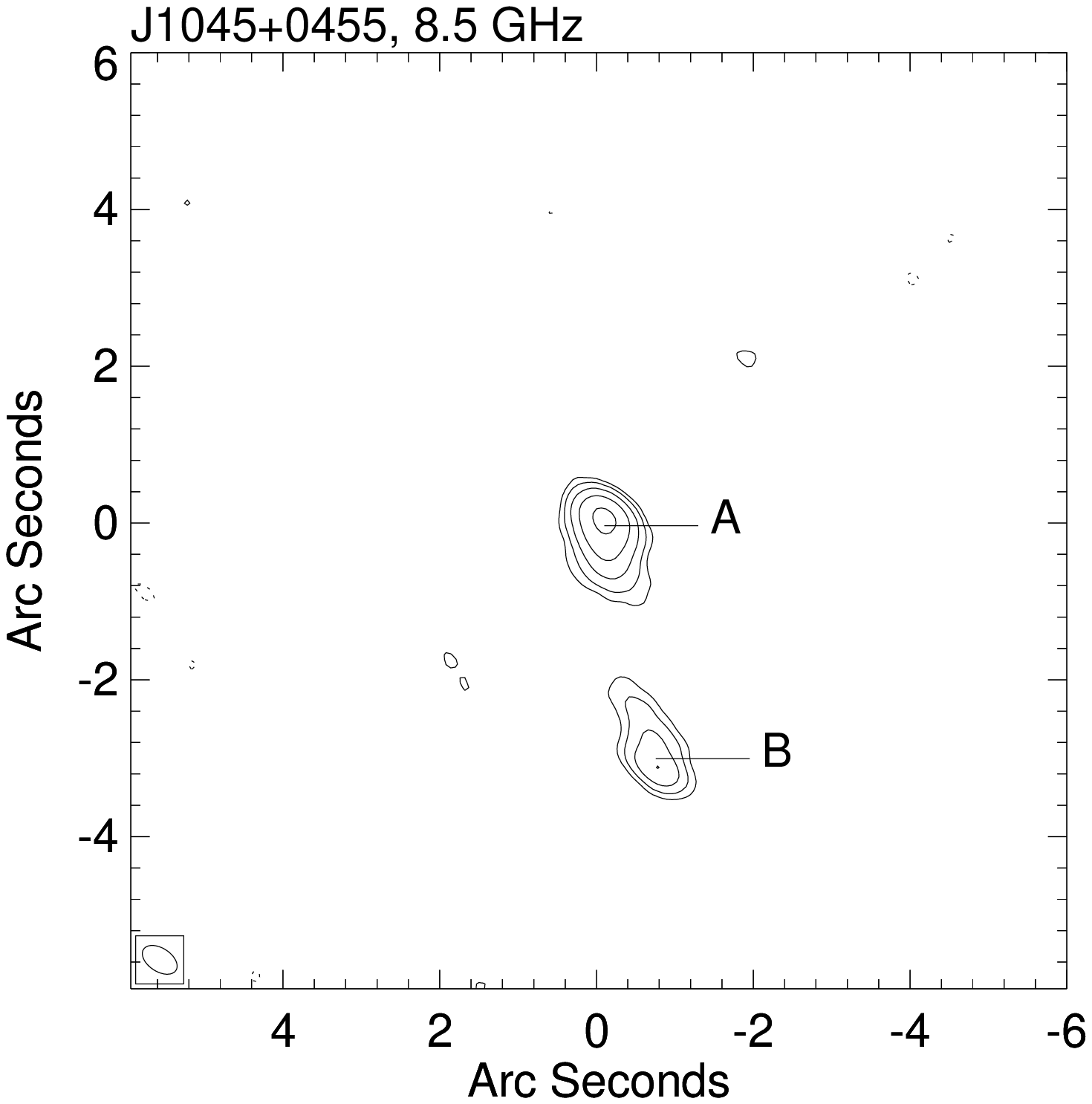}	&		\\
\includegraphics[scale=0.25]{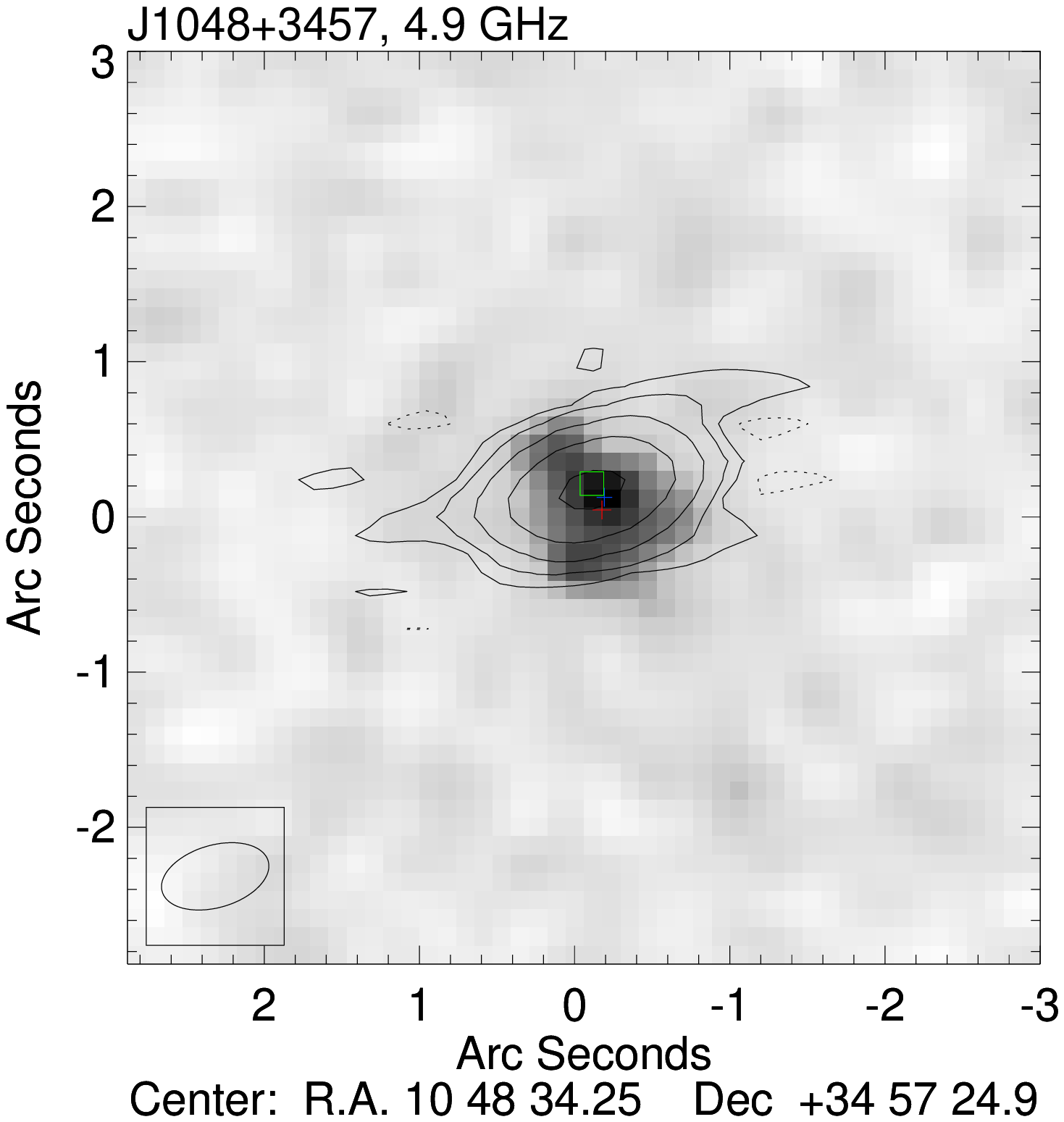}	&		&	\includegraphics[scale=0.25]{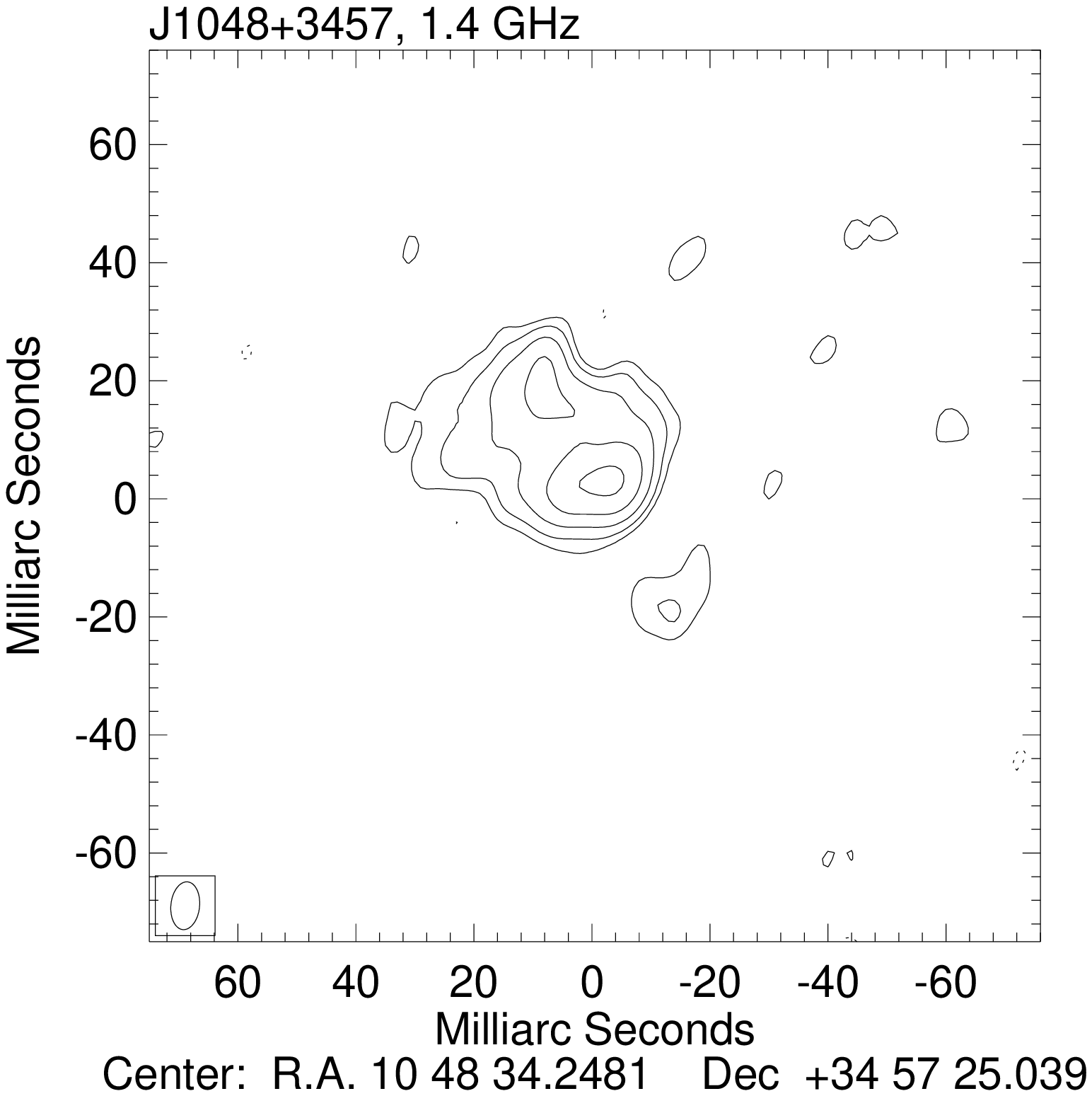}	\\
\includegraphics[scale=0.25]{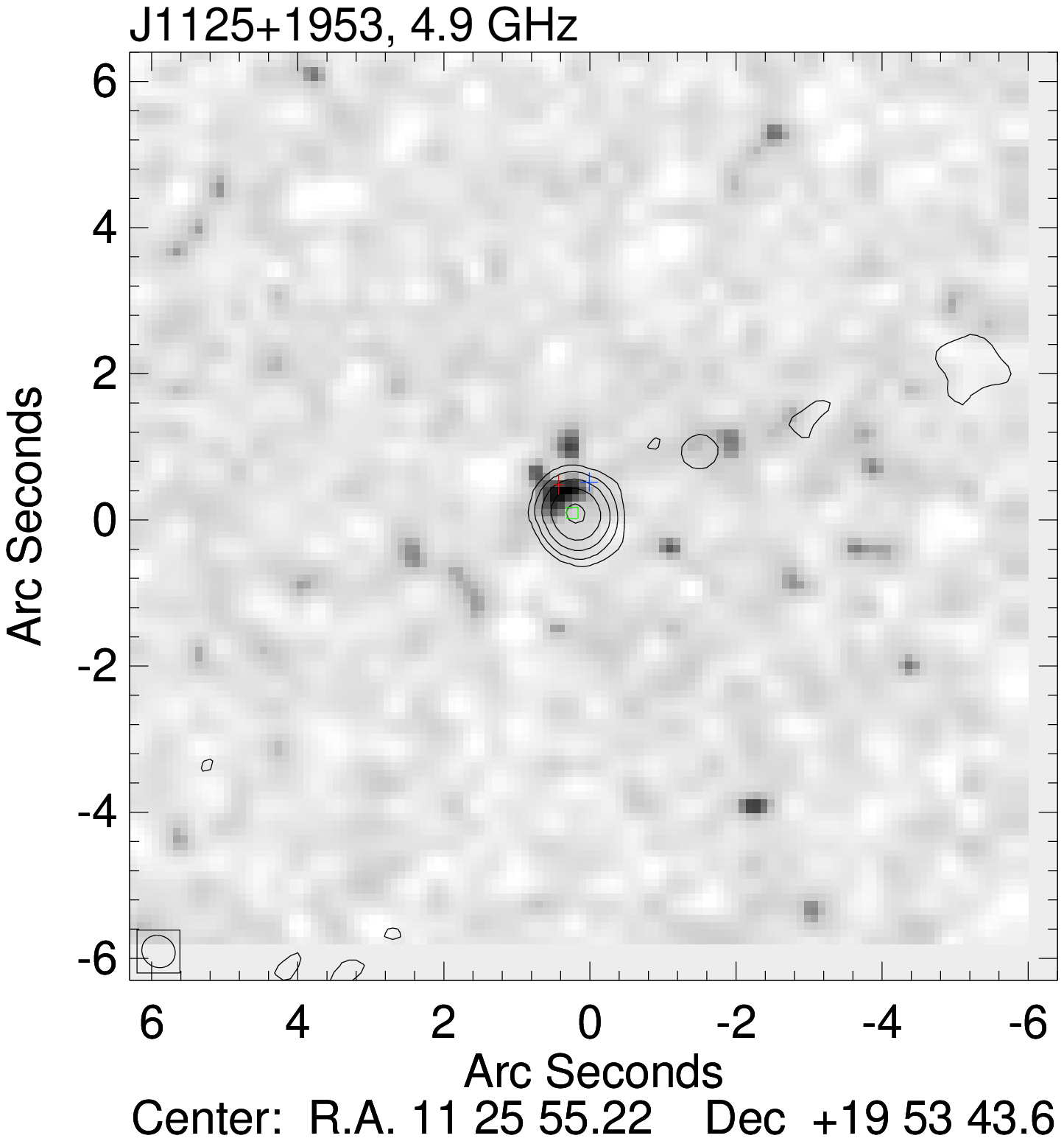}	&		&	\includegraphics[scale=0.25]{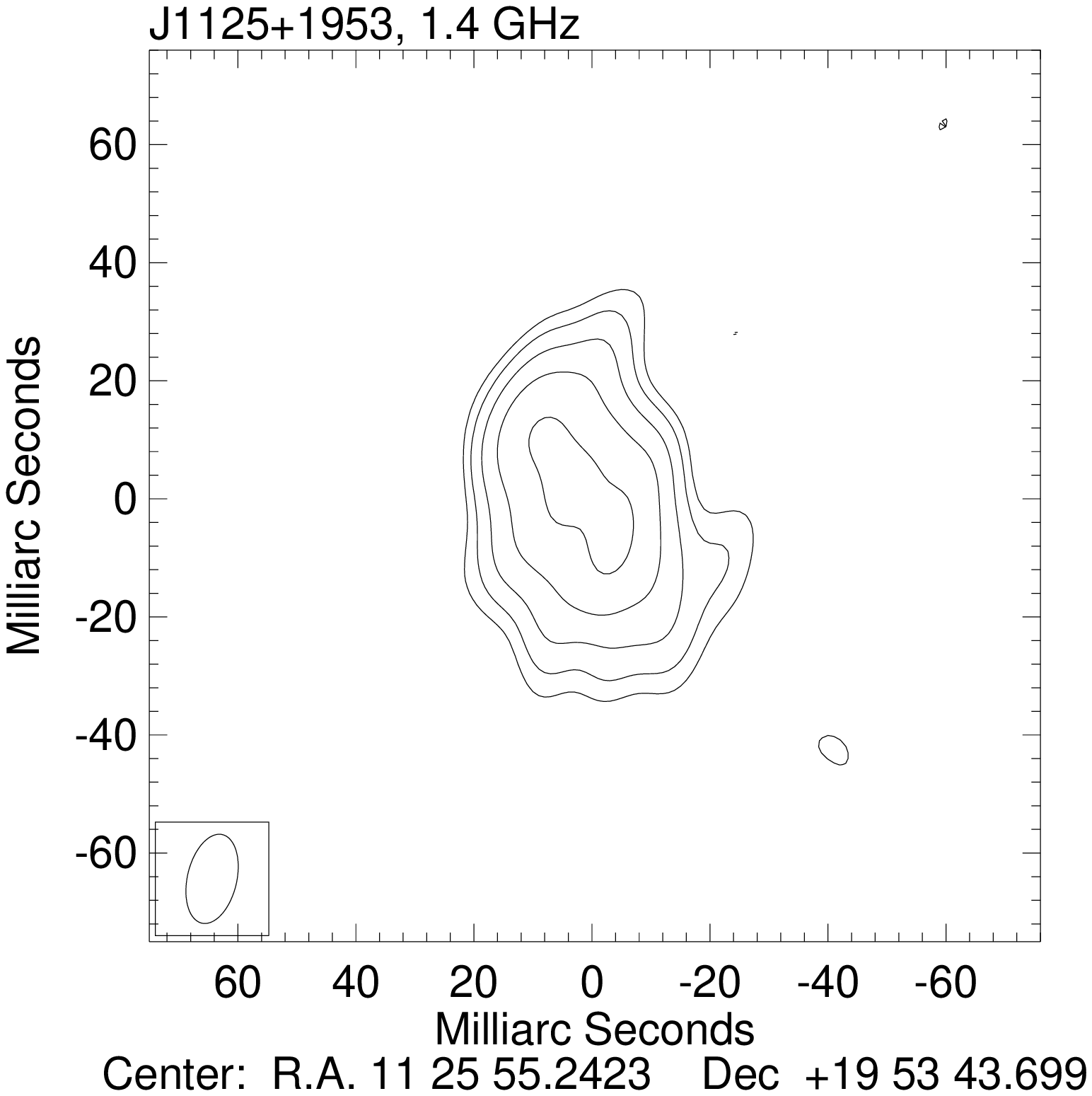}	\\

\end{tabular}
\end{figure}

\begin{figure}[htdp]
\begin{tabular}{lll}

\includegraphics[scale=0.25]{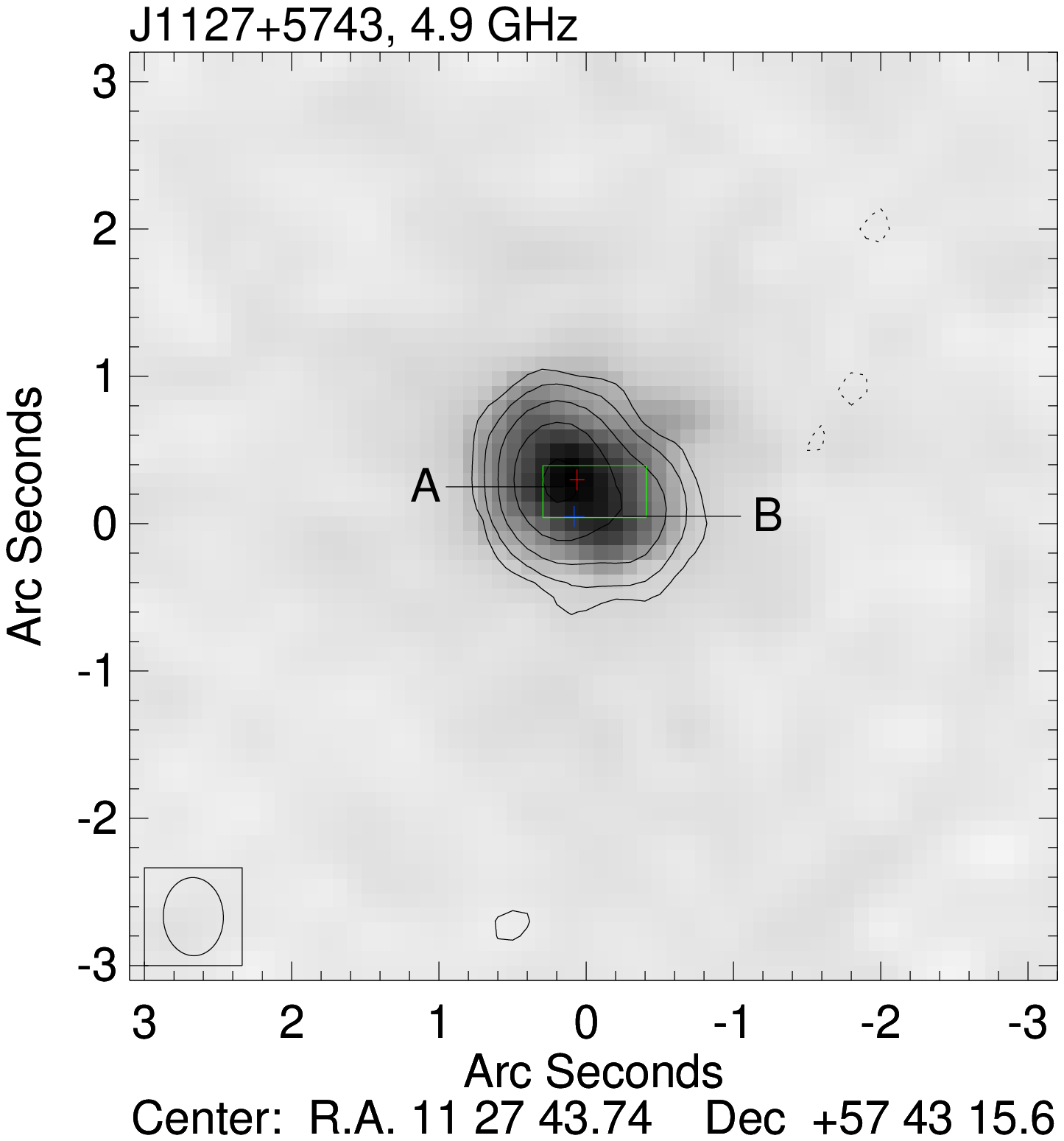}	&	\includegraphics[scale=0.25]{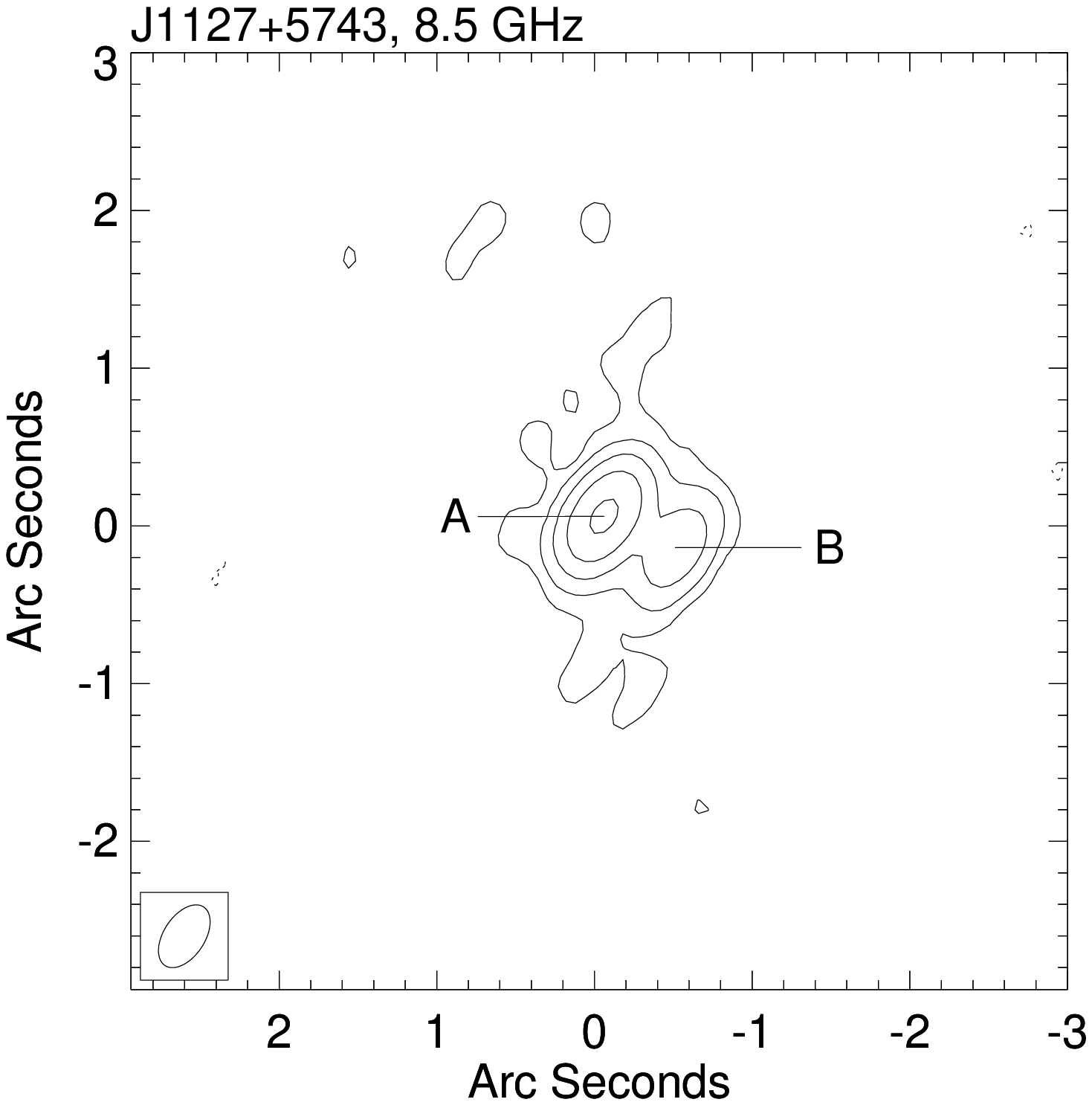}	&	\includegraphics[scale=0.25,bb=50 0 904 453]{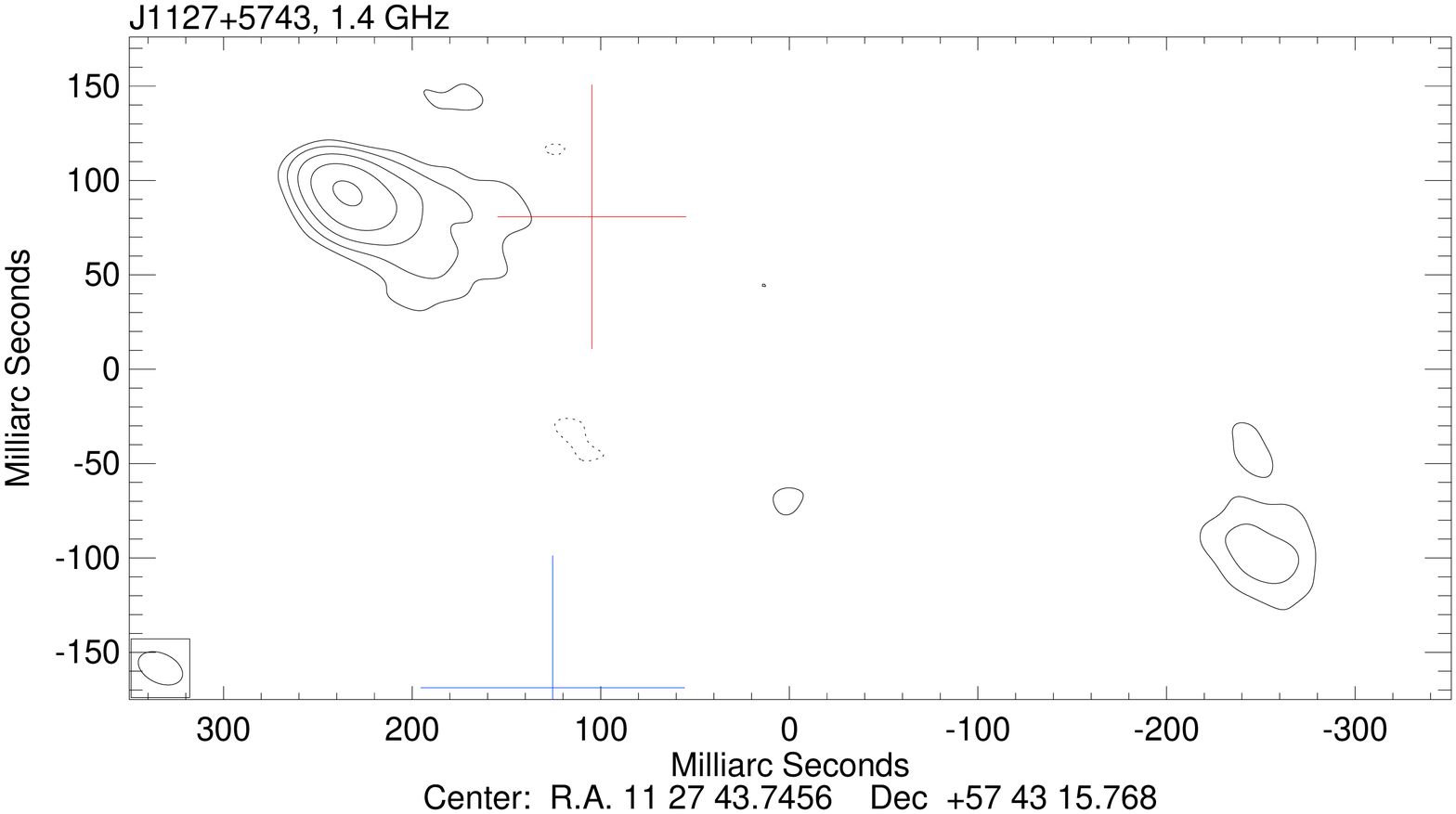}	\\
\includegraphics[scale=0.25]{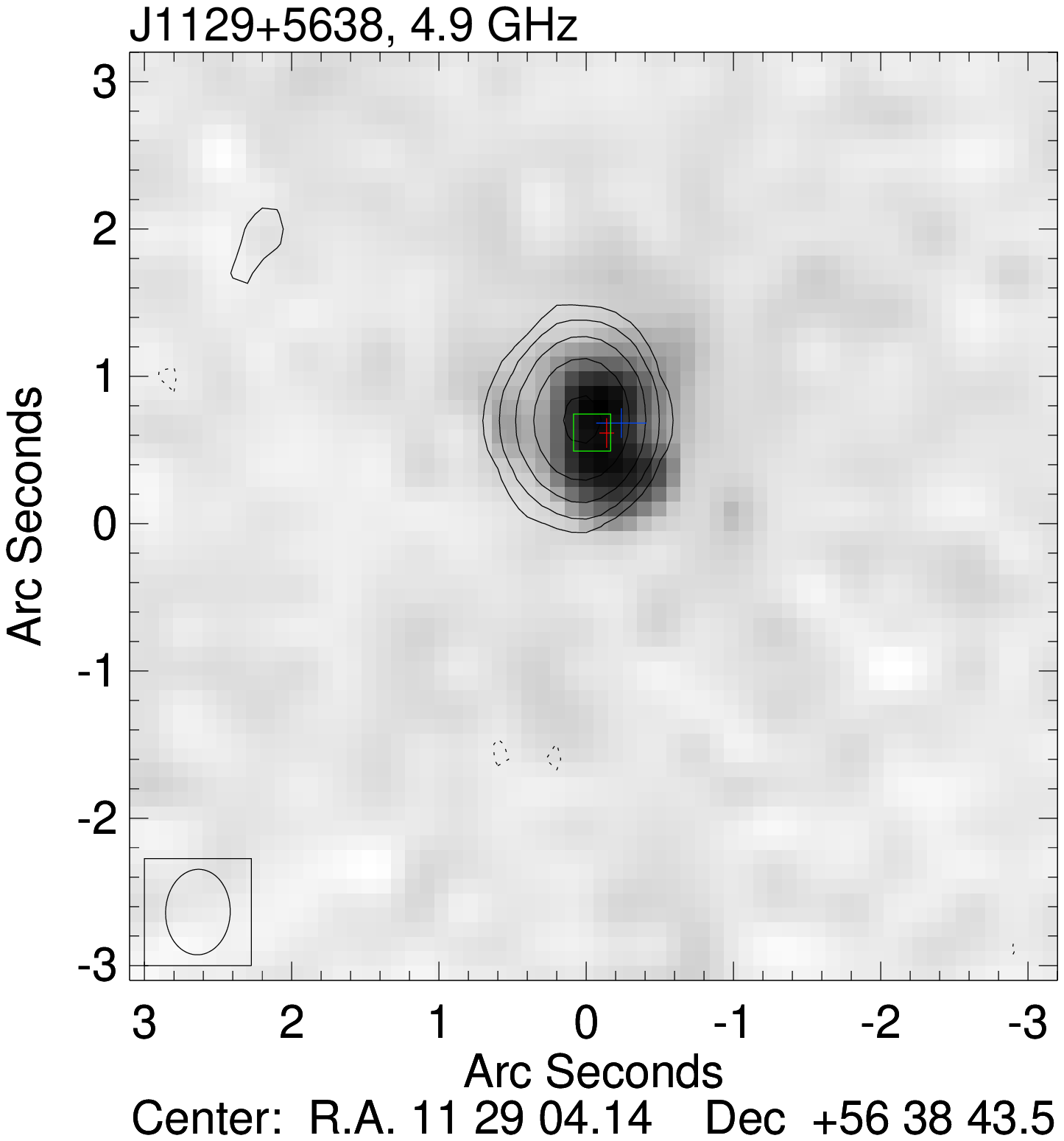}	&	\includegraphics[scale=0.25]{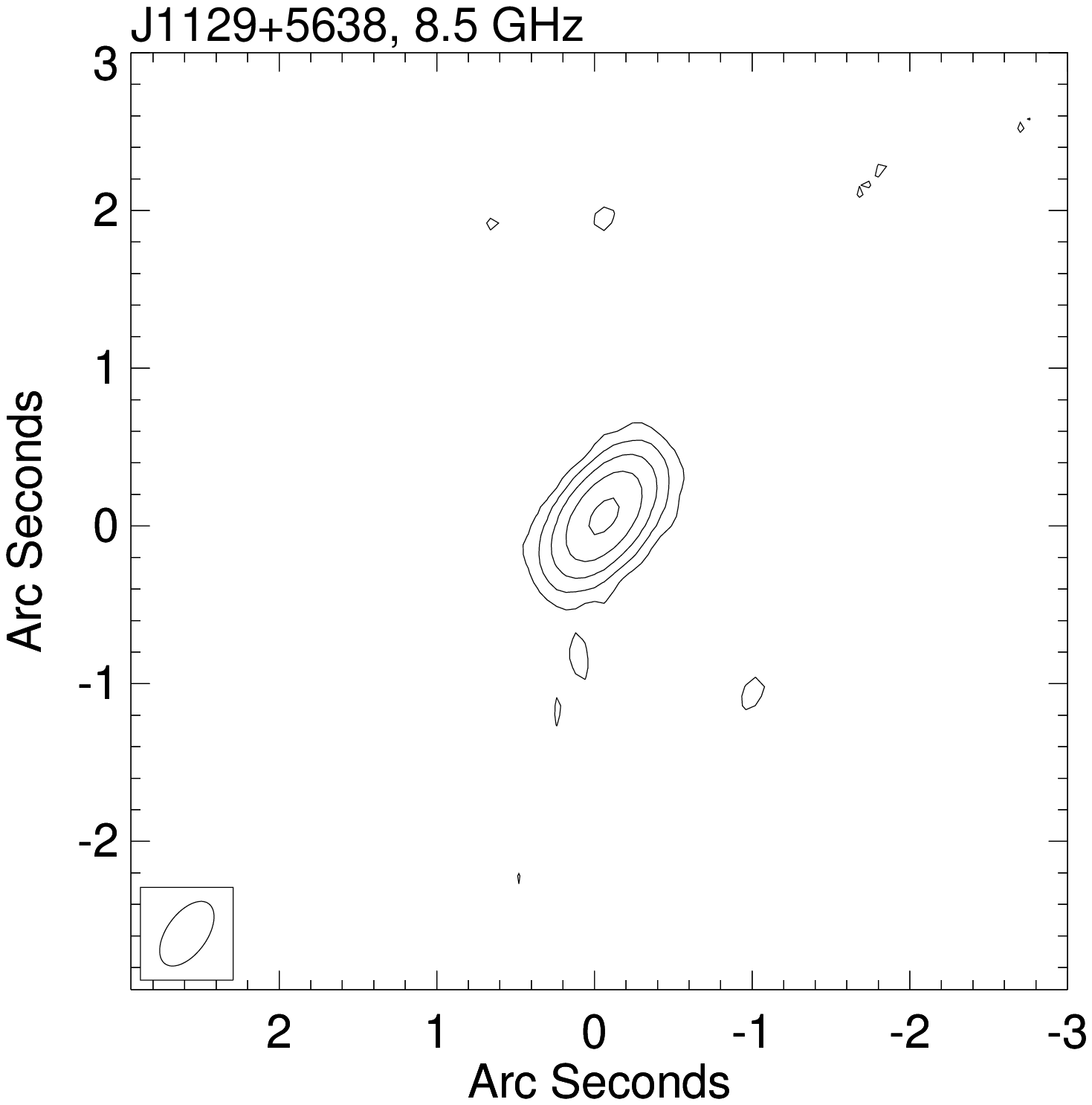}	&	\includegraphics[scale=0.25]{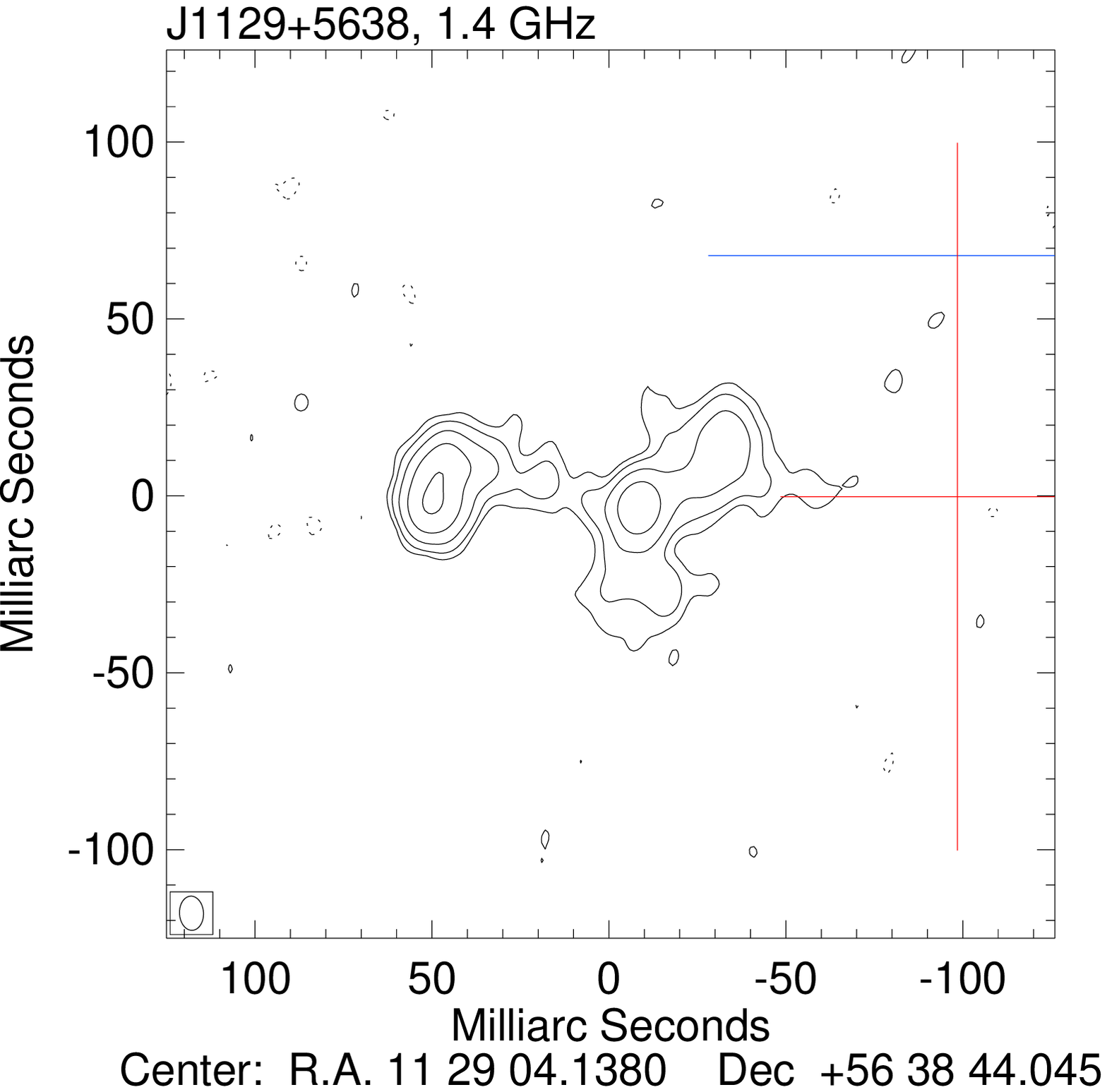}	\\
\includegraphics[scale=0.25]{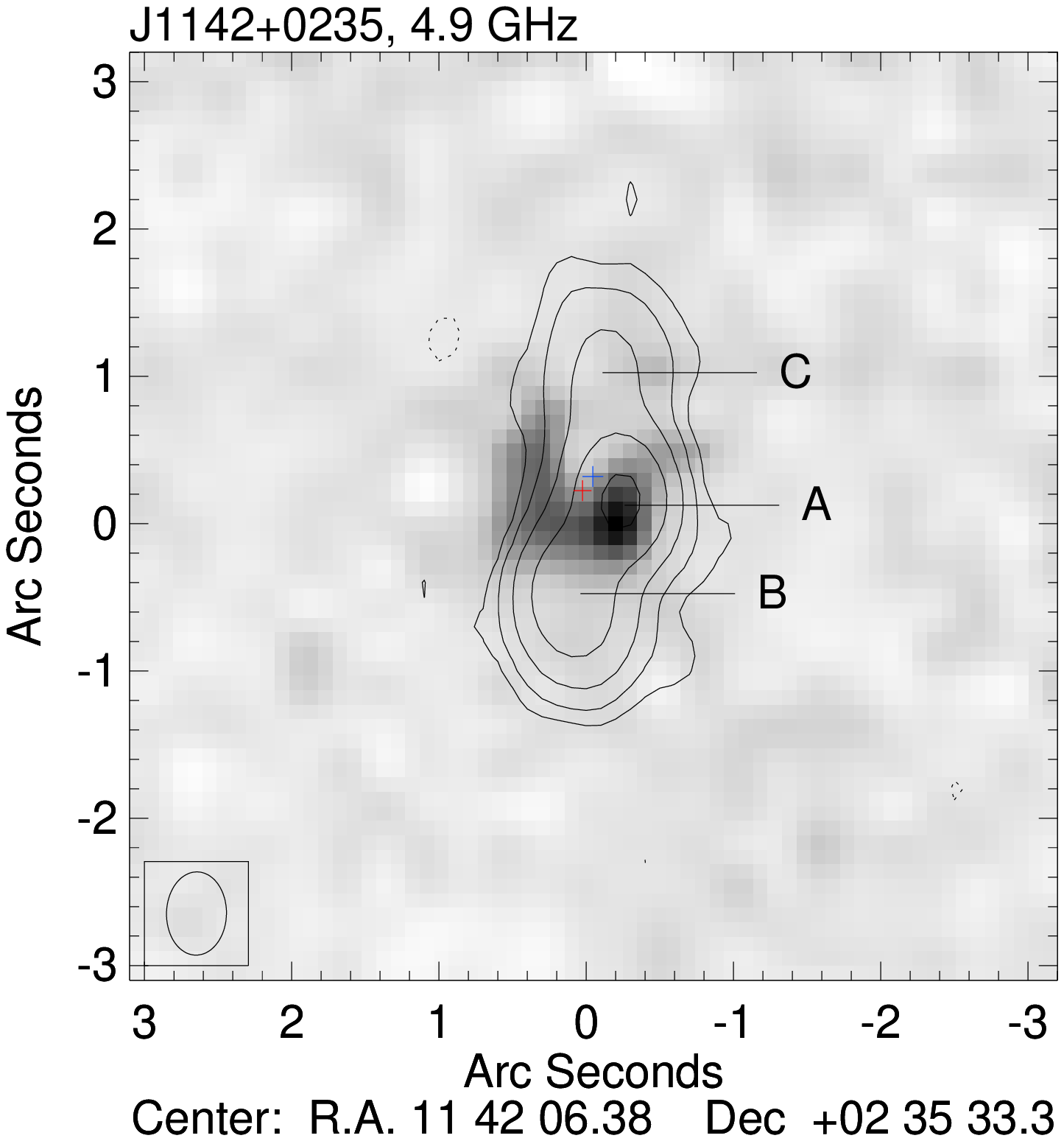}	&	\includegraphics[scale=0.25]{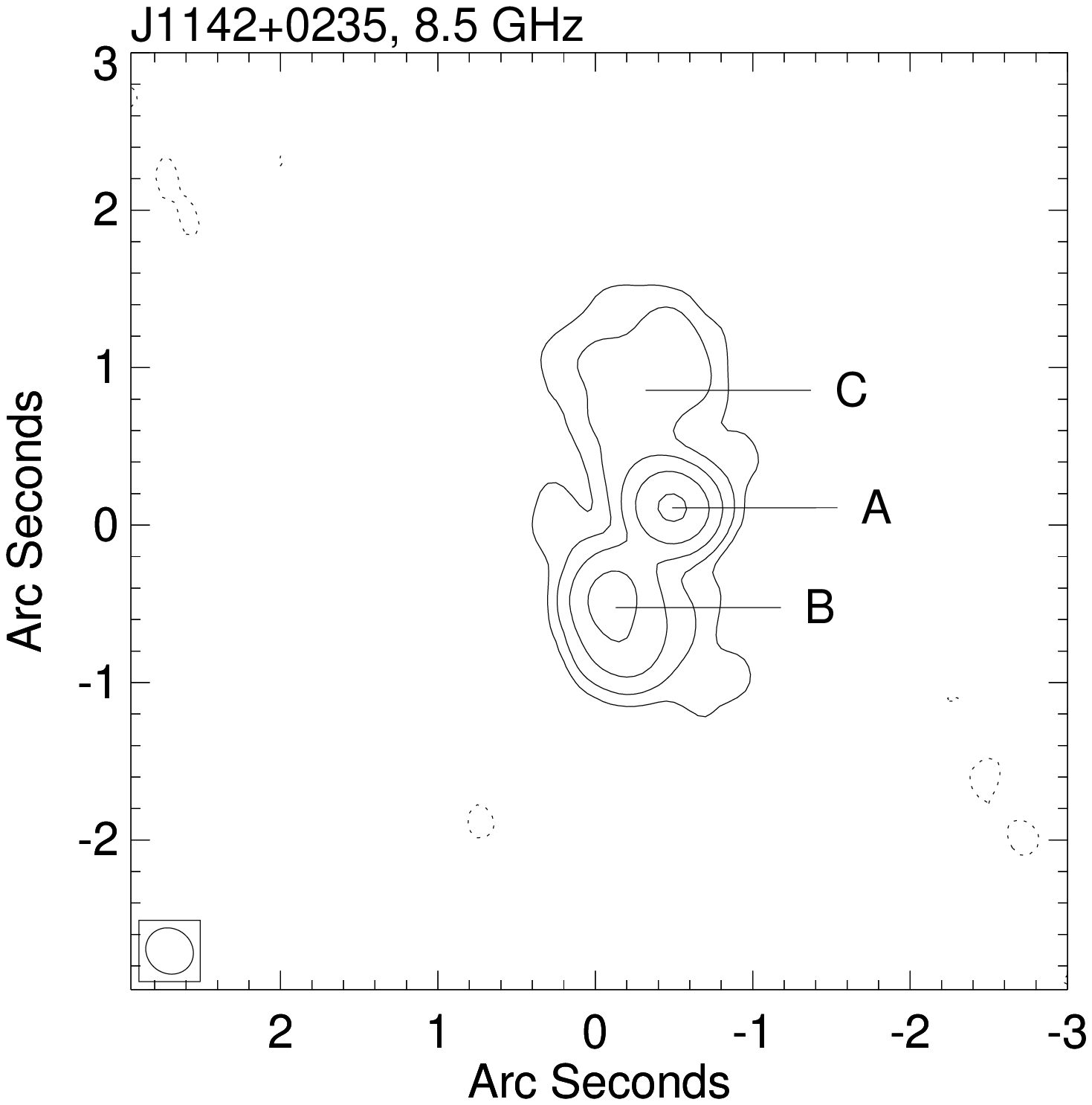}	&		\\
\includegraphics[scale=0.25]{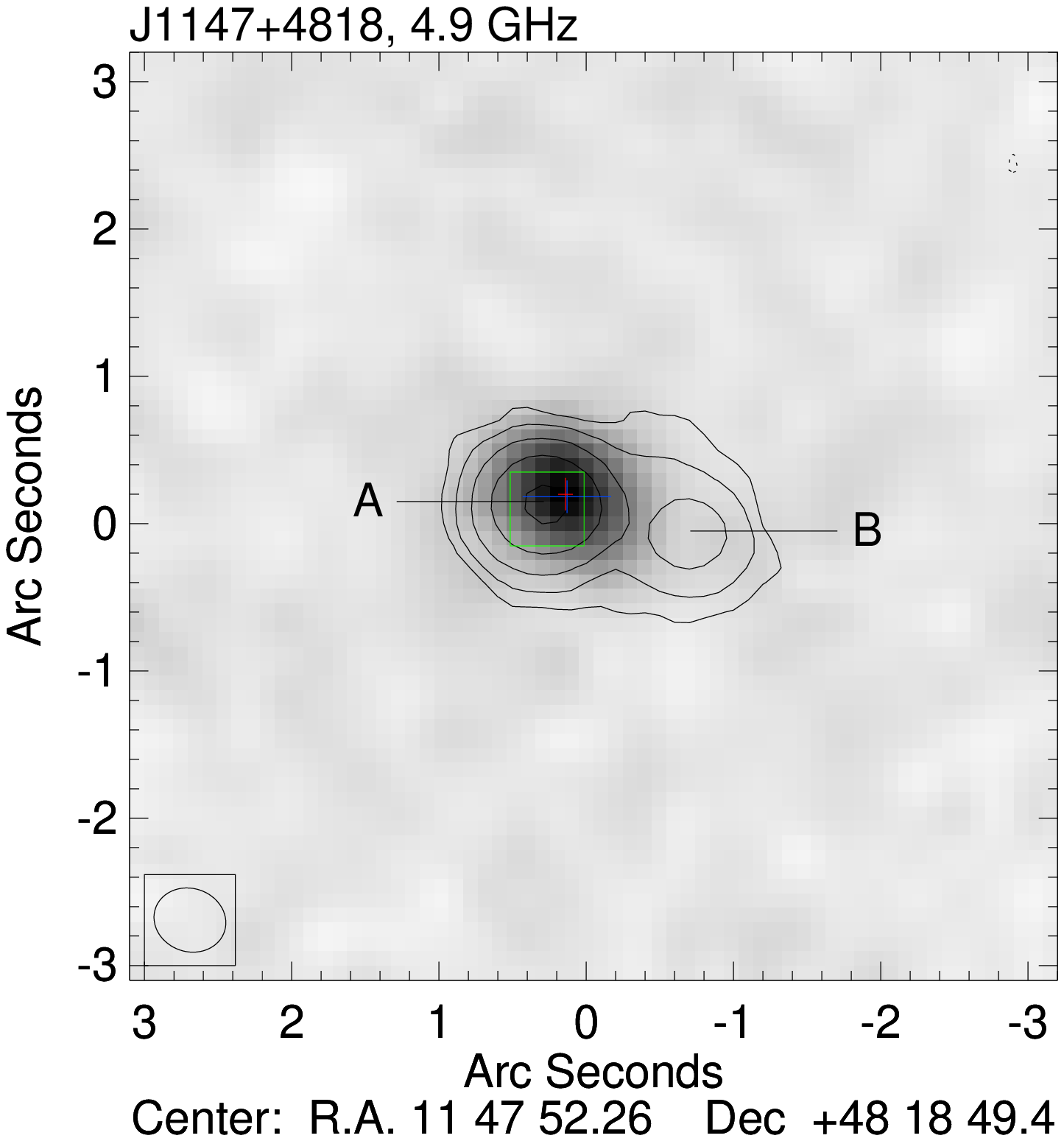}	&	\includegraphics[scale=0.25]{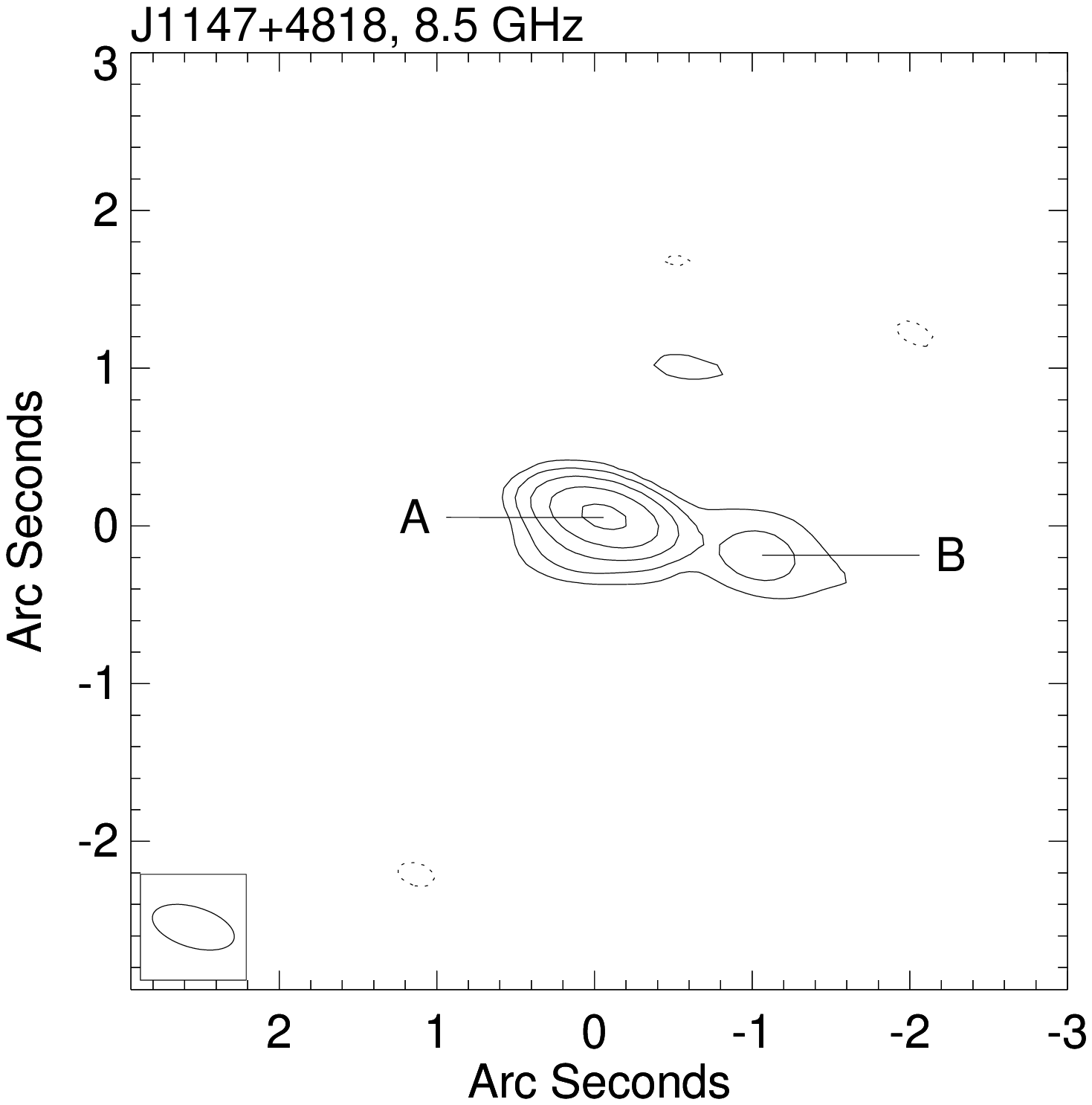}	&	\includegraphics[scale=0.25]{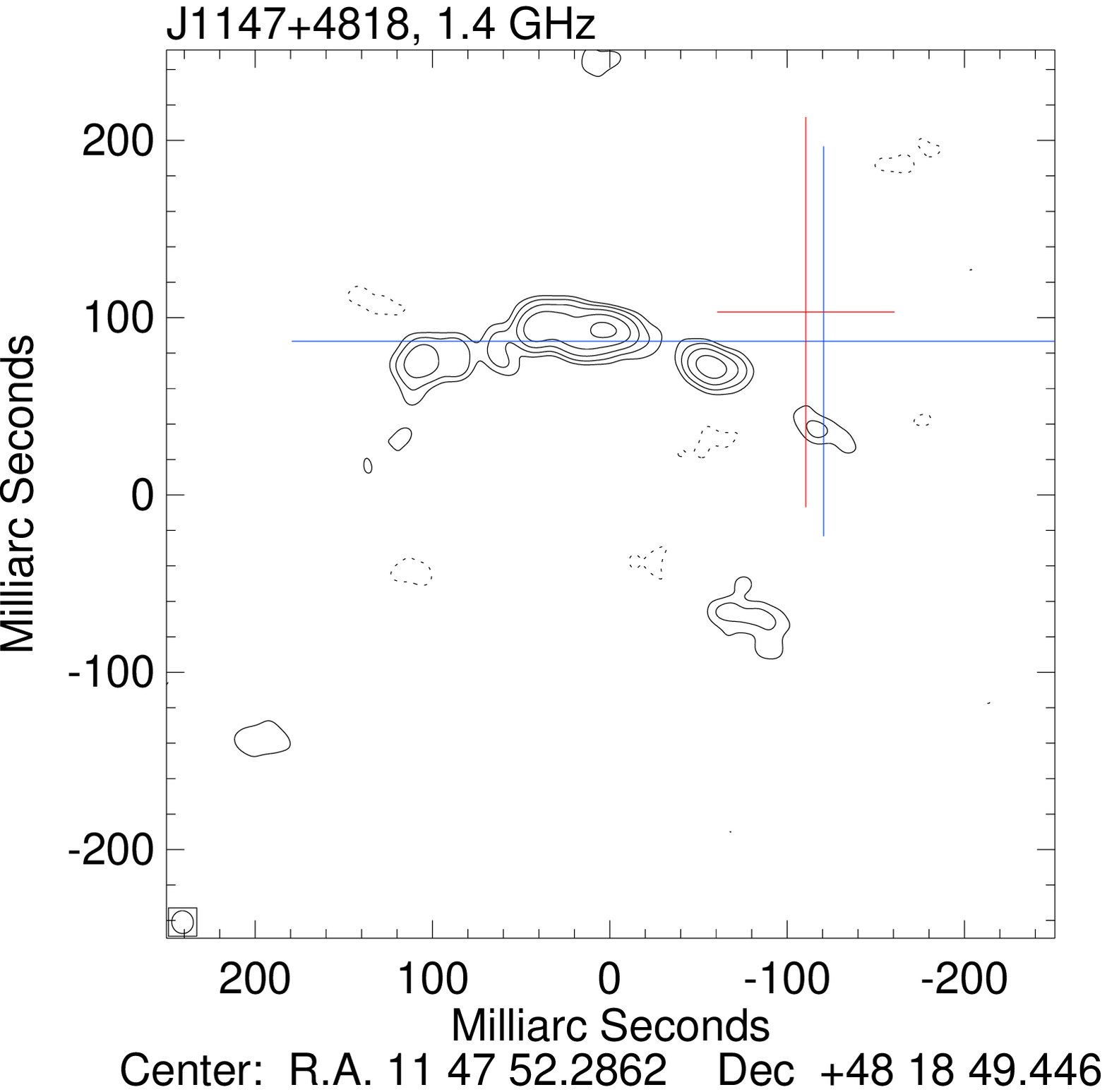}	\\

\end{tabular}
\end{figure}

\begin{figure}[htdp]
\begin{tabular}{lll}

\includegraphics[scale=0.25]{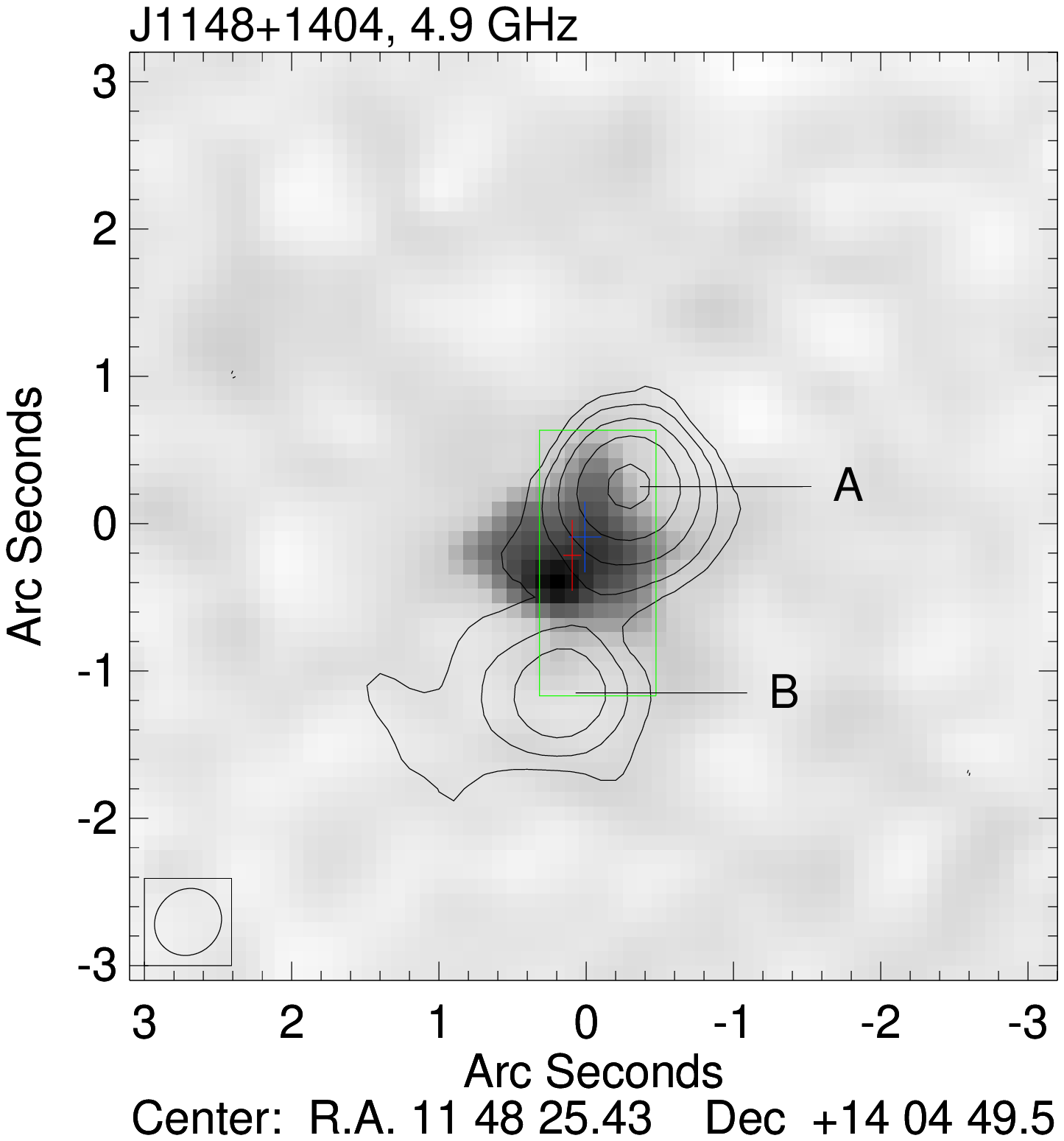}	&	\includegraphics[scale=0.25]{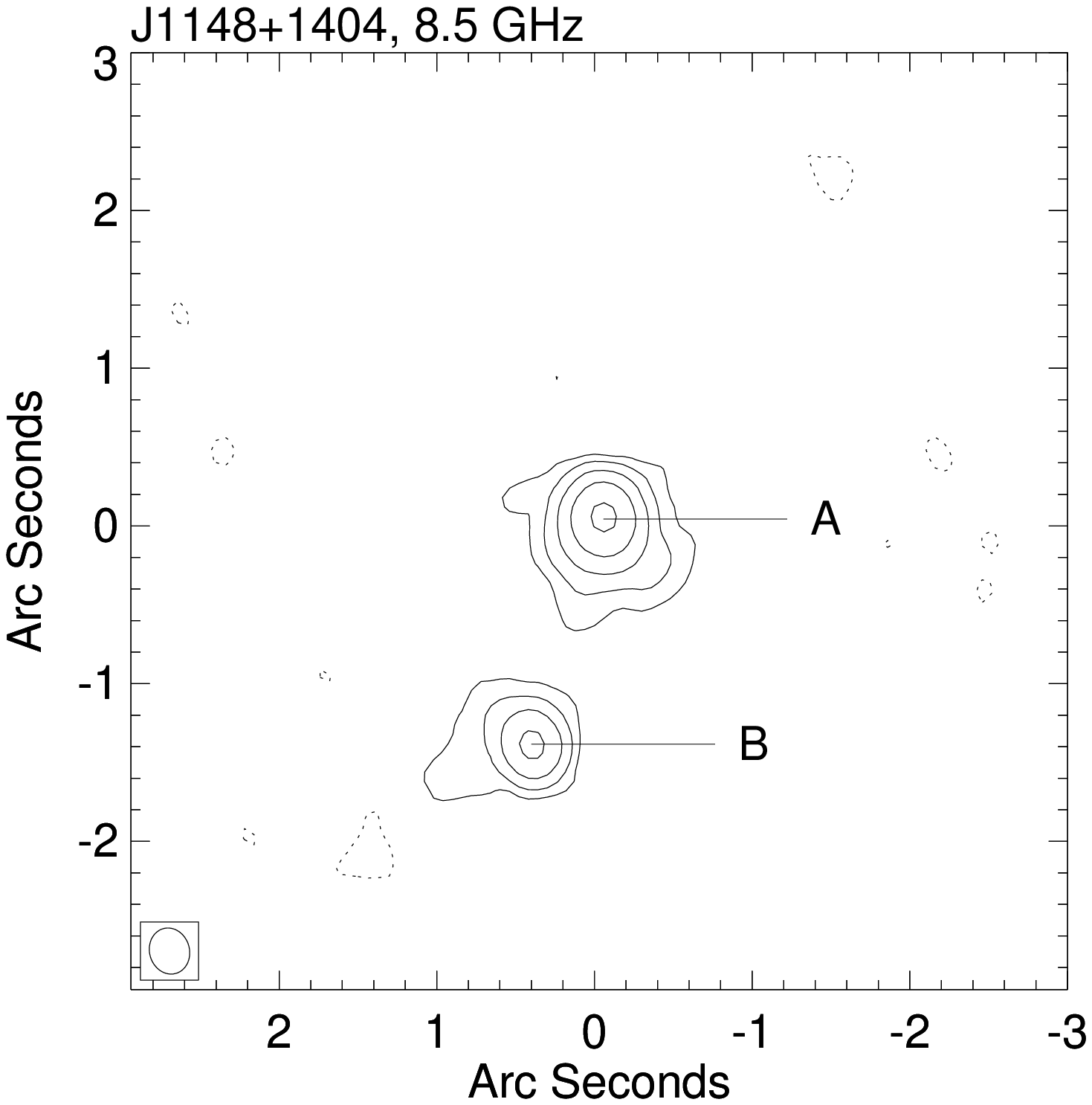}	&	\includegraphics[scale=0.25,bb=0 80 453 1031]{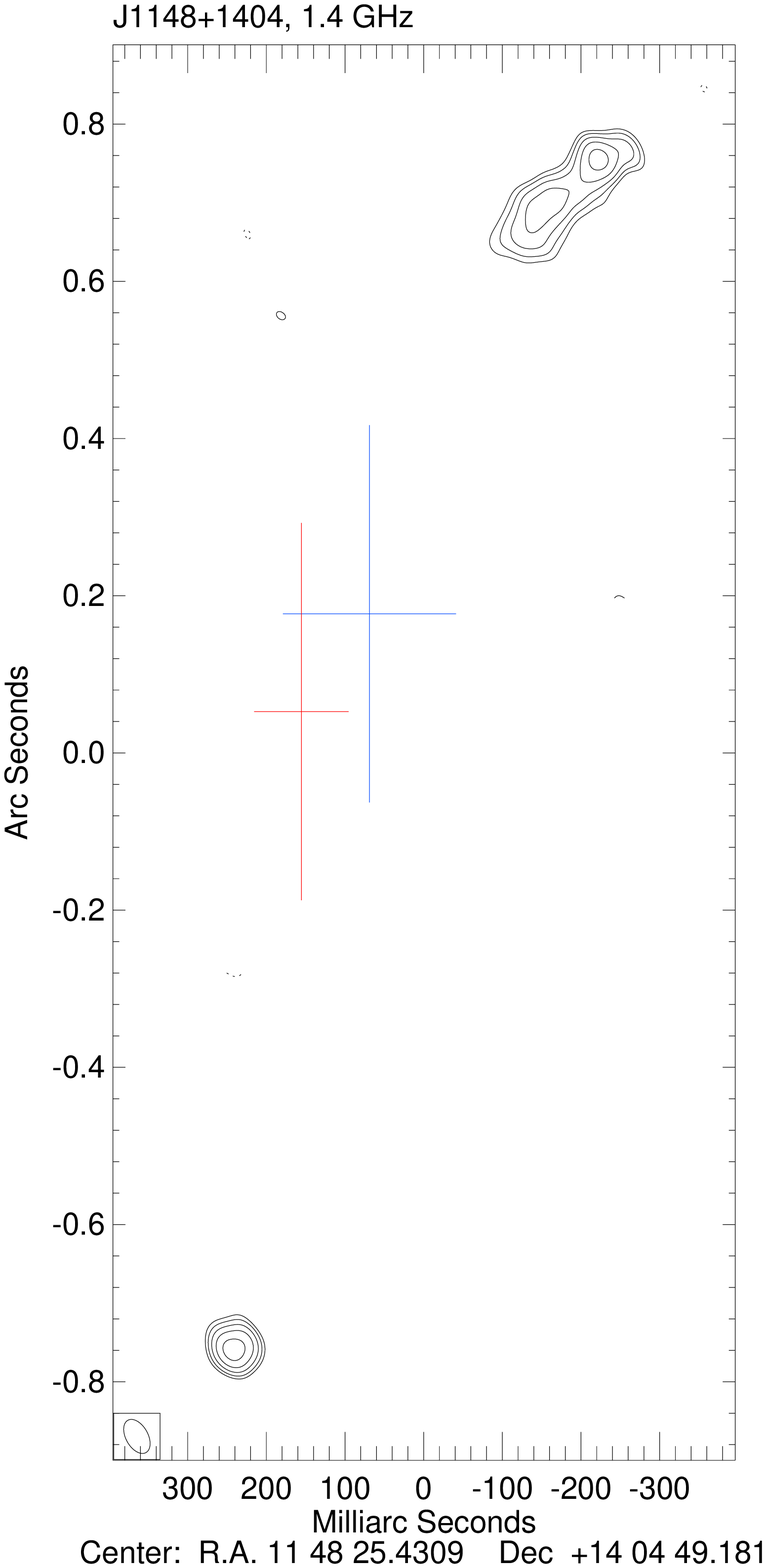}	\\
\includegraphics[scale=0.25]{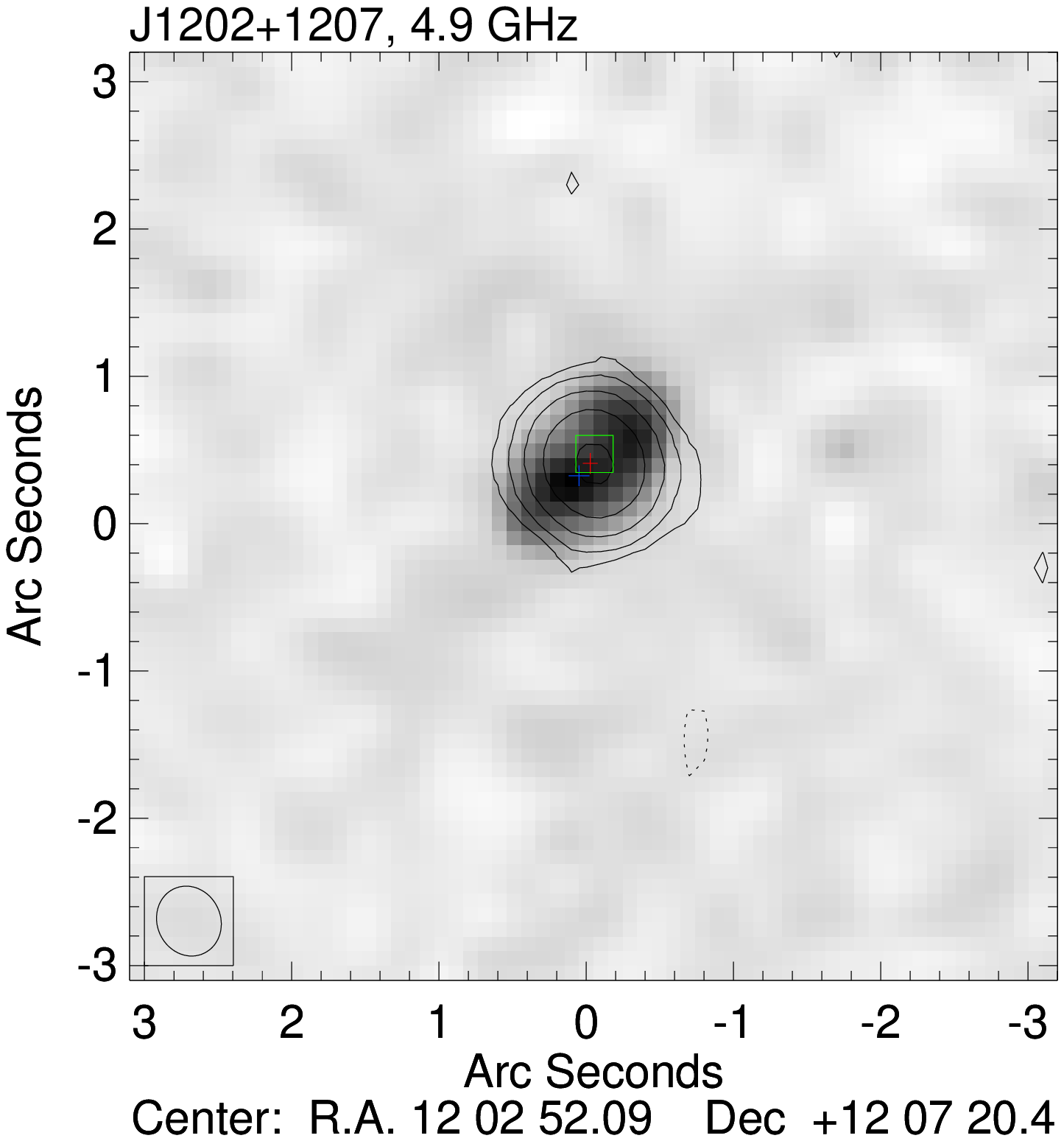}	&	\includegraphics[scale=0.25]{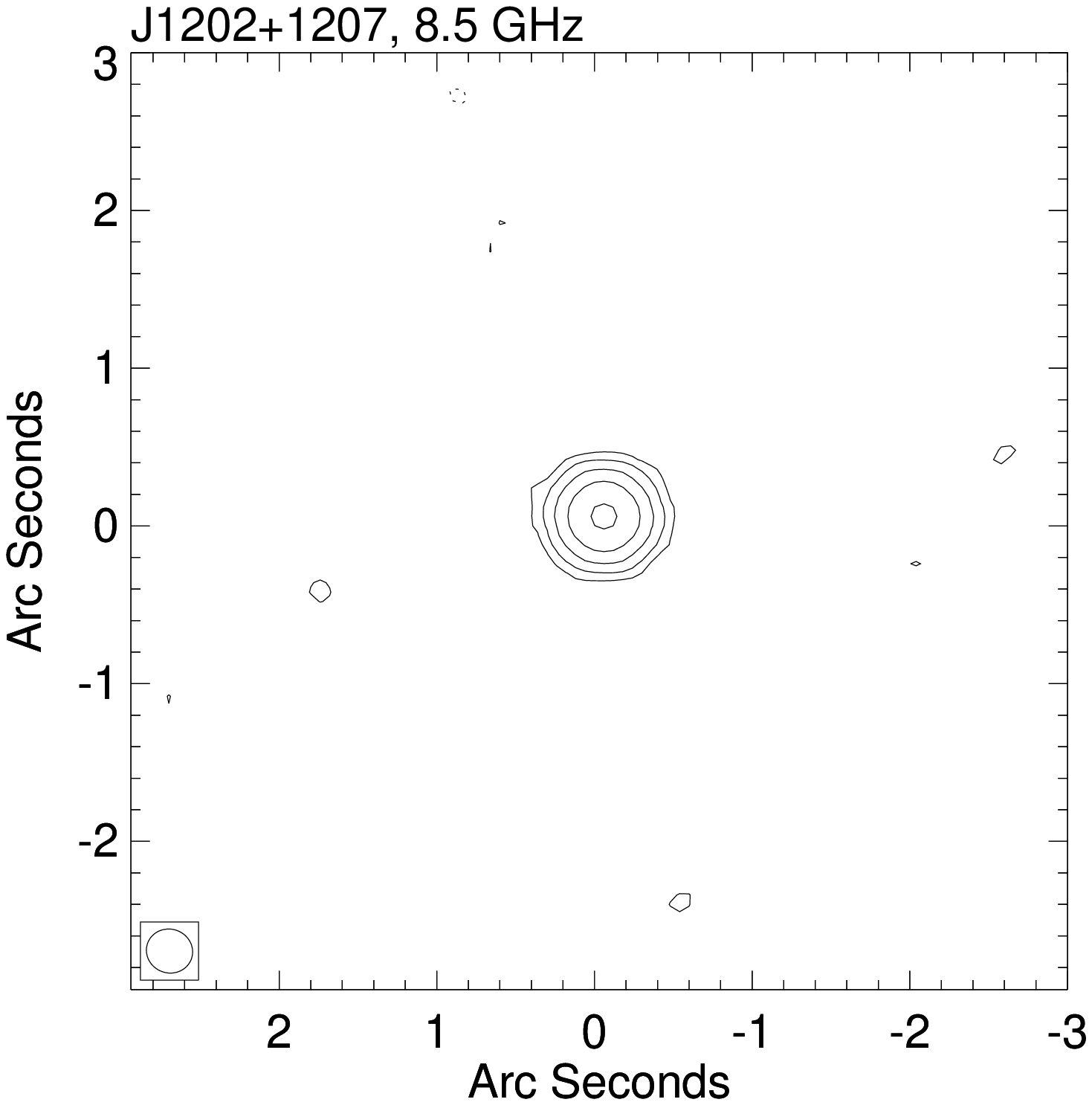}	&	\includegraphics[scale=0.25]{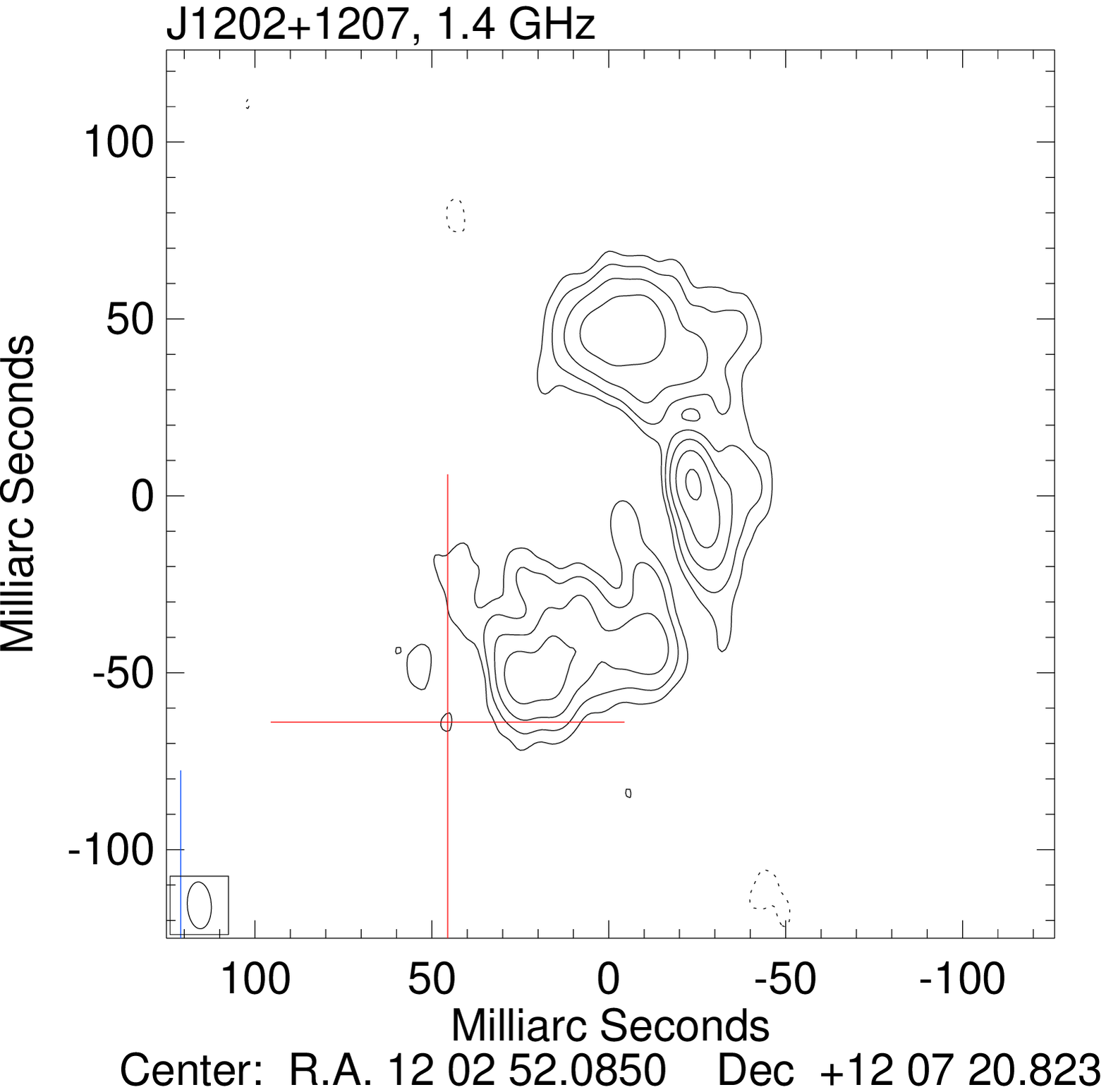}	\\
\includegraphics[scale=0.25]{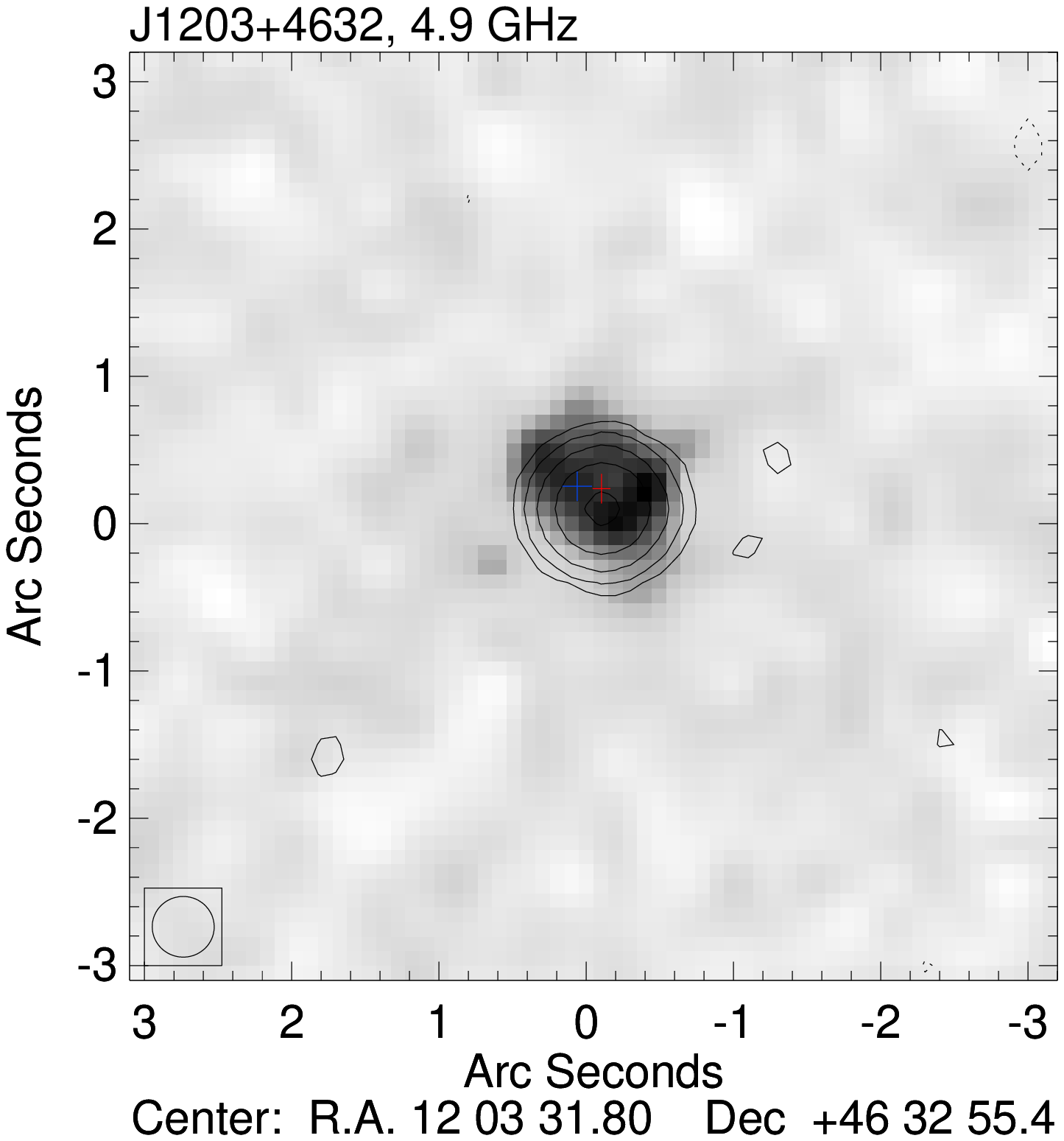}	&		&		\\
\includegraphics[scale=0.25]{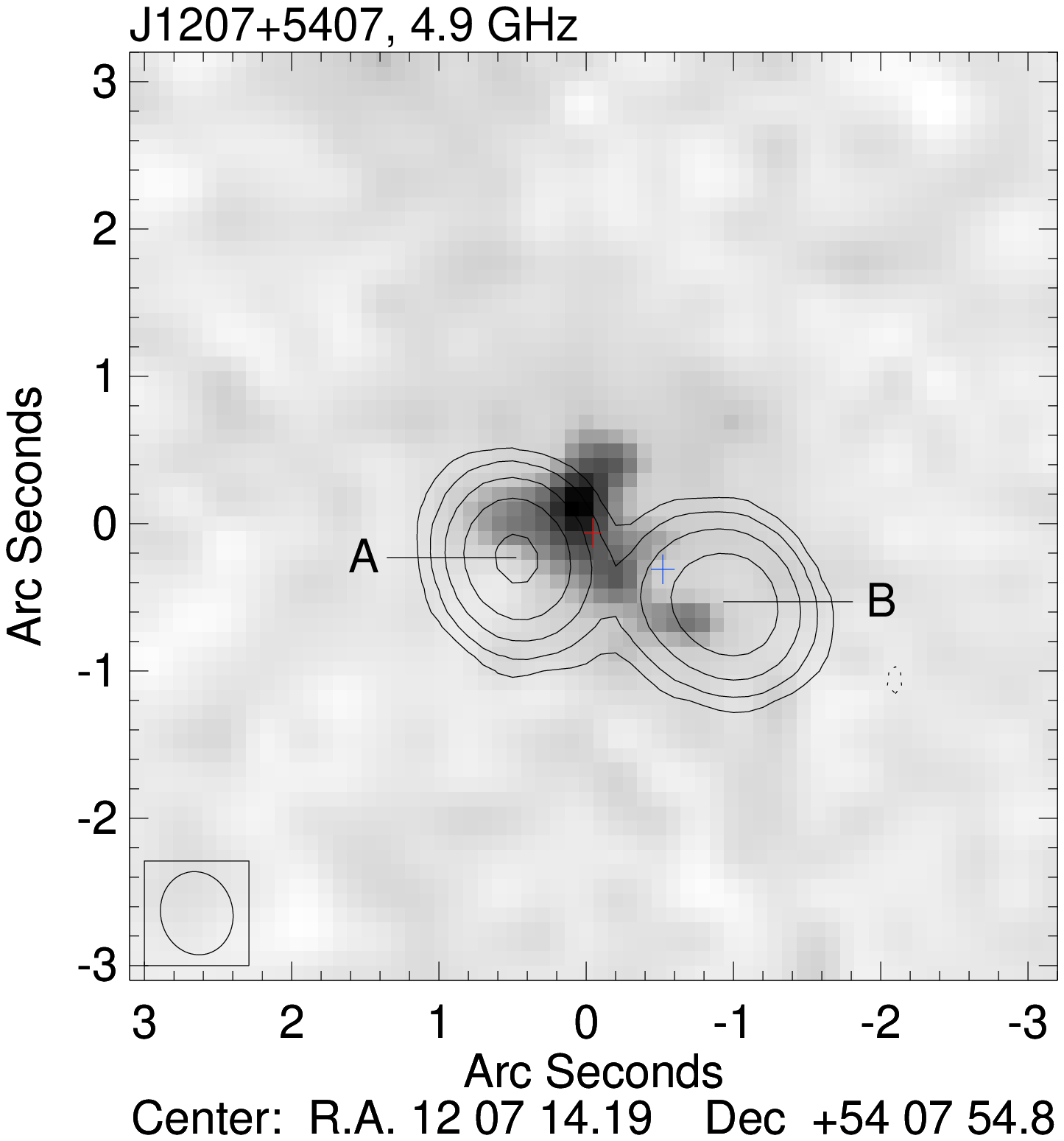}	&	\includegraphics[scale=0.25]{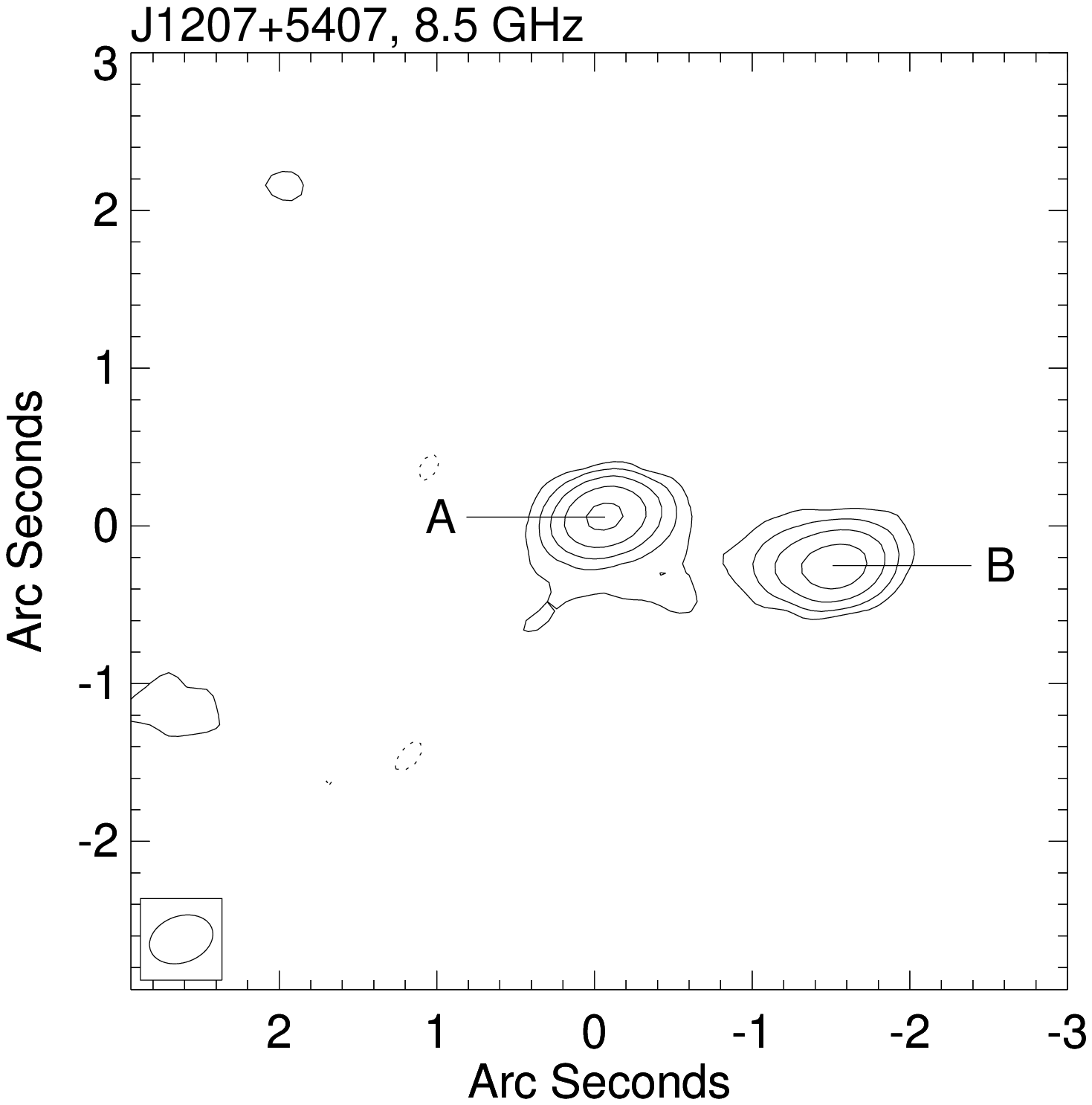}	&		\\

\end{tabular}
\end{figure}

\begin{figure}[htdp]
\begin{tabular}{lll}

\includegraphics[scale=0.25]{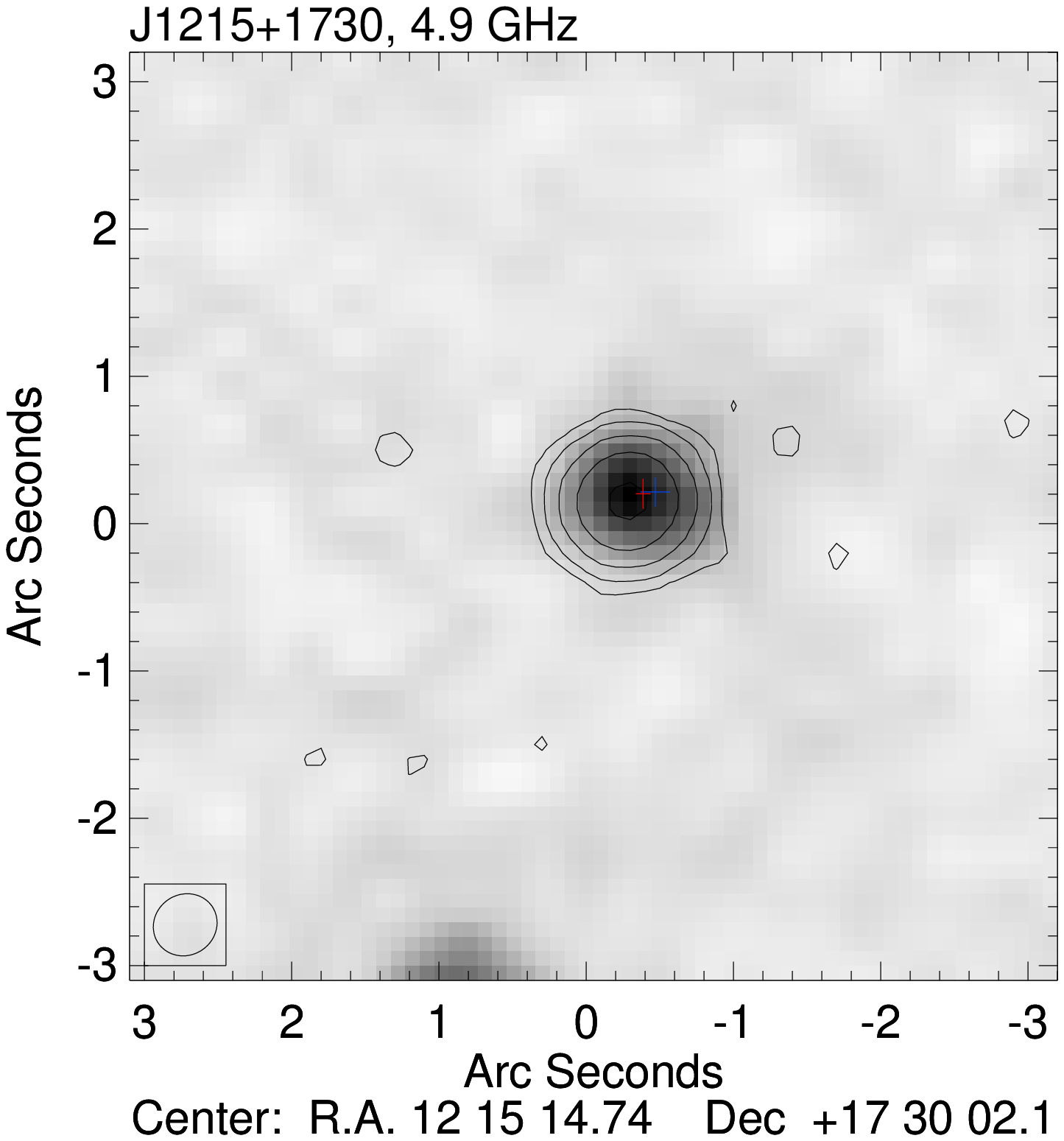}	&		&		\\
\includegraphics[scale=0.25]{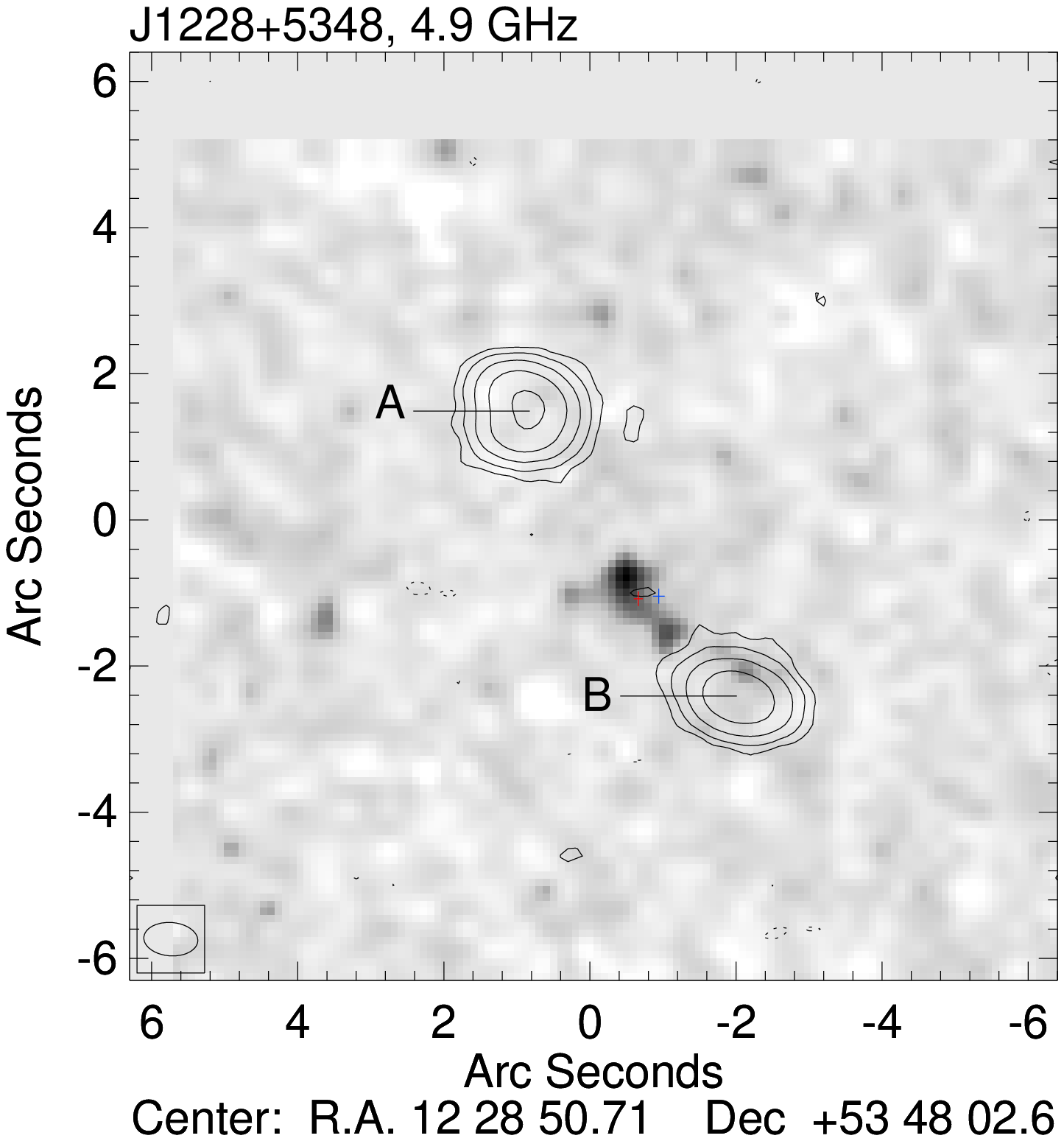}	&	\includegraphics[scale=0.25]{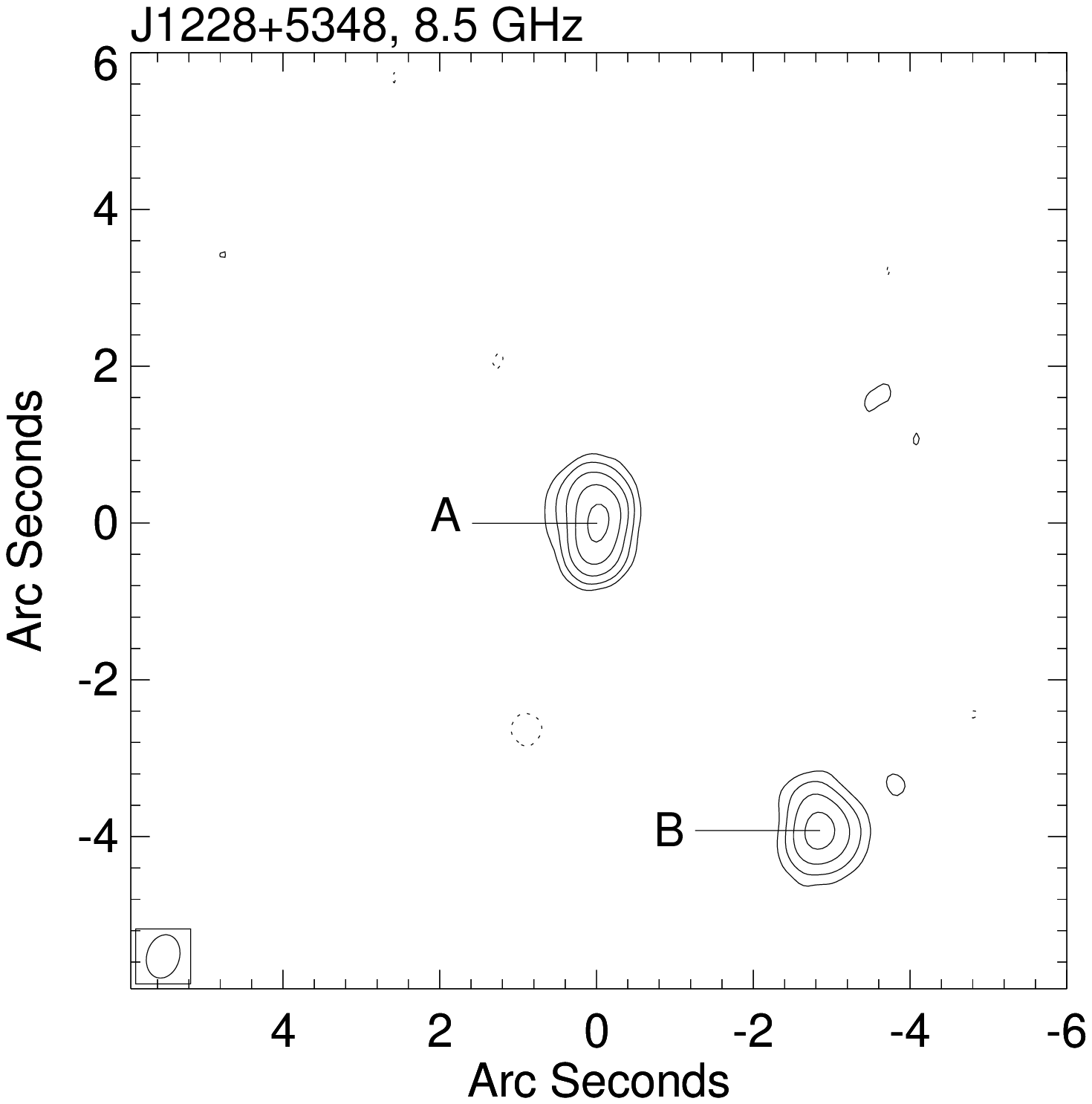}	&		\\
\includegraphics[scale=0.25]{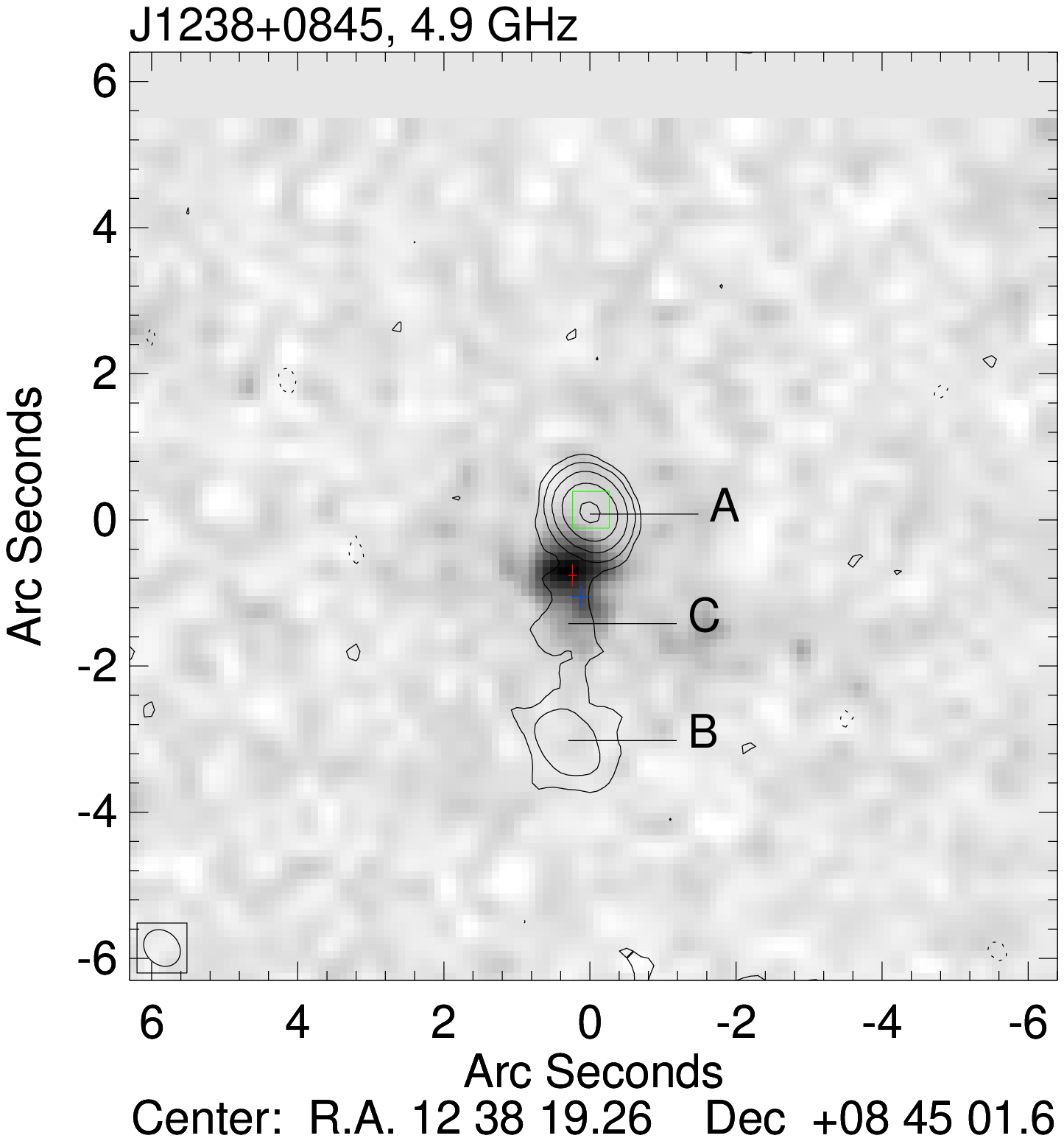}	&	\includegraphics[scale=0.25]{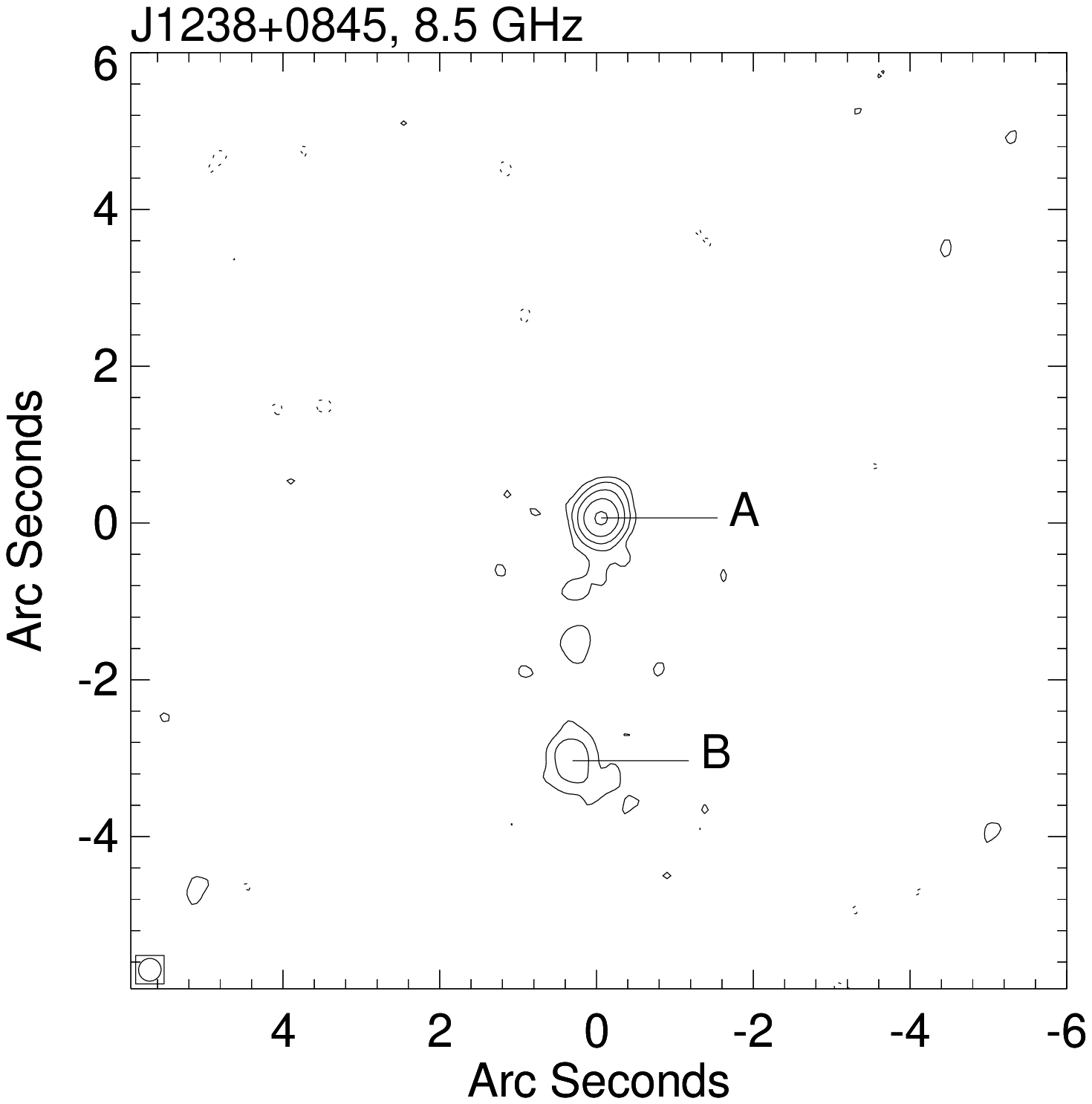}	&	\includegraphics[scale=0.25]{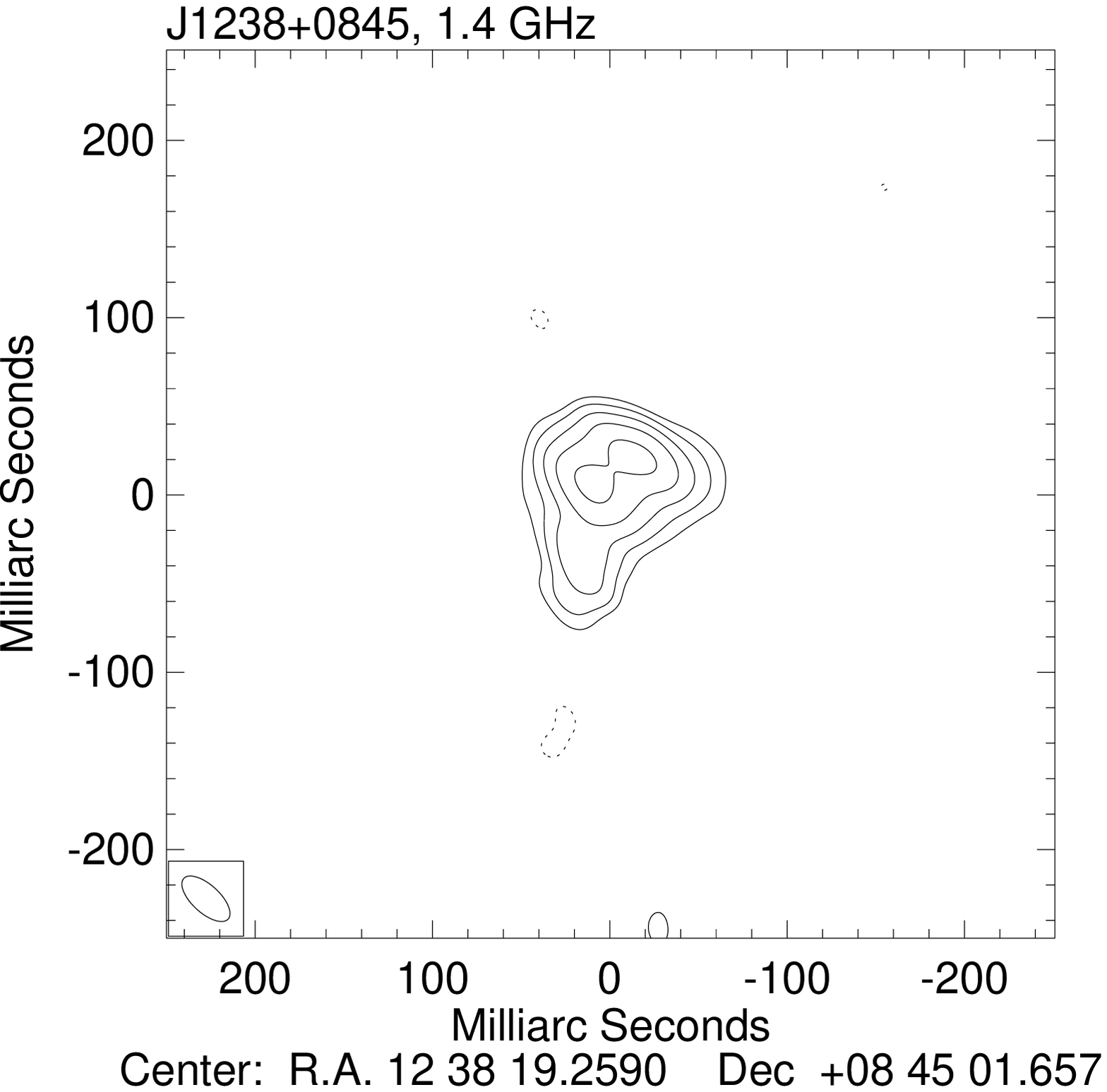}	\\
\includegraphics[scale=0.25]{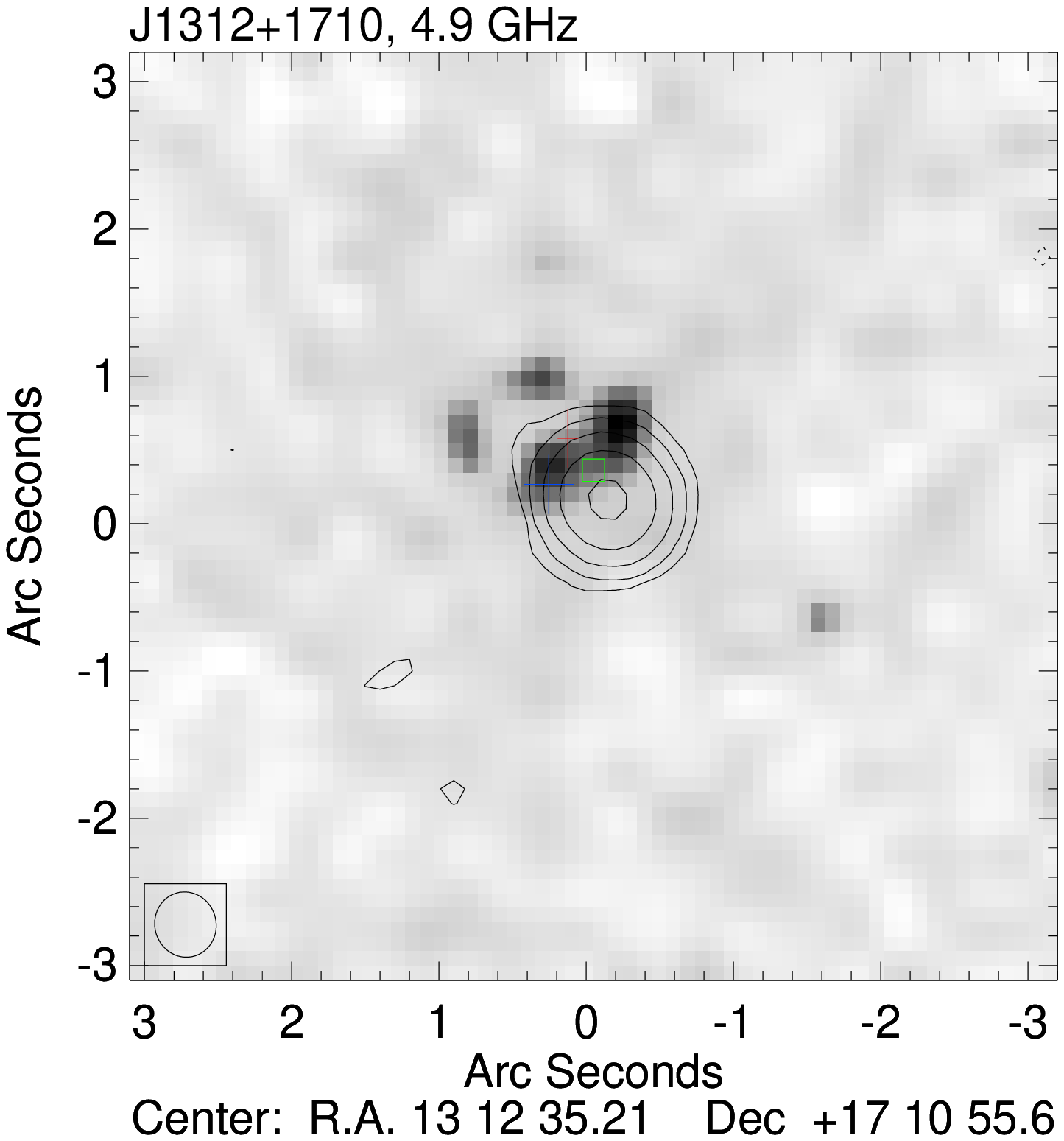}	&	\includegraphics[scale=0.25]{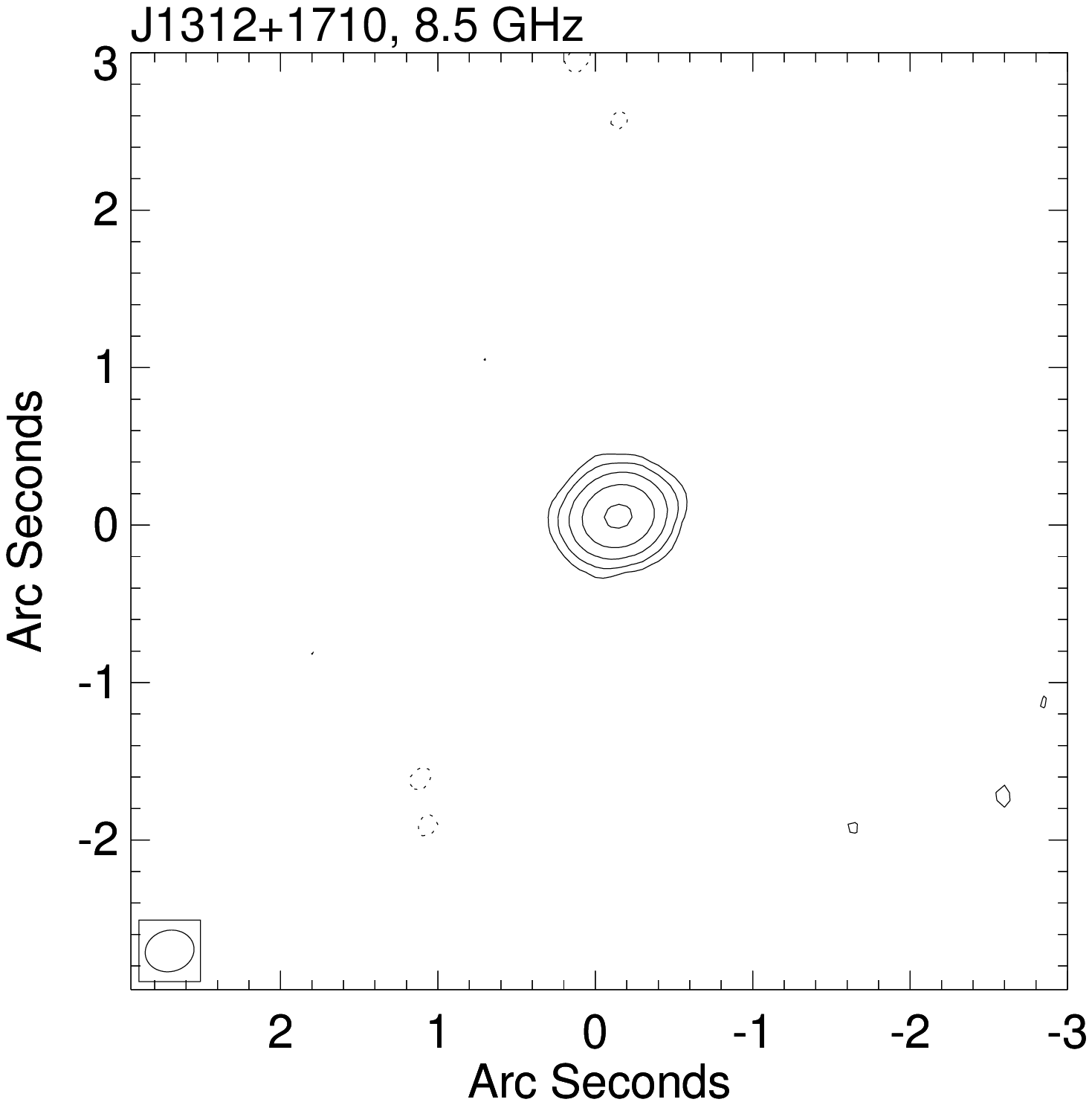}	&	\includegraphics[scale=0.25]{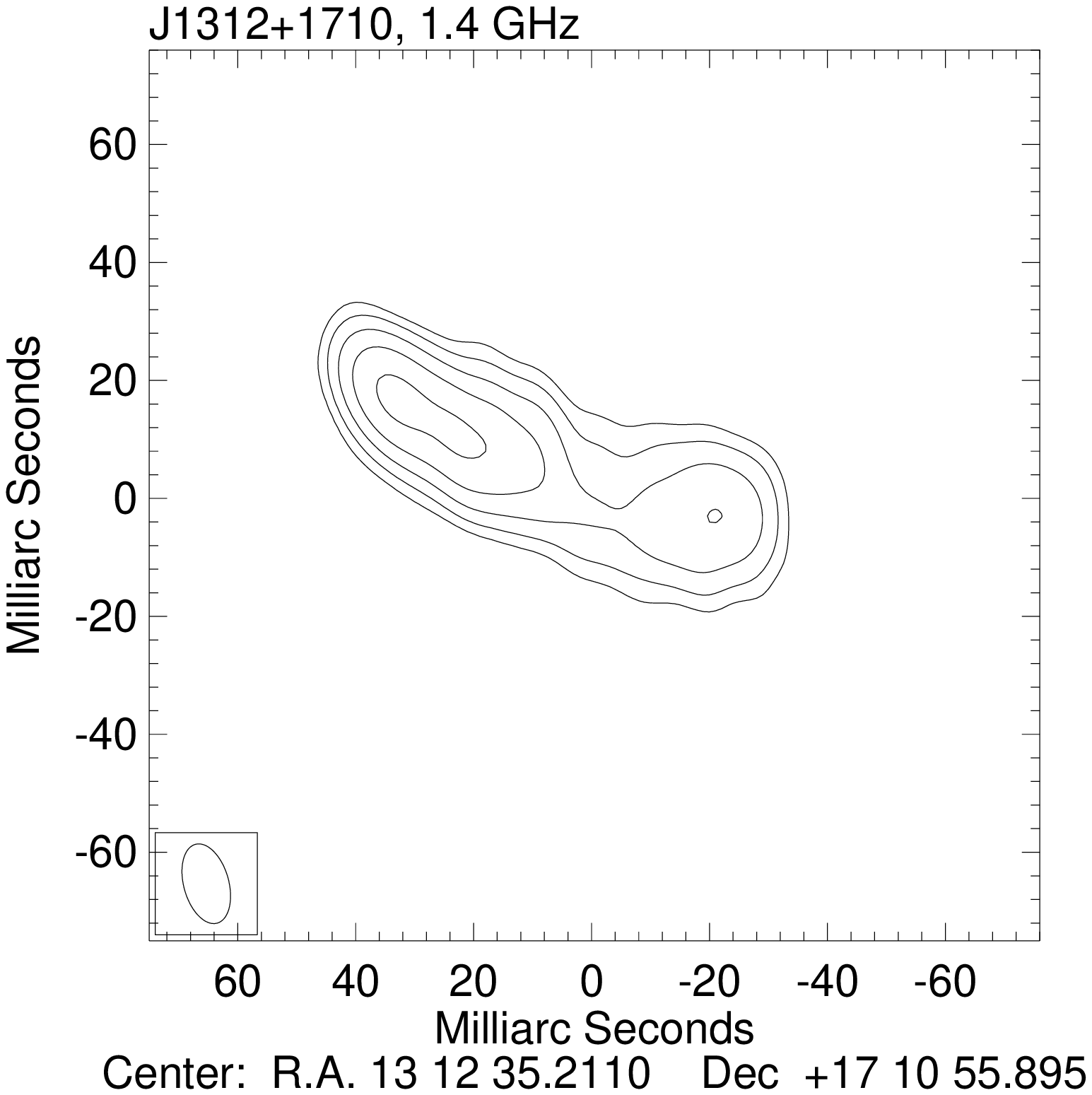}	\\
\includegraphics[scale=0.25]{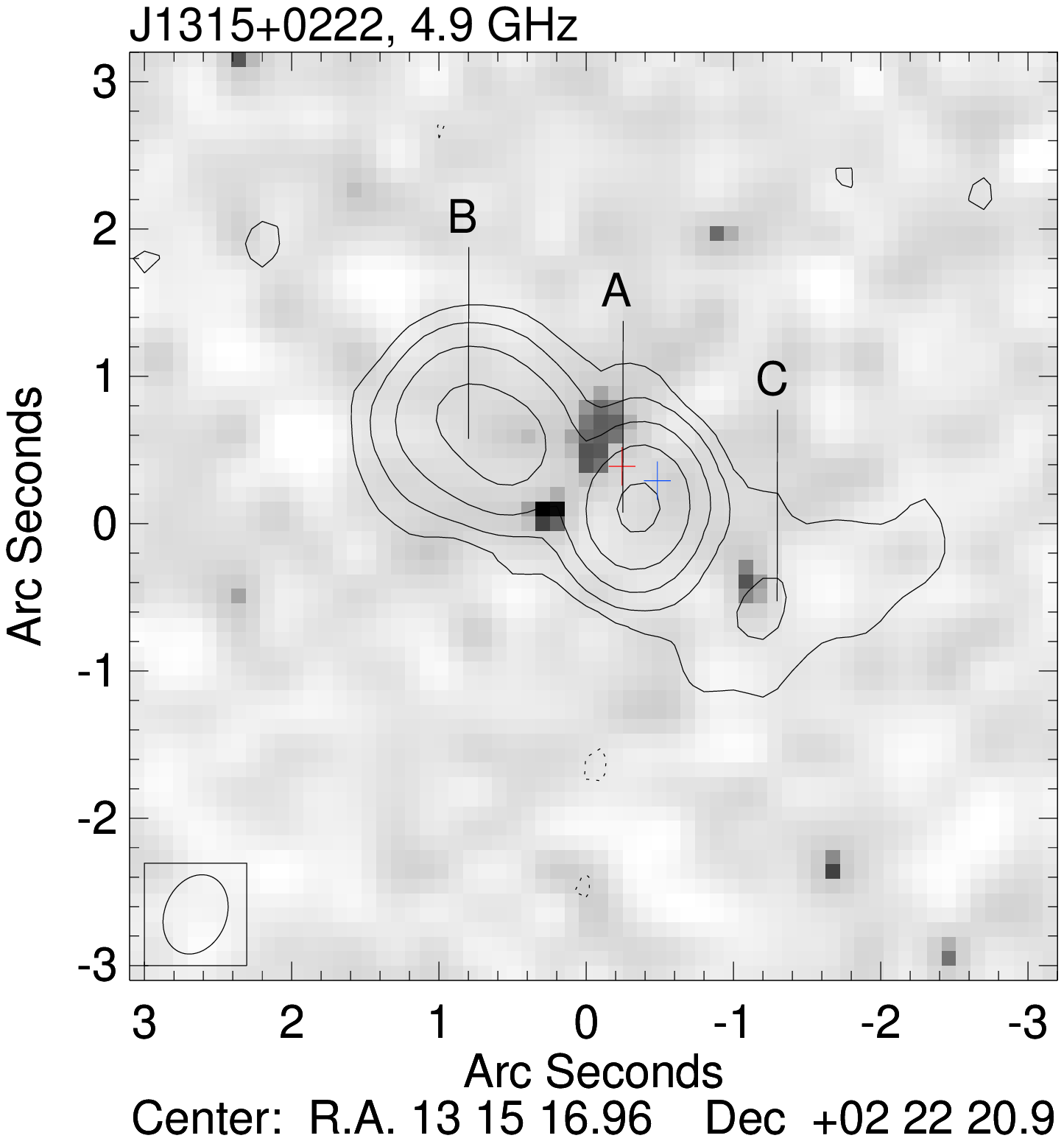}	&	\includegraphics[scale=0.25]{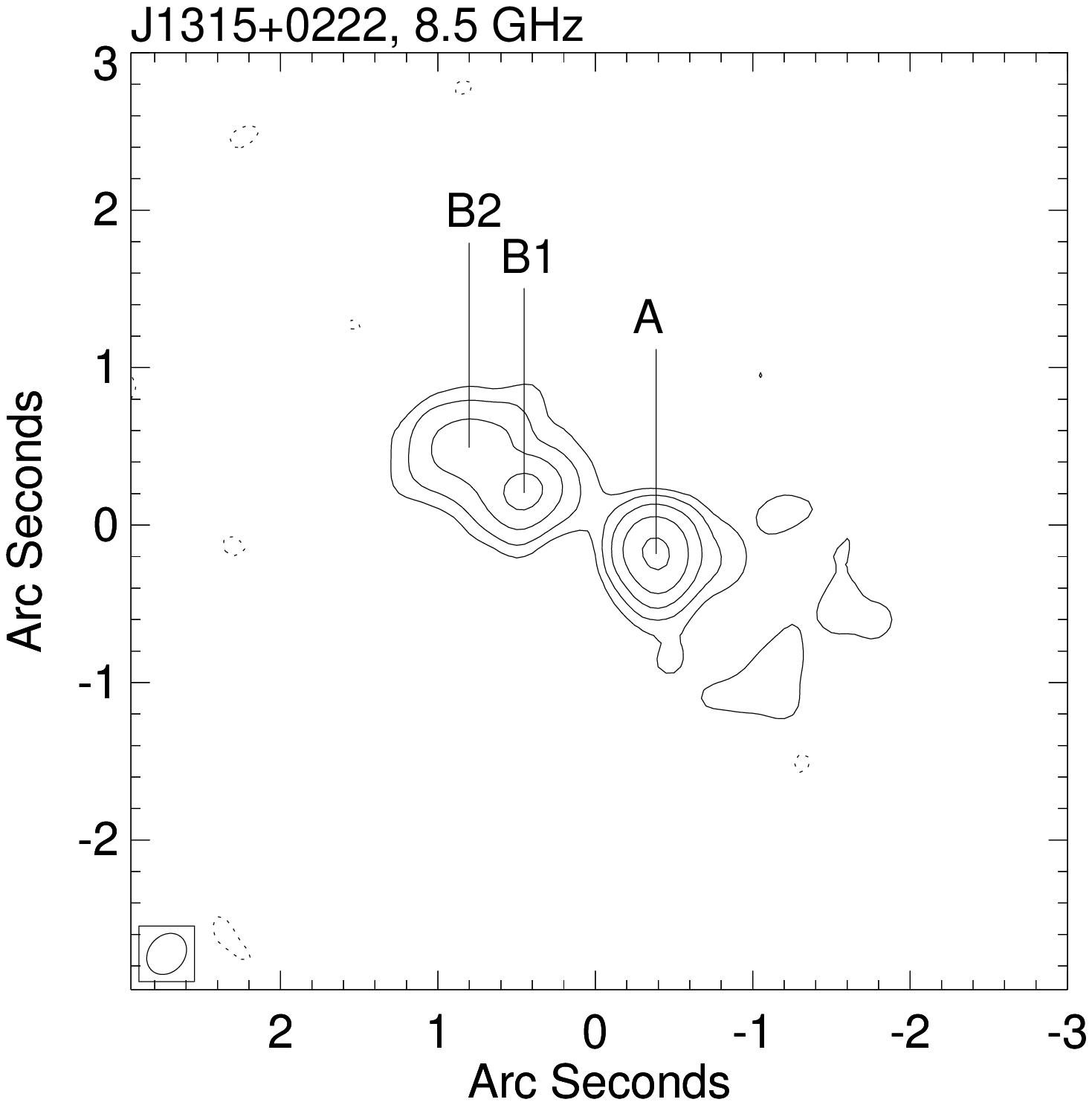}	&		\\

\end{tabular}
\end{figure}

\begin{figure}[htdp]
\begin{tabular}{lll}

\includegraphics[scale=0.25]{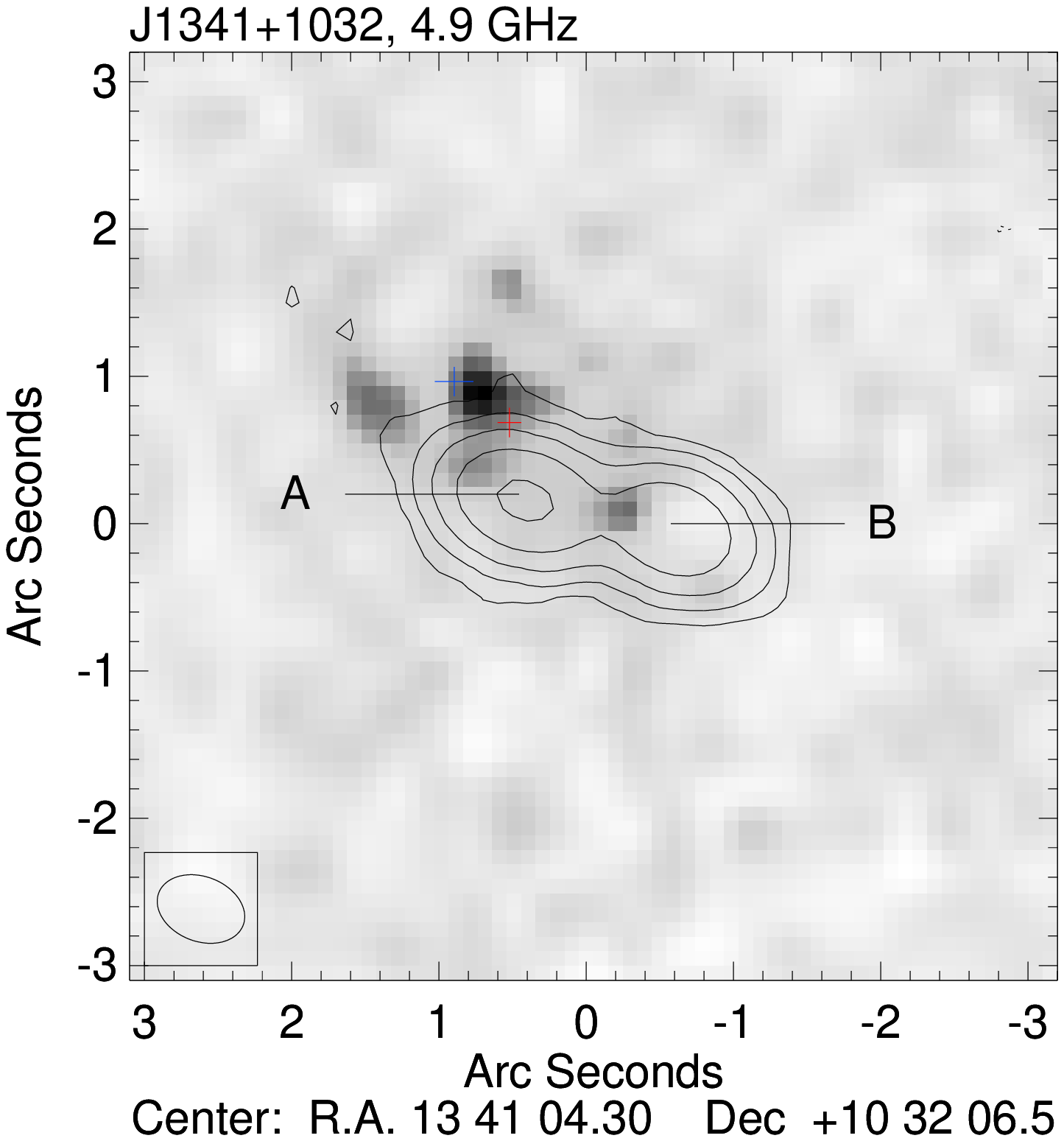}	&	\includegraphics[scale=0.25]{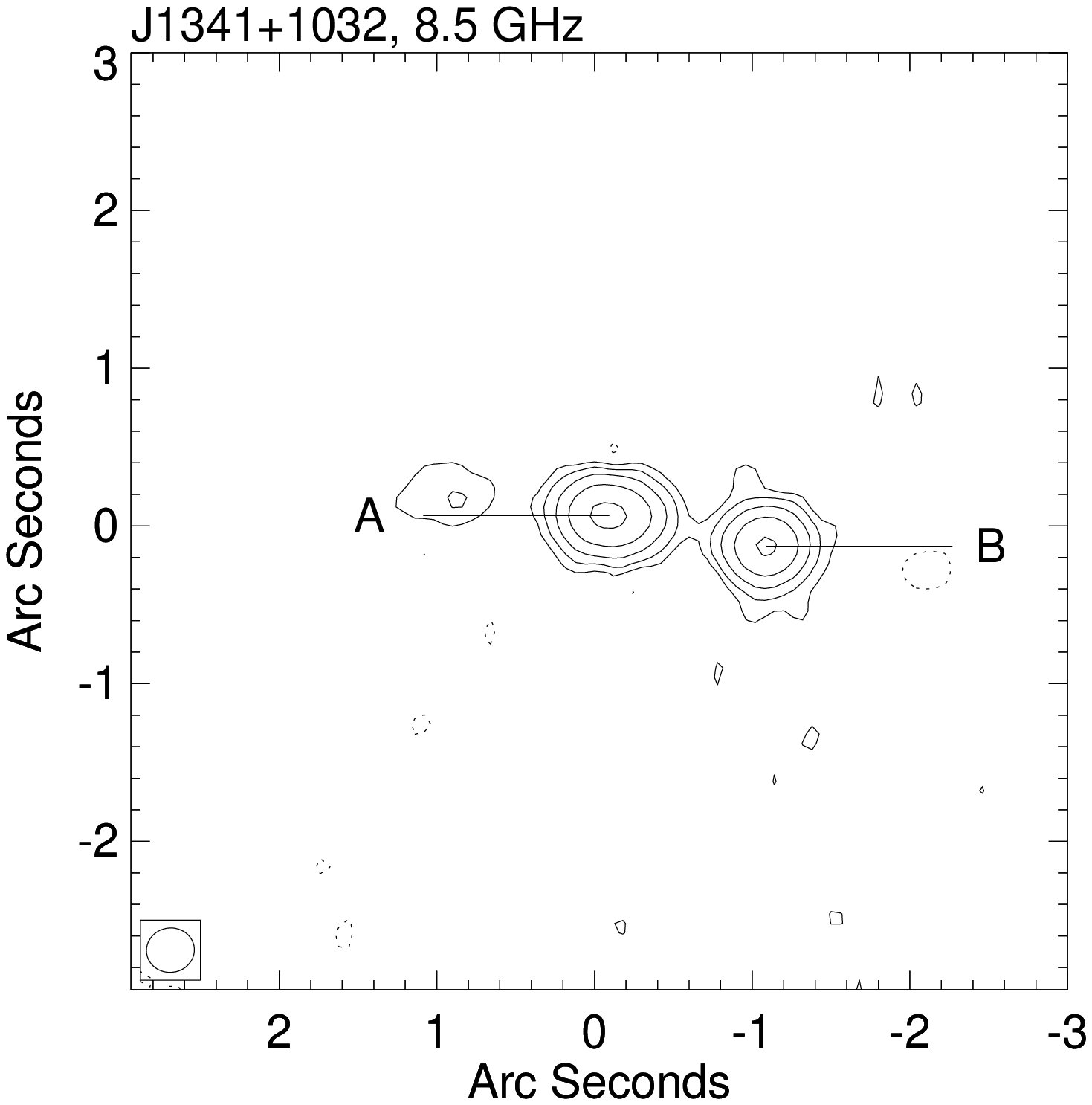}	&		\\
\includegraphics[scale=0.25]{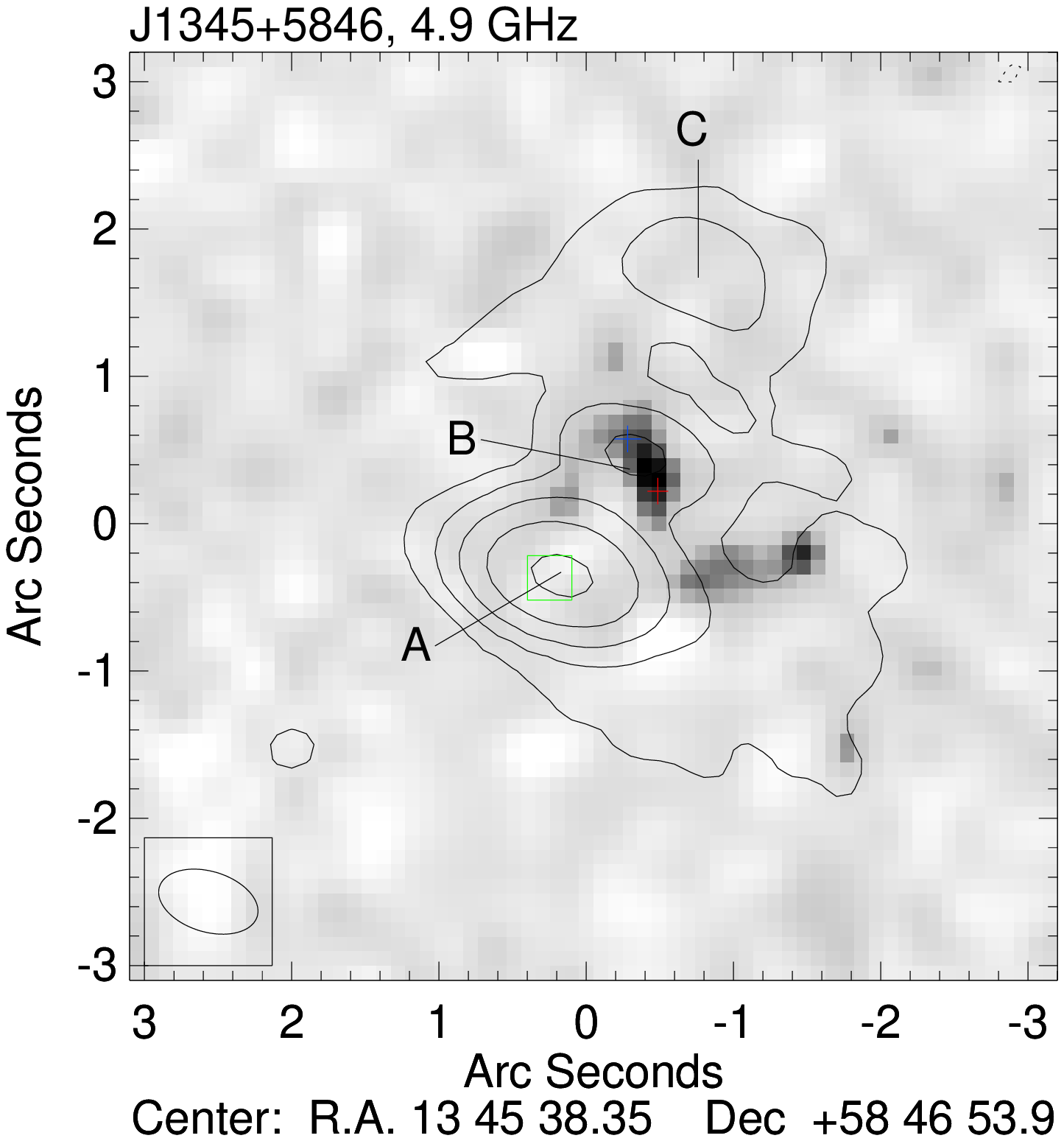}	&	\includegraphics[scale=0.25]{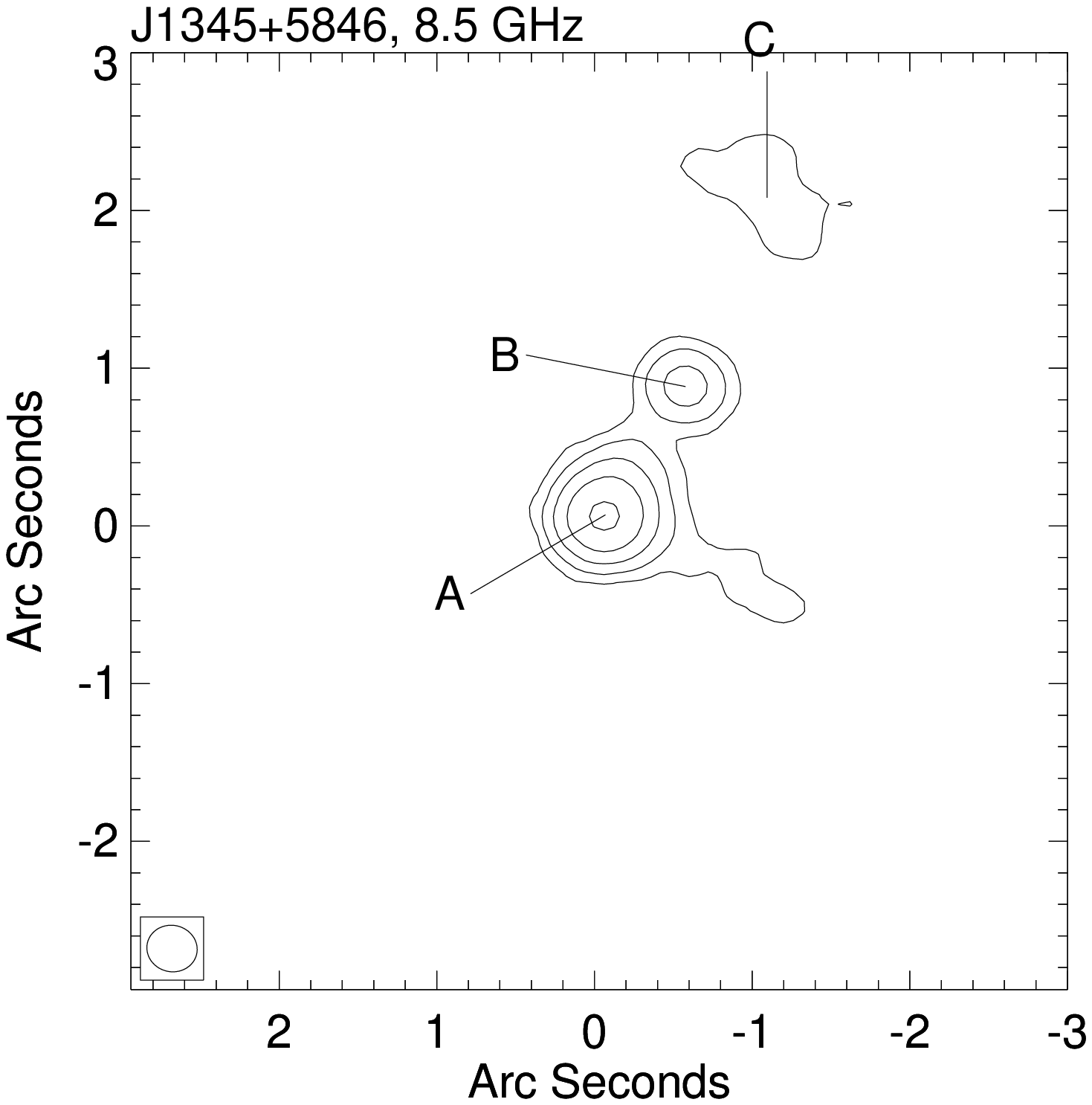}	&	\includegraphics[scale=0.25]{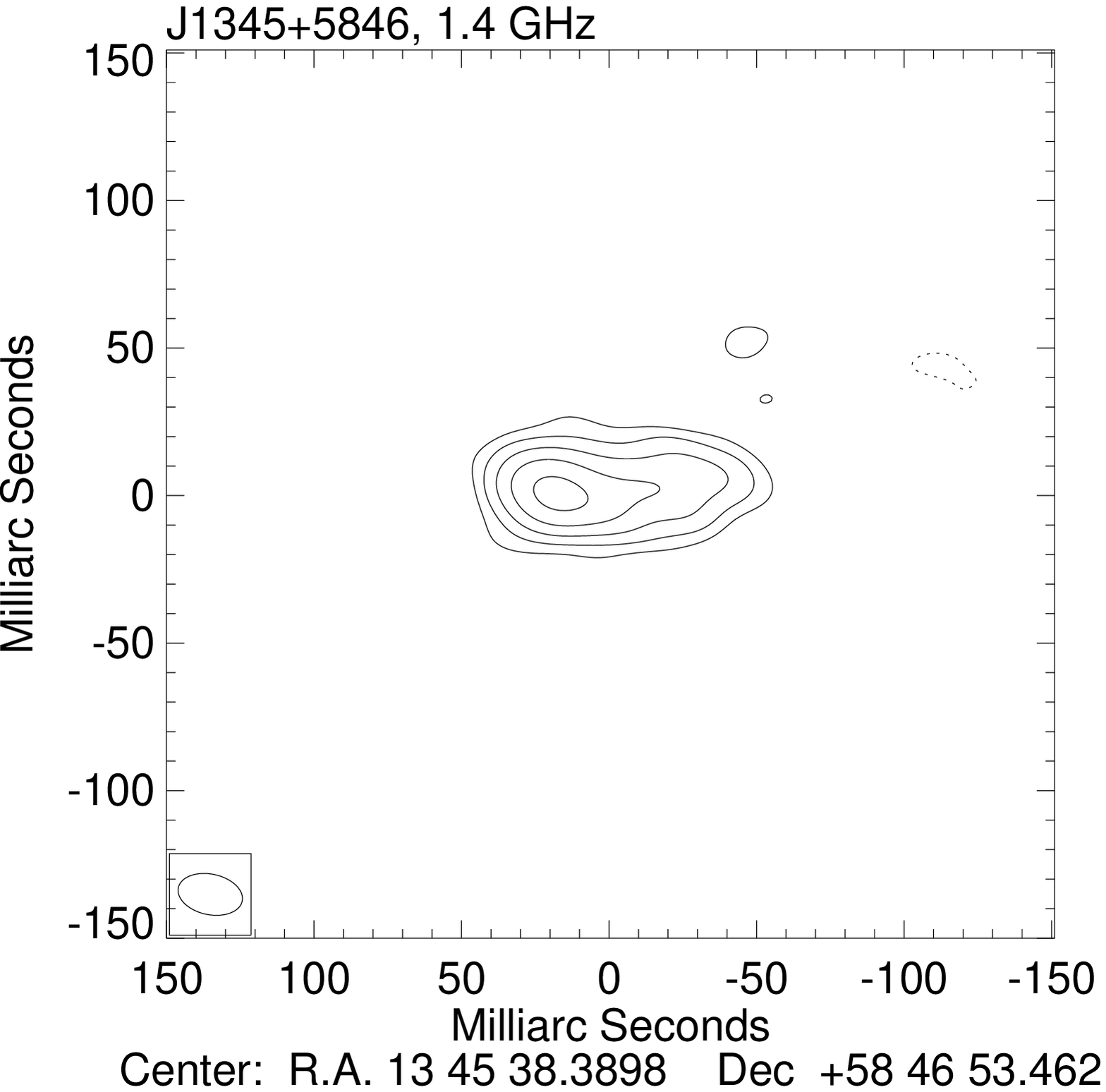}	\\
\includegraphics[scale=0.25]{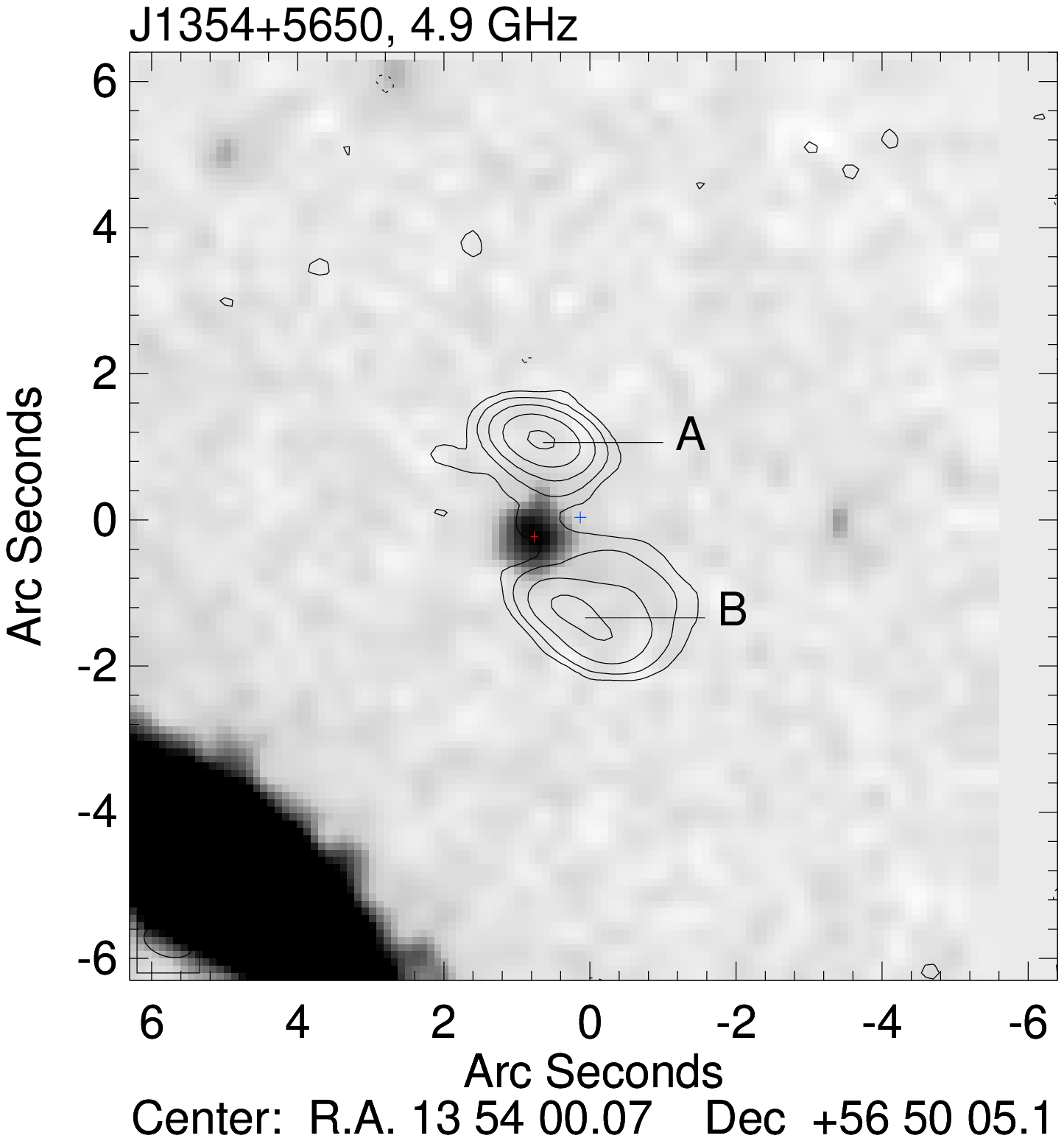}	&	\includegraphics[scale=0.25]{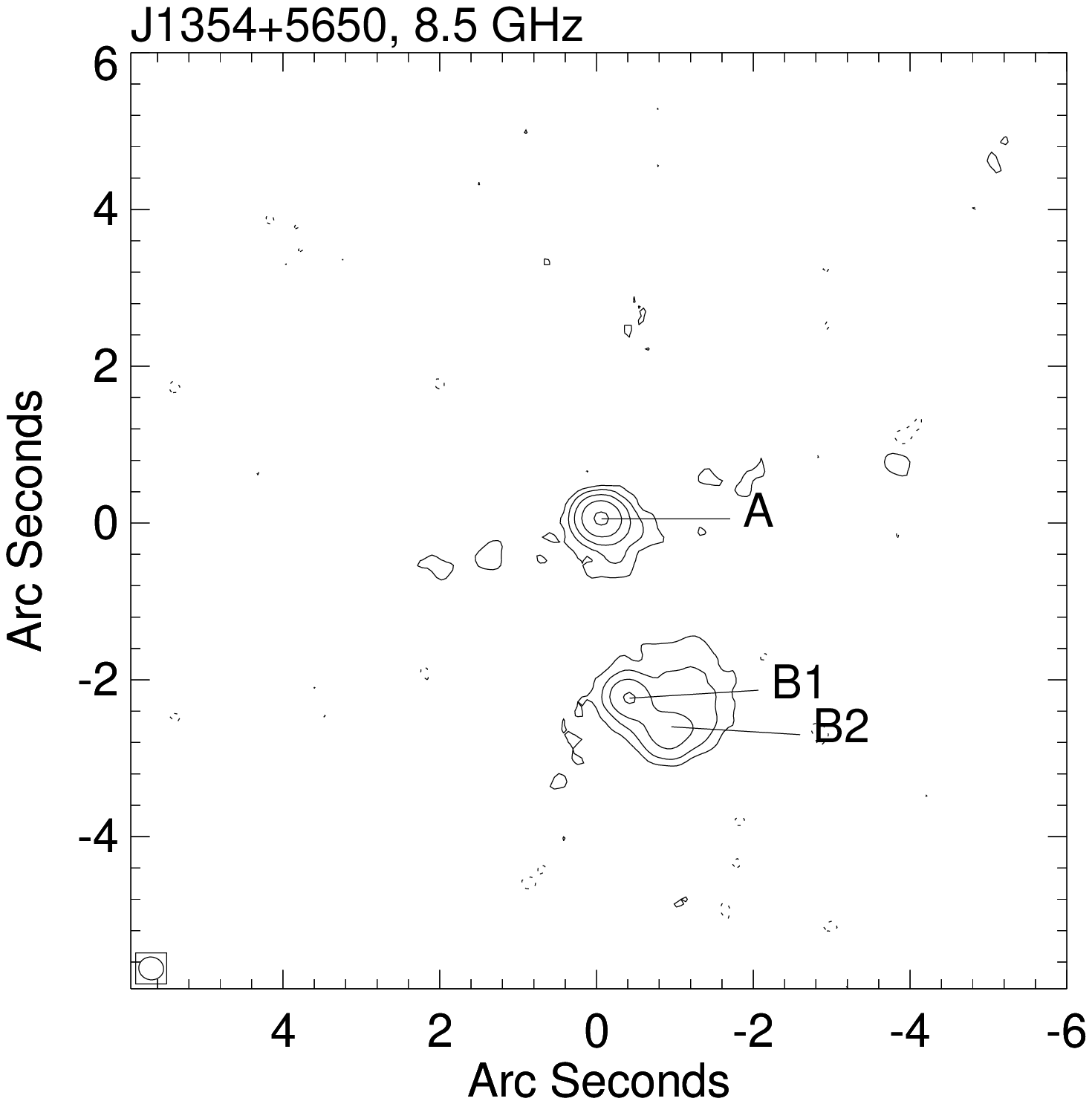}	&		\\
\includegraphics[scale=0.25]{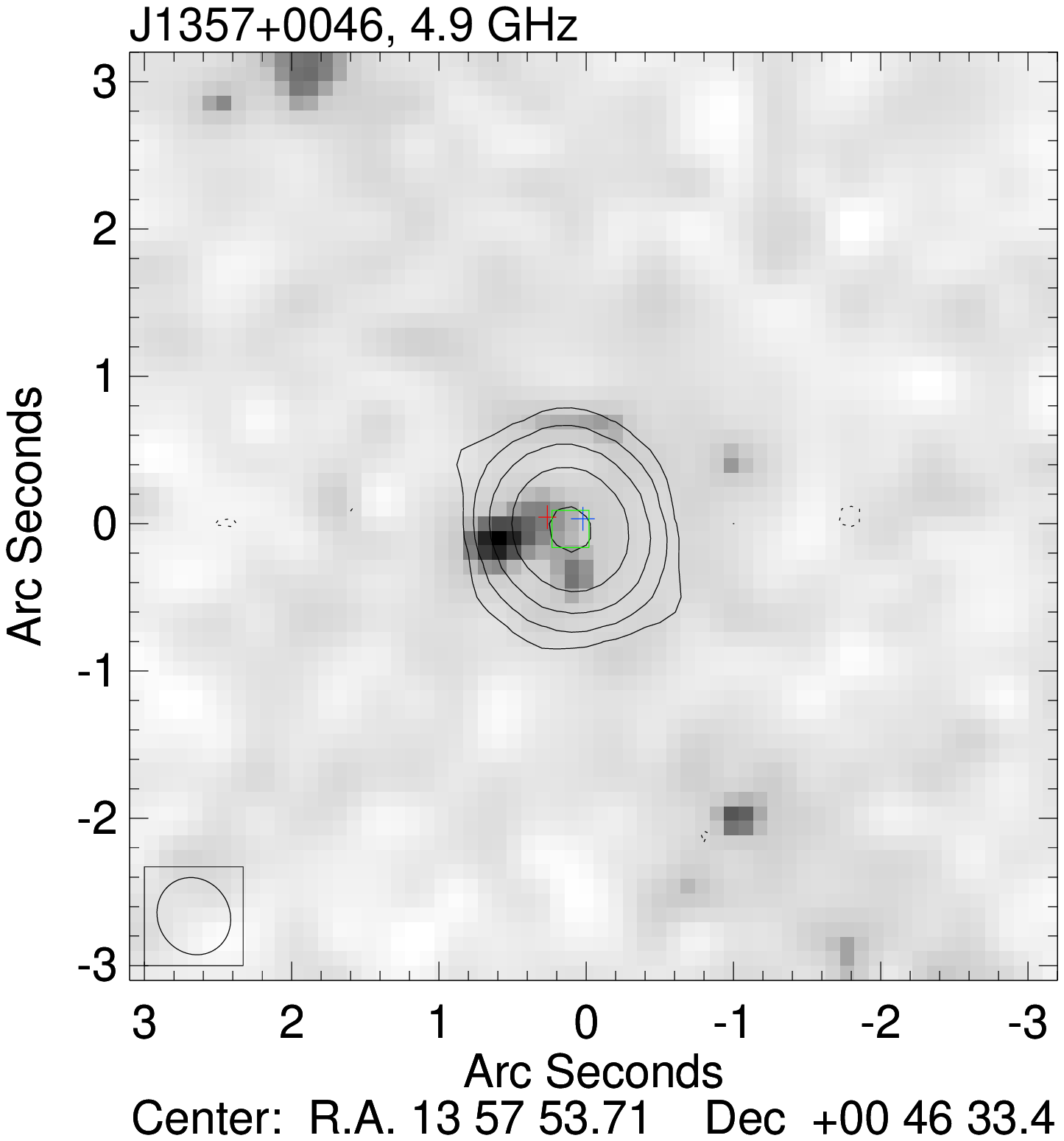}	&	\includegraphics[scale=0.25]{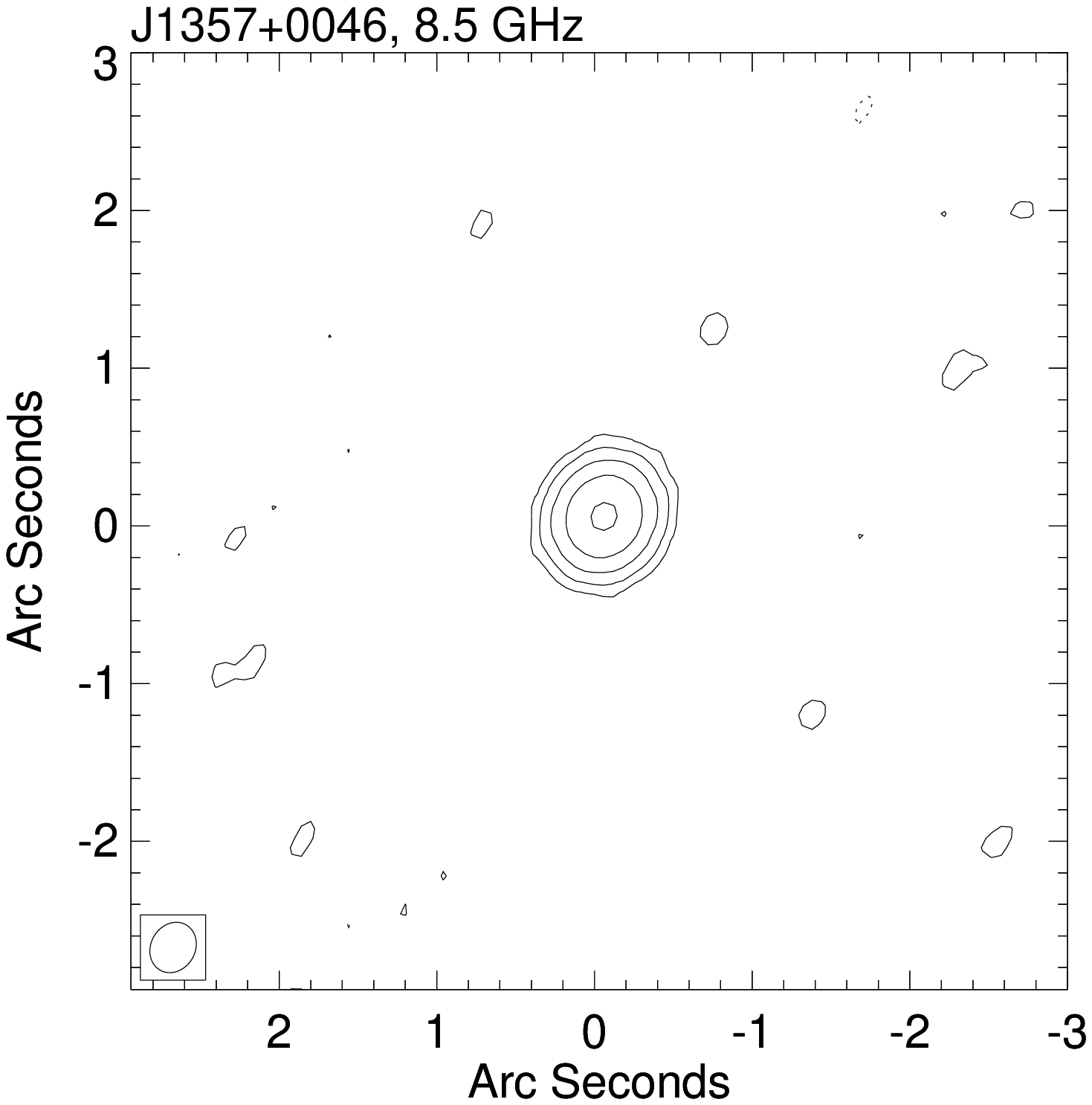}	&	\includegraphics[scale=0.25]{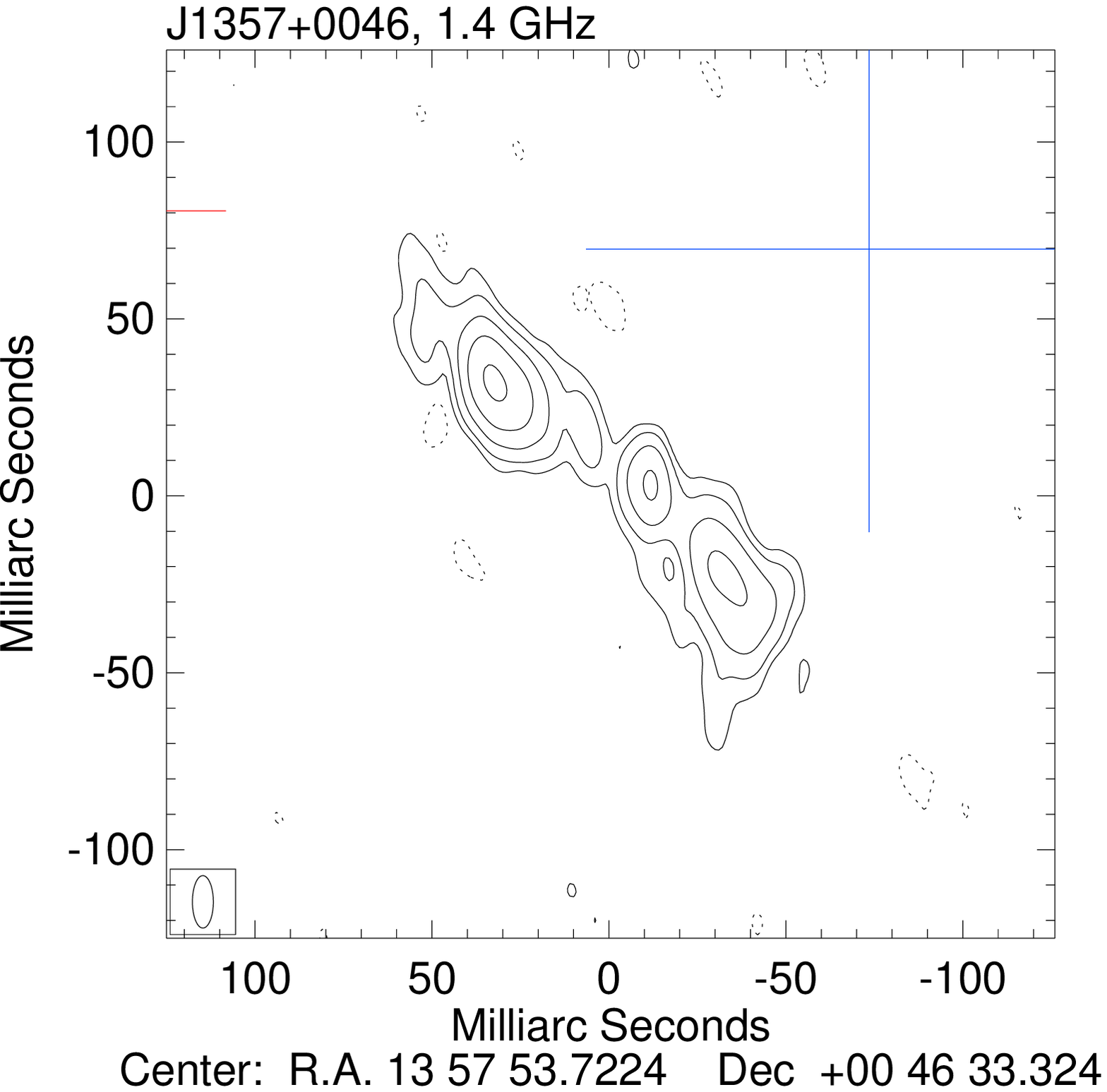}	\\
\includegraphics[scale=0.25]{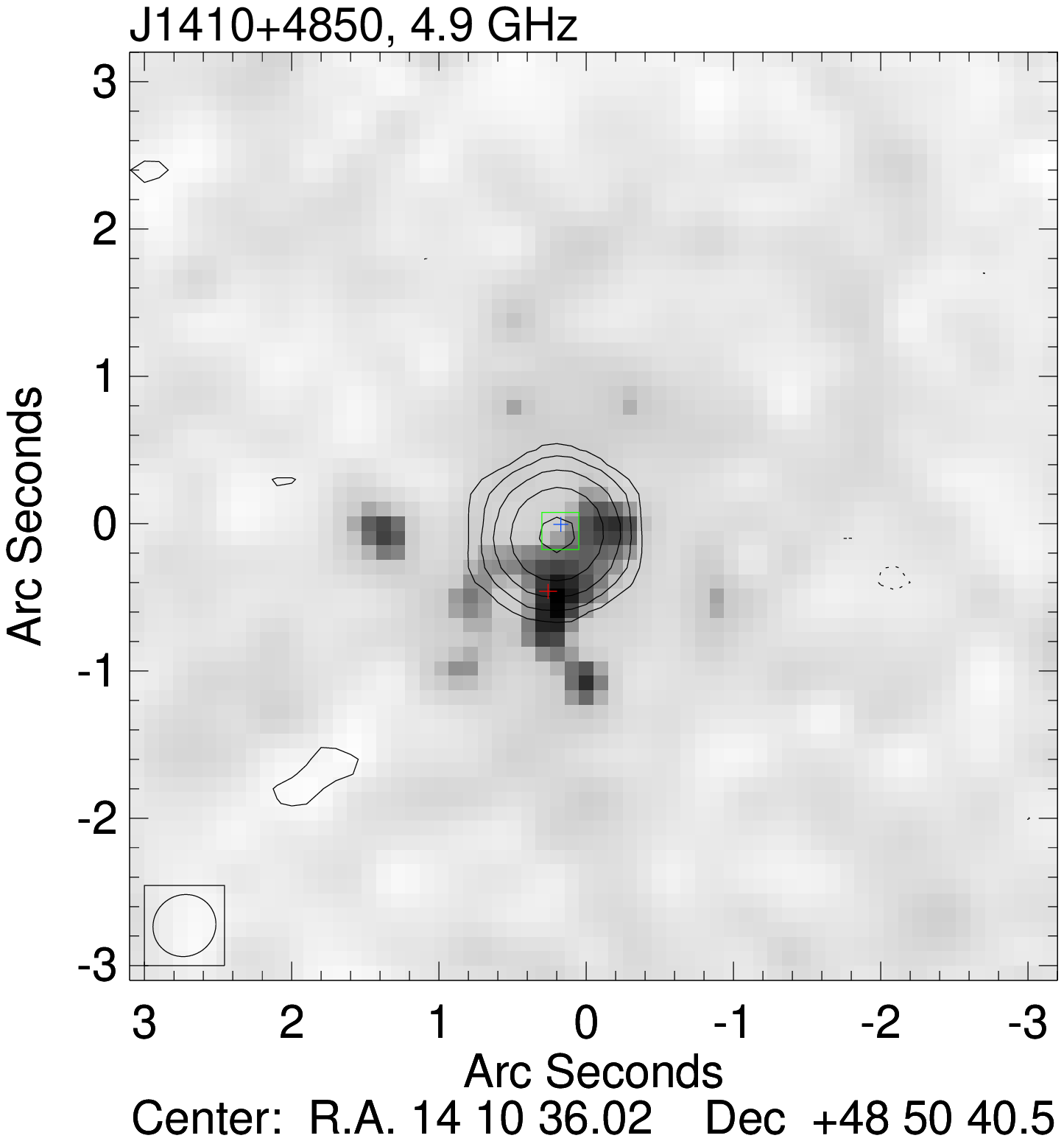}	&	\includegraphics[scale=0.25]{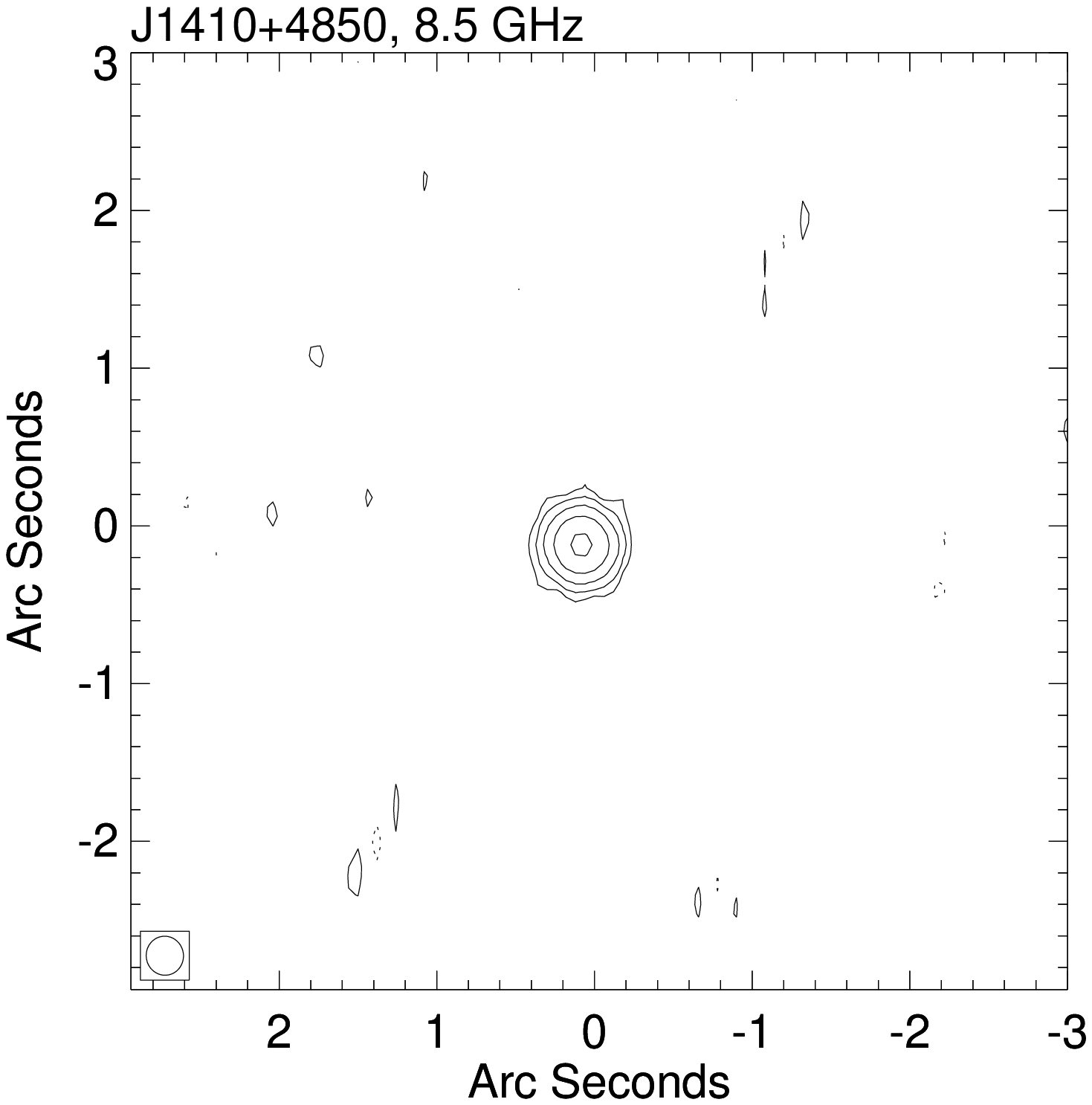}	&	\includegraphics[scale=0.25]{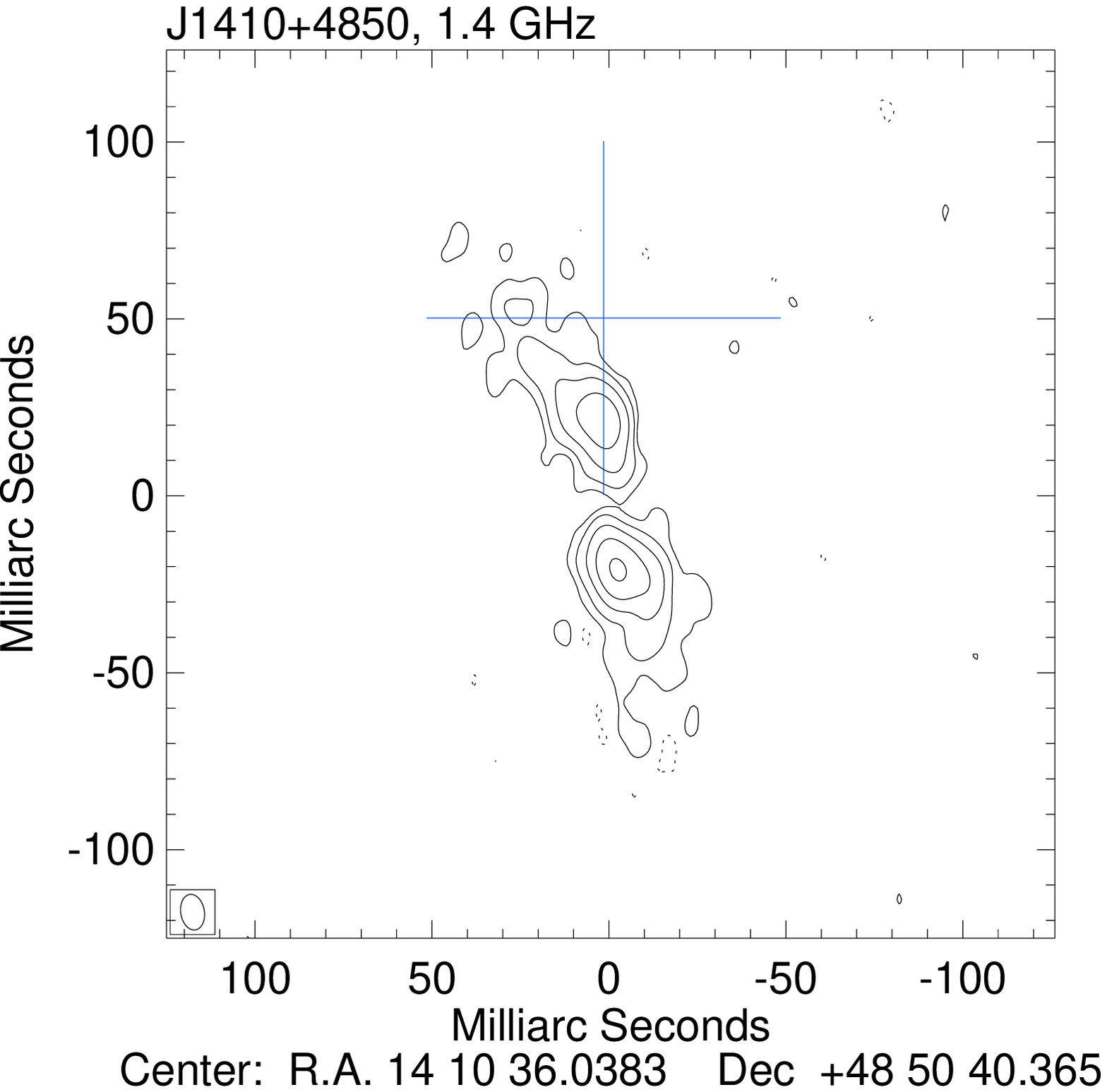}	\\

\end{tabular}
\end{figure}

\begin{figure}[htdp]
\begin{tabular}{lll}

\includegraphics[scale=0.25]{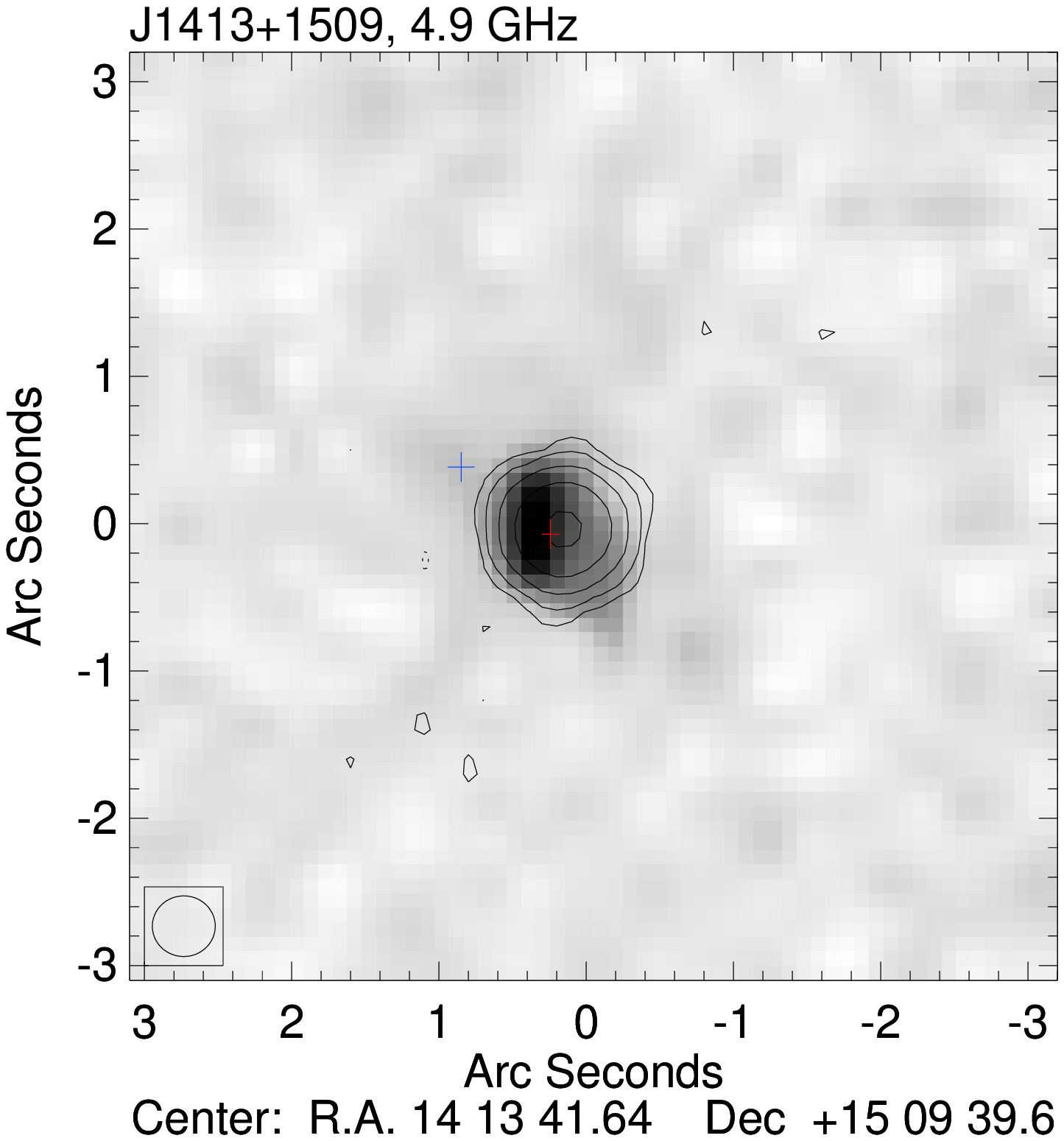}	&		&		\\
\includegraphics[scale=0.25]{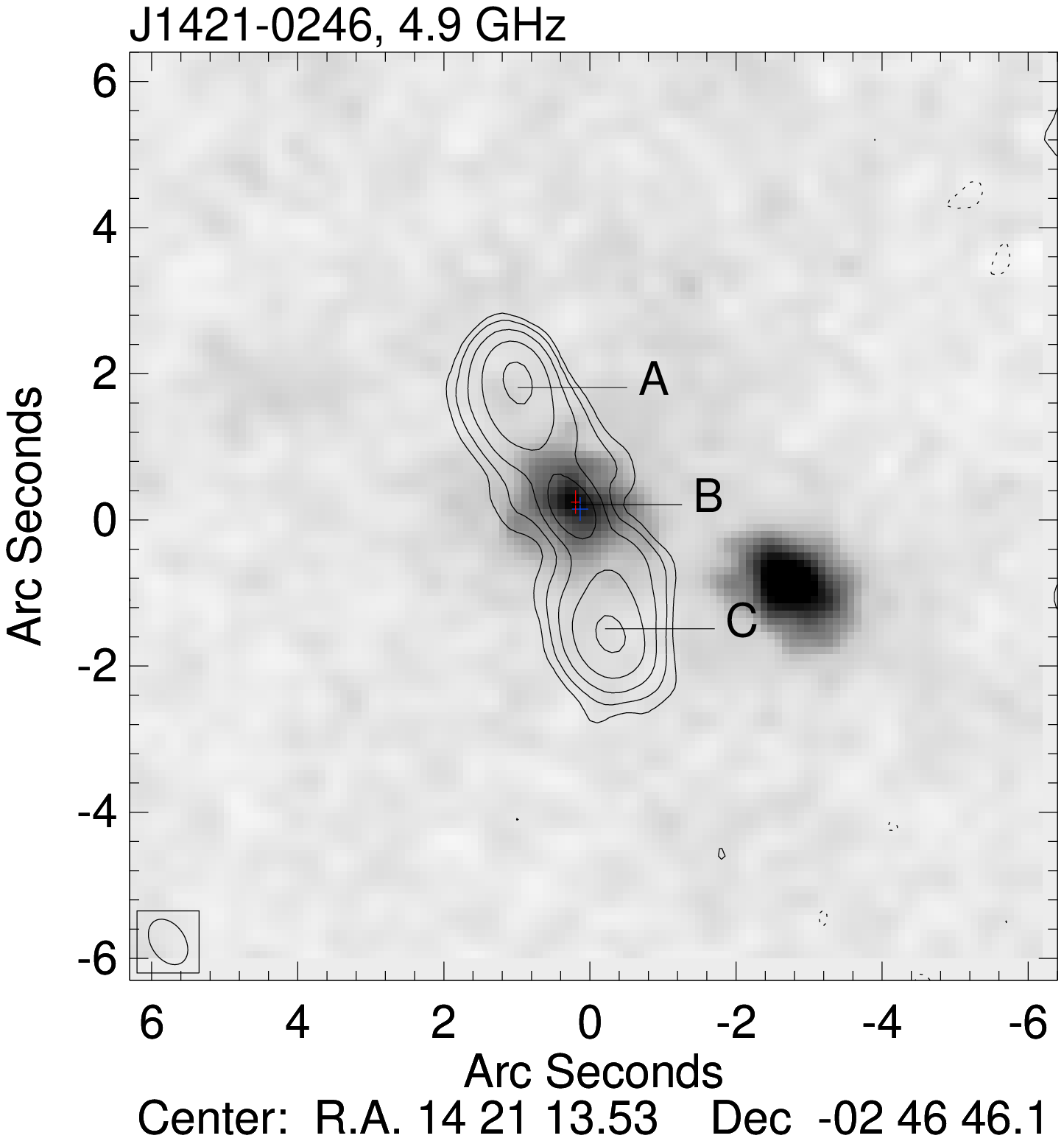}	&	\includegraphics[scale=0.25]{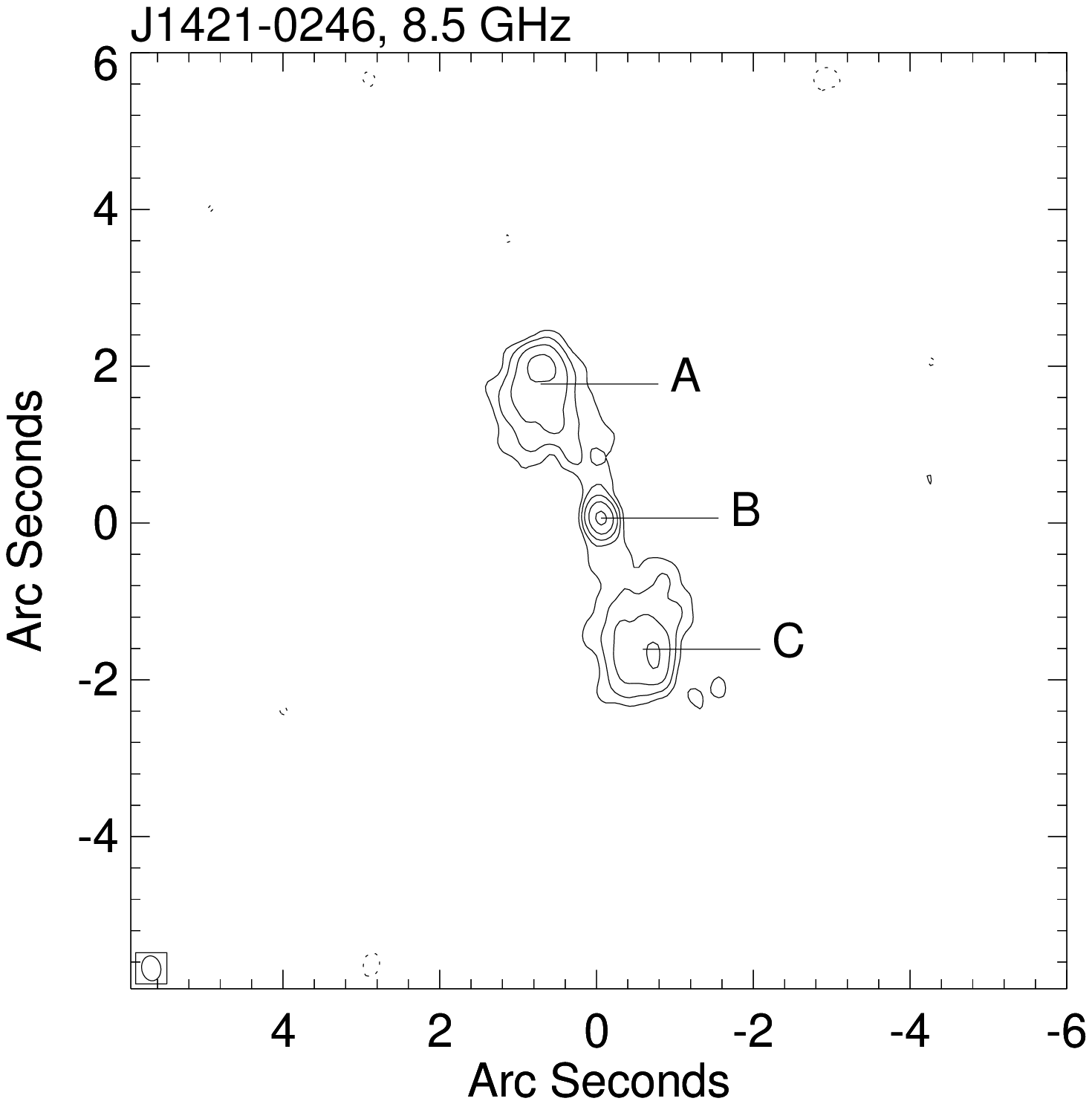}	&		\\
\includegraphics[scale=0.25]{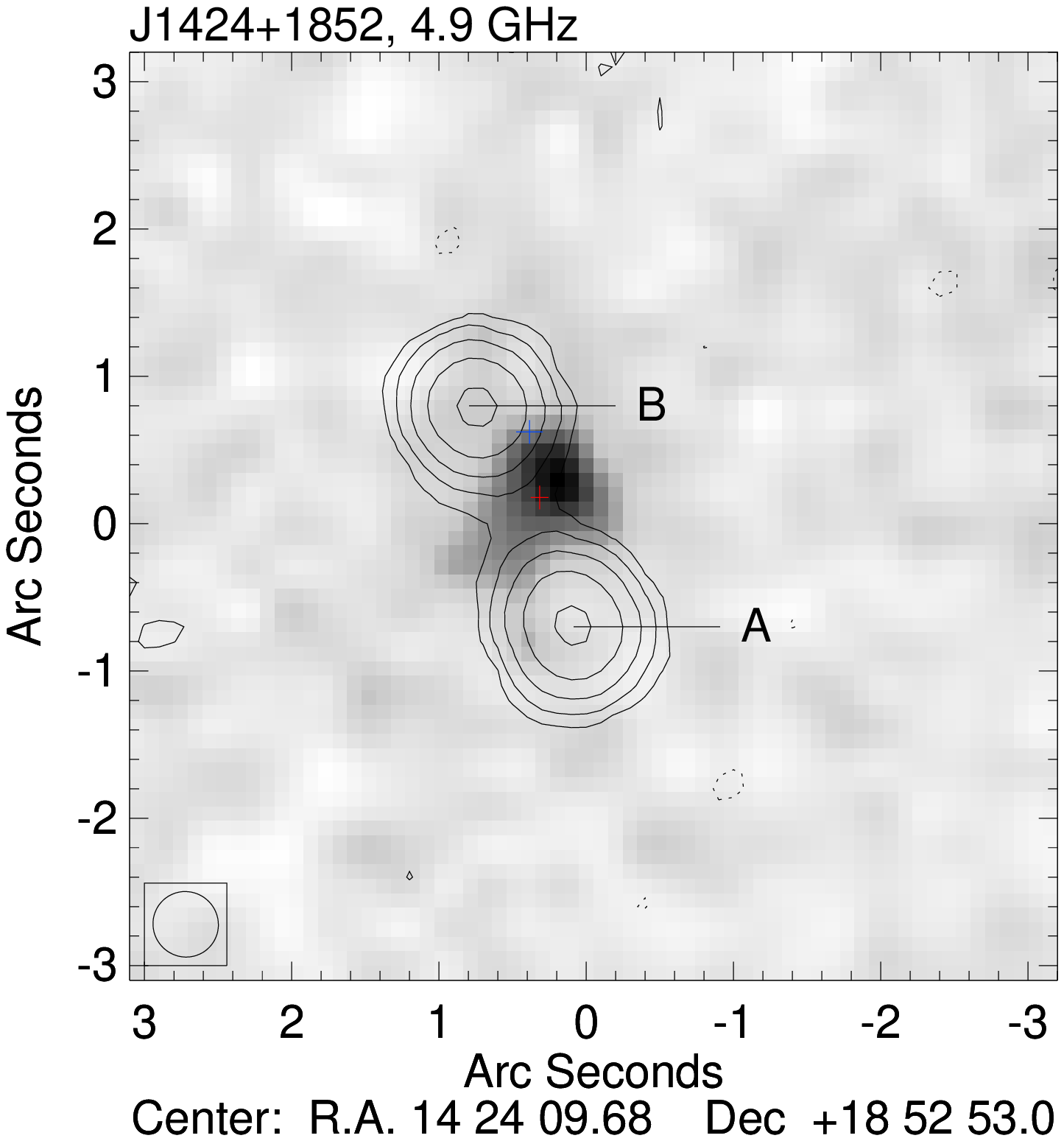}	&	\includegraphics[scale=0.25]{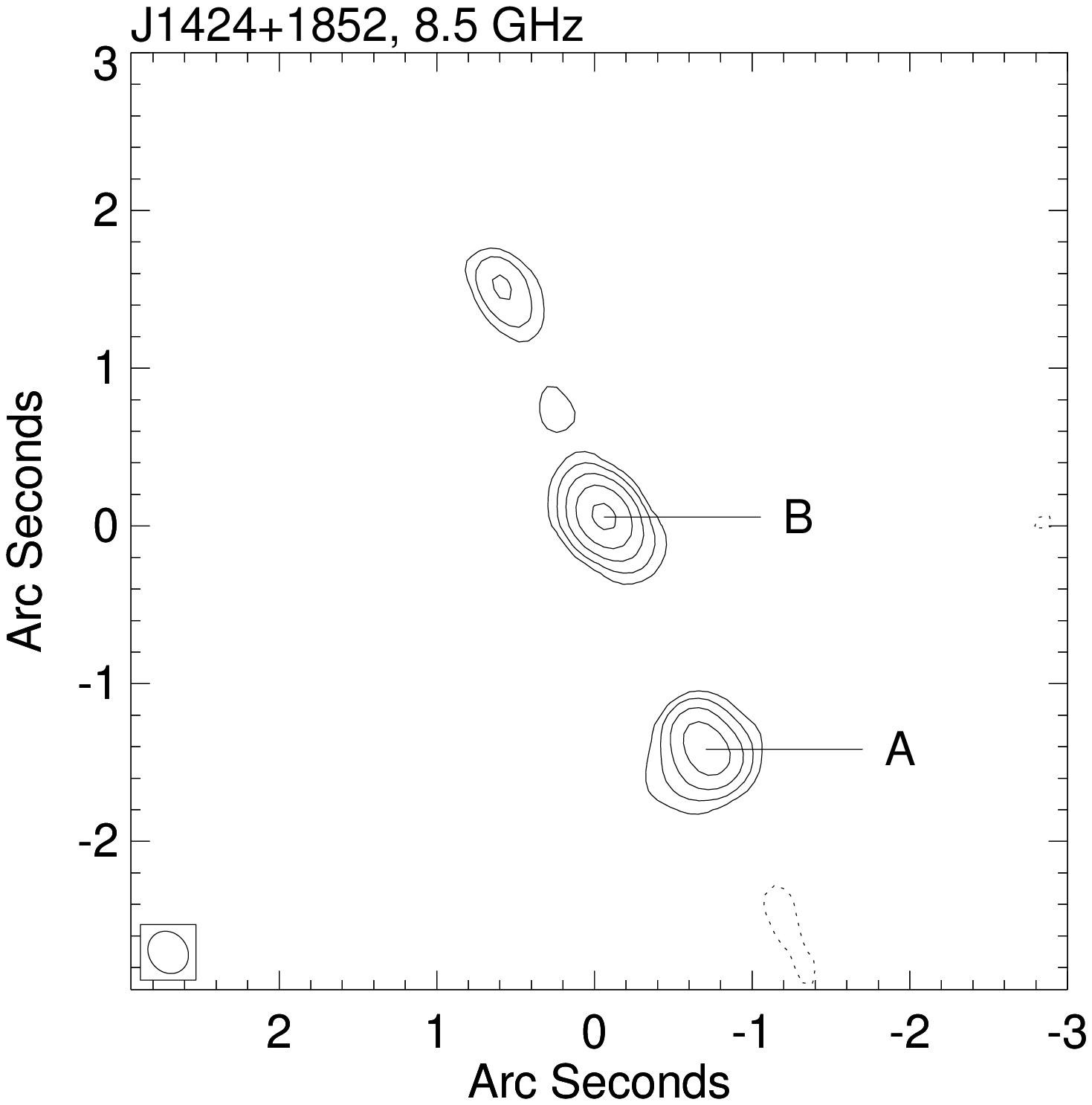}	&		\\
\includegraphics[scale=0.25]{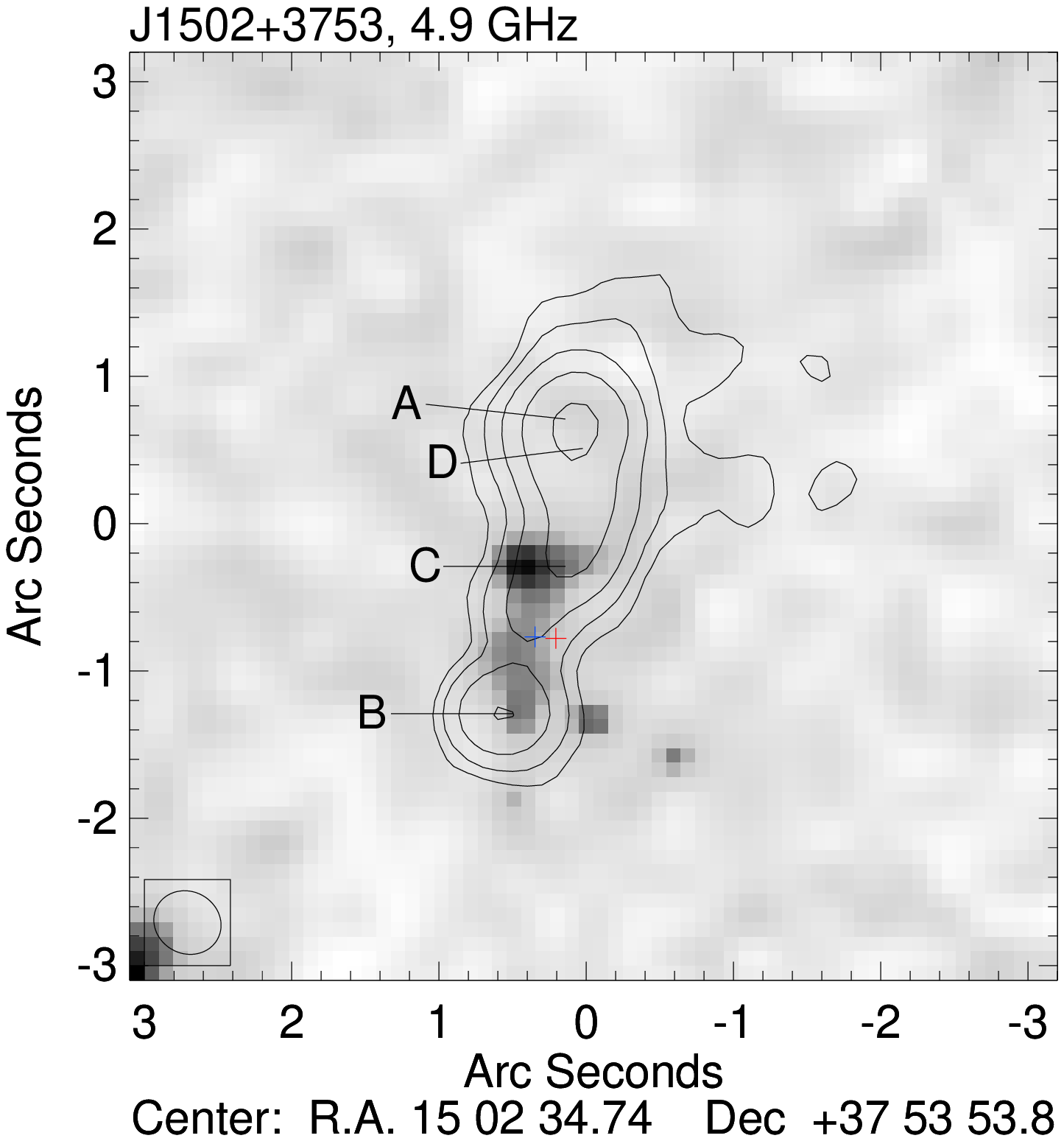}	&	\includegraphics[scale=0.25]{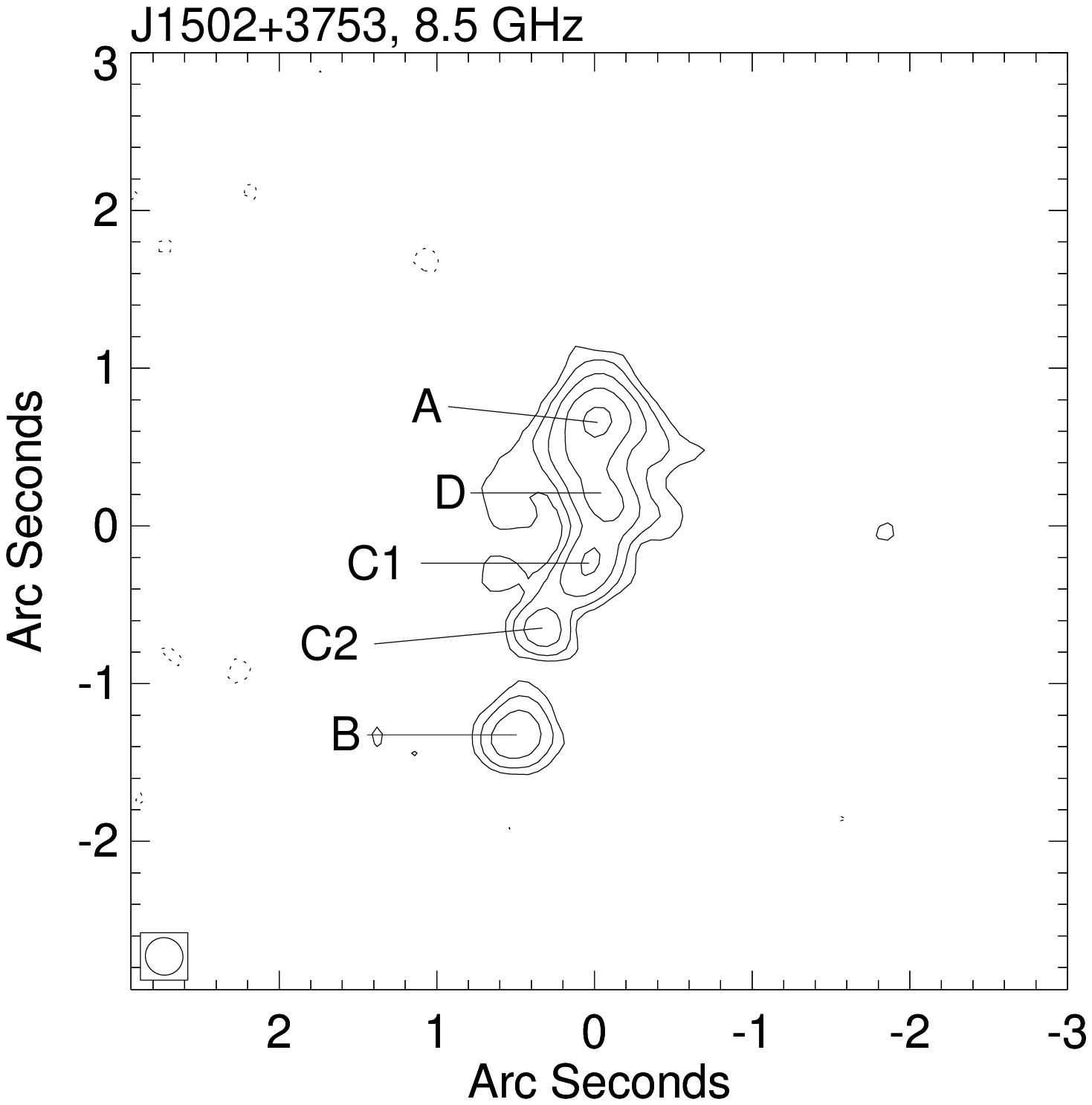}	&		\\
\includegraphics[scale=0.25]{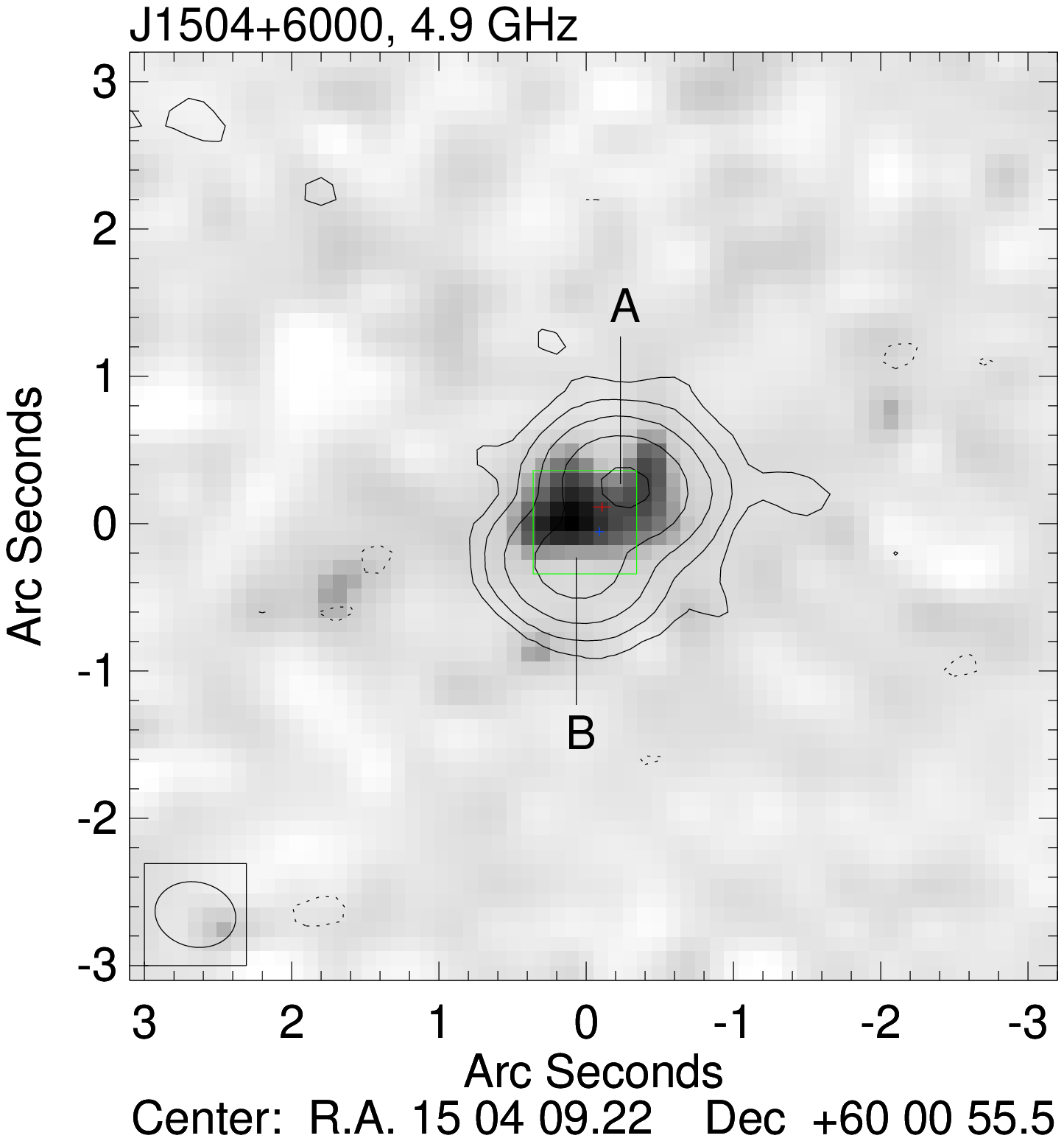}	&	\includegraphics[scale=0.25]{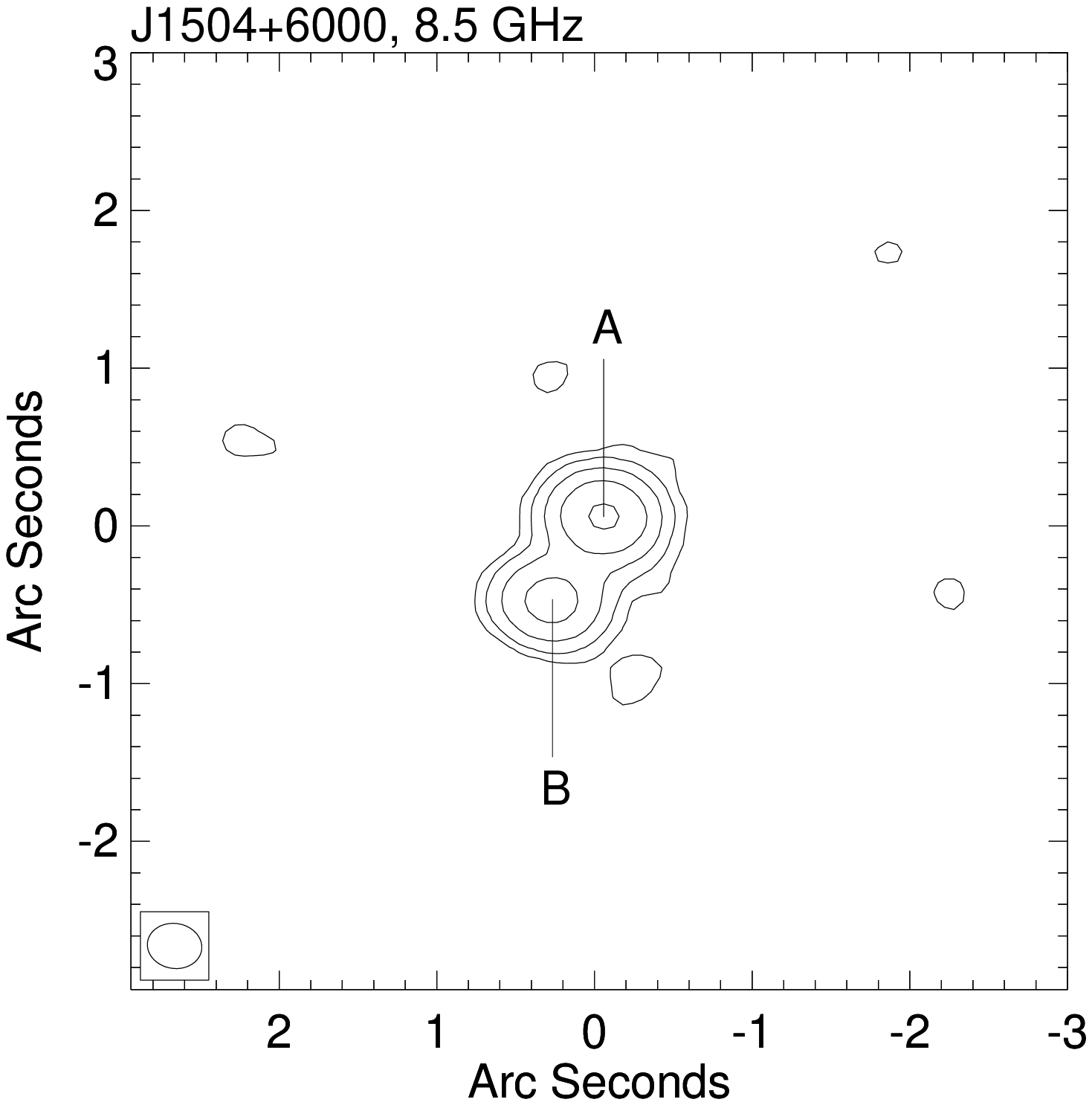}	&	\includegraphics[scale=0.25]{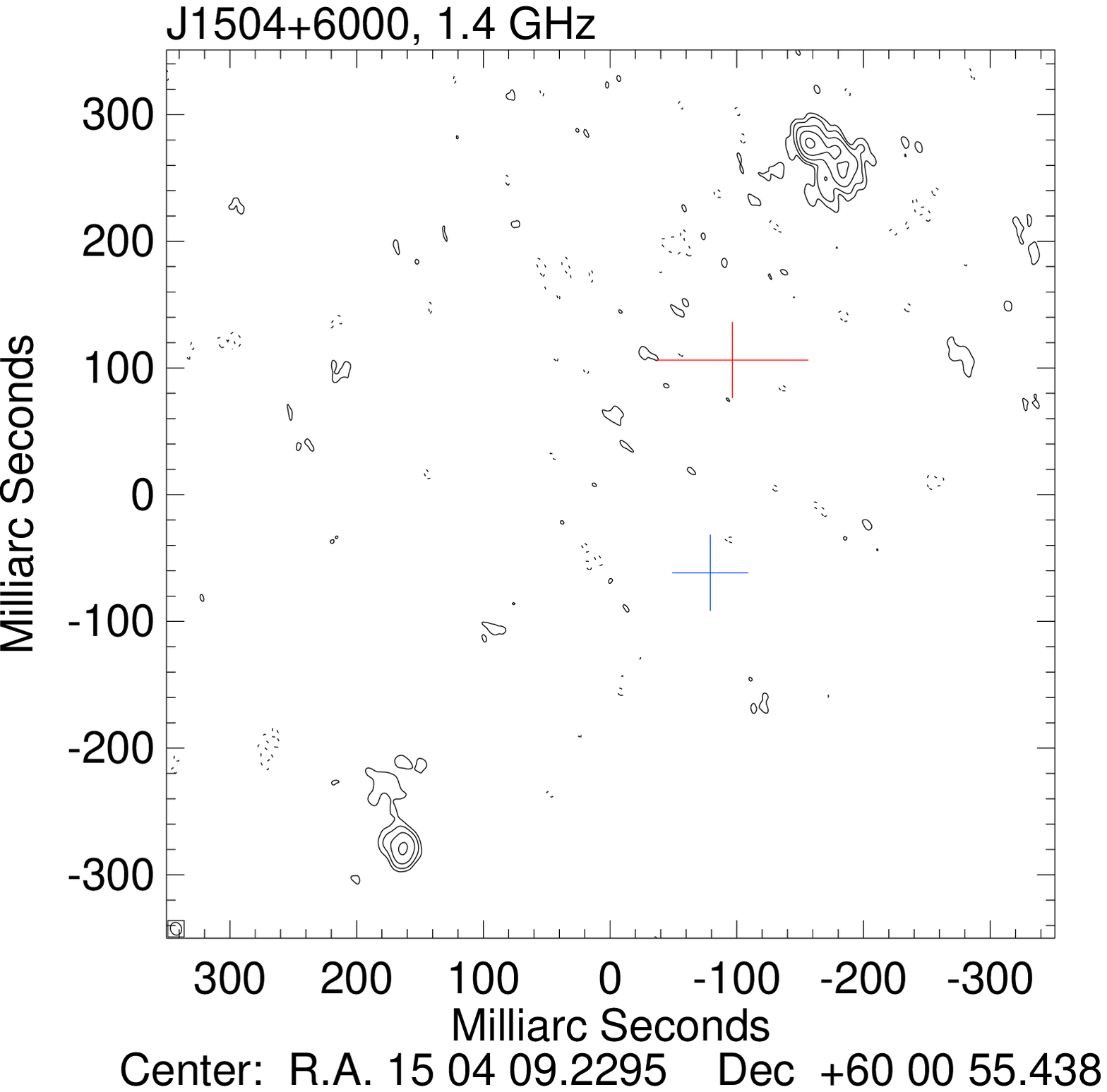}	\\

\end{tabular}
\end{figure}

\begin{figure}[htdp]
\begin{tabular}{lll}

\includegraphics[scale=0.25]{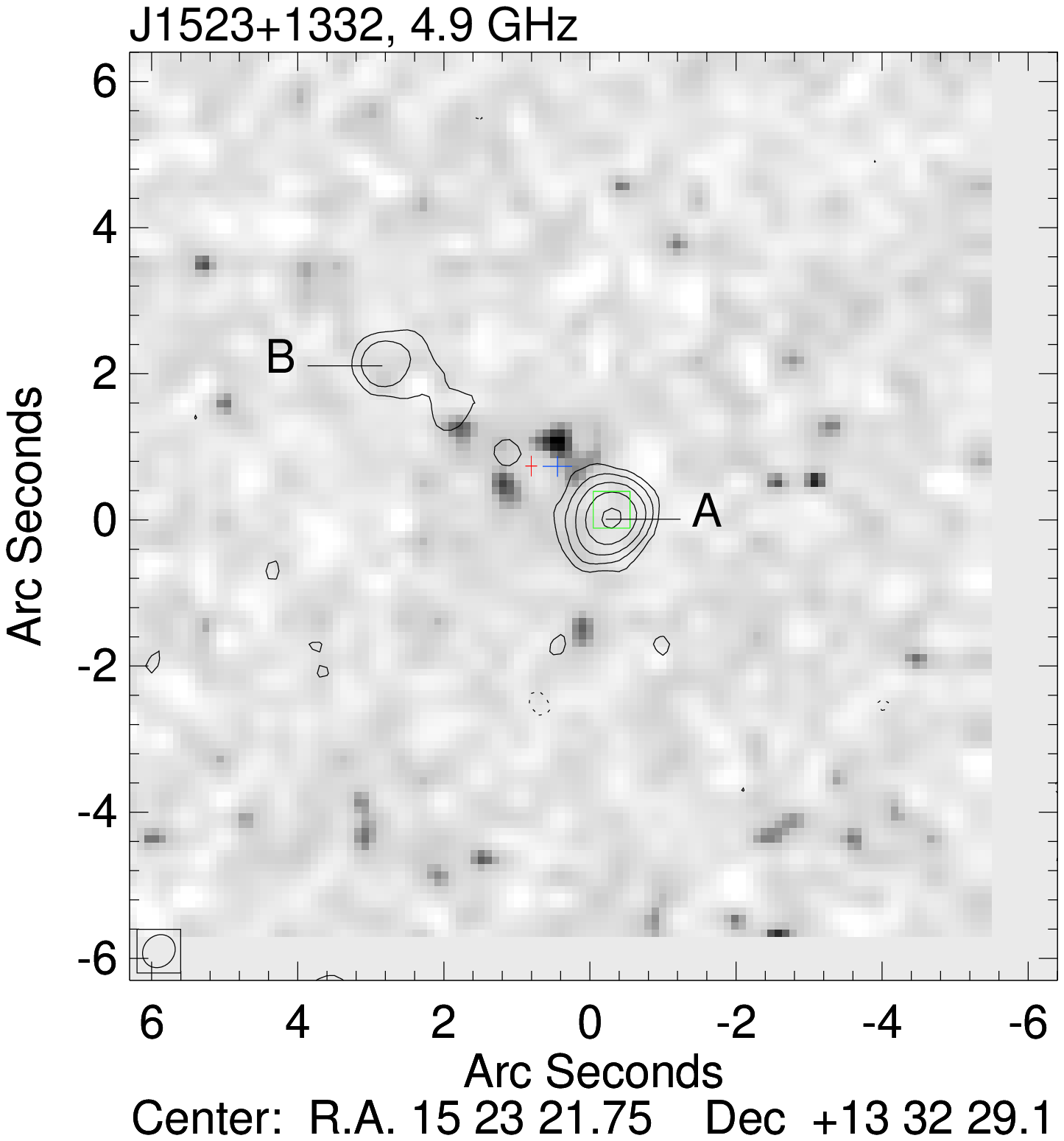}	&	\includegraphics[scale=0.25]{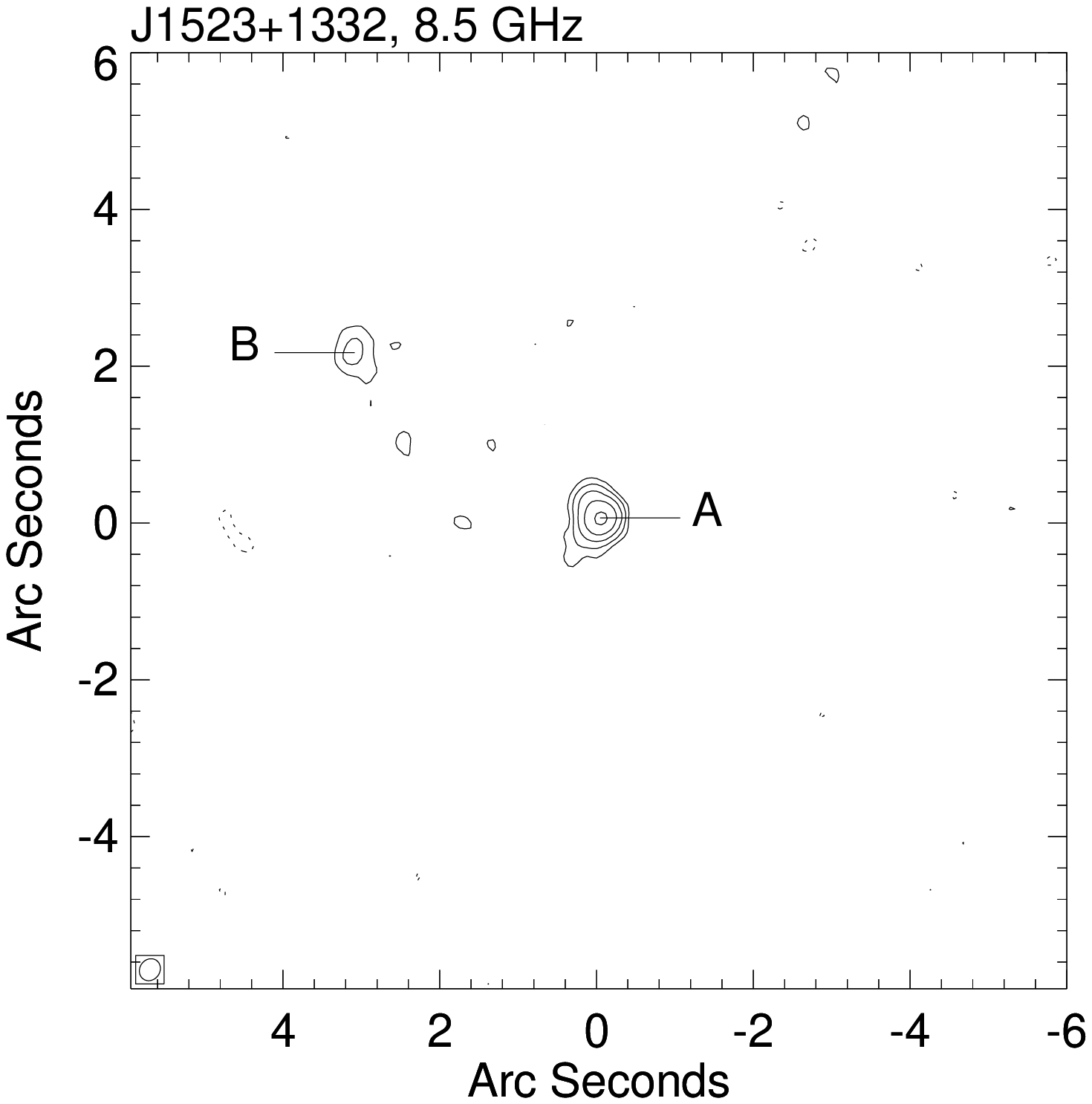}	&	\includegraphics[scale=0.25]{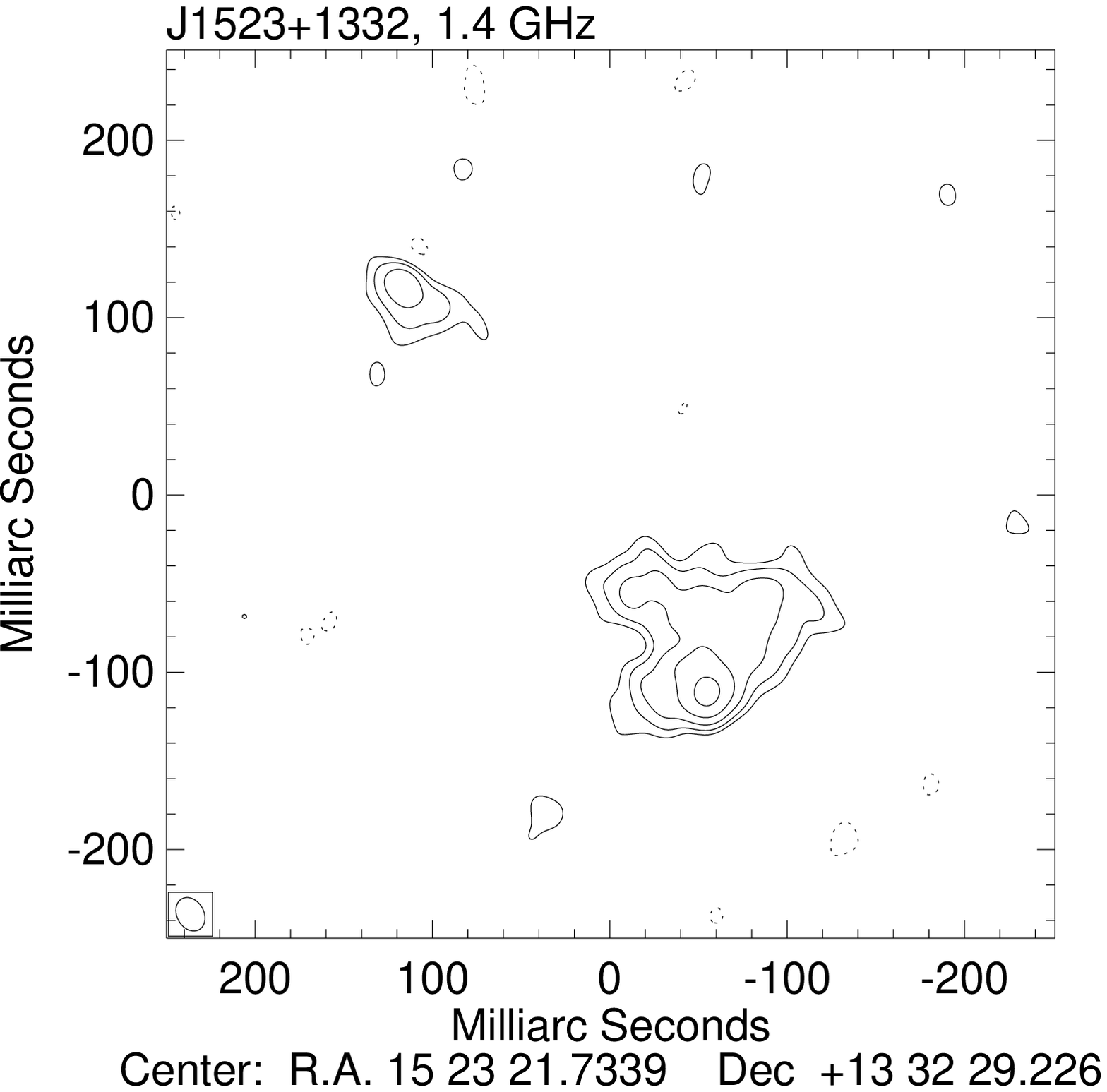}	\\
	&	\includegraphics[scale=0.25]{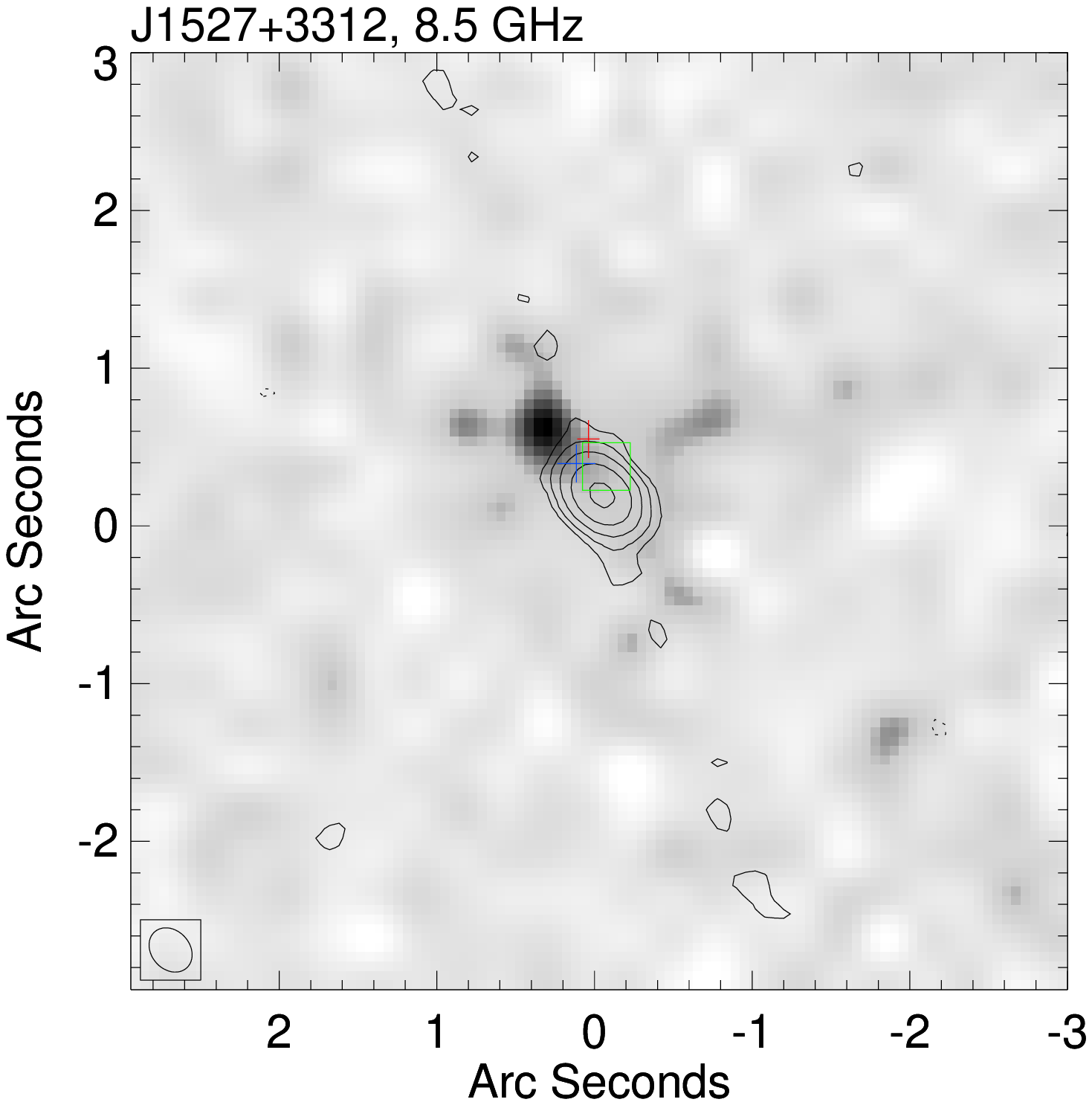}	&	\includegraphics[scale=0.25]{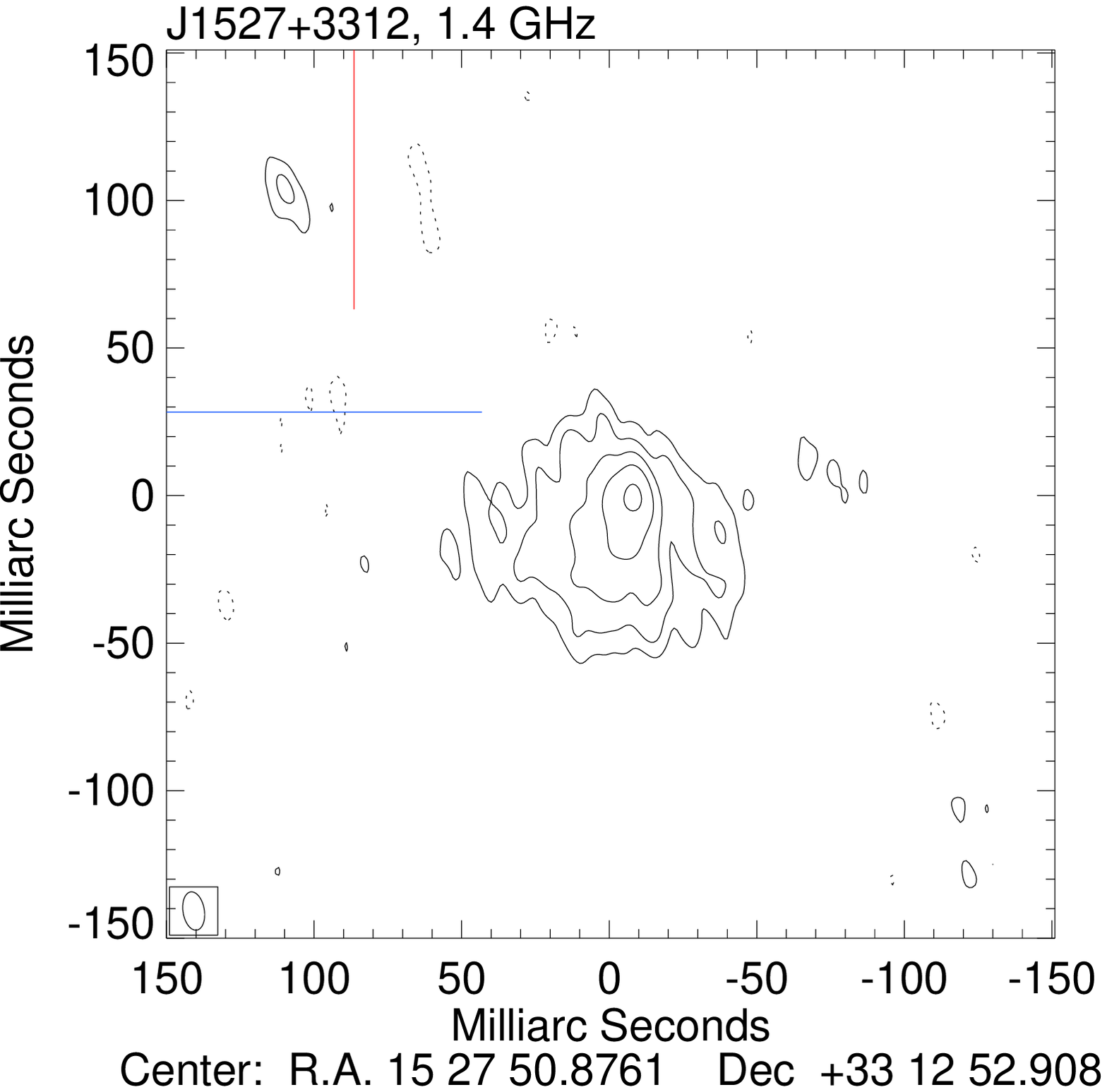}	\\
\includegraphics[scale=0.25]{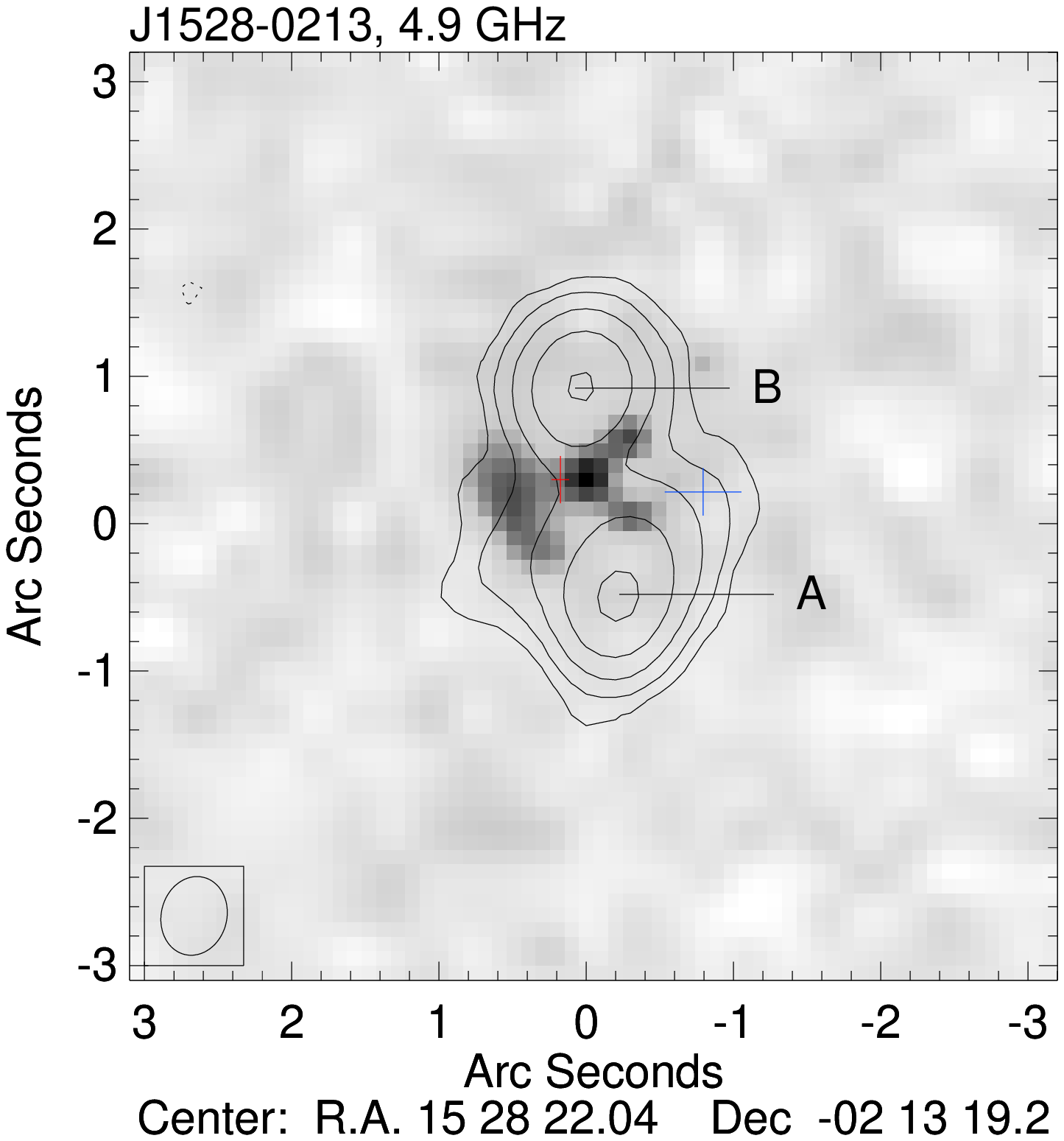}	&	\includegraphics[scale=0.25]{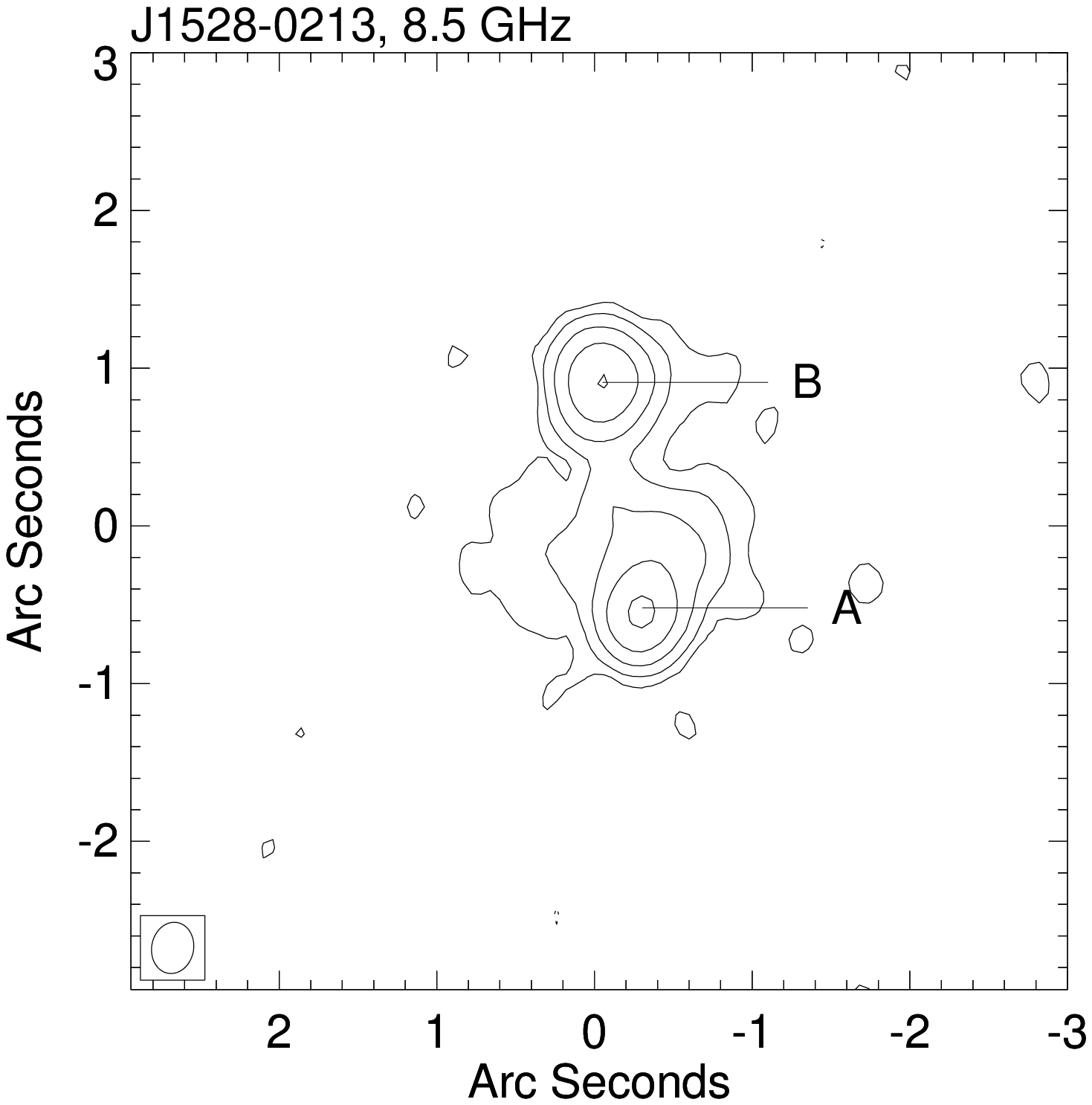}	&		\\
\includegraphics[scale=0.25]{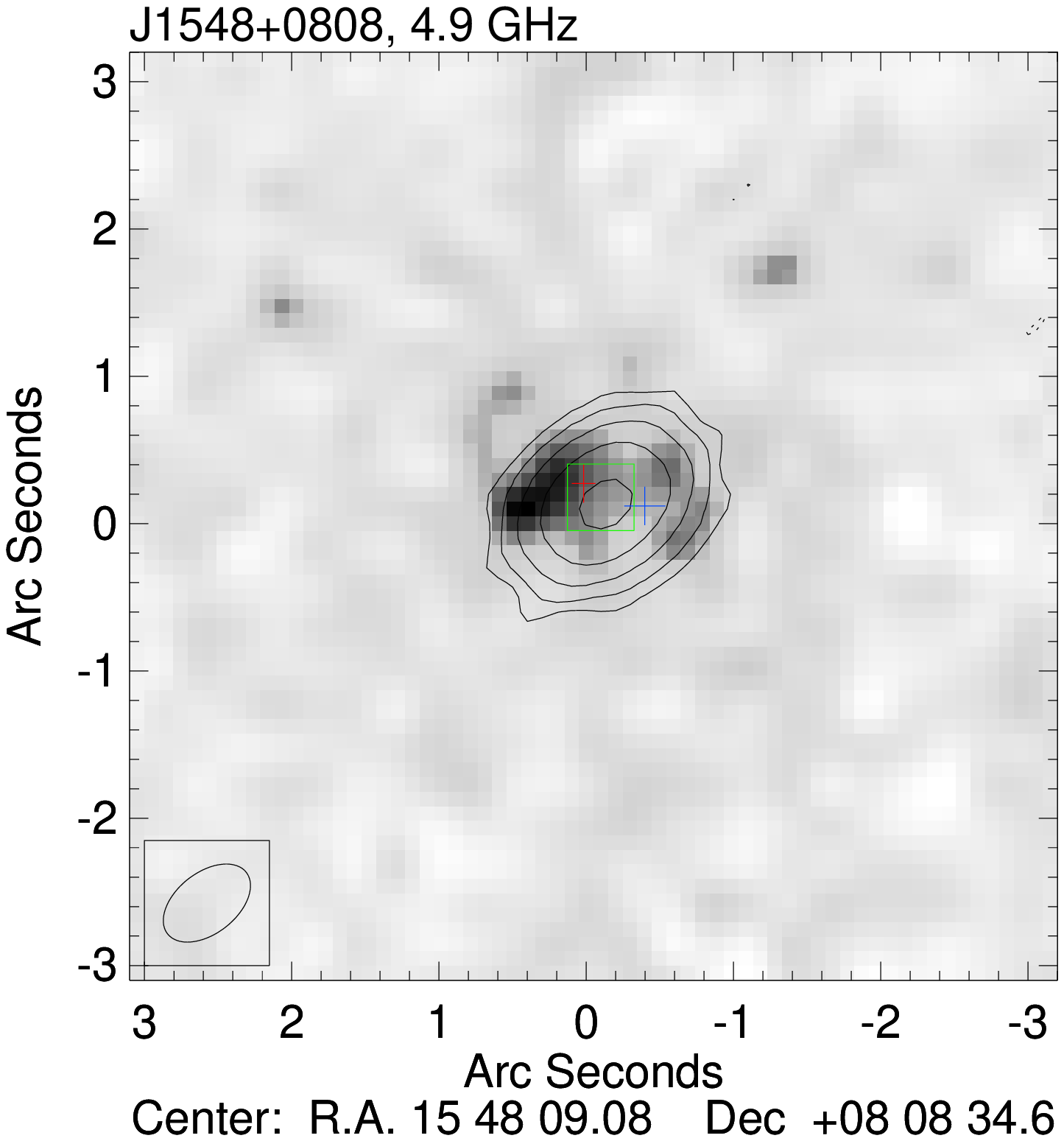}	&	\includegraphics[scale=0.25]{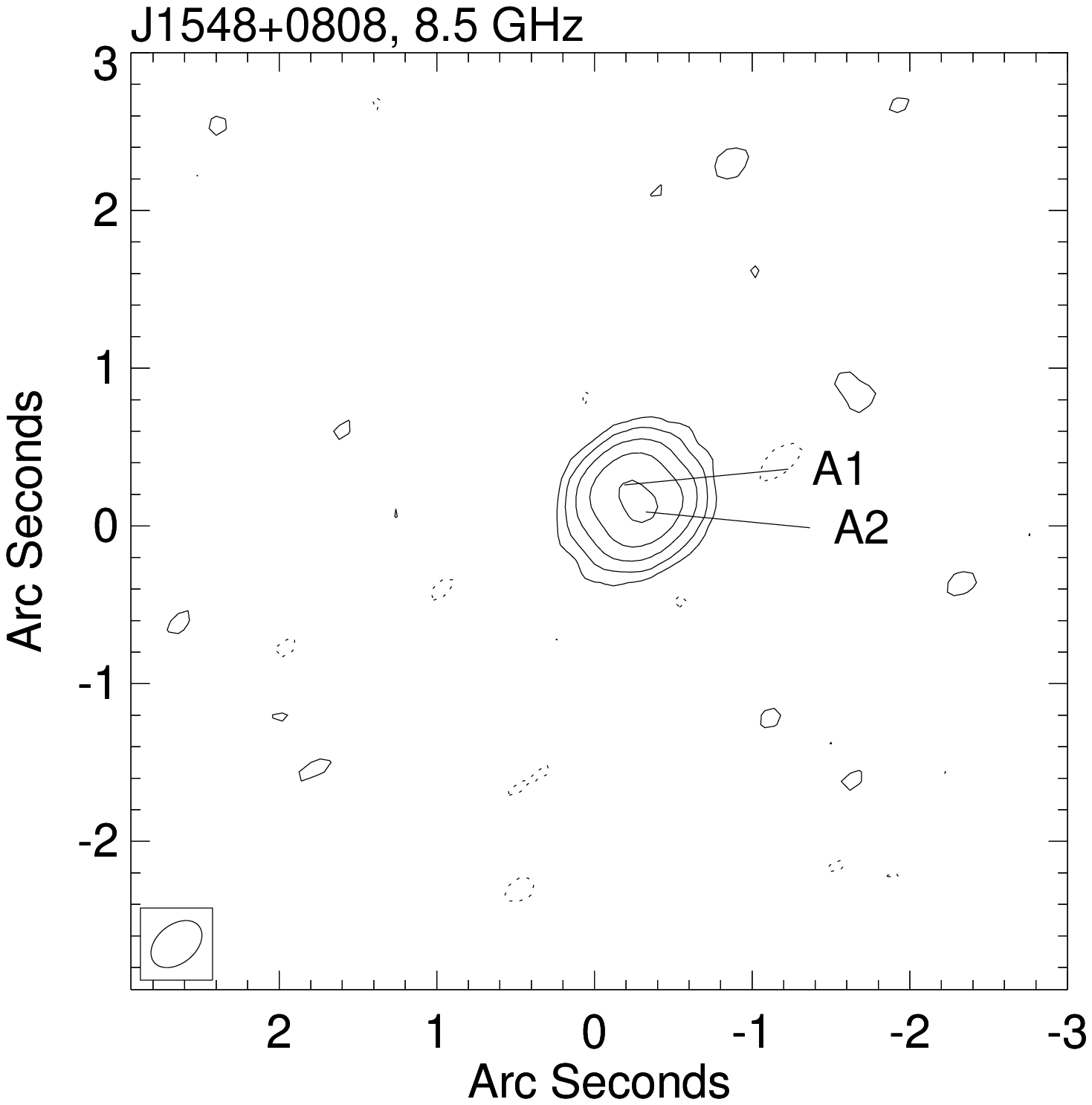}	&	\includegraphics[scale=0.25]{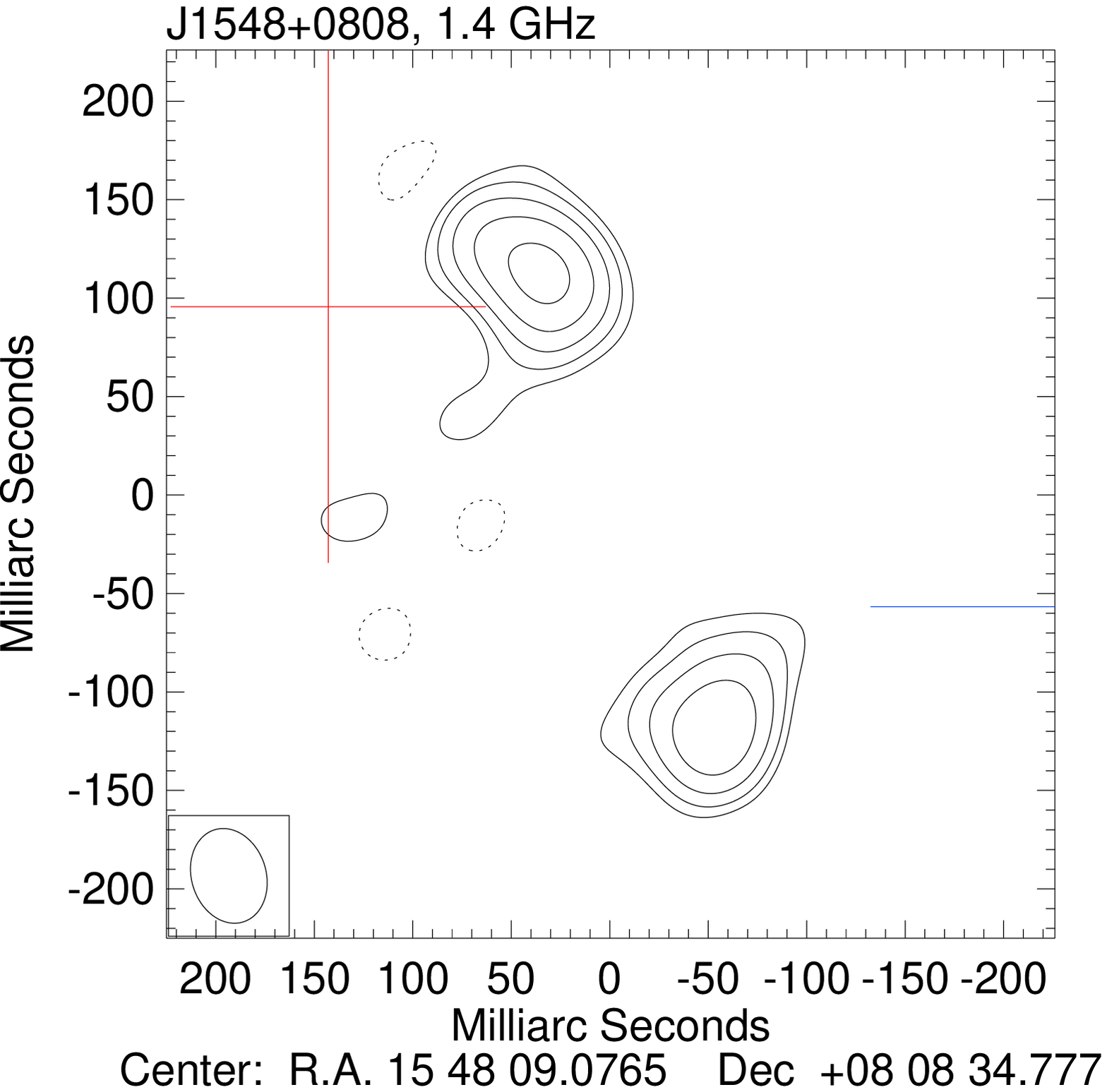}	\\
\includegraphics[scale=0.25]{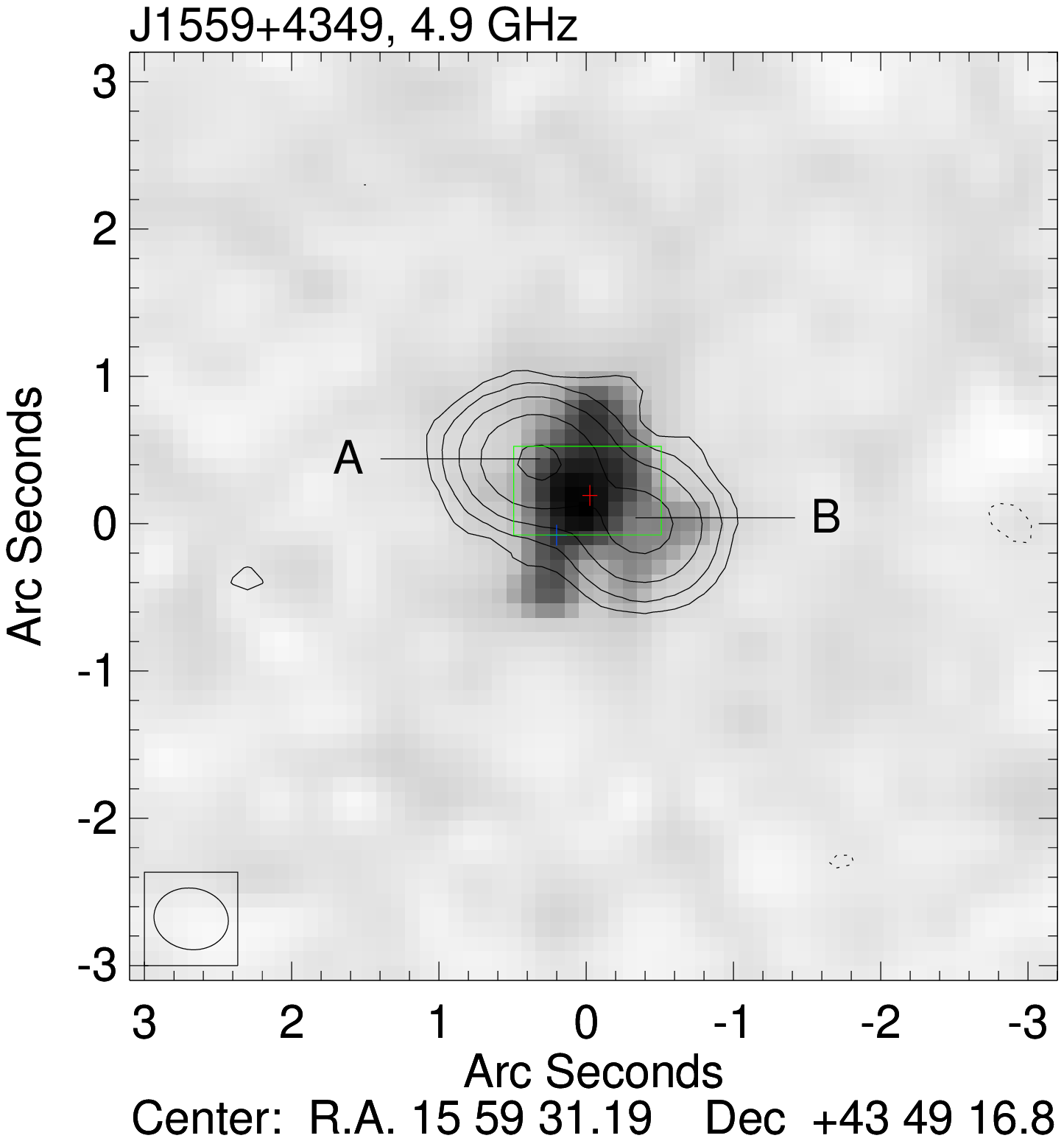}	&	\includegraphics[scale=0.25]{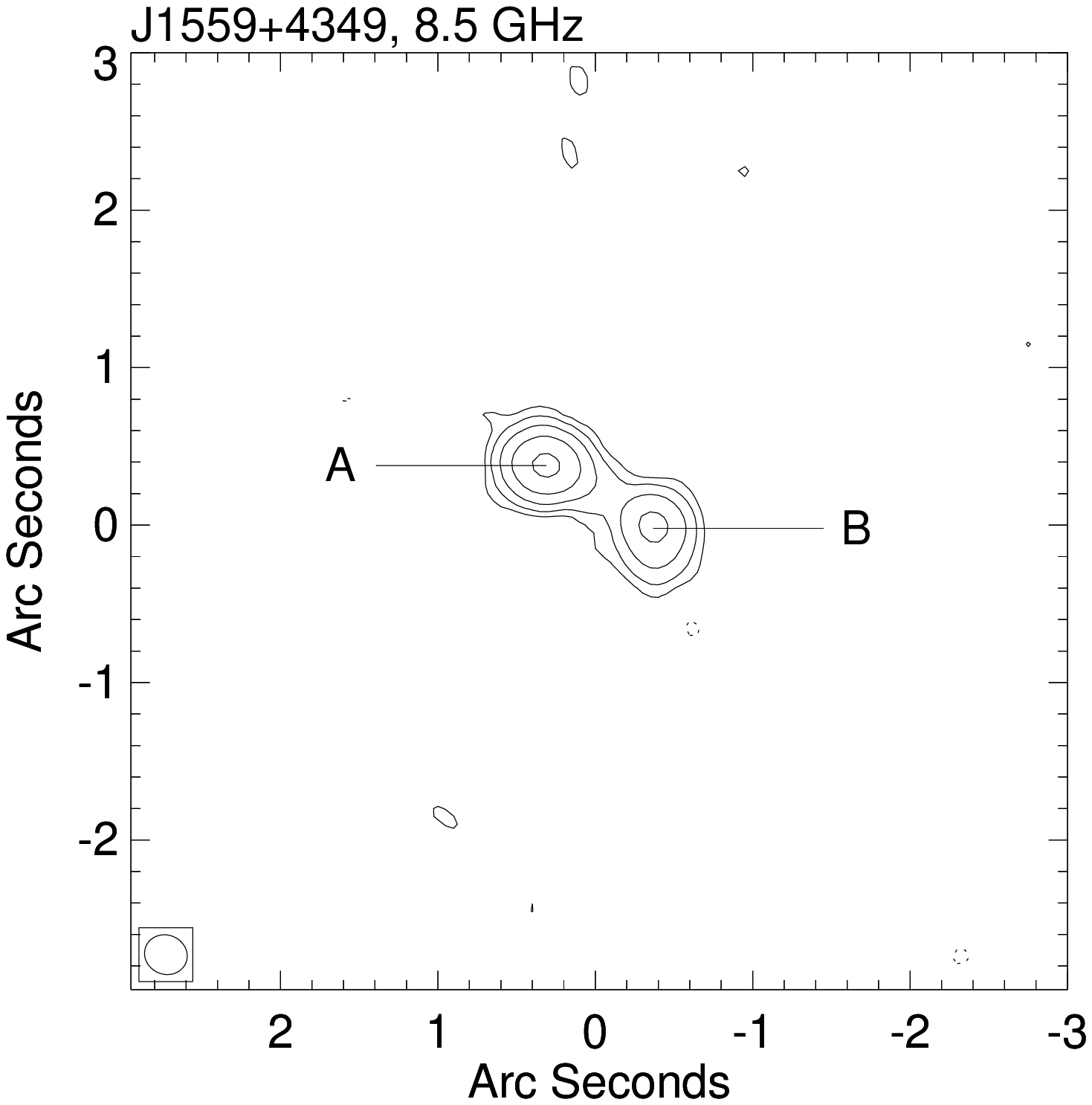}	&	\includegraphics[scale=0.25,bb=50 0 754 453]{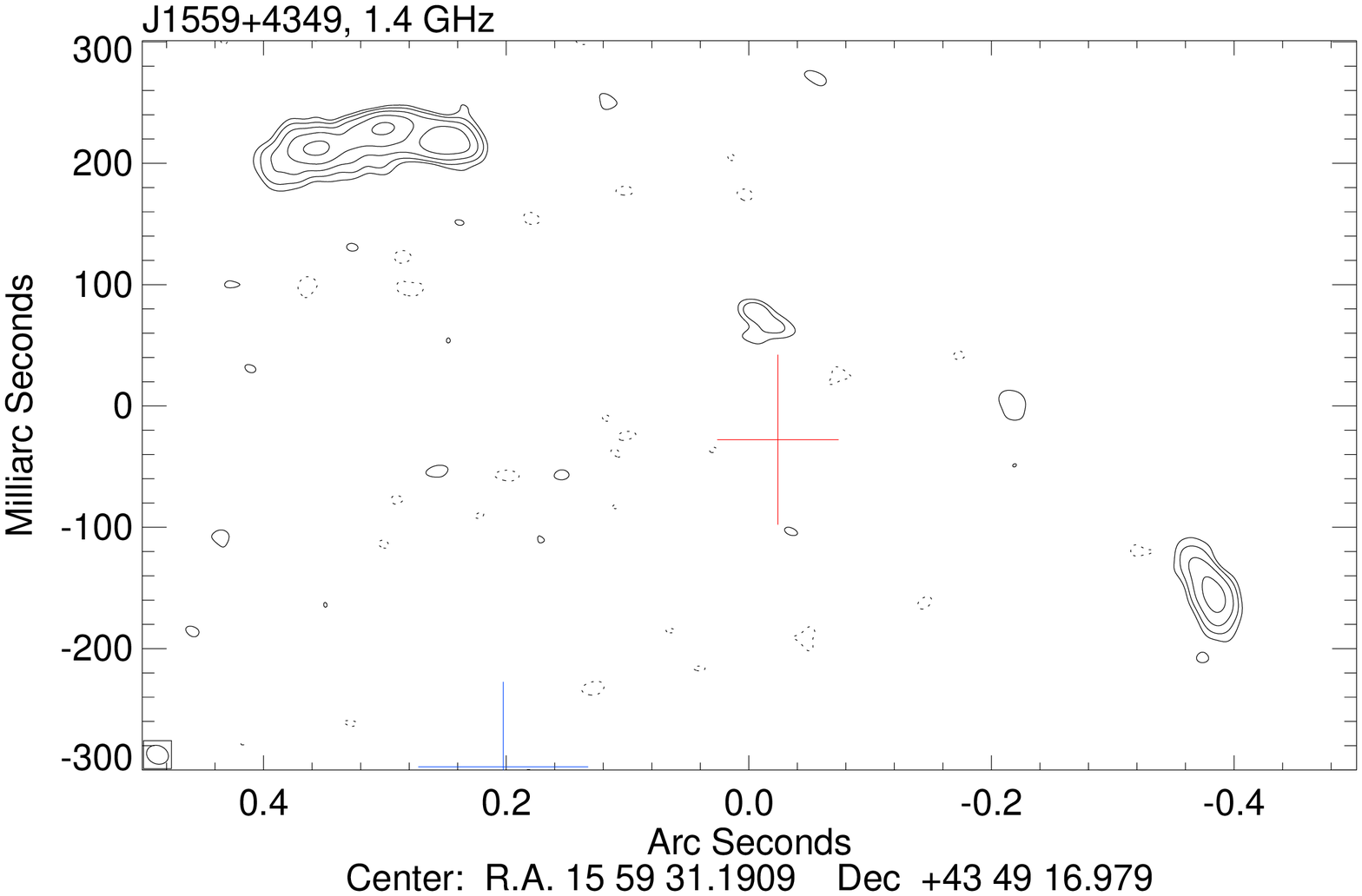}	\\

\end{tabular}
\end{figure}

\begin{figure}[htdp]
\begin{tabular}{lll}

\includegraphics[scale=0.25]{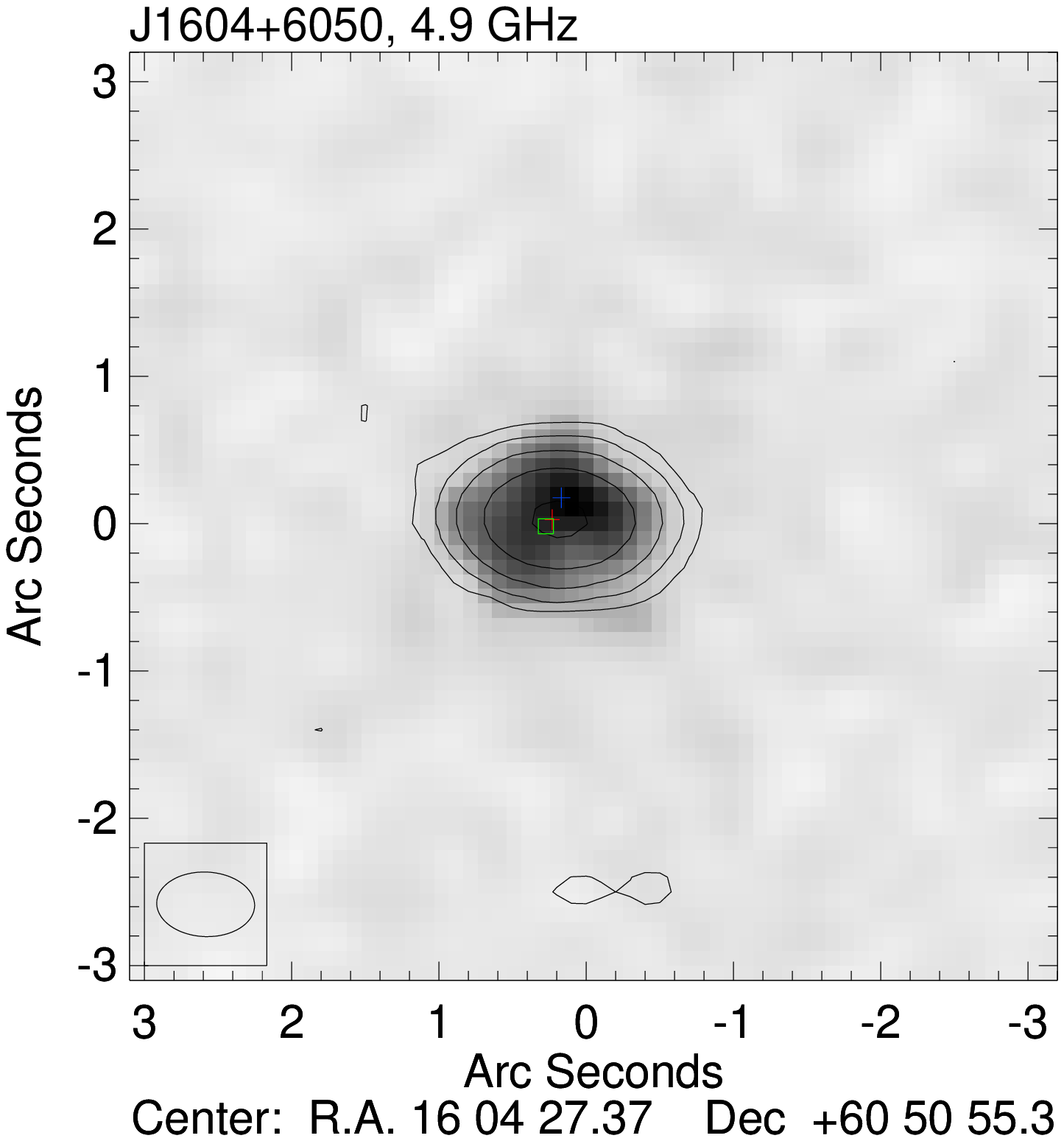}	&	\includegraphics[scale=0.25]{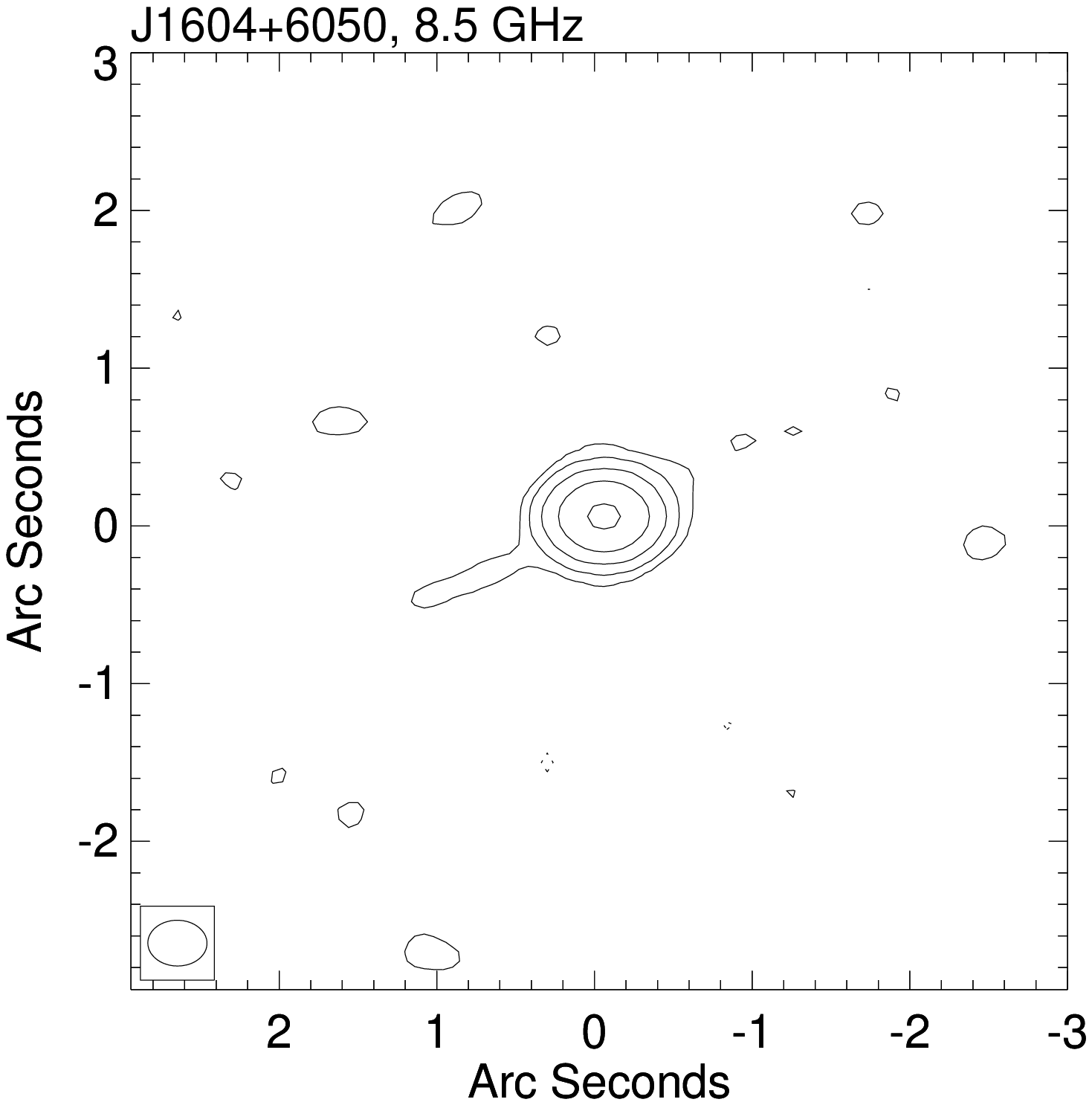}	&	\includegraphics[scale=0.25]{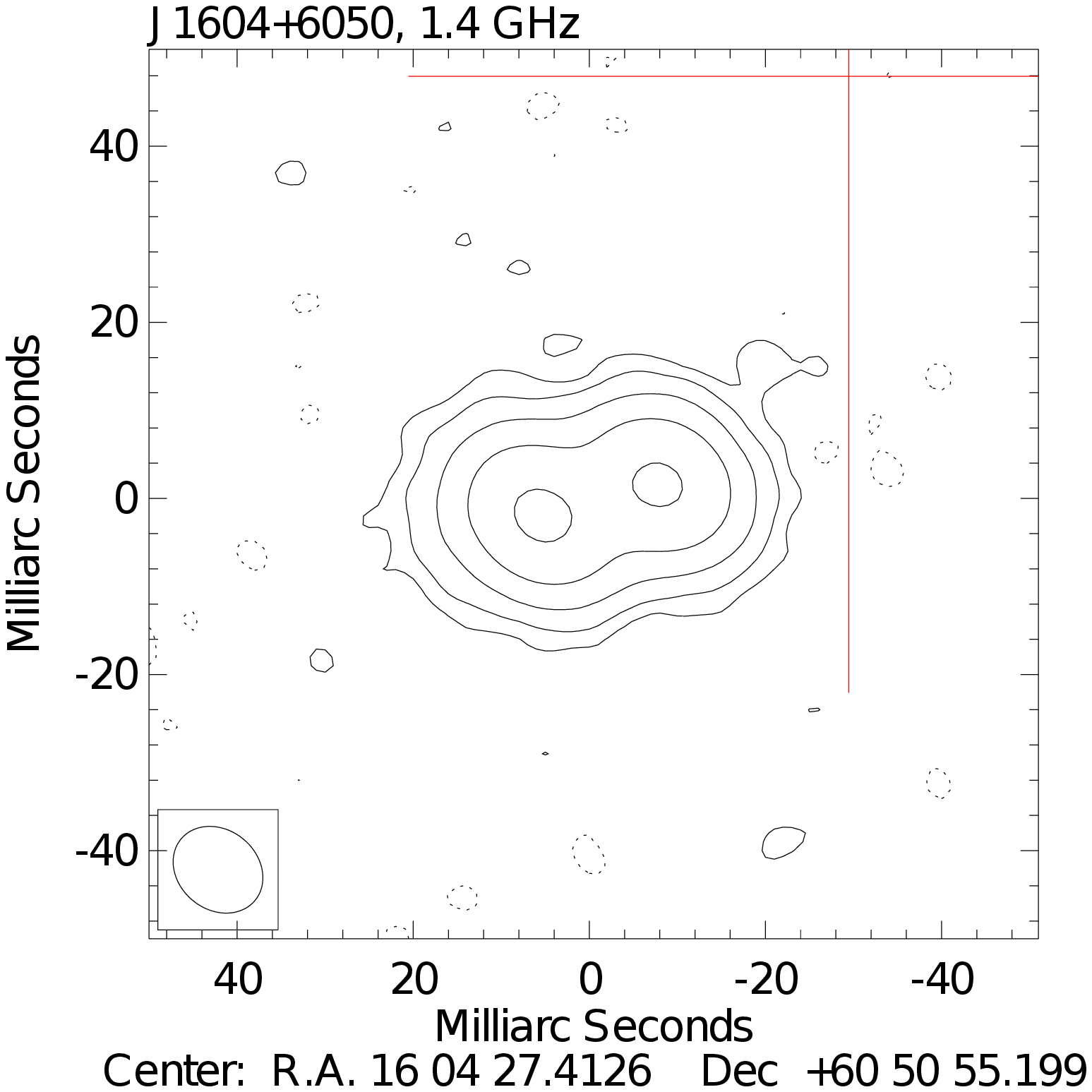}	\\
\includegraphics[scale=0.25]{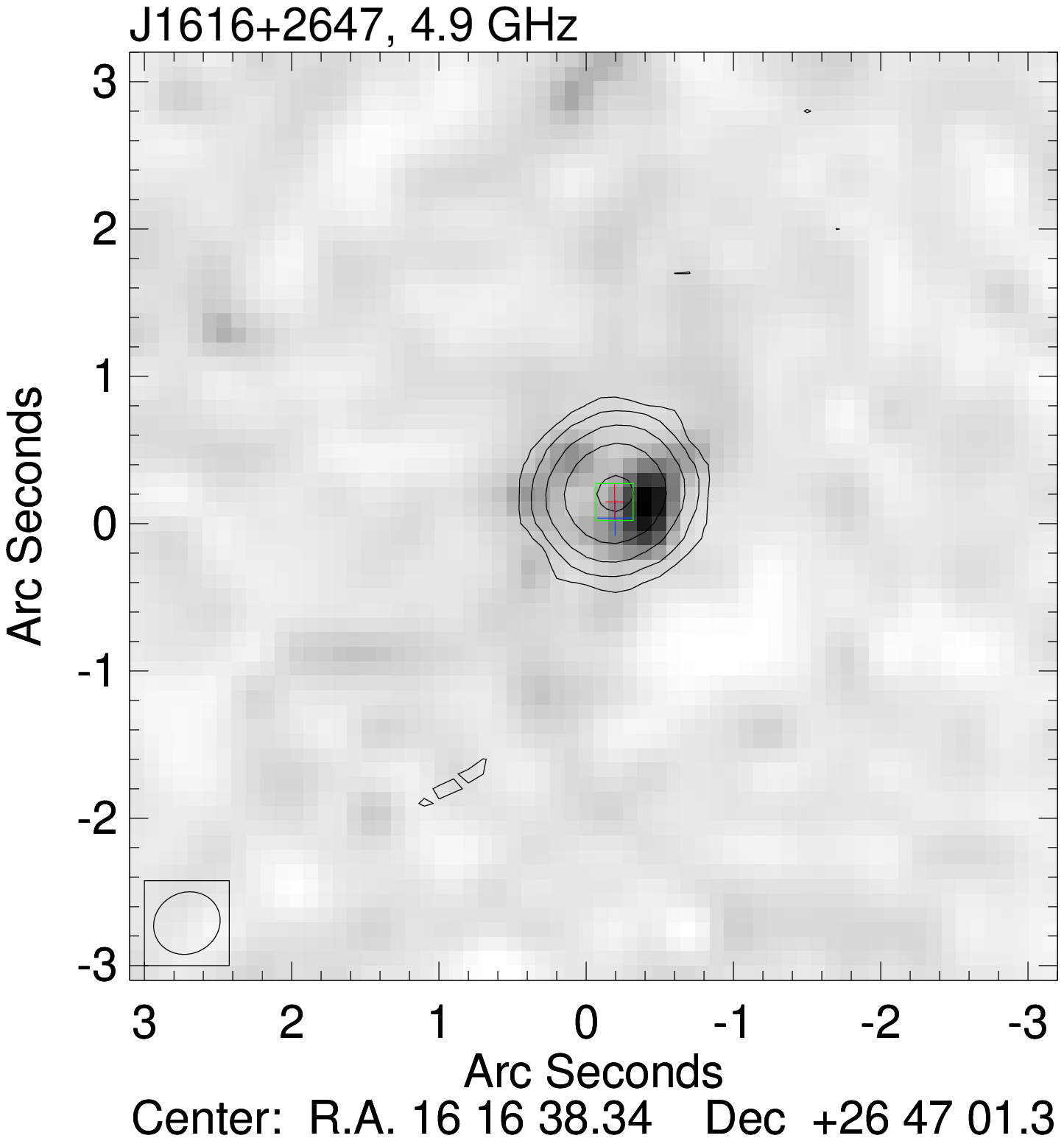}	&		&	\includegraphics[scale=0.25]{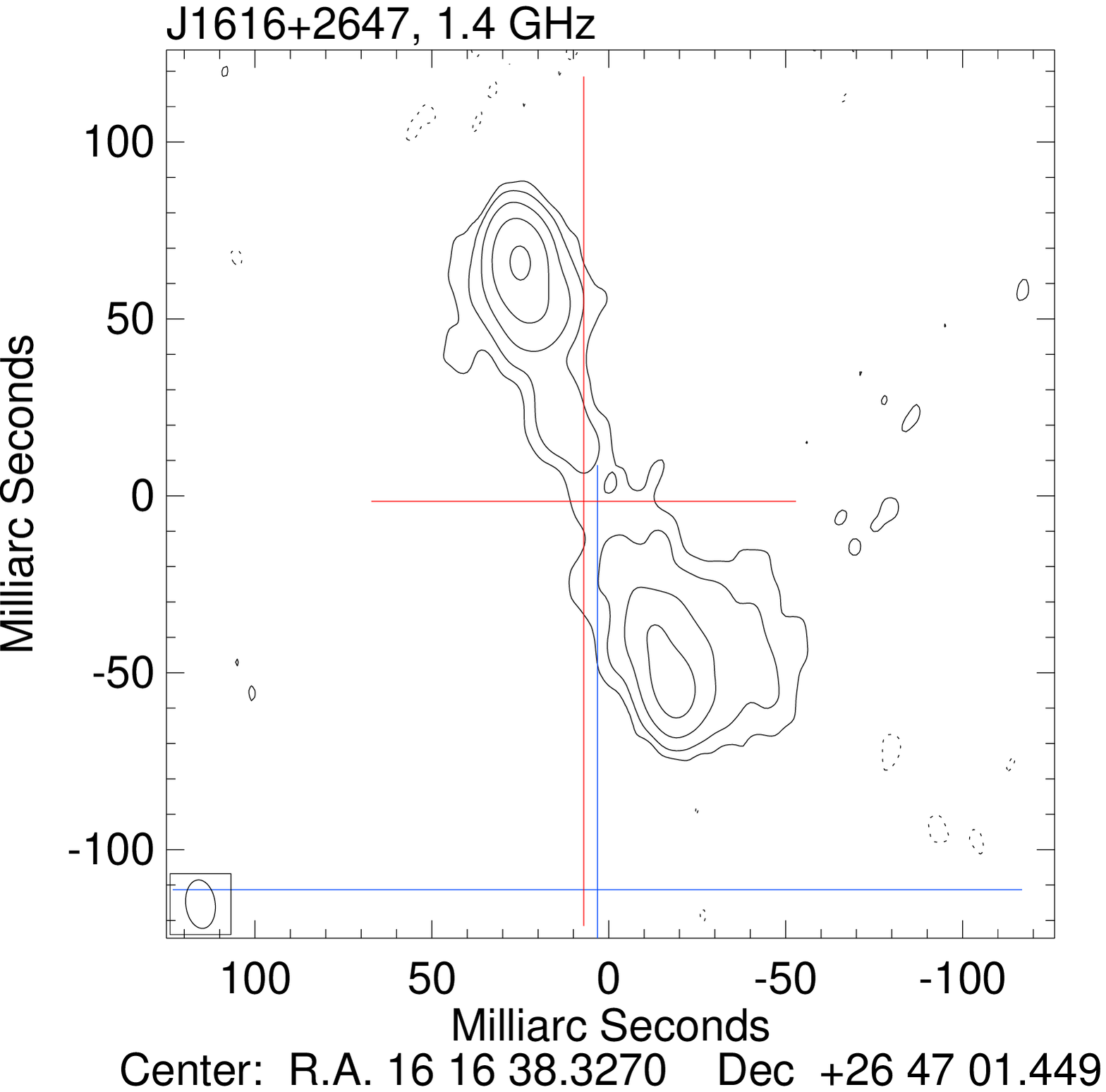}	\\
\includegraphics[scale=0.25]{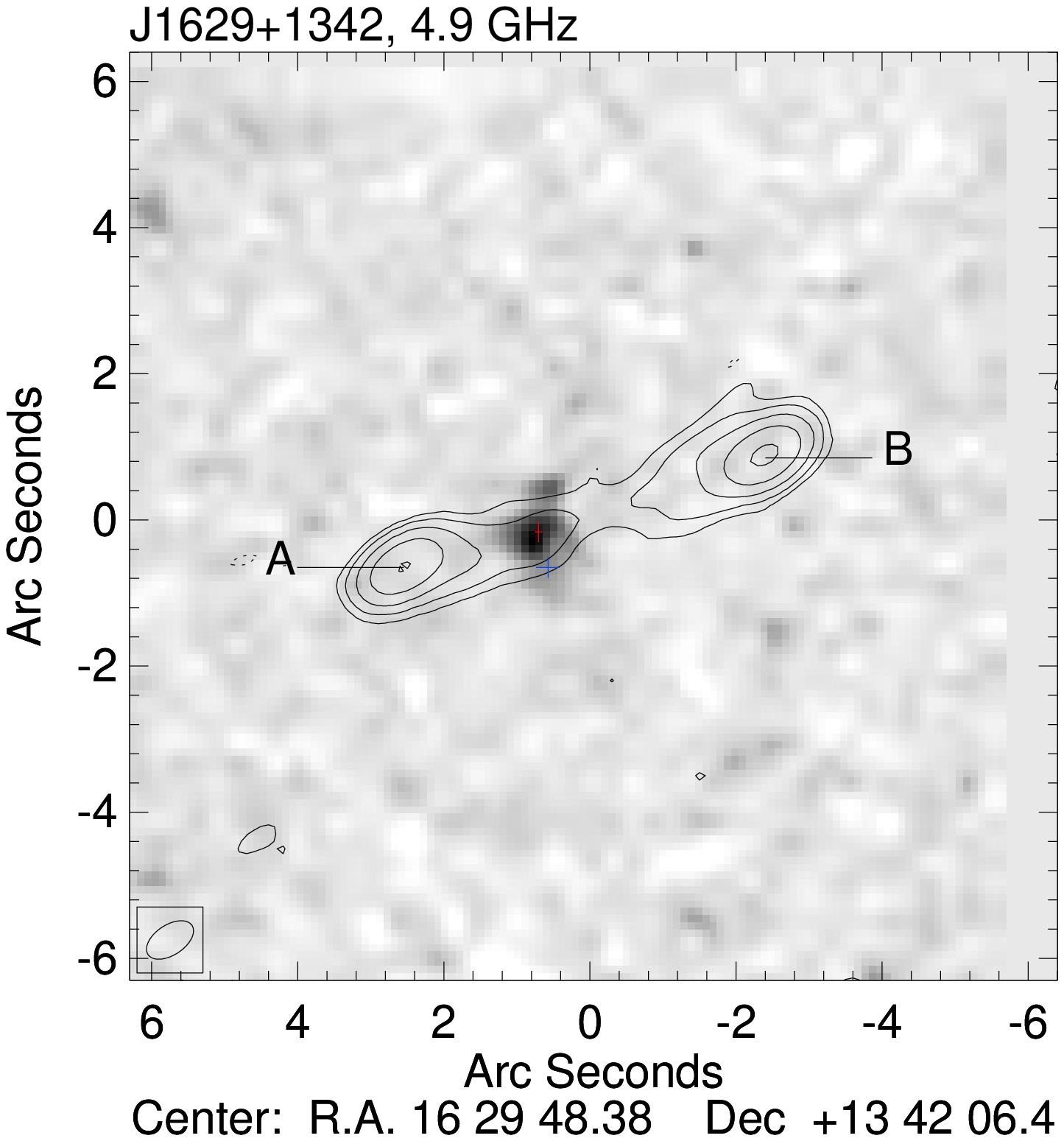}	&	\includegraphics[scale=0.25]{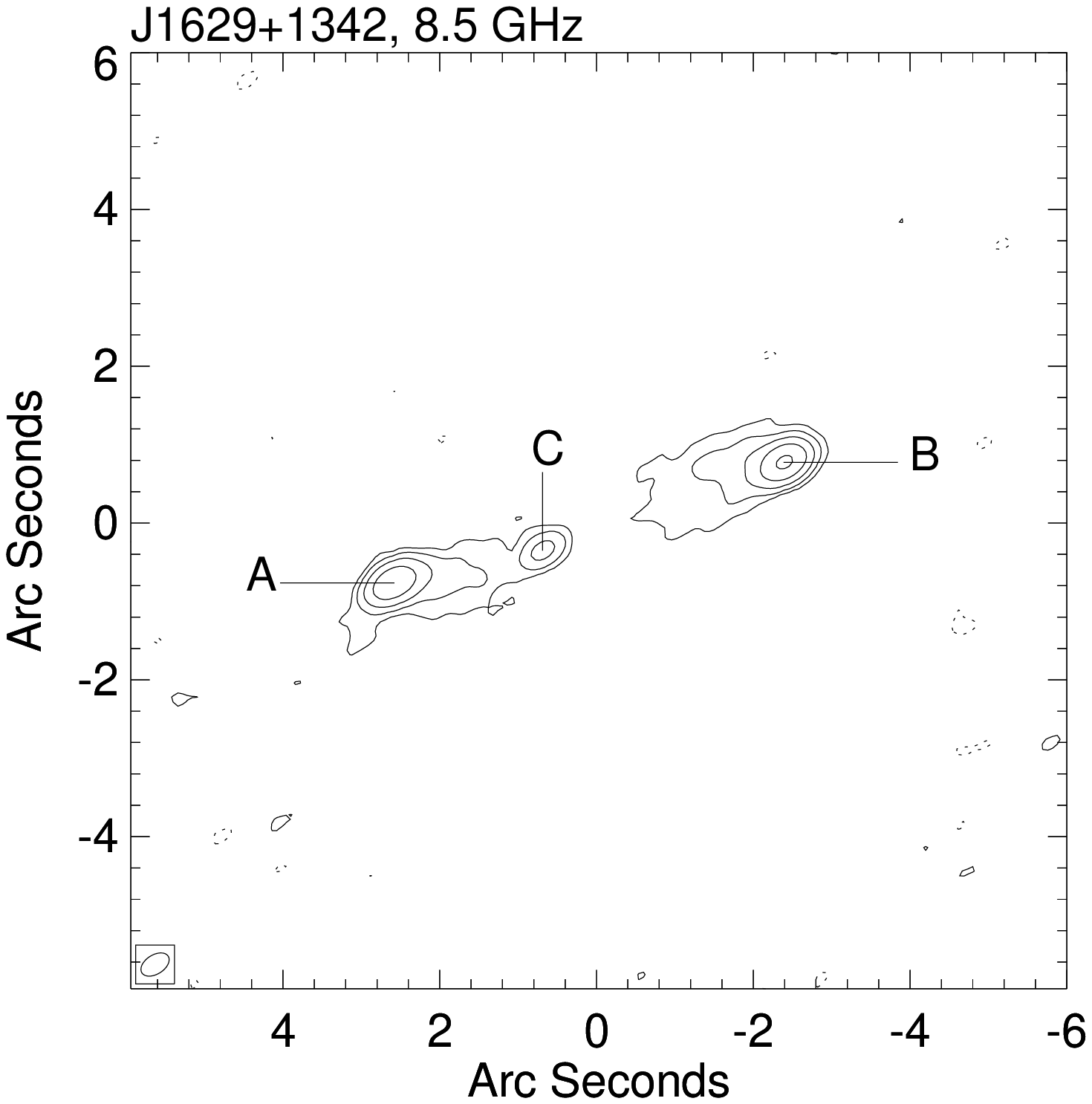}	&		\\
\includegraphics[scale=0.25]{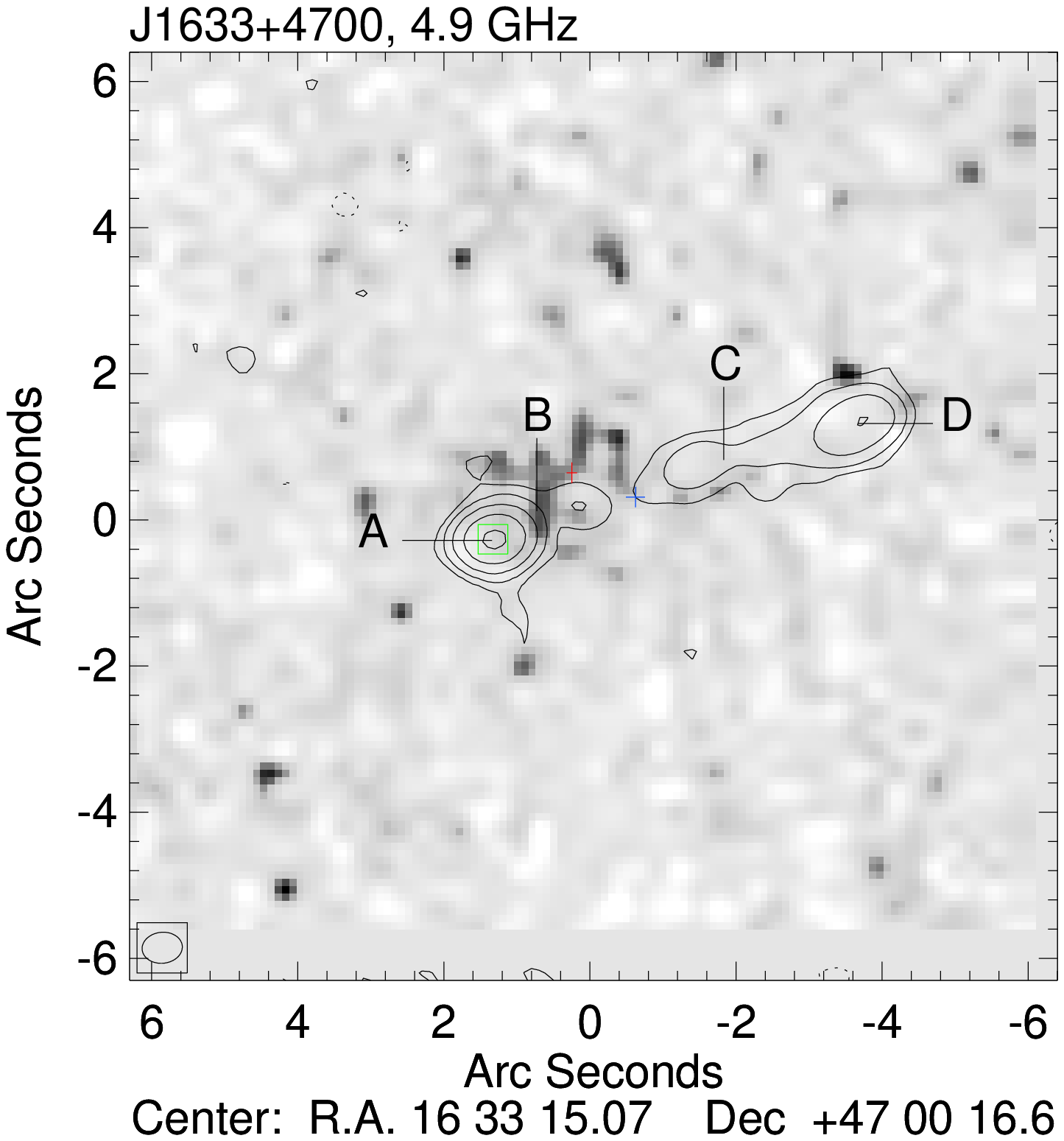}	&	\includegraphics[scale=0.25]{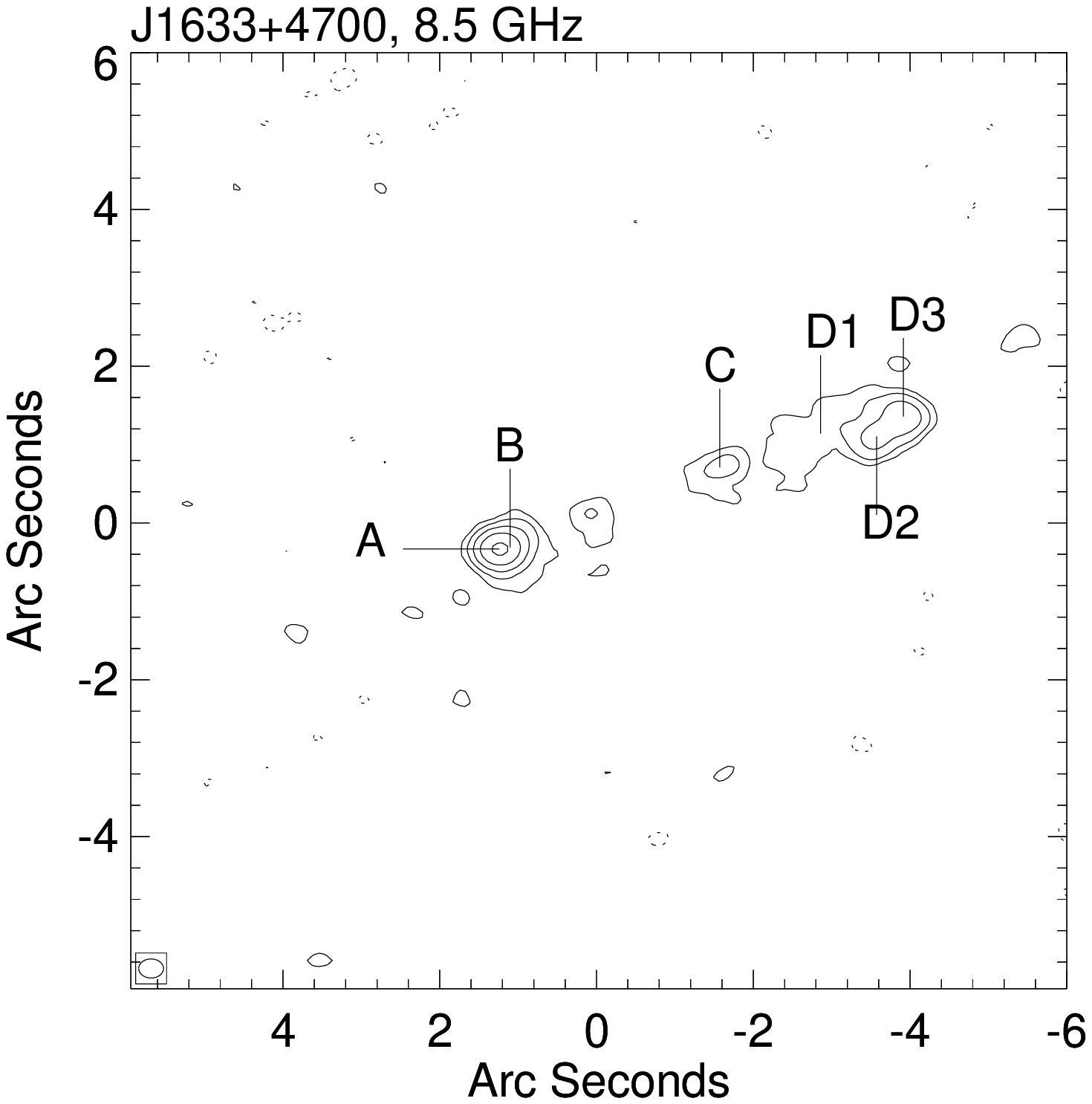}	&	\includegraphics[scale=0.25]{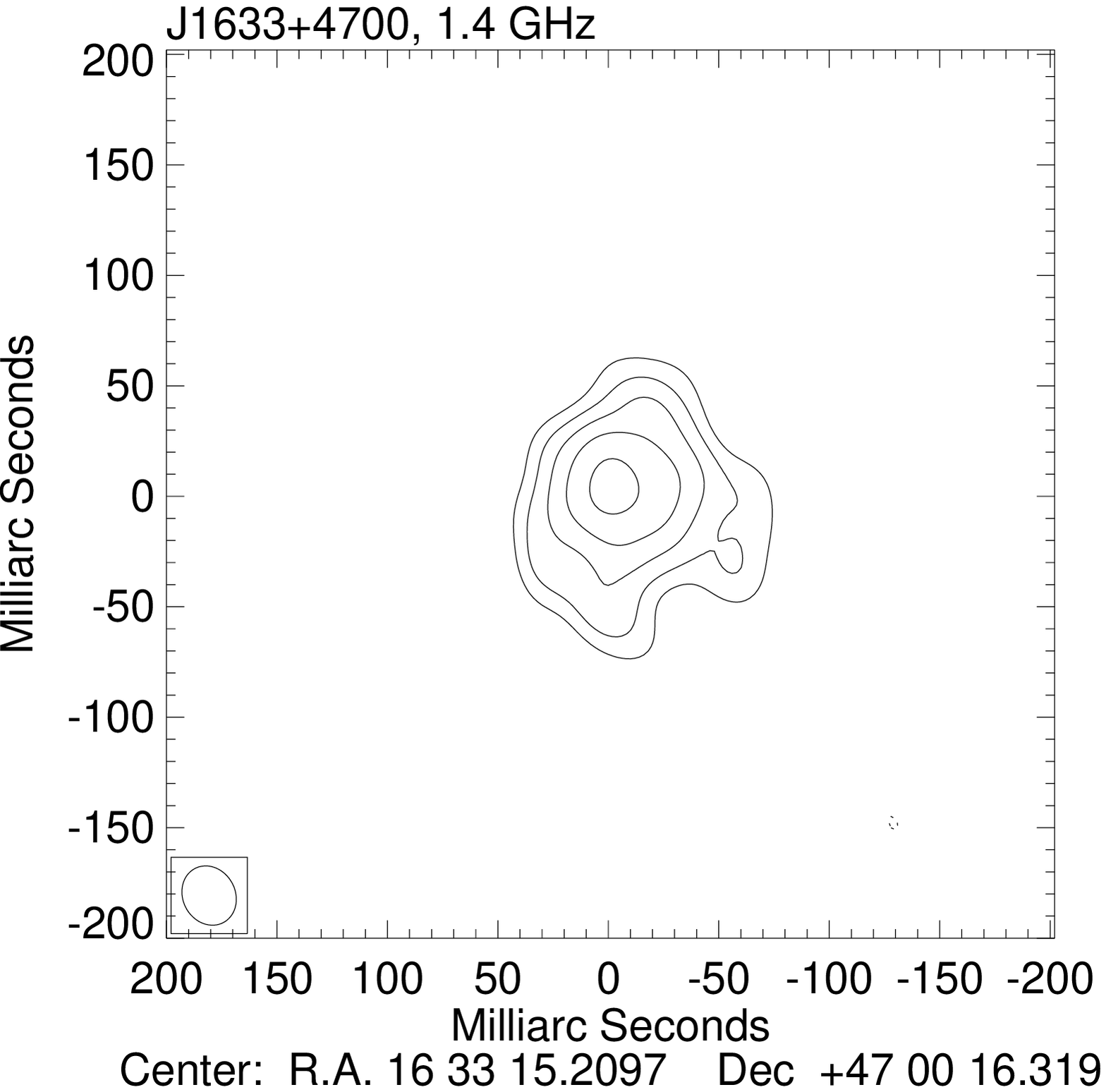}	\\
\includegraphics[scale=0.25]{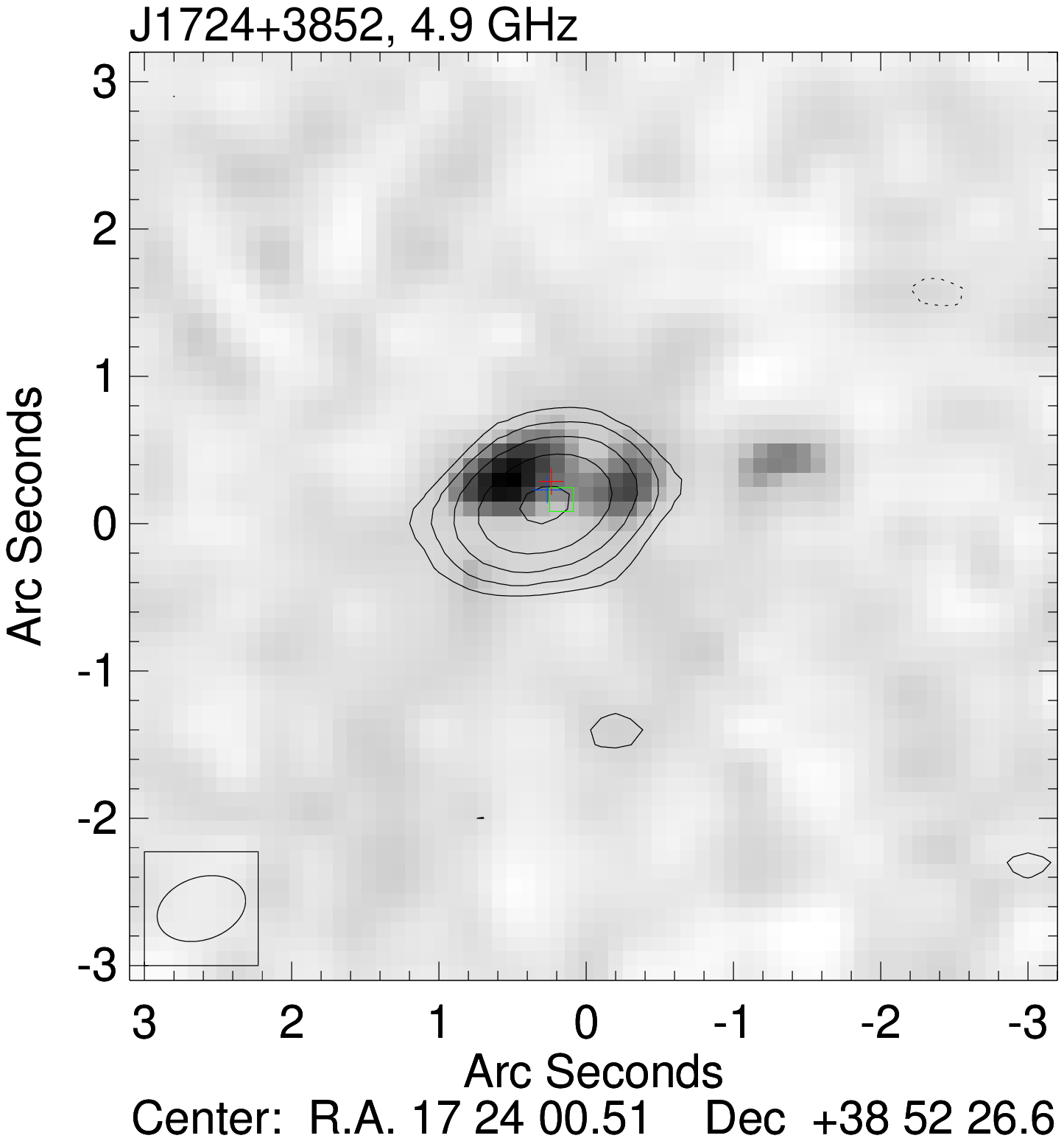}	&	\includegraphics[scale=0.25]{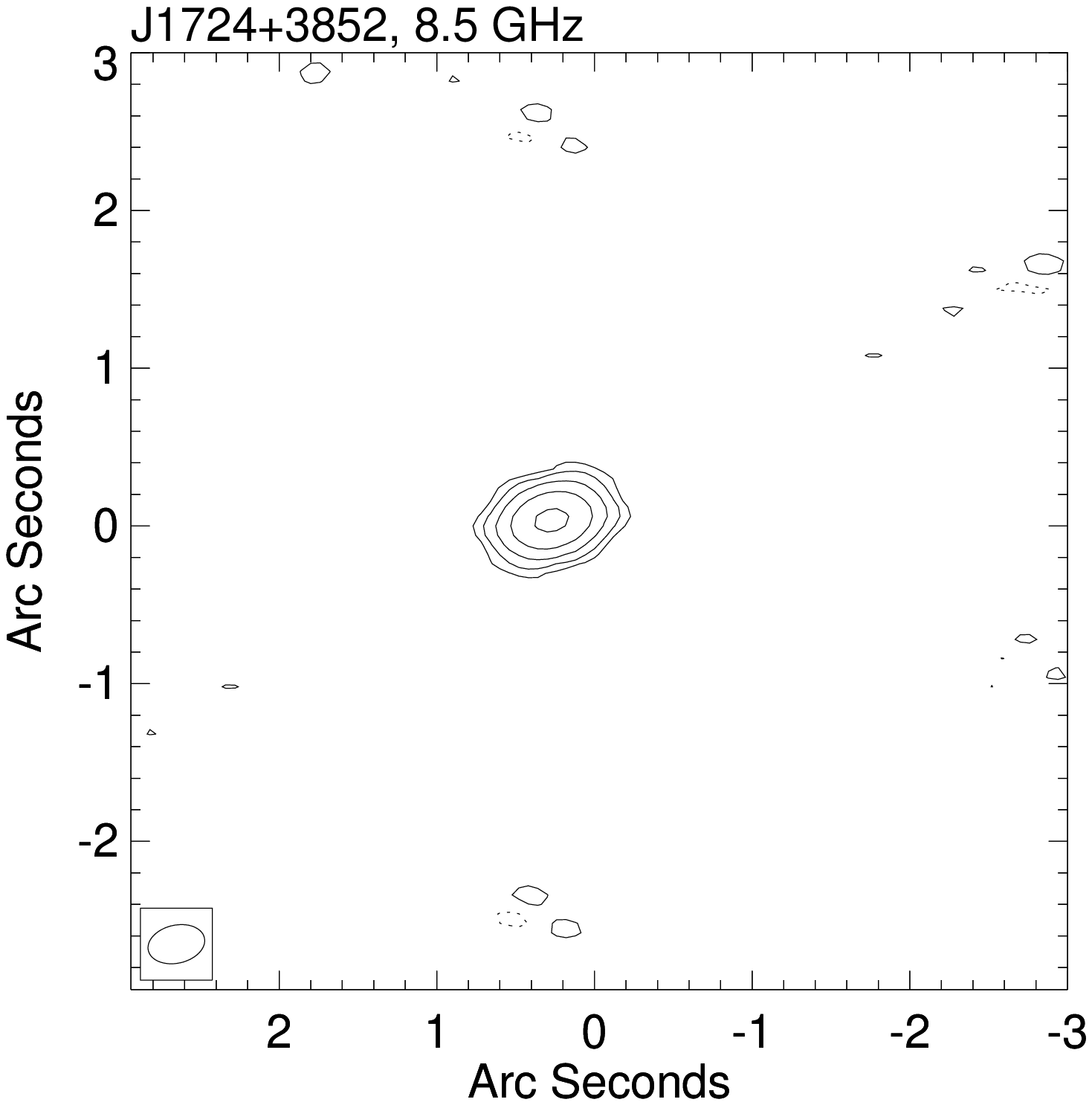}	&	\includegraphics[scale=0.25]{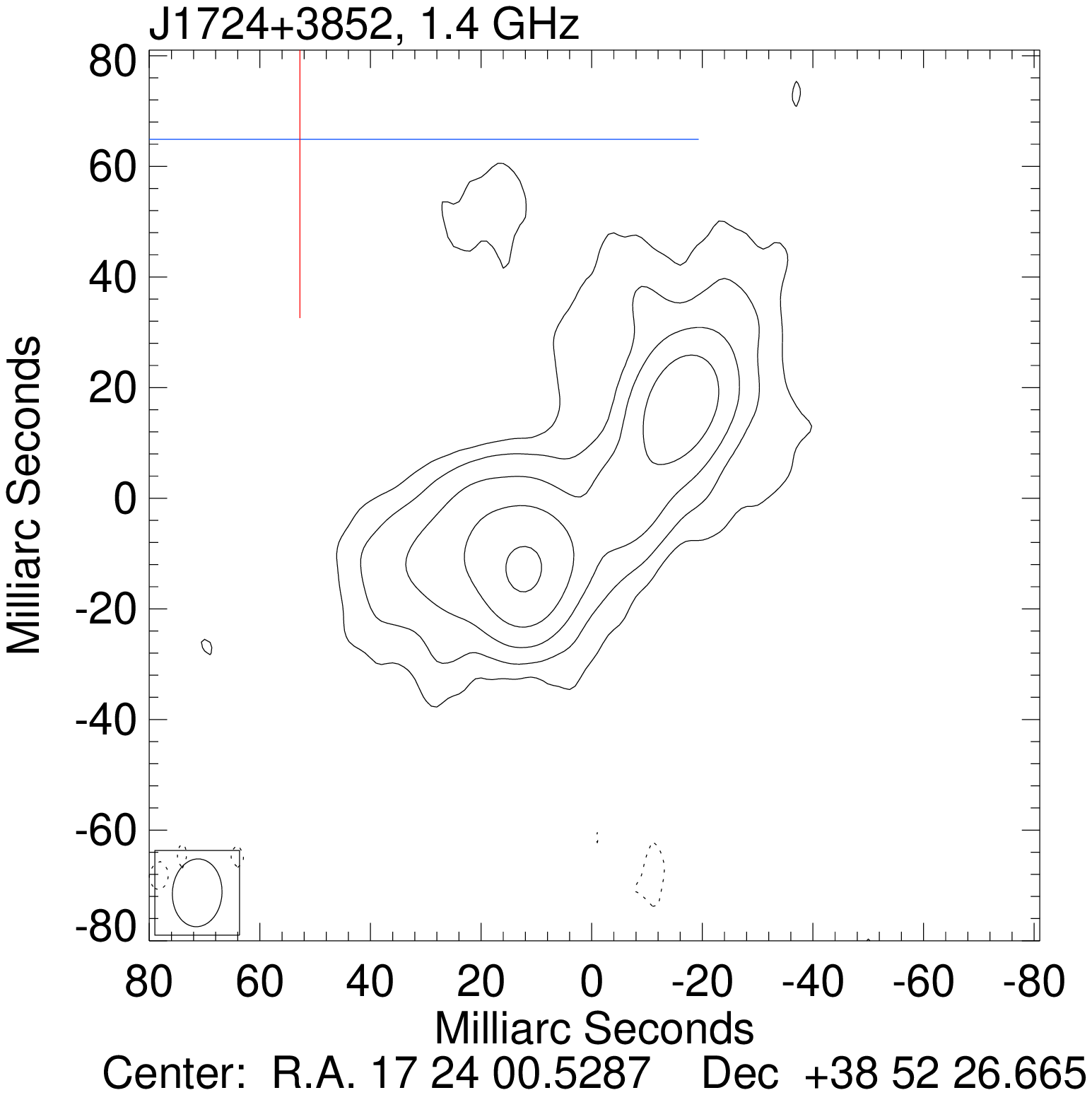}	\\

\end{tabular}
\end{figure}

\begin{figure}[htdp]
\begin{tabular}{lll}

\includegraphics[scale=0.25]{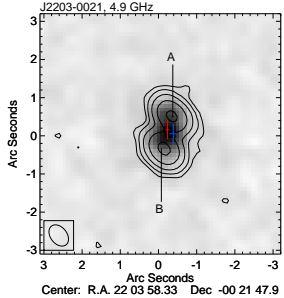}	&		&		\\
\end{tabular}

\caption{VLA 4.9 GHz, VLA 8.5 GHz and VLBA 1.4 GHz images (left to right columns) for all sources observed in this study. 
The VLA and VLBA maps obtained for each individual source occupy one row in this Figure. Each image has five contour levels 
ranging from 3$\sigma$ to 0.8$F_\mathrm{m}$ and logarithmically spaced, where $F_\mathrm{m}$ is the maximum of pixel counts, 
i.e., assuming $R=0.8F_\mathrm{m}/3\sigma$, the contour levels are (1, $R^{1/4}$, $R^{2/4}$, $R^{3/4}$, R)$\times 3\sigma$. 
Where present, a negative 3$\sigma$ value is shown as a dashed line. In the 4.9 GHz and 1.4 GHz images, the optical centroid 
position in the $r$-band is shown as a blue cross with uncertainties. The red cross is the NIR centroid position (usually 
$K_\mathrm{s}$-band) with uncertainties. A green rectangle in a 4.9 GHz image represents the boundary of the object's 1.4 GHz 
VLBA image where available (usually very small; see VLBA map field-of-view in right-hand column). Therefore, even if an object's 
NIR and/or optical position crosses are completely or partly outside of its 1.4 GHz image, their relative positions can be 
inferred from the crosses and rectangle in the 4.9 GHz image. The absolute coordinates of the map center of the 4.9 GHz and 
1.4 GHz VLBA images are shown at the bottom of each plot. Relative RA and Dec positions are labeled on the X and Y axes 
respectively. We note that absolute 8.5 GHz astrometry is problematic (see Section~\ref{sec:obs}) and the central coordinates 
of the 8.5 GHz images are not given for this reason. The 4.9 GHz images are overlaid with NIR images from the APO 3.5m
telescope (see Paper \Rmnum{1}). NIR images have been re-gridded and smoothed by cubic interpolation to match the astrometry 
of the 4.9 GHz images. Nevertheless, in some cases, for the most diffuse sources, their NIR images become fractured by the gridding in this
display. See Paper~\Rmnum{1} and \citet{yan:2013} for better quality NIR images where required. Pixel counts are displayed by 
grayscale between  0 (white) to 1 (dark) where 1 represents the maximum pixel count of the object; 0 and 0.2 represent the
$-$2.5$\sigma$ and 2.5$\sigma$ noise of the sky background respectively; between 0.2 to 1 the gray shade 
scales linearly with pixel counts between these extremes. The restored beam is shown at the lower left corner of each image. 
J0834$+$1700 and J1527$+$3312 have no 4.9 GHz images so we have overlaid their NIR images on the 8.5 GHz images. For those two cases,
we label the optical and NIR centroid positions on their 8.5 GHz images as well as the field-of-view of the 1.4 GHz VLBA images. \label{fig:image}}

\end{figure}

\clearpage

\begin{landscape}
\begin{deluxetable}{lccrcrccrcrccrcr}
\tablewidth{20cm}
\tabletypesize{\tiny}
\tablecaption{Basic Observing Results \label{tab:obs}} 

  \tablehead{ \colhead{Object} &\multicolumn{5}{|c|}{VLA 4.9 GHz} &
    \multicolumn{5}{|c|}{VLA 8.5 GHz} & \multicolumn{5}{|c|}{VLBA 1.4
      GHz} \\ \hline \colhead{} & \colhead{Maj.} & \colhead{Min.} &
    \colhead{P.A.} & \colhead{RMS} & \colhead{$S_\mathrm{m}$} &
    \colhead{Maj.} & \colhead{Min.} & \colhead{P.A.} & \colhead{RMS} &
    \colhead{$S_\mathrm{m}$} & \colhead{Maj.} & \colhead{Min.} &
    \colhead{P.A.} & \colhead{RMS} & \colhead{$S_\mathrm{m}$}
    \\
    \colhead{} & \colhead{($\arcsec$)} & \colhead{($\arcsec$)} &
    \colhead{($\degr$)} & \colhead{(mJy/} & \colhead{(mJy/} &
    \colhead{($\arcsec$)} & \colhead{($\arcsec$)} &
    \colhead{($\degr$)} & \colhead{(mJy/} & \colhead{(mJy/} &
    \colhead{(mas)} & \colhead{(mas)} & \colhead{($\degr$)} &
    \colhead{(mJy/} & \colhead{(mJy/}
    \\
    \colhead{} & \colhead{} & \colhead{} & \colhead{} &
    \colhead{beam)} & \colhead{beam)} & \colhead{} & \colhead{} &
    \colhead{} & \colhead{beam)} & \colhead{beam)} & \colhead{} &
    \colhead{} & \colhead{} & \colhead{beam)} & \colhead{beam)} }

\startdata
	J0000$-$1054	&	0.60	&	0.44	&	-6.3	&	0.15		&	34.9		&	0.35	&	0.25	&	174.3	&	0.11		&	17.6		&	\nodata	&	\nodata	&	\nodata	&	\nodata	&	\nodata	\\
	J0003$-$1053	&	0.60	&	0.45	&	-3.3	&	0.17		&	31.1		&	0.40	&	0.32	&	32.7	&	0.15		&	22.4		&	\nodata	&	\nodata	&	\nodata	&	\nodata	&	\nodata	\\
	J0134$+$0003	&	0.54	&	0.44	&	-22.7	&	0.57	*	&	541.5	*	&	\nodata	&	\nodata	&	\nodata	&	\nodata		&	\nodata		&	\nodata	&	\nodata	&	\nodata	&	\nodata	&	\nodata	\\
	J0249$-$0759	&	0.72	&	0.46	&	-39.8	&	0.33	*	&	117.9	*	&	0.44	&	0.32	&	-44.6	&	0.71	*	&	137.7	*	&	\nodata	&	\nodata	&	\nodata	&	\nodata	&	\nodata	\\
	J0736$+$2954	&	\nodata	&	\nodata	&	\nodata	&	\nodata		&	\nodata		&	\nodata	&	\nodata	&	\nodata	&	\nodata		&	\nodata		&	\nodata	&	\nodata	&	\nodata	&	\nodata	&	\nodata	\\
	J0747$+$4618	&	\nodata	&	\nodata	&	\nodata	&	\nodata		&	\nodata		&	\nodata	&	\nodata	&	\nodata	&	\nodata		&	\nodata		&	\nodata	&	\nodata	&	\nodata	&	\nodata	&	\nodata	\\
	J0749$+$2129	&	0.81	&	0.46	&	39.8	&	0.14		&	45.0		&	0.33	&	0.27	&	-19.3	&	0.23		&	10.0		&	\nodata	&	\nodata	&	\nodata	&	\nodata	&	\nodata	\\
	J0751$+$2716	&	\nodata	&	\nodata	&	\nodata	&	\nodata		&	\nodata		&	\nodata	&	\nodata	&	\nodata	&	\nodata		&	\nodata		&	\nodata	&	\nodata	&	\nodata	&	\nodata	&	\nodata	\\
	J0759$+$5312	&	0.74	&	0.48	&	61.0	&	0.15		&	97.5		&	0.33	&	0.28	&	1.5	&	0.18		&	50.8		&	11.93	&	7.02	&	1.4	&	0.34	&	66.7	\\
	J0805$+$1614	&	0.61	&	0.46	&	64.0	&	0.19		&	114.0		&	0.28	&	0.23	&	37.5	&	0.29		&	89.5		&	14.43	&	8.48	&	9.3	&	1.00	&	97.2	\\
	J0807$+$5327	&	0.75	&	0.49	&	59.5	&	0.15		&	81.0		&	0.33	&	0.28	&	1.4	&	0.12		&	39.0		&	61.82	&	55.91	&	-33.9	&	2.74	&	45.3	\\
	J0824$+$5413	&	0.75	&	0.49	&	58.6	&	0.15		&	62.9		&	0.33	&	0.28	&	3.9	&	0.11		&	30.1		&	\nodata	&	\nodata	&	\nodata	&	\nodata	&	\nodata	\\
	J0834$+$1700	&	\nodata	&	\nodata	&	\nodata	&	\nodata		&	\nodata		&	0.26	&	0.24	&	-171.1	&	0.49		&	226.7		&	21.78	&	14.14	&	66.6	&	0.17	&	65.2	\\
	J0839$+$2403	&	0.43	&	0.41	&	-43.6	&	0.25		&	97.5		&	\nodata	&	\nodata	&	\nodata	&	\nodata		&	\nodata		&	\nodata	&	\nodata	&	\nodata	&	\nodata	&	\nodata	\\
	J0843$+$4215	&	\nodata	&	\nodata	&	\nodata	&	\nodata		&	\nodata		&	\nodata	&	\nodata	&	\nodata	&	\nodata		&	\nodata		&	\nodata	&	\nodata	&	\nodata	&	\nodata	&	\nodata	\\
	J0901$+$0304	&	0.62	&	0.42	&	50.4	&	0.35		&	160.5		&	\nodata	&	\nodata	&	\nodata	&	\nodata		&	\nodata		&	14.56	&	6.21	&	-1.3	&	0.30	&	184.9	\\
	J0903$+$5012	&	0.63	&	0.49	&	52.2	&	0.20		&	230.0		&	0.31	&	0.27	&	-13.3	&	0.07		&	39.4		&	\nodata	&	\nodata	&	\nodata	&	\nodata	&	\nodata	\\
	J0905$+$4128	&	\nodata	&	\nodata	&	\nodata	&	\nodata		&	\nodata		&	\nodata	&	\nodata	&	\nodata	&	\nodata		&	\nodata		&	\nodata	&	\nodata	&	\nodata	&	\nodata	&	\nodata	\\
	J0907$+$0413	&	0.58	&	0.47	&	-41.5	&	0.25		&	81.8		&	\nodata	&	\nodata	&	\nodata	&	\nodata		&	\nodata		&	\nodata	&	\nodata	&	\nodata	&	\nodata	&	\nodata	\\
	J0910$+$2419	&	0.44	&	0.41	&	-59.6	&	0.19		&	97.1		&	\nodata	&	\nodata	&	\nodata	&	\nodata		&	\nodata		&	\nodata	&	\nodata	&	\nodata	&	\nodata	&	\nodata	\\
	J0915$+$1018	&	0.54	&	0.40	&	-9.7	&	0.21		&	51.6		&	\nodata	&	\nodata	&	\nodata	&	\nodata		&	\nodata		&	\nodata	&	\nodata	&	\nodata	&	\nodata	&	\nodata	\\
	J0917$+$4725	&	0.75	&	0.48	&	71.0	&	0.19		&	63.0		&	0.31	&	0.27	&	-19.4	&	0.09		&	24.4		&	51.39	&	48.50	&	-11.1	&	0.56	&	7.5	\\
	J0920$+$1753	&	0.49	&	0.43	&	-87.6	&	0.29		&	177.2		&	\nodata	&	\nodata	&	\nodata	&	\nodata		&	\nodata		&	\nodata	&	\nodata	&	\nodata	&	\nodata	&	\nodata	\\
	J0920$+$2714	&	0.54	&	0.44	&	73.8	&	0.17		&	91.7		&	0.35	&	0.25	&	78.3	&	0.22		&	42.2		&	34.52	&	15.91	&	46.7	&	0.25	&	34.6	\\
	J0939$+$0304	&	0.68	&	0.41	&	-40.5	&	0.26		&	198.0		&	\nodata	&	\nodata	&	\nodata	&	\nodata		&	\nodata		&	15.38	&	6.90	&	1.2	&	0.24	&	160.6	\\
	J0945$+$2640	&	0.97	&	0.47	&	61.2	&	0.15		&	162.0		&	\nodata	&	\nodata	&	\nodata	&	\nodata		&	\nodata		&	13.66	&	10.10	&	8.2	&	0.11	&	64.6	\\
	J0951$+$1154	&	0.49	&	0.44	&	-3.0	&	0.18		&	26.3		&	0.74	&	0.39	&	-46.7	&	0.15		&	15.9		&	\nodata	&	\nodata	&	\nodata	&	\nodata	&	\nodata	\\
	J1008$+$2401	&	0.47	&	0.41	&	-47.5	&	0.16		&	58.7		&	\nodata	&	\nodata	&	\nodata	&	\nodata		&	\nodata		&	\nodata	&	\nodata	&	\nodata	&	\nodata	&	\nodata	\\
	J1010$+$4159	&	\nodata	&	\nodata	&	\nodata	&	\nodata		&	\nodata		&	\nodata	&	\nodata	&	\nodata	&	\nodata		&	\nodata		&	\nodata	&	\nodata	&	\nodata	&	\nodata	&	\nodata	\\
	J1019$+$4408	&	\nodata	&	\nodata	&	\nodata	&	\nodata		&	\nodata		&	\nodata	&	\nodata	&	\nodata	&	\nodata		&	\nodata		&	\nodata	&	\nodata	&	\nodata	&	\nodata	&	\nodata	\\
	J1023$+$0424	&	0.51	&	0.42	&	-24.4	&	0.21		&	71.5		&	0.35	&	0.25	&	30.0	&	0.09		&	38.6		&	14.69	&	6.62	&	5.6	&	0.28	&	43.7	\\
	J1033$+$3935	&	\nodata	&	\nodata	&	\nodata	&	\nodata		&	\nodata		&	\nodata	&	\nodata	&	\nodata	&	\nodata		&	\nodata		&	\nodata	&	\nodata	&	\nodata	&	\nodata	&	\nodata	\\
	J1034$+$1112	&	0.54	&	0.46	&	51.4	&	0.17		&	123.7		&	0.31	&	0.27	&	71.7	&	0.28		&	53.7		&	\nodata	&	\nodata	&	\nodata	&	\nodata	&	\nodata	\\
	J1043$+$0537	&	0.53	&	0.44	&	38.7	&	0.12		&	46.7		&	\nodata	&	\nodata	&	\nodata	&	\nodata		&	\nodata		&	\nodata	&	\nodata	&	\nodata	&	\nodata	&	\nodata	\\
	J1045$+$0455	&	0.55	&	0.46	&	42.3	&	0.14		&	52.0		&	0.49	&	0.30	&	56.9	&	0.20		&	23.1		&	\nodata	&	\nodata	&	\nodata	&	\nodata	&	\nodata	\\
	J1048$+$3457	&	\nodata	&	\nodata	&	\nodata	&	\nodata		&	\nodata		&	\nodata	&	\nodata	&	\nodata	&	\nodata		&	\nodata		&	8.12	&	4.86	&	-5.9	&	0.60	&	73.7	\\
	J1120$+$2327	&	\nodata	&	\nodata	&	\nodata	&	\nodata		&	\nodata		&	\nodata	&	\nodata	&	\nodata	&	\nodata		&	\nodata		&	\nodata	&	\nodata	&	\nodata	&	\nodata	&	\nodata	\\
	J1125$+$1953	&	0.47	&	0.43	&	54.8	&	0.21		&	101.9		&	\nodata	&	\nodata	&	\nodata	&	\nodata		&	\nodata		&	15.40	&	8.37	&	-13.1	&	0.20	&	18.2	\\
	J1127$+$5743	&	0.53	&	0.41	&	2.0	&	0.21		&	147.3		&	0.44	&	0.26	&	-33.4	&	0.20		&	79.3		&	24.87	&	15.68	&	65.1	&	0.36	&	219.9	\\
	J1129$+$5638	&	0.58	&	0.44	&	-2.0	&	0.28		&	108.3		&	0.47	&	0.26	&	-35.9	&	0.18		&	54.0		&	9.67	&	6.78	&	3.0	&	0.14	&	80.5	\\
	J1142$+$0235	&	0.57	&	0.41	&	-1.3	&	0.23		&	64.6		&	0.29	&	0.25	&	-175.8	&	0.46		&	40.9		&	\nodata	&	\nodata	&	\nodata	&	\nodata	&	\nodata	\\
	J1147$+$4818	&	0.49	&	0.43	&	71.3	&	0.20		&	93.2		&	0.54	&	0.26	&	73.4	&	0.33		&	50.8		&	12.83	&	12.19	&	8.6	&	0.45	&	33.3	\\
	J1148$+$1404	&	0.47	&	0.44	&	-44.8	&	0.25		&	57.8		&	0.29	&	0.25	&	15.7	&	0.12		&	31.4		&	47.11	&	28.02	&	29.3	&	0.57	&	11.7	\\
	J1202$+$1207	&	0.48	&	0.43	&	24.3	&	0.17		&	130.4		&	0.29	&	0.28	&	70.8	&	0.10		&	88.7		&	13.23	&	6.76	&	3.4	&	0.15	&	32.3	\\
	J1203$+$4632	&	0.42	&	0.41	&	87.3	&	0.25		&	184.2		&	\nodata	&	\nodata	&	\nodata	&	\nodata		&	\nodata		&	\nodata	&	\nodata	&	\nodata	&	\nodata	&	\nodata	\\
	J1207$+$5407	&	0.57	&	0.49	&	12.6	&	0.19		&	71.4		&	0.41	&	0.29	&	109.6	&	0.33		&	33.0		&	\nodata	&	\nodata	&	\nodata	&	\nodata	&	\nodata	\\
	J1215$+$1730	&	0.44	&	0.41	&	-55.5	&	0.40		&	356.1		&	\nodata	&	\nodata	&	\nodata	&	\nodata		&	\nodata		&	\nodata	&	\nodata	&	\nodata	&	\nodata	&	\nodata	\\
	J1228$+$5348	&	0.74	&	0.46	&	85.9	&	0.17		&	47.6		&	0.56	&	0.42	&	-16.1	&	0.35		&	30.8		&	\nodata	&	\nodata	&	\nodata	&	\nodata	&	\nodata	\\
	J1238$+$0845	&	0.55	&	0.45	&	45.0	&	0.13		&	86.6		&	0.29	&	0.29	&	-27.9	&	0.07		&	41.6		&	33.93	&	15.65	&	47.5	&	0.64	&	30.5	\\
	J1300$+$5029	&	\nodata	&	\nodata	&	\nodata	&	\nodata		&	\nodata		&	\nodata	&	\nodata	&	\nodata	&	\nodata		&	\nodata		&	\nodata	&	\nodata	&	\nodata	&	\nodata	&	\nodata	\\
	J1312$+$1710	&	0.44	&	0.42	&	14.1	&	0.21		&	94.3		&	0.26	&	0.25	&	-46.7	&	0.07		&	56.3		&	13.86	&	7.64	&	15.5	&	0.40	&	36.7	\\
	J1315$+$0222	&	0.56	&	0.42	&	-22.2	&	0.20		&	68.6		&	0.30	&	0.25	&	-26.2	&	0.11		&	33.9		&	\nodata	&	\nodata	&	\nodata	&	\nodata	&	\nodata	\\
	J1341$+$1032	&	0.61	&	0.44	&	68.9	&	0.16		&	80.8		&	0.30	&	0.28	&	96.6	&	0.17		&	34.5		&	\nodata	&	\nodata	&	\nodata	&	\nodata	&	\nodata	\\
	J1345$+$5846	&	0.69	&	0.41	&	73.2	&	0.15		&	72.7		&	0.32	&	0.29	&	77.4	&	0.12		&	37.1		&	22.12	&	13.74	&	78.2	&	0.50	&	20.5	\\
	J1347$+$1217	&	\nodata	&	\nodata	&	\nodata	&	\nodata		&	\nodata		&	\nodata	&	\nodata	&	\nodata	&	\nodata		&	\nodata		&	\nodata	&	\nodata	&	\nodata	&	\nodata	&	\nodata	\\
	J1348$+$2415	&	\nodata	&	\nodata	&	\nodata	&	\nodata		&	\nodata		&	\nodata	&	\nodata	&	\nodata	&	\nodata		&	\nodata		&	\nodata	&	\nodata	&	\nodata	&	\nodata	&	\nodata	\\
	J1354$+$5650	&	0.68	&	0.40	&	73.2	&	0.14		&	141.9		&	0.32	&	0.29	&	77.8	&	0.11		&	76.2		&	\nodata	&	\nodata	&	\nodata	&	\nodata	&	\nodata	\\
	J1357$+$0046	&	0.54	&	0.49	&	30.7	&	0.36		&	475.1		&	0.33	&	0.28	&	-30.2	&	0.12		&	229.0		&	14.81	&	5.91	&	0.1	&	0.65	&	401.1	\\


	J1410$+$4850	&	0.43	&	0.41	&	-54.1	&	0.20		&	115.9		&	0.25	&	0.24	&	6.8	&	0.14		&	63.9		&	10.15	&	6.69	&	9.4	&	0.13	&	94.1	\\
	J1413$+$1509	&	0.43	&	0.41	&	-89.4	&	0.29		&	221.2		&	\nodata	&	\nodata	&	\nodata	&	\nodata		&	\nodata		&	\nodata	&	\nodata	&	\nodata	&	\nodata	&	\nodata	\\
	J1414$+$4554	&	\nodata	&	\nodata	&	\nodata	&	\nodata		&	\nodata		&	\nodata	&	\nodata	&	\nodata	&	\nodata		&	\nodata		&	\nodata	&	\nodata	&	\nodata	&	\nodata	&	\nodata	\\
	J1415$+$1320	&	\nodata	&	\nodata	&	\nodata	&	\nodata		&	\nodata		&	\nodata	&	\nodata	&	\nodata	&	\nodata		&	\nodata		&	\nodata	&	\nodata	&	\nodata	&	\nodata	&	\nodata	\\
	J1421$-$0246	&	0.68	&	0.46	&	33.7	&	0.13		&	21.6		&	0.32	&	0.24	&	10.1	&	0.11		&	11.9		&	\nodata	&	\nodata	&	\nodata	&	\nodata	&	\nodata	\\
	J1424$+$1852	&	0.45	&	0.44	&	42.0	&	0.18		&	69.8		&	0.28	&	0.24	&	36.5	&	0.27		&	32.8		&	\nodata	&	\nodata	&	\nodata	&	\nodata	&	\nodata	\\
	J1502$+$3753	&	0.47	&	0.42	&	61.4	&	0.19		&	32.5		&	0.24	&	0.24	&	46.7	&	0.11		&	11.8		&	\nodata	&	\nodata	&	\nodata	&	\nodata	&	\nodata	\\
	J1504$+$5438	&	\nodata	&	\nodata	&	\nodata	&	\nodata		&	\nodata		&	\nodata	&	\nodata	&	\nodata	&	\nodata		&	\nodata		&	\nodata	&	\nodata	&	\nodata	&	\nodata	&	\nodata	\\
	J1504$+$6000	&	0.55	&	0.44	&	77.0	&	0.41		&	297.1		&	0.35	&	0.29	&	81.3	&	0.15		&	175.1		&	10.29	&	8.68	&	19.1	&	0.60	&	132.5	\\
	J1523$+$1332	&	0.48	&	0.42	&	-43.0	&	0.15		&	63.1		&	0.29	&	0.26	&	-35.0	&	0.11		&	30.1		&	19.93	&	15.12	&	29.2	&	0.36	&	18.2	\\
	J1527$+$3312	&	\nodata	&	\nodata	&	\nodata	&	\nodata		&	\nodata		&	0.31	&	0.24	&	41.5	&	0.38		&	73.7		&	13.11	&	7.23	&	9.2	&	0.19	&	59.9	\\
	J1528$-$0213	&	0.54	&	0.45	&	-13.8	&	0.18		&	60.4		&	0.33	&	0.27	&	-10.7	&	0.09		&	31.4		&	\nodata	&	\nodata	&	\nodata	&	\nodata	&	\nodata	\\
	J1548$+$0808	&	0.68	&	0.41	&	-52.0	&	0.42		&	149.2		&	0.37	&	0.24	&	-51.0	&	0.09		&	59.0		&	49.00	&	37.85	&	17.3	&	3.50	&	113.6	\\
	J1551$+$6405	&	\nodata	&	\nodata	&	\nodata	&	0.25		&	\nodata		&	\nodata	&	\nodata	&	\nodata	&	\nodata		&	\nodata		&	\nodata	&	\nodata	&	\nodata	&	\nodata	&	\nodata	\\
	J1559$+$4349	&	0.51	&	0.42	&	80.6	&	0.19		&	134.3		&	\nodata	&	\nodata	&	\nodata	&	\nodata		&	\nodata		&	18.49	&	15.02	&	66.5	&	0.33	&	30.0	\\
	J1604$+$6050	&	0.66	&	0.44	&	87.8	&	0.16		&	165.6		&	0.38	&	0.29	&	89.3	&	0.09		&	86.1		&	10.92	&	9.03	&	50.2	&	0.28	&	222.3	\\
	J1616$+$2647	&	0.46	&	0.41	&	-62.8	&	0.77		&	578.7		&	0.43	&	0.24	&	19.0	&	0.42	*	&	605.1	*	&	13.76	&	8.31	&	7.0	&	0.40	&	366.6	\\
	J1625$+$4134	&	\nodata	&	\nodata	&	\nodata	&	\nodata		&	\nodata		&	\nodata	&	\nodata	&	\nodata	&	\nodata		&	\nodata		&	\nodata	&	\nodata	&	\nodata	&	\nodata	&	\nodata	\\
	J1629$+$1342	&	0.72	&	0.41	&	-56.5	&	0.14		&	79.8		&	0.40	&	0.24	&	122.3	&	0.07		&	41.2		&	\nodata	&	\nodata	&	\nodata	&	\nodata	&	\nodata	\\
	J1633$+$4700	&	0.55	&	0.42	&	-84.4	&	0.13		&	73.9		&	0.32	&	0.24	&	89.4	&	0.06		&	35.8		&	27.64	&	23.72	&	27.9	&	0.65	&	26.7	\\
	J1724$+$3852	&	0.62	&	0.42	&	-71.0	&	0.19		&	126.1		&	0.36	&	0.24	&	102.8	&	0.20		&	68.7		&	12.26	&	8.96	&	-3.2	&	0.15	&	89.4	\\
	J2203$-$0021	&	0.62	&	0.43	&	38.1	&	0.33	*	&	131.4	*	&	\nodata	&	\nodata	&	\nodata	&	\nodata		&	\nodata		&	\nodata	&	\nodata	&	\nodata	&	\nodata	&	\nodata	\\

\enddata

\tablecomments{Flux densities designated with ``*'' lack accurate flux
  calibrations. See Table~\ref{tab:log} for details.}

\end{deluxetable}

\end{landscape}

\begin{deluxetable}{lcrrrrrrr}
\tablewidth{0pt}
  \tablecaption{VLA 4.9 GHz Gaussian Fit Results.\label{tab:vlac}}

  \tablehead{ \colhead{Object} & \colhead{ID} & \colhead{R.A.} &
    \colhead{Dec.} & \colhead{$S_{p}$ (mJy} & \colhead{$S_{tot}$} &
    \colhead{Maj.} & \colhead{Min.} &
    \colhead{P.A.} \\
    \colhead{} & \colhead{} &\colhead{(hh mm ss.sss)} & \colhead{(dd
      mm ss.ss)} &\colhead{/beam)} & \colhead{(mJy)} &
    \colhead{($\arcsec$)} & \colhead{($\arcsec$)} &
    \colhead{($\degr$)} \\
    \colhead{(1)} & \colhead{(2)} & \colhead{(3)} & \colhead{(4)} &
    \colhead{(5)} & \colhead{(6)} & \colhead{(7)} & \colhead{(8)} &
    \colhead{(9)}}

\startdata
J0000$-$1054	&	A	&	00 00 57.647	&	$-$10 54 32.87	&	35.3		&	42.6		&	0.28	&	0.17	&	40.2	\\
	&	B	&	00 00 57.690	&	$-$10 54 31.50	&	18.8		&	23.8		&	0.36	&	0.14	&	32.0	\\
J0003$-$1053	&	A	&	00 03 56.254	&	$-$10 53 02.49	&	30.7		&	39.6		&	0.30	&	0.23	&	96.1	\\
	&	B	&	00 03 56.299	&	$-$10 53 02.47	&	30.5		&	40.5		&	0.40	&	0.07	&	60.0	\\
	&	C	&	00 03 56.332	&	$-$10 53 01.92	&	22.3		&	23.0		&	0.11	&	0.07	&	140.9	\\
	&	D	&	00 03 56.381	&	$-$10 53 01.06	&	24.3		&	27.8		&	0.21	&	0.17	&	48.7	\\
J0134$+$0003	&		&	01 34 12.704	&	$+$00 03 45.14	&	548.3	*	&	553.9	*	&	0.05	&	0.05	&	106.0	\\
J0249$-$0759	&	A	&	02 49 35.370	&	$-$07 59 21.92	&	118.9	*	&	136.1	*	&	0.23	&	0.18	&	5.9	\\
	&	B	&	02 49 35.426	&	$-$07 59 20.21	&	55.6	*	&	67.1	*	&	0.29	&	0.23	&	149.8	\\
J0749$+$2129	&	A	&	07 49 48.708	&	$+$21 29 33.47	&	44.7		&	63.5		&	0.68	&	0.18	&	27.6	\\
	&	B	&	07 49 48.813	&	$+$21 29 32.87	&	28.2		&	36.0		&	0.45	&	0.22	&	27.5	\\
	&	C	&	07 49 48.697	&	$+$21 29 34.90	&	8.6		&	10.3		&	0.34	&	0.21	&	43.5	\\
J0759$+$5312	&	A	&	07 59 06.465	&	$+$53 12 47.95	&	97.8		&	100.1		&	0.11	&	0.07	&	40.5	\\
	&	B	&	07 59 06.598	&	$+$53 12 47.23	&	3.7		&	4.5		&	0.34	&	0.24	&	79.9	\\
J0805$+$1614	&	A	&	08 05 02.180	&	$+$16 14 05.10	&	107.1		&	114.2		&	0.18	&	0.10	&	72.6	\\
	&	B	&	08 05 02.180	&	$+$16 14 04.60	&	69.2		&	73.0		&	0.16	&	0.09	&	72.4	\\
J0807$+$5327	&	A	&	08 07 40.746	&	$+$53 27 38.63	&	81.9		&	86.3		&	0.15	&	0.10	&	152.9	\\
	&	B	&	08 07 40.714	&	$+$53 27 37.60	&	22.1		&	24.9		&	0.27	&	0.16	&	35.9	\\
J0824$+$5413	&	A	&	08 24 25.427	&	$+$54 13 48.06	&	62.5		&	79.4		&	0.36	&	0.23	&	112.3	\\
	&	B	&	08 24 25.568	&	$+$54 13 49.37	&	35.2		&	48.0		&	0.39	&	0.29	&	139.0	\\
J0839$+$2403	&	A	&	08 39 57.851	&	$+$24 03 11.27	&	100.8		&	108.0		&	0.12	&	0.10	&	50.9	\\
	&	B	&	08 39 57.997	&	$+$24 03 12.52	&	33.9		&	36.9		&	0.15	&	0.09	&	58.2	\\
	&	C	&	08 39 57.828	&	$+$24 03 10.57	&	8.2		&	21.9		&	0.70	&	0.39	&	55.5	\\
	&	D	&	08 39 58.024	&	$+$24 03 13.72	&	4.5		&	18.6		&	1.00	&	0.52	&	120.9	\\
J0901$+$0304	&		&	09 01 50.981	&	$+$03 04 22.66	&	166.5		&	172.5		&	0.14	&	0.06	&	44.9	\\
J0903$+$5012	&	A	&	09 03 49.973	&	$+$50 12 36.55	&	233.3		&	268.4		&	0.26	&	0.15	&	105.6	\\
	&	B	&	09 03 49.675	&	$+$50 12 34.73	&	11.2		&	22.7		&	0.76	&	0.38	&	81.7	\\
J0907$+$0413	&	A	&	09 07 50.770	&	$+$04 13 36.30	&	82.5		&	106.9		&	0.38	&	0.14	&	8.6	\\
	&	B	&	09 07 50.720	&	$+$04 13 38.80	&	64.0		&	89.7		&	0.41	&	0.23	&	7.5	\\
J0910$+$2419	&	A	&	09 10 22.376	&	$+$24 19 19.48	&	97.5		&	118.3		&	0.22	&	0.17	&	91.1	\\
	&	B	&	09 10 22.684	&	$+$24 19 19.48	&	51.8		&	78.3		&	0.33	&	0.28	&	109.0	\\
J0915$+$1018	&	A	&	09 15 12.960	&	$+$10 18 27.20	&	50.2		&	67.5		&	0.31	&	0.21	&	113.1	\\
	&	B	&	09 15 12.960	&	$+$10 18 28.00	&	28.3		&	40.6		&	0.34	&	0.27	&	135.1	\\
	&	C	&	09 15 12.980	&	$+$10 18 26.10	&	19.9		&	28.5		&	0.36	&	0.23	&	120.2	\\
J0917$+$4725	&		&	09 17 27.032	&	$+$47 25 24.07	&	60.9		&	95.7		&	0.58	&	0.29	&	29.3	\\
J0920$+$1753	&	A	&	09 20 11.143	&	$+$17 53 25.12	&	167.8		&	181.6		&	0.15	&	0.10	&	24.7	\\
	&	B	&	09 20 11.142	&	$+$17 53 24.70	&	106.2		&	113.6		&	0.13	&	0.10	&	159.4	\\
J0920$+$2714	&	A	&	09 20 45.142	&	$+$27 14 05.58	&	93.0		&	105.0		&	0.20	&	0.15	&	39.1	\\
	&	B	&	09 20 45.060	&	$+$27 14 02.90	&	5.7		&	6.9		&	0.23	&	0.20	&	166.1	\\
J0939$+$0304	&		&	09 37 09.249	&	$+$03 18 03.33	&	199.9		&	216.0		&	0.17	&	0.12	&	111.6	\\
J0945$+$2640	&	A	&	09 45 30.950	&	$+$26 40 53.90	&	161.2		&	171.8		&	0.16	&	0.13	&	154.3	\\
	&	B	&	09 45 31.000	&	$+$26 40 48.60	&	7.9		&	16.1		&	0.72	&	0.51	&	170.1	\\
	&	C	&	09 45 30.970	&	$+$26 40 53.10	&	3.6		&	11.3		&	1.68	&	0.41	&	29.2	\\
J0951$+$1154	&	A	&	09 51 33.780	&	$+$11 54 58.20	&	26.0		&	44.3		&	0.46	&	0.32	&	46.7	\\
	&	B	&	09 51 33.880	&	$+$11 55 00.70	&	25.6		&	39.8		&	0.37	&	0.32	&	127.3	\\
	&	C	&	09 51 33.850	&	$+$11 54 59.70	&	1.4		&	14.7		&	2.10	&	0.95	&	48.5	\\
J1008$+$2401	&	A	&	10 08 32.681	&	$+$24 01 19.77	&	57.7		&	84.1		&	0.38	&	0.19	&	77.8	\\
	&	B	&	10 08 32.548	&	$+$24 01 18.93	&	10.5		&	21.3		&	0.46	&	0.43	&	44.0	\\
J1023$+$0424	&	A	&	10 23 37.550	&	$+$04 24 14.20	&	71.3		&	74.4		&	0.12	&	0.07	&	5.5	\\
	&	B	&	10 23 37.560	&	$+$04 24 13.60	&	29.0		&	31.7		&	0.20	&	0.00	&	164.8	\\
J1034$+$1112	&	A	&	10 34 05.014	&	$+$11 12 30.58	&	120.3		&	143.9		&	0.24	&	0.19	&	153.9	\\
	&	B	&	10 34 05.299	&	$+$11 12 33.07	&	41.8		&	53.5		&	0.36	&	0.17	&	71.1	\\
	&	C	&	10 34 05.216	&	$+$11 12 32.35	&	16.8		&	61.7		&	1.44	&	0.37	&	61.8	\\
	&	D	&	10 34 05.132	&	$+$11 12 31.57	&	6.1		&	9.4		&	0.58	&	0.10	&	80.2	\\
	&	E	&	10 34 05.024	&	$+$11 12 31.08	&	15.3		&	56.8		&	1.09	&	0.59	&	95.1	\\
J1043$+$0537	&	A	&	10 43 40.425	&	$+$05 37 13.01	&	43.5		&	78.4		&	0.58	&	0.28	&	71.5	\\
	&	B	&	10 43 39.984	&	$+$05 37 12.28	&	23.8		&	53.0		&	0.62	&	0.44	&	91.5	\\
J1045$+$0455	&	A	&	10 45 51.764	&	$+$04 55 52.34	&	51.2		&	92.5		&	0.61	&	0.29	&	2.6	\\
	&	B	&	10 45 51.723	&	$+$04 55 49.52	&	14.9		&	32.6		&	0.83	&	0.29	&	16.2	\\
J1125$+$1953	&		&	11 25 55.240	&	$+$19 53 43.68	&	101.4		&	113.9		&	0.20	&	0.09	&	2.0	\\
J1127$+$5743	&	A	&	11 27 43.773	&	$+$57 43 15.84	&	147.8		&	152.1		&	0.09	&	0.06	&	60.0	\\
	&	B	&	11 27 43.718	&	$+$57 43 15.63	&	25.4		&	27.7		&	0.16	&	0.10	&	128.2	\\
J1129$+$5638	&		&	11 29 04.147	&	$+$56 38 44.14	&	109.6		&	112.9		&	0.11	&	0.03	&	88.6	\\
J1142$+$0235	&	A	&	11 42 06.370	&	$+$02 35 33.40	&	64.9		&	72.3		&	0.18	&	0.14	&	38.3	\\
	&	B	&	11 42 06.390	&	$+$02 35 32.80	&	41.2		&	65.5		&	0.49	&	0.27	&	9.6	\\
	&	C	&	11 42 06.380	&	$+$02 35 34.30	&	8.7		&	21.0		&	0.73	&	0.45	&	175.4	\\
J1147$+$4818	&	A	&	11 47 52.288	&	$+$48 18 49.48	&	94.9		&	107.0		&	0.19	&	0.12	&	128.1	\\
	&	B	&	11 47 52.193	&	$+$48 18 49.29	&	11.0		&	18.0		&	0.39	&	0.33	&	21.9	\\
J1148$+$1404	&	A	&	11 48 25.410	&	$+$14 04 49.70	&	57.9		&	77.2		&	0.28	&	0.24	&	15.4	\\
	&	B	&	11 48 25.440	&	$+$14 04 48.30	&	15.3		&	21.5		&	0.32	&	0.25	&	83.1	\\
J1202$+$1207	&		&	12 02 52.086	&	$+$12 07 20.76	&	133.9		&	140.2		&	0.12	&	0.07	&	105.5	\\
J1203$+$4632	&		&	12 03 31.795	&	$+$46 32 55.48	&	184.3		&	187.2		&	0.07	&	0.03	&	99.9	\\
J1207$+$5407	&	A	&	12 07 14.249	&	$+$54 07 54.49	&	72.6		&	76.7		&	0.16	&	0.08	&	44.7	\\
	&	B	&	12 07 14.088	&	$+$54 07 54.18	&	49.3		&	61.3		&	0.35	&	0.10	&	79.4	\\
J1215$+$1730	&		&	12 15 14.724	&	$+$17 30 02.23	&	369.4		&	388.8		&	0.12	&	0.06	&	50.1	\\
J1228$+$5348	&	A	&	12 28 50.815	&	$+$53 48 04.00	&	49.6		&	91.9		&	0.67	&	0.24	&	2.4	\\
	&	B	&	12 28 50.487	&	$+$53 48 00.08	&	31.9		&	50.9		&	0.54	&	0.33	&	48.3	\\
J1238$+$0845	&	A	&	12 38 19.260	&	$+$08 45 01.63	&	86.3		&	94.2		&	0.19	&	0.08	&	171.8	\\
	&	B	&	12 38 19.282	&	$+$08 44 58.47	&	4.7		&	8.3		&	0.56	&	0.33	&	41.0	\\
	&	C	&	12 38 19.277	&	$+$08 45 00.08	&	0.8		&	4.3		&	1.93	&	0.49	&	11.9	\\
J1312$+$1710	&		&	13 10 07.820	&	$+$17 26 49.70	&	98.8		&	109.4		&	0.17	&	0.11	&	0.1	\\
J1315$+$0222	&	A	&	13 15 16.950	&	$+$02 22 20.90	&	70.1		&	86.1		&	0.27	&	0.17	&	30.9	\\
	&	B	&	13 15 17.020	&	$+$02 22 21.40	&	31.1		&	74.1		&	0.74	&	0.34	&	58.2	\\
	&	C	&	13 15 16.880	&	$+$02 22 20.30	&	1.9		&	9.5		&	1.30	&	0.66	&	107.9	\\
J1341$+$1032	&	A	&	13 41 04.327	&	$+$10 32 06.56	&	83.0		&	92.3		&	0.25	&	0.11	&	80.3	\\
	&	B	&	13 41 04.259	&	$+$10 32 06.36	&	62.2		&	65.3		&	0.13	&	0.10	&	48.4	\\
J1345$+$5846	&	A	&	13 45 38.379	&	$+$58 46 53.48	&	74.5		&	87.9		&	0.22	&	0.21	&	96.8	\\
	&	B	&	13 45 38.319	&	$+$58 46 54.24	&	5.6		&	10.5		&	0.54	&	0.39	&	140.2	\\
	&	C	&	13 45 38.261	&	$+$58 46 55.54	&	3.0		&	7.8		&	0.84	&	0.56	&	78.0	\\
J1354$+$5650	&	A	&	13 54 00.153	&	$+$56 50 06.13	&	141.1		&	157.1		&	0.21	&	0.15	&	57.0	\\
	&	B	&	13 54 00.078	&	$+$56 50 03.70	&	38.5		&	99.4		&	0.98	&	0.42	&	57.5	\\
J1357$+$0046	&		&	13 57 53.723	&	$+$00 46 33.32	&	483.0		&	495.6		&	0.10	&	0.06	&	18.4	\\
J1410$+$4850	&		&	14 10 36.041	&	$+$48 50 40.35	&	117.0		&	118.7		&	0.07	&	0.01	&	11.9	\\
J1413$+$1509	&		&	14 13 41.657	&	$+$15 09 39.51	&	233.7		&	236.4		&	0.06	&	0.01	&	151.3	\\
J1421$-$0246	&	A	&	14 21 13.600	&	$-$02 46 44.33	&	21.0		&	52.9		&	0.87	&	0.53	&	7.1	\\
	&	B	&	14 21 13.550	&	$-$02 46 45.94	&	12.6		&	27.3		&	1.01	&	0.31	&	32.0	\\
	&	C	&	14 21 13.516	&	$-$02 46 47.63	&	20.2		&	53.3		&	0.87	&	0.57	&	176.2	\\
J1424$+$1852	&	A	&	14 24 09.691	&	$+$18 52 52.21	&	68.0		&	85.3		&	0.29	&	0.14	&	20.6	\\
	&	B	&	14 24 09.736	&	$+$18 52 53.69	&	71.2		&	79.8		&	0.18	&	0.13	&	40.6	\\
J1502$+$3753	&	A	&	15 02 34.755	&	$+$37 53 54.49	&	14.4		&	19.9		&	0.44	&	0.00	&	102.4	\\
	&	B	&	15 02 34.794	&	$+$37 53 52.50	&	10.5		&	12.4		&	0.25	&	0.06	&	127.0	\\
	&	C	&	15 02 34.758	&	$+$37 53 53.54	&	9.0		&	22.0		&	0.92	&	0.09	&	142.8	\\
	&	D	&	15 02 34.755	&	$+$37 53 54.27	&	21.0		&	46.9		&	0.79	&	0.17	&	175.1	\\
J1504$+$6000	&	A	&	15 04 09.195	&	$+$60 00 55.68	&	306.8		&	323.7		&	0.12	&	0.11	&	13.4	\\
	&	B	&	15 04 09.240	&	$+$60 00 55.16	&	128.8		&	138.9		&	0.16	&	0.11	&	119.1	\\
J1523$+$1332	&	A	&	15 23 21.735	&	$+$13 32 29.11	&	62.5		&	76.5		&	0.23	&	0.20	&	171.0	\\
	&	B	&	15 23 21.946	&	$+$13 32 31.23	&	1.9		&	3.0		&	0.38	&	0.26	&	80.7	\\
J1528$-$0213	&	A	&	15 28 22.030	&	$-$02 13 19.67	&	56.4		&	91.3		&	0.47	&	0.31	&	165.7	\\
	&	B	&	15 28 22.047	&	$-$02 13 18.30	&	52.1		&	67.5		&	0.29	&	0.25	&	9.1	\\
J1548$+$0808	&		&	15 48 09.076	&	$+$08 08 34.73	&	152.4		&	187.4		&	0.29	&	0.05	&	36.8	\\
J1559$+$4349	&	A	&	15 59 31.221	&	$+$43 49 17.18	&	135.3		&	143.7		&	0.16	&	0.06	&	82.2	\\
	&	B	&	15 59 31.158	&	$+$43 49 16.77	&	50.0		&	59.5		&	0.25	&	0.11	&	16.5	\\
J1604$+$6050	&		&	16 04 27.402	&	$+$60 50 55.25	&	168.4		&	168.6		&	0.03	&	0.00	&	121.5	\\
J1616$+$2647	&		&	16 16 38.327	&	$+$26 47 01.50	&	579.3		&	610.2		&	0.14	&	0.00	&	107.6	\\
J1629$+$1342	&	A	&	16 29 48.558	&	$+$13 42 05.71	&	65.8		&	81.5		&	0.38	&	0.16	&	102.4	\\
	&	B	&	16 29 48.222	&	$+$13 42 07.23	&	77.5		&	96.6		&	0.36	&	0.19	&	101.7	\\
J1633$+$4700	&	A	&	16 33 15.206	&	$+$47 00 16.31	&	74.6		&	82.8		&	0.18	&	0.13	&	138.9	\\
	&	B	&	16 33 15.148	&	$+$47 00 16.74	&	1.3		&	3.9		&	1.30	&	0.25	&	82.3	\\
	&	C	&	16 33 14.898	&	$+$47 00 17.42	&	2.7		&	13.2		&	1.95	&	0.38	&	104.6	\\
	&	D	&	16 33 14.722	&	$+$47 00 17.89	&	17.2		&	32.4		&	0.64	&	0.28	&	124.0	\\
J1724$+$3852	&		&	17 24 00.537	&	$+$38 52 26.63	&	127.7		&	130.3		&	0.09	&	0.06	&	112.7	\\
J2203$-$0021	&	A	&	22 03 58.311	&	$-$00 21 47.46	&	127.7	*	&	144.8	*	&	0.21	&	0.15	&	85.6	\\
	&	B	&	22 03 58.325	&	$-$00 21 48.31	&	134.4	*	&	152.9	*	&	0.22	&	0.13	&	107.0	\\
\enddata

\tablecomments{Flux densities designated with ``*'' lack accurate flux
  calibrations. See Table~\ref{tab:log} for details.}

\end{deluxetable}

\begin{deluxetable}{lcrrrrrr}
\tablewidth{0pt}
  \tablecaption{VLA 8.5 GHz Gaussian Fit Results.\label{tab:vlax}} 

  \tablehead{ \colhead{Object} & \colhead{ID} & \colhead{$S_{p}$ (mJy}
    & \colhead{$S_{tot}$} & \colhead{Maj.} & \colhead{Min.} &
    \colhead{P.A.} & \colhead{$\alpha$} \\
    \colhead{} & \colhead{} &\colhead{/beam)} & \colhead{(mJy)} &
    \colhead{($\arcsec$)} & \colhead{($\arcsec$)} &
    \colhead{($\degr$)} & \colhead{} \\
    \colhead{(1)} & \colhead{(2)} & \colhead{(3)} & \colhead{(4)} &
    \colhead{(5)} & \colhead{(6)} & \colhead{(7)} & \colhead{(8)}}

\startdata
J0000$-$1054	&	A	&	17.5		&	23.3		&	0.19	&	0.13	&	66.4	&	-1.1	\\
	&	B	&	7.9		&	11.9		&	0.27	&	0.14	&	33.4	&	-1.2	\\
J0003$-$1053	&	A	&	13.1		&	22.7		&	0.35	&	0.24	&	99.6	&	-1.0	\\
	&	B	&	16.7		&	22.7		&	0.31	&	0.09	&	68.9	&	-1.0	\\
	&	C	&	22.3		&	22.5		&	0.06	&	0.00	&	2.8	&	0.0	\\
	&	D	&	13.2		&	16.1		&	0.19	&	0.15	&	49.1	&	-1.0	\\
J0249$-$0759	&	A	&	138.3	*	&	172.8	*	&	0.20	&	0.16	&	13.9	&	\nodata	\\
	&	B	&	67.7	*	&	89.1	*	&	0.23	&	0.19	&	118.3	&	\nodata	\\
J0749$+$2129	&	A	&	18.3		&	31.7		&	0.61	&	0.14	&	26.8	&	-1.2	\\
	&	B	&	13.1		&	17.4		&	0.33	&	0.18	&	21.8	&	-1.3	\\
	&	C	&	3.5		&	4.7		&	0.35	&	0.19	&	43.1	&	-1.4	\\
J0759$+$5312	&	A	&	53.4		&	55.1		&	0.08	&	0.05	&	67.1	&	-1.1	\\
	&	B	&	1.4		&	2.0		&	0.34	&	0.13	&	81.9	&	-1.5	\\
J0805$+$1614	&	A	&	89.1		&	93.4		&	0.06	&	0.05	&	74.0	&	-0.4	\\
	&	B	&	54.5		&	57.8		&	0.07	&	0.05	&	136.9	&	-0.4	\\
J0807$+$5327	&	A	&	41.1		&	45.3		&	0.13	&	0.09	&	139.7	&	-1.2	\\
	&	B	&	11.8		&	15.2		&	0.21	&	0.18	&	57.3	&	-0.9	\\
J0824$+$5413	&	A1	&	32.8		&	40.8		&	0.19	&	0.16	&	157.8	&	-1.2	\\
	&	A2	&	2.7		&	6.7		&	0.48	&	0.41	&	136.5	&	\nodata	\\
	&	B	&	16.2		&	26.2		&	0.33	&	0.23	&	122.6	&	-1.1	\\
J0834$+$1700	&		&	219.0		&	299.4		&	0.18	&	0.13	&	53.7	&	\nodata	\\
J0903$+$5012	&	A	&	118.9		&	134.3		&	0.13	&	0.09	&	102.5	&	-1.2	\\
	&	B	&	2.9		&	9.1		&	0.73	&	0.25	&	85.3	&	-1.6	\\
J0917$+$4725	&		&	21.6		&	47.8		&	0.41	&	0.22	&	31.1	&	-1.2	\\
J0920$+$2714	&	A	&	40.8		&	50.0		&	0.18	&	0.09	&	50.8	&	-1.3	\\
	&	B	&	2.6		&	4.3		&	0.38	&	0.10	&	68.3	&	-0.9	\\
J0951$+$1154	&	A	&	13.2		&	21.7		&	0.44	&	0.33	&	15.5	&	-1.1	\\
	&	B	&	15.6		&	22.1		&	0.35	&	0.25	&	53.5	&	-1.2	\\
	&	C	&	1.2		&	7.7		&	2.04	&	0.55	&	24.9	&	-1.2	\\
J1023$+$0424	&	A	&	38.5		&	39.2		&	0.04	&	0.03	&	68.0	&	-1.2	\\
	&	B	&	16.6		&	17.8		&	0.11	&	0.04	&	67.7	&	-1.0	\\
J1034$+$1112	&	A	&	51.0		&	83.0		&	0.25	&	0.21	&	174.4	&	-1.0	\\
	&	B	&	17.8		&	28.2		&	0.30	&	0.15	&	78.0	&	-1.2	\\
	&	C	&	4.7		&	41.9		&	1.69	&	0.34	&	63.3	&	-0.7	\\
	&	D	&	2.7		&	4.9		&	0.42	&	0.10	&	55.7	&	-1.2	\\
	&	E	&	7.7		&	10.1		&	0.20	&	0.13	&	64.6	&	3.1	\\
J1045$+$0455	&	A	&	22.3		&	50.6		&	0.55	&	0.26	&	3.3	&	-1.1	\\
	&	B	&	6.9		&	15.9		&	0.64	&	0.22	&	16.3	&	-1.3	\\
J1127$+$5743	&	A	&	78.7		&	82.5		&	0.08	&	0.06	&	172.5	&	-1.1	\\
	&	B	&	14.3		&	18.6		&	0.24	&	0.14	&	156.3	&	-0.7	\\
J1129$+$5638	&		&	53.8		&	56.5		&	0.08	&	0.07	&	121.2	&	-1.2	\\
J1142$+$0235	&	A	&	37.6		&	43.5		&	0.12	&	0.11	&	146.8	&	-0.9	\\
	&	B	&	15.0		&	34.6		&	0.45	&	0.24	&	6.3	&	-1.1	\\
	&	C	&	1.9		&	13.1		&	0.85	&	0.61	&	162.7	&	-0.8	\\
J1147$+$4818	&	A	&	50.0		&	55.5		&	0.14	&	0.09	&	99.6	&	-1.2	\\
	&	B	&	3.6		&	7.2		&	0.48	&	0.29	&	83.9	&	-1.6	\\
J1148$+$1404	&	A	&	30.3		&	44.4		&	0.20	&	0.17	&	166.8	&	-1.0	\\
	&	B	&	10.8		&	13.2		&	0.14	&	0.12	&	36.2	&	-0.9	\\
J1202$+$1207	&		&	88.6		&	93.9		&	0.09	&	0.05	&	150.9	&	-0.7	\\
J1207$+$5407	&	A	&	32.4		&	36.1		&	0.16	&	0.06	&	78.4	&	-1.4	\\
	&	B	&	19.0		&	26.4		&	0.32	&	0.11	&	86.9	&	-1.5	\\
J1228$+$5348	&	A	&	30.6		&	54.5		&	0.64	&	0.25	&	2.1	&	-0.9	\\
	&	B	&	15.1		&	27.2		&	0.49	&	0.36	&	27.8	&	-1.1	\\
J1238$+$0845	&	A	&	40.9		&	48.3		&	0.16	&	0.08	&	169.4	&	-1.2	\\
	&	B	&	1.9		&	4.4		&	0.43	&	0.25	&	14.4	&	-1.1	\\
J1312$+$1710	&		&	56.5		&	60.5		&	0.09	&	0.06	&	157.2	&	-1.1	\\
J1315$+$0222	&	A	&	30.3		&	50.5		&	0.26	&	0.14	&	26.3	&	-1.0	\\
	&	B1	&	11.8		&	19.2		&	0.23	&	0.16	&	92.9	&	-1.3	\\
	&	B2	&	5.3		&	17.2		&	0.46	&	0.31	&	98.1	&	\nodata	\\
J1341$+$1032	&	A	&	35.1		&	47.3		&	0.25	&	0.07	&	83.5	&	-1.2	\\
	&	B	&	31.2		&	33.6		&	0.09	&	0.07	&	54.9	&	-1.2	\\
J1345$+$5846	&	A	&	36.2		&	47.2		&	0.20	&	0.13	&	147.5	&	-1.1	\\
	&	B	&	5.3		&	6.1		&	0.14	&	0.10	&	136.8	&	-1.0	\\
	&	C	&	0.6		&	5.8		&	1.07	&	0.77	&	57.8	&	-0.5	\\
J1354$+$5650	&	A	&	75.0		&	88.0		&	0.14	&	0.11	&	66.9	&	-1.0	\\
	&	B1	&	18.6		&	24.3		&	0.20	&	0.13	&	38.0	&	-1.1	\\
	&	B2	&	8.2		&	31.1		&	0.59	&	0.43	&	118.4	&	\nodata	\\
J1357$+$0046	&		&	229.0		&	240.0		&	0.09	&	0.02	&	51.5	&	-1.3	\\
J1410$+$4850	&		&	65.5		&	66.4		&	0.04	&	0.00	&	13.2	&	-1.0	\\
J1421$-$0246	&	A	&	4.7		&	34.2		&	0.88	&	0.56	&	179.2	&	-0.8	\\
	&	B	&	11.6		&	12.9		&	0.14	&	0.04	&	7.9	&	-1.3	\\
	&	C	&	4.1		&	31.9		&	0.84	&	0.63	&	177.9	&	-0.9	\\
J1424$+$1852	&	A	&	20.1		&	32.4		&	0.25	&	0.17	&	26.7	&	-1.7	\\
	&	B	&	32.1		&	42.8		&	0.22	&	0.08	&	34.8	&	-1.1	\\
J1502$+$3753	&	A	&	11.4		&	24.3		&	0.28	&	0.23	&	145.1	&	0.4	\\
	&	B	&	3.8		&	5.7		&	0.20	&	0.14	&	128.0	&	-1.4	\\
	&	C1	&	4.7		&	7.9		&	0.31	&	0.04	&	152.8	&	-1.2	\\
	&	C2	&	3.2		&	3.6		&	0.11	&	0.04	&	13.8	&	\nodata	\\
	&	D	&	5.7		&	13.1		&	0.40	&	0.15	&	46.4	&	-2.3	\\
J1504$+$6000	&	A	&	173.7		&	183.7		&	0.09	&	0.05	&	143.4	&	-1.0	\\
	&	B	&	59.3		&	71.5		&	0.19	&	0.08	&	135.1	&	-1.2	\\
J1523$+$1332	&	A	&	28.9		&	38.6		&	0.18	&	0.12	&	31.1	&	-1.2	\\
	&	B	&	1.3		&	3.1		&	0.38	&	0.25	&	13.0	&	0.1	\\
J1527$+$3312	&		&	73.8		&	81.0		&	0.13	&	0.03	&	46.6	&	\nodata	\\
J1528$-$0213	&	A	&	28.6		&	48.1		&	0.30	&	0.19	&	164.0	&	-1.2	\\
	&	B	&	25.6		&	39.4		&	0.22	&	0.21	&	28.5	&	-1.0	\\
J1548$+$0808	&	A1	&	42.5		&	43.9		&	0.06	&	0.04	&	166.8	&	-1.1	\\
	&	A2	&	51.9		&	55.7		&	0.09	&	0.05	&	15.1	&	\nodata	\\
J1559$+$4349	&	A	&	68.4		&	79.6		&	0.15	&	0.05	&	85.4	&	-1.1	\\
	&	B	&	22.2		&	30.2		&	0.20	&	0.10	&	13.7	&	-1.2	\\
J1604$+$6050	&		&	86.1		&	86.2		&	0.03	&	0.00	&	126.9	&	-1.2	\\
J1616$+$2647	&		&	650.6	*	&	717.8	*	&	0.14	&	0.08	&	29.2	&	\nodata	\\
J1629$+$1342	&	A	&	28.1		&	39.8		&	0.27	&	0.14	&	108.6	&	-1.3	\\
	&	B	&	39.9		&	52.1		&	0.20	&	0.14	&	96.0	&	-1.1	\\
	&	C	&	4.2		&	4.8		&	0.14	&	0.09	&	127.4	&	\nodata	\\
J1633$+$4700	&	A	&	34.6		&	37.6		&	0.09	&	0.06	&	150.1	&	-1.4	\\
	&	B	&	4.0		&	9.5		&	0.42	&	0.21	&	154.2	&	1.6	\\
	&	C	&	1.1		&	1.8		&	0.36	&	0.04	&	119.5	&	-3.6	\\
	&	D1	&	0.4		&	6.8		&	1.99	&	0.52	&	122.6	&	-0.6	\\
	&	D2	&	3.8		&	7.4		&	0.32	&	0.22	&	119.1	&	\nodata	\\
	&	D3	&	6.2		&	9.1		&	0.22	&	0.15	&	44.0	&	\nodata	\\
J1724$+$3852	&		&	71.6		&	72.8		&	0.06	&	0.01	&	140.9	&	-1.0	\\
\enddata

\tablecomments{Flux densities designated with ``*'' lack accurate flux
  calibrations. See Table~\ref{tab:log} for details.}

\end{deluxetable}

\begin{deluxetable}{lcrrrrrrr}
\tablewidth{0pt}
  \tablecaption{VLBA 1.4 GHz Gaussian Fit Results.\label{tab:vlba}}

  \tablehead{ \colhead{Object} & \colhead{ID} & \colhead{R.A.} &
    \colhead{Dec.} & \colhead{$S_{p}$ (mJy} & \colhead{$S_{tot}$} &
    \colhead{Maj.} & \colhead{Min.} &
    \colhead{P.A.} \\
    \colhead{} & \colhead{} &\colhead{(hh mm ss.ssss)} & \colhead{(dd
      mm ss.sss)} &\colhead{/beam)} & \colhead{(mJy)} &
    \colhead{(mas)} & \colhead{(mas)} & \colhead{($\degr$)} \\
    \colhead{(1)} & \colhead{(2)} & \colhead{(3)} & \colhead{(4)} &
    \colhead{(5)} & \colhead{(6)} & \colhead{(7)} & \colhead{(8)} &
    \colhead{(9)}}

\startdata
J0759$+$5312	&	a	&	07 59 06.4744	&	$+$53 12 47.943	&	8.4	&	20.9	&	13	&	8	&	97.9	\\
	&	b	&	07 59 06.4792	&	$+$53 12 47.917	&	1.5	&	7.3	&	21	&	15	&	119.3	\\
	&	c	&	07 59 06.4664	&	$+$53 12 47.956	&	62.7	&	132.3	&	11	&	7	&	51.0	\\
	&	d	&	07 59 06.4638	&	$+$53 12 47.938	&	8.8	&	43.2	&	26	&	12	&	32.3	\\
J0805$+$1614	&	a	&	08 05 02.1877	&	$+$16 14 05.223	&	92.7	&	175.5	&	11	&	9	&	118.5	\\
	&	b	&	08 05 02.1832	&	$+$16 14 04.772	&	37.4	&	87.6	&	13	&	11	&	95.7	\\
J0807$+$5327	&		&	08 07 40.7482	&	$+$53 27 38.624	&	43.3	&	103.1	&	109	&	34	&	125.1	\\
J0834$+$1700	&	a	&	08 34 48.2174	&	$+$17 00 42.594	&	10.8	&	69.0	&	51	&	31	&	107.5	\\
	&	b	&	08 34 48.2126	&	$+$17 00 42.608	&	61.8	&	184.1	&	31	&	18	&	100.7	\\
	&	c	&	08 34 48.2102	&	$+$17 00 42.634	&	11.3	&	57.6	&	52	&	23	&	91.6	\\
J0901$+$0304	&		&	09 01 50.9780	&	$+$03 04 22.709	&	181.8	&	295.0	&	11	&	5	&	1.2	\\
J0917$+$4725	&		&	09 17 27.0198	&	$+$47 25 23.865	&	7.4	&	24.7	&	90	&	63	&	110.2	\\
J0920$+$2714	&		&	09 20 45.1428	&	$+$27 14 05.620	&	32.1	&	157.2	&	52	&	36	&	149.7	\\
J0939$+$0304	&	a	&	09 39 45.1424	&	$+$03 04 26.491	&	137.9	&	165.3	&	4	&	3	&	101.0	\\
	&	b	&	09 39 45.1429	&	$+$03 04 26.491	&	38.1	&	94.3	&	15	&	5	&	98.7	\\
	&	c	&	09 39 45.1459	&	$+$03 04 26.485	&	2.6	&	49.9	&	79	&	20	&	84.5	\\
	&	d	&	09 39 45.1577	&	$+$03 04 26.472	&	9.2	&	38.5	&	21	&	14	&	118.4	\\
J0945$+$2640	&		&	09 45 30.9555	&	$+$26 40 54.063	&	52.0	&	205.8	&	27	&	13	&	67.1	\\
J1023$+$0424	&	a	&	10 23 37.5518	&	$+$04 24 14.332	&	29.9	&	124.4	&	24	&	10	&	135.0	\\
	&	b	&	10 23 37.5508	&	$+$04 24 13.757	&	14.1	&	57.9	&	19	&	15	&	172.1	\\
J1048$+$3457	&	a	&	10 48 34.2480	&	$+$34 57 25.042	&	69.4	&	203.0	&	12	&	4	&	104.8	\\
	&	b	&	10 48 34.2487	&	$+$34 57 25.055	&	27.3	&	203.5	&	18	&	13	&	73.6	\\
J1125$+$1953	&		&	11 25 55.2424	&	$+$19 53 43.699	&	19.4	&	100.0	&	31	&	16	&	26.2	\\
J1127$+$5743	&	a	&	11 27 43.7751	&	$+$57 43 15.861	&	187.3	&	243.5	&	13	&	8	&	18.8	\\
	&	b	&	11 27 43.7732	&	$+$57 43 15.852	&	56.7	&	158.4	&	31	&	22	&	86.7	\\
	&	c	&	11 27 43.7144	&	$+$57 43 15.669	&	7.0	&	21.9	&	33	&	24	&	41.3	\\
J1129$+$5638	&	a	&	11 29 04.1439	&	$+$56 38 44.047	&	80.7	&	237.7	&	16	&	8	&	158.4	\\
	&	b	&	11 29 04.1370	&	$+$56 38 44.042	&	50.4	&	96.2	&	9	&	6	&	139.8	\\
	&	c	&	11 29 04.1343	&	$+$56 38 44.056	&	12.7	&	61.5	&	21	&	12	&	145.3	\\
J1147$+$4818	&	a	&	11 47 52.2965	&	$+$48 18 49.522	&	9.8	&	38.2	&	29	&	14	&	103.2	\\
	&	b	&	11 47 52.2900	&	$+$48 18 49.541	&	24.0	&	61.6	&	21	&	10	&	108.7	\\
	&	c	&	11 47 52.2868	&	$+$48 18 49.538	&	31.5	&	98.6	&	27	&	10	&	93.9	\\
	&	d	&	11 47 52.2804	&	$+$48 18 49.518	&	20.4	&	40.7	&	18	&	6	&	72.0	\\
	&	e	&	11 47 52.2744	&	$+$48 18 49.484	&	2.9	&	11.9	&	44	&	6	&	57.4	\\
	&	f	&	11 47 52.2783	&	$+$48 18 49.375	&	3.6	&	20.8	&	37	&	19	&	67.6	\\
J1148$+$1404	&	a	&	11 48 25.4153	&	$+$14 04 49.937	&	9.8	&	20.4	&	51	&	4	&	119.4	\\
	&	b	&	11 48 25.4204	&	$+$14 04 49.869	&	7.9	&	37.6	&	106	&	34	&	130.8	\\
	&	c	&	11 48 25.4474	&	$+$14 04 48.424	&	12.1	&	19.6	&	35	&	10	&	119.5	\\
J1202$+$1207	&	a	&	12 02 52.0847	&	$+$12 07 20.869	&	20.4	&	103.0	&	22	&	14	&	83.7	\\
	&	b	&	12 02 52.0833	&	$+$12 07 20.823	&	28.4	&	58.0	&	18	&	4	&	18.1	\\
	&	c	&	12 02 52.0844	&	$+$12 07 20.781	&	7.4	&	21.3	&	13	&	11	&	100.0	\\
	&	d	&	12 02 52.0863	&	$+$12 07 20.774	&	17.0	&	73.6	&	22	&	12	&	124.0	\\
J1238$+$0845	&	a	&	12 38 19.2580	&	$+$08 45 01.677	&	29.2	&	78.9	&	41	&	21	&	71.3	\\
	&	b	&	12 38 19.2596	&	$+$08 45 01.660	&	24.4	&	53.4	&	36	&	17	&	57.1	\\
	&	c	&	12 38 19.2601	&	$+$08 45 01.624	&	11.3	&	37.8	&	48	&	18	&	172.4	\\
J1312$+$1710	&	a	&	13 12 35.2134	&	$+$17 10 55.914	&	23.0	&	27.3	&	8	&	0	&	72.2	\\
	&	b	&	13 12 35.2125	&	$+$17 10 55.905	&	31.0	&	103.4	&	24	&	7	&	66.2	\\
	&	c	&	13 12 35.2098	&	$+$17 10 55.892	&	12.9	&	43.6	&	20	&	10	&	97.9	\\
J1345$+$5846	&		&	13 45 38.3909	&	$+$58 46 53.462	&	16.1	&	70.9	&	53	&	18	&	91.4	\\
J1357$+$0046	&	a	&	13 57 53.7245	&	$+$00 46 33.354	&	390.7	&	929.4	&	15	&	5	&	46.8	\\
	&	b	&	13 57 53.7216	&	$+$00 46 33.326	&	107.2	&	153.1	&	6	&	3	&	56.5	\\
	&	c	&	13 57 53.7201	&	$+$00 46 33.298	&	117.1	&	449.5	&	23	&	9	&	34.0	\\
J1410$+$4850	&	a	&	14 10 36.0386	&	$+$48 50 40.386	&	52.5	&	106.2	&	11	&	6	&	38.4	\\
	&	b	&	14 10 36.0381	&	$+$48 50 40.344	&	88.1	&	147.0	&	9	&	4	&	52.3	\\
J1504$+$6000	&	a	&	15 04 09.2075	&	$+$60 00 55.711	&	78.3	&	528.7	&	38	&	12	&	57.3	\\
	&	b	&	15 04 09.2514	&	$+$60 00 55.158	&	45.7	&	131.1	&	15	&	11	&	175.6	\\
J1523$+$1332	&	a	&	15 23 21.7418	&	$+$13 32 29.339	&	6.2	&	17.0	&	32	&	15	&	58.2	\\
	&	b	&	15 23 21.7302	&	$+$13 32 29.113	&	14.1	&	27.7	&	19	&	14	&	112.6	\\
	&	c	&	15 23 21.7321	&	$+$13 32 29.171	&	4.9	&	29.1	&	58	&	23	&	89.9	\\
	&	d	&	15 23 21.7298	&	$+$13 32 29.139	&	6.7	&	87.0	&	92	&	37	&	126.6	\\
J1527$+$3312	&		&	15 27 50.8757	&	$+$33 12 52.903	&	45.1	&	187.3	&	23	&	12	&	170.3	\\
J1548$+$0808	&	a	&	15 48 09.0791	&	$+$08 08 34.890	&	112.5	&	180.4	&	43	&	22	&	60.7	\\
	&	b	&	15 48 09.0730	&	$+$08 08 34.661	&	85.5	&	135.6	&	44	&	16	&	135.6	\\
J1559$+$4349	&	a	&	15 59 31.2239	&	$+$43 49 17.191	&	27.1	&	99.9	&	40	&	17	&	99.5	\\
	&	b	&	15 59 31.2188	&	$+$43 49 17.206	&	27.3	&	82.8	&	35	&	14	&	102.5	\\
	&	c	&	15 59 31.2140	&	$+$43 49 17.197	&	23.5	&	66.3	&	31	&	15	&	91.3	\\
	&	d	&	15 59 31.1898	&	$+$43 49 17.051	&	4.2	&	9.1	&	31	&	8	&	57.1	\\
	&	e	&	15 59 31.1556	&	$+$43 49 16.825	&	15.5	&	54.2	&	42	&	13	&	22.6	\\
J1604$+$6050	&	a	&	16 04 27.4134	&	$+$60 50 55.197	&	219.9	&	257.3	&	5	&	4	&	74.0	\\
	&	b	&	16 04 27.4116	&	$+$60 50 55.201	&	208.8	&	234.5	&	4	&	2	&	108.9	\\
J1616$+$2647	&	a	&	16 16 38.3289	&	$+$26 47 01.516	&	315.3	&	416.0	&	8	&	5	&	170.4	\\
	&	b	&	16 16 38.3288	&	$+$26 47 01.505	&	93.6	&	255.3	&	17	&	11	&	25.2	\\
	&	c	&	16 16 38.3258	&	$+$26 47 01.399	&	148.4	&	408.3	&	22	&	9	&	19.4	\\
J1633$+$4700	&		&	16 33 15.2092	&	$+$47 00 16.321	&	22.6	&	107.0	&	54	&	45	&	152.7	\\
J1724$+$3852	&	a	&	17 24 00.5298	&	$+$38 52 26.652	&	82.5	&	159.1	&	10	&	9	&	76.5	\\
	&	b	&	17 24 00.5274	&	$+$38 52 26.680	&	49.2	&	97.9	&	17	&	4	&	150.0	\\
\enddata

\end{deluxetable}


\acknowledgements

This work was support financially by National Science Foundation grant
AST-0707480 (JTS, PI). In addition, TY acknowledges financial support
for this work from an NRAO studentship and an observing grant from the
{\it Spitzer Space Telescope} under project number 1439221. This work is
a portion of the PhD dissertation of TY presented to the Astrophysical \&
Planetary Sciences Department at the University of Colorado, Boulder.

{\it Facilities:} \facility{VLA}, \facility{VLBA}, \facility{Green Bank Telescope}, \facility{Arecibo},
\facility{GMRT}, \facility{ARC}, \facility{Gemini}.

\bibliographystyle{apj}
\bibliography{ting_refs.bib}

\end{document}